\documentclass[10pt]{article}

\usepackage{amsmath}

\usepackage{array}

\usepackage{appendix}

\usepackage{tocloft}                   

\usepackage{graphicx}

\usepackage{amsfonts}

\usepackage{amssymb}

\usepackage{mathrsfs}

\usepackage{yfonts}

\usepackage{euscript}

\usepackage{centernot}                 

\usepackage{ifsym}                     

\usepackage{upgreek}

\usepackage{mathtools}

\usepackage{color}

\usepackage{slantsc}
\usepackage{calligra}

\usepackage{bbold}          

\usepackage[T1]{fontenc}

\usepackage{epsf}

\usepackage{latexsym}

\usepackage{tipa}

\usepackage{makeidx}

\makeindex

\def\b{\begin{equation}}

\def\e{\begin{equation}}

\def\be{\begin{equation}}              

\def\ee{\end{equation}}

\def\beq{\begin{equation}}

\def\eeq{\end{equation}}

\def\bea{\begin{eqnarray}}

\def\eea{\end{eqnarray}}

\def\m{\mbox{ }}

\def\mma {\m , \m \m }

\def\!{\hspace{-1.6667em}}

\def\Proof{{\n{\u{Proof}}}\m}

\def\c{\cite}

\def\l{\label}

\def\r{\ref}

\def\n{\noindent}

\def\f{\footnote}

\def\u{\underline}

\def\w{\widetilde}

\def\s{\stackrel}

\def\slTheta{\mathit{\Theta}}                     

\def\slPhi{\mathit{\Phi}}                         

\def\uiR{\u{R}}

\def\uiq{\u{q}}

\def\uir{\u{r}} 

\def\birho{\mbox{\boldmath$\rho$}}

\def\brho{\birho}                                   

\def\mB{\mbox{B}}  

\def\mC{\mbox{C}}                        

\def\mD{\mbox{D}}                        

\def\mF{\mbox{F}}

\def\mG{\mbox{G}}

\def\mH{\mbox{H}} 

\def\mK{\mbox{K}}

\def\mM{\mbox{M}}                        

\def\mO{\mbox{O}}

\def\mP{\mbox{P}}

\def\mR{\mbox{R}}                        

\def\mS{\mbox{S}}                        

\def\mU{\mbox{U}}                        

\def\mV{\mbox{V}}

\def\mW{\mbox{W}}

\def\mb{\mbox{b}}

\def\md{\mbox{d}} 

\def\me{\mbox{e}}

\def\mg{\mbox{g}}

\def\mh{\mbox{h}}

\def\mi{\mbox{i}}

\def\ml{\mbox{l}}  

\def\mm{\mbox{m}}  

\def\mn{\mbox{n}}  

\def\mo{\mbox{o}}

\def\mp{\mbox{p}}

\def\ms{\mbox{s}}

\def\uR{\u{\mbox{R}}}

\def\uq{\u{\mbox{q}}} 

\def\ur{\u{\mbox{r}}}

\def\ux{\u{\mbox{x}}}

\def\bh{\u{\u{\mbox{h}}}  }            

\def\urho{{\u{\rho}}}

\def\bC{\mbox{\bf C}}                    

\def\bG{\mbox{\bf G}}                     

\def\bM{\mbox{\bf M}}

\def\bg{\mbox{\bf g}}

\def\bh{\mbox{\bf h}}

\def\bm{\mbox{\bf m}}

\def\bq{\mbox{\bf q}}

\def\buu{\mbox{\bf u}}             

\def\bupSigma{\mbox{\boldmath$\Sigma$}}                 

\def\fA{\mbox{\sffamily A}}           

\def\fB{\mbox{\sffamily B}}

\def\fO{\mbox{\sffamily O}}

\def\fP{\mbox{\sffamily P}}

\def\fQ{\mbox{\sffamily Q}}

\def\fT{\mbox{\sffamily T}}

\def\fX{\mbox{\sffamily X}}

\def\sa{\mbox{\scriptsize a}}

\def\scc{\mbox{\scriptsize c}}

\def\se{\mbox{\scriptsize e}}

\def\sh{\mbox{\scriptsize h}} 

\def\si{\mbox{\scriptsize i}}

\def\sll{\mbox{\scriptsize l}}  

\def\sm{\mbox{\scriptsize m}}

\def\sn{\mbox{\scriptsize n}} 

\def\so{\mbox{\scriptsize o}} 

\def\sp{\mbox{\scriptsize p}}

\def\sr{\mbox{\scriptsize r}}

\def\sss{\mbox{\scriptsize s}}  

\def\st{\mbox{\scriptsize t}}

\def\sv{\mbox{\scriptsize v}}

\def\sx{\mbox{\scriptsize x}}

\def\sC{\mbox{\scriptsize C}}

\def\sD{\mbox{\scriptsize D}}

\def\sF{\mbox{\scriptsize F}}

\def\sG{\mbox{\scriptsize G}}

\def\sI{\mbox{\scriptsize I}}

\def\sJ{\mbox{\scriptsize J}}

\def\sL{\mbox{\scriptsize L}} 

\def\sM{\mbox{\scriptsize M}}

\def\sO{\mbox{\scriptsize O}}

\def\sS{\mbox{\scriptsize S}}

\def\sT{\mbox{\scriptsize T}}

\def\sU{\mbox{\scriptsize U}}

\def\sY{\mbox{\scriptsize Y}}

\def\sfP{\mbox{\sffamily{\scriptsize P}}}      

\def\sfT{\mbox{\sffamily{\scriptsize T}}}      

\def\sfX{\mbox{\sffamily{\scriptsize X}}}      

\def\sbm{\mbox{{\bf \scriptsize m}}}

\def\tO{\mbox{\tiny O}}

\def\Thomas{\mbox{\textcircled{$\rightarrow$}}}

\def\sumi2{\sum\mbox{}_{\mbox{}_{\mbox{\scriptsize $i$=1}}}^2}

\def\sumi3{\sum\mbox{}_{\mbox{}_{\mbox{\scriptsize $i$=1}}}^3}

\def\sumABcycles3{\sum\mbox{}_{\mbox{}_{\mbox{\scriptsize cycles $A,B$=1}}}^{3}}

\def\sumCDcycles3{\sum\mbox{}_{\mbox{}_{\mbox{\scriptsize cycles $C,D$=1}}}^{3}}

\def\sumj3{\sum\mbox{}_{\mbox{}_{\mbox{\scriptsize $j$=1}}}^3}

\def\sumk3{\sum\mbox{}_{\mbox{}_{\mbox{\scriptsize $k$=1}}}^3}

\def\prodiA1{\prod\mbox{}_{\mbox{}_{\mbox{\scriptsize $i$=1}}}^{A - 1}}

\def\bigtimes{\mbox{\Large $\times$}}

\def\sumpn{\sum\mbox{}_{\mbox{}_{\mbox{\scriptsize $p$ = 2}}}^{n - 1}}

\def\d{\textrm{d}}                                                  

\def\pa{\partial}                                                   

\def\es{\m = \m}

\def\:={\m := \m}

\def\=:{\m =: \m}

\def\leqs{\m \leq \m}

\def\geqs{\m \geq \m}

\def\ls{\m < \m}

\def\Abs{\mbox{\Large $\mathfrak{a}$}\mb\ms}                         

\def\FrI{\mbox{$\mathfrak{I}$}}                                

\def\FrA{\mbox{$\mathfrak{A}$}}                                

\def\FrT{\mathfrak{T}}                                         

\def\FrC{\mbox{$\mathfrak{C}$}}                                
                                                       
\def\FrS{\mbox{\Large $\mathfrak{s}$}}                         
\def\sFrS{\mbox{\large$\mathfrak{s}$}}                         
   
\def\tFrS{\mbox{\footnotesize$\mathfrak{s}$}} 
\def\FrU{\mbox{$\mathfrak{U}$}}                                
\def\FrM{\mbox{$\mathfrak{M}$}}                                
\def\nFrg{\mbox{\large$\mathfrak{g}$}}                         
\def\FrT{\mbox{\boldmath$\mathfrak{T}$}}                       
\def\bFrPP{\mbox{\boldmath$\mathfrak{P}$}}                     
 
\def\FrG{\mathfrak{G}}                                         
\def\Hilb{\mbox{{\boldmath$\mathfrak{H}$}ilb}}                 

\def\Space{\bFrS\mbox{pace}}                                         

\def\FrQ{\mbox{\Large $\mathfrak{q}$}}                               
\def\bFrC{\mbox{\boldmath$\mathfrak{C}$}}                            
\def\bFrL{\mbox{\boldmath$\mathfrak{L}$}}                            

\def\Phase{\mbox{{\boldmath$\mathfrak{P}$}hase}}                     

\def\bFrR{\mbox{\boldmath$\mathfrak{R}$}}                            
\def\Rig-Phase{\bFrR\mbox{ig-}\Phase}                                
                                                                													   
\def\lFrr{\mbox{\Large $\mathfrak{r}$}}                              

\def\FrP{\mbox{\Large $\mathfrak{p}$}}                                 
\def\FrR{\mbox{\boldmath$\mathfrak{R}$}}                             

\def\bFrM{\mbox{\boldmath${\mathfrak{M}}$}}                             

\def\bFrU{\mbox{\boldmath$\mathfrak{U}$}}                            

\def\diam{\delta iam}                                                

\def\Imat{\u{\u{I}}}                                                 

\def\1mat{\u{\u{1}}}                                                 

\def\Lmat{\u{\u{L}}}                                                 

\def\Positive-Modespace{\mbox{{\boldmath$\mathfrak{M}$}odespace$^+$}}

\def\POSITIVE-MODESPACE{\mbox{{\boldmath$\mathfrak{M}$}ODESPACE$^+$}}
                                                                                                                             														
\def\bFrS{\mbox{\Large $\mathfrak{s}$}}                              
			
\def\bFrG{\mbox{ $\mathfrak{G}$}}                                    %
\def\Riem{\bFrR\mbox{iem}}                                           
\def\CRiem{\bFrC\Riem}                                               

\def\Superspace{\bFrS\mbox{uperspace}}                               
\def\CS{\bFrC\bFrS}                                                  
\def\bFrT{\mbox{\boldmath$\mathfrak{T}$}}                            %
\def\TrueSpace{\bFrT\mbox{ruespace}}                                 

\def\Sym{\bFrS\mbox{ym}}                                             
\def\Leib{\bFrL\mbox{eib}}                                           

\def\PRiem{\bFrPP\bFrR\mbox{iem}}                                    
\def\Superspacetime{\bFrS\mbox{uperspacetime}}                       
\def\Co{\bFrC\mbox{o}}                                               

\def\Uni{\bFrU\mbox{ni}}                                             

\def\CoM{\bFrC\mbox{oM}}                                             

\def\Merger{\bFrM\mbox{erger}}                                       

\def\FrO{\mbox{$\mathfrak{O}$}}                                      

\def\Top{\FrT\mo\mp}

\def\GrandRiem{\bFrG\mbox{rand-}\Riem}                                           

\def\GrandSuperspace{\bFrG\mbox{rand-}\Superspace}                               

\def\Kin-Hilb{\mbox{{\boldmath$\mathfrak{K}$}in-\Hilb}}                     

\def\Mid-Hilb{\mbox{{\boldmath$\mathfrak{M}$}id-\Hilb}}                     

\def\Dyn-Hilb{\mbox{{\boldmath$\mathfrak{D}$}yn-\Hilb}}                     

\def\5Star{\mbox{\Large$\star$}}              

\def\Frr{\mbox{$\mathfrak{r}$}}

\def\upi{\mbox{$\pi$}}                              

\def\bigiota{\mbox{\Large $\iota$}}

\begin{document}

\begin{titlepage}

\begin{center}

\large{\bf THE SMALLEST SHAPE SPACES} \normalsize

\vspace{0.1in}

\large{\bf I. Shape Theory Posed, with Example of 3 Points on the Line} \normalsize

\vspace{0.1in}

\normalsize

\vspace{0.1in}

{\large \bf Edward Anderson$^*$}

\vspace{.2in}

\end{center}

\begin{abstract}

This treatise concerns shapes in the sense of constellations of points with various automorphisms quotiented out: 
continuous translations, rotations and dilations, and also discrete mirror image identification and labelling indistinguishability of the points.  
We consider in particular the corresponding configuration spaces, which include shape spaces and shape-and-scale spaces.
This is a substantial model arena for developing concepts of Background Independence, with many analogies to General Relativity and Quantum Gravity; 
it also has many applications to Dynamics, Quantization, Probability and Statistics.     
We also explain the necessity of working within the shape-theoretic Aufbau Principle: 
only considering larger particle number $N$, spatial dimension $d$ and continuous group of automorphisms $G$ when all the relatively smaller cases have been considered.  

\m 

\n We show that topological shape spaces are graphs,  
opening up hitherto untapped combinatorial foundations both for these and for topological features of the more usually-considered spaces of metric shapes.
We give a conceptual analysis of {\sl inhomogeneous} Background Independence's clustering and uniformness aspects.    
We also consider the fate of shape spaces' (similarity) Killing vectors upon performing the mirror image and particle indistinguishability quotientings;   
this is crucial for dynamical and quantization considerations. 
For now, in Part I, we illustrate all these topological, combinatorial, differential-geometric and inhomogeneity innovations with the example of 3 points in 1-$d$.
Papers II, III and IV then extend the repertoire of examples to 4 points in 1-$d$, 
                                                     triangles (3 points in 2- and 3-$d$) 
											and quadrilaterals (4 points in 2-$d$) respectively. 
The quadrilateral is a minimal requirement prior to most implementations of the third part of the shape-theoretic Aufbau Principle: 
adding further generators to the automorphism group $G$.  

\end{abstract}

\n PACS: 04.20.Cv, Physics keywords: background independence, inhomogeneity, configuration spaces, Killing vectors, dynamical and quantization aspects of General Relativity.

\m 

\n Mathematics keywords: shapes, spaces of shapes, Shape Geometry, Shape Statistics, Applied Topology, Applied Geometry, applications of Graph Theory to Shape Theory.

\vspace{0.1in}
  
\n $^*$ Dr.E.Anderson.Maths.Physics@protonmail.com

\vspace{0.1in}
 
{            \begin{figure}[!ht]
\centering
\includegraphics[width=0.5\textwidth]{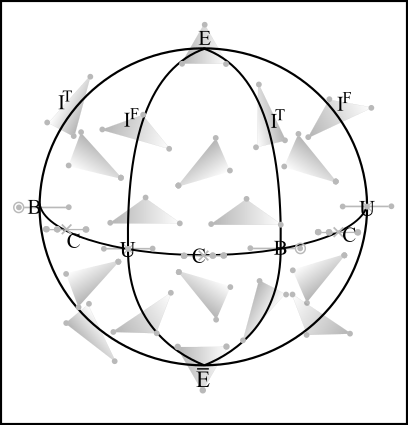}
\caption[Text der im Bilderverzeichnis auftaucht]{        \footnotesize{The triangleland sphere.  
Its poles E are each of the two labelling orientations of equilateral triangles, 
the meridians marked I are isosceles triangles and the equator C is where collinear configurations are located.
The points labelled B are binary coincidences, for which two of the triangle's three points are superposed. 
$\sfX$ denotes here the centre of mass of the outermost 2 points-or-particles.  
} }
\l{S(3, 2)-Intro} \end{figure}          }
 
\end{titlepage}

\section{Introduction}\l{Intro-I}

\subsection{Example of what these articles means by shape, shape space and Shape Theory}
%
Let us first illustrate what this article means mean by shape and shape spaces 
by outlining statistician David Kendall's pioneering example \c{Kendall89, Kendall}.
This involves triangular shapes, in the sense of triples of labelled points in flat 2-$d$ space, 
for which configurations related by a transformation belonging to the flat 2-$d$ similarity group, $Sim(2)$, are identified.  
The information content of each triangle can be taken to be one ratio of sides and one relative angle between sides.
The configuration space consisting of all these triangles is
\beq
\w{\FrS}(3, 2) \es  \frac{\mathbb{R}^6}{Sim(2)}  \m  ,
\eeq 
denoting  the {\it shape space} of 3 particles in 2-$d$.
This turns out to be a sphere 
\beq
\w{\FrS}(3, 2) = \mathbb{S}^2
\eeq 
at the topological level, and equipped with the standard spherical metric \c{Kendall}.  
This is a very standard geometry, 
yet one in which the polar and azimuthal angles carry an unusual shape-theoretic significance as per Figure \r{S(3, 2)-Intro}.

\m 

\n  Suppose instead that similar triangles of different sizes are taken to be distinct. 
Now configurations related by a transformation belonging to the flat 2-$d$ Euclidean group, $Eucl(2)$, are identified.  
The configuration space of these shape-and-scale triangles is  
\be
\w{\cal R}(3, 2) \es  \frac{\mathbb{R}^6}{Eucl(2)}  \m  ,
\ee 
denoting the {\it relational space} of 3 particles in 2-$d$.
This turns out to be 
\be
\w{\cal R}(3, 2) = \mathbb{R}^3
\ee
at the topological level, by being the topological cone over $\w{\FrS}(3, 2) = \mathbb{S}^2$.  
The corresponding metric, however, turns out not to be the flat metric, though it is in a sense conformally flat \c{Iwai87, FileR}.  

\m 

\n The current treatise lays out how the above approach readily conceptually generalizes to other spatial dimensions $d$, particle numbers $N$ and groups $G$, 
with a large number of such remaining topologically and geometrically tractable, and the following rich fields of applications ensuing.

\subsection{Motivation for the Shape Theory subject}

\n{\bf Subject Motivation 1} Shape Theory offers further insights and results 
to the classic subject of the Geometry of figures in flat space, by use of topological and differential-geometric techniques 
rooted in the nature of the corresponding shape(-and-scale) spaces.
See Papers III and IV and \c{Small, MIT, III, IV, 2-Herons, Ineq, A-Pillow, Max-Angle-Flow, A-Sylvester, A-Coolidge, 
A-Quad-Ineq, Affine-Shape-1, Affine-Shape-2} for examples of such results for triangles and quadilaterals.
The flat space in question can moreover be interpreted not only as (an approximate model of) physical space, but as an abstract carrier space (Sec 2.1).  
This generalization substantially increases the number of applications.

\m 

\n{\bf Subject Motivation 2} If the flat spaces are interpreted as sample spaces for location data in the sense of Probability and Statistics, 
one can then furthermore consider the shape(-and-scale) data content of $N$-tuples of location data, corresponding to sampling in groups of $N$ points.  
This can be studied through its living on the corresponding shape spaces of known topology and geometry, 
upon which measures and probability distributions can consequently also be constructed.  
This was Kendall's own motivation for establishing Shape Theory. 
For instance, he studied \c{Kendall89, Kendall} sampling points in the plane in threes 
using Probability and Statistics defined on the aforementioned shape sphere and spherical blackboard examples of shape spaces.
He used this to quantify whether a given a set of data points in a plane contain more than just a random amount of approximate collinearities in threes: 
a seminal prototype of pattern-detecting methods which properly respects the topology and geometry of the underlying configuration space.   
Applications of this are widely interdisciplinary, from distributions of ancient monuments in archaeological study \c{Kendall84} 
to those of astrophysical objects \c{Kendall89, PE16}, 
to many instances of biological modelling involving quantifying the shapes of skulls and other organs \c{Small, JM00, Bhatta, PE16}.
See e.g.\ \c{Bookstein, Grenander96, Grenander07} for further approaches.

\m 

\n{\bf Subject Motivation 3} Shape Theory gives a topological and geometrical approach to the $N$-body problem via consideration of the configuration space, 
as used in the following.    

\m 

\n a) The Celestial Mechanics branch of Dynamics \c{ArchRat, Montgomery2}.  

\m 

\n b) Molecular Physics \c{Zick-2, ACG86, PP87, Iwai87, LR95, LR97}.    

\m 

\n c) In providing concrete models of Mechanics which are relational \c{BB82, B03, FORD, QuadI, FileR, AMech, ABook, Project-1, A-Generic} as regards the foundationally interesting 
relational `Leibniz--Mach' \c{L, L2, M} side of the absolute versus relational motion debate whose absolute side goes back (at least \c{DoD, Buckets}) to Newton \cite{Newton}.    

\m 

\n The connection between c) and Shape Theory lies in observing, firstly \c{FORD}, 
that the similarity-invariant shape mechanics of \c{B03} has Kendall-type shape spaces as its reduced configuration spaces.  
Secondly \c{Kendall, Cones}, that the Euclidean-invariant relational Mechanics' \c{BB82} reduced configuration spaces are the cones over these shape spaces.  

\m 

\n In a) and the classical part of c), one considers motion as a geodesic \c{Lanczos, DeWitt70, Arnol'd, B94I}
(or conformally-related parageodesic \c{Magic} in cases with nontrivial potential) on shape(-and-scale) space.  

\m  

\n{\bf Subject Motivation 4} Mechanics can moreover be quantized. 
There are obvious reasons for entertaining this for b)'s Molecular Physics;   
we shall moreover give reasons below why quantizing c) \c{AF, FileR, QuadII, ABook, QLS, Quantum-Triangles} is of considerable foundational interest.  
A useful reminder at this point is that $N$-body Quantum Mechanics (QM) unfolds on configuration space, rather than on space, 
so in the present context it unfolds on such as shape space or relational space.  
Another useful reminder is that in studying ordinary QM, 
it has become customary to accept the position, momentum and angular momentum operators afforded by $\mathbb{R}^d$.
However, in setting up a quantum theory on a more general topology and geometry, one needs to ascertain {\sl what replaces} these notions 
prior to considering the dynamical part of quantization such as what the Schr\"{o}dinger equation is and how to solve it. 
This preliminary replacement step is known as {\it kinematical quantization} \c{I84} (and the Conclusion of Part IV), 
and is observed to be sensitive to configuration space topology.
Shape(-and-scale) spaces then represent an interesting extension of the scope of both kinematical and dynamical quantization:  can conventional QM cope with modelling shapes? 

\m 

\n{\bf Extension 1} Shape(-and-Scale) Theory is moreover very flexible as regards which groups of transformations $G$ are held to be irrelevant.  
A given (absolute) space can have quite a number of distinct notions of automorphism, each of which gives a different notion of shape(-and-scale).  
Such include affine, equiareal, projective, conformal and supersymmetric geometries (and further consistent combinations that these afford \c{AMech}).  
This means that Motivation 1) extends to the further foundationally interesting consideration of how relational mechanics modelling 
is affected under varying what geometrical structure is attributed to space \c{AMech, ABook, Project-1, Affine-Shape-1, Affine-Shape-2}.
On the other hand, affine and projective analogues of Kendall's notion of shape can be taken to underly much of 
Image Analysis and Computer Vision \c{Sparr98, Bhatta, PE16}.  
In this way, e.g.\ medical, biological and astrophysical {\sl images} give further applications in each of these subjects.  

\m  

\n{\bf Extension 2} Shape(-and-Scale) Theory can also be posed for carrier spaces of different topology and/or geometry than $\mathbb{R}^n$.  
This enters some of the more practical applications, such as the {\it celestial sphere} of astronomical observations, 
for which Kendall and collaborators also provided shape-theoretic modelling \c{Kendall}.
Another is {\it directional statistics} \c{Watson, JM00}, used for instance for modelling bird migration or rock magnetization.   
See Sec \r{Abs} for further examples.  

\m 

\n{\bf Extension 3} Shape(-and-Scale)  Theory moreover has a counterpart involving much larger spaces, 
in which the shapes are not $N$-point figures but rather in general inhomogenous $d$-dimensional manifolds.  

\m 

\n{\bf Example 1} General Relativity (GR) is a subject in which this has long been established to occur.  
Upon viewing GR dynamically, one is dealing with evolving 3-metrics, or, better, 3-geometries: 
3-metrics modulo diffeomorphisms.
These form the configuration space of 3-metrics on a fixed spatial topological manifold $\bupSigma$, $\Riem(\bupSigma)$, 
and then Wheeler's \c{Battelle, DeWitt67, Giu09} 
\be
\Superspace(\bupSigma) \es  \frac{\Riem(\bupSigma)}{Diff(\bupSigma)} \m  .  
\l{Intro-Superspace}
\ee
Note the analogy with Kendall's triangle shape space; there are in fact very many more analogies (and some differences) between shape(-and-scale) mechanics 
and this Geometrodynamics formulation of GR, as exposited in \c{FileR, ABook}; see Sec \r{Gdyn} for a partial summary.  

\m 

\n Some also well-known restricted GR configuration spaces are also shape(-and-scale) spaces: 
the minisuperspace \c{Misner-70, Magic, Ryan} and anisotropyspace \c{Misner-Ani, ABook} of homogeneous spatial geometries, 
and the inhomogeneous yet symmetric midisuperspaces, whether full \c{Kuchar71, Kuchar94, Krasinski, MacCallum, Gowdy} 
or as perturbations about minisuperspace \c{Bardeen, Cos, HallHaw}.  
Many of these are familiar from Cosmology and from studies of anisotropy and/or inhomogeneity in GR (which can indeed be recast as shape problems).  

\m 

\n{\bf Example 2} A first more practical subject in which similar diffeomorphism modelling has more recently begun is Anatomy, 
in the form of `Comparative Anatomy and Morphometrics' \c{Younes10, Younes, Morphometrics, Diff-Med-Imaging-1, Diff-Med-Imaging-2}.  

\m 

\n{\bf Example 3} A second is the theory of search engines, in the specific direction of manifold-level matching \c{Geom-Search-Engine}.  

\m 

\n{\bf Subject Motivation 5} The absolute versus relational motion debate mentioned in Motivation 3.c) moreover does not just concern the foundations of Mechanics.  
It is additionally played a part in GR's inception \c{Einstein1, Einstein2}, 
and to its subsequent interpretation as a dynamical and Background Independent theory \c{ADM, BSW, A64, A67, BB82, Kuchar92, I93, B94I, RWR, ABFKO, APoT2, ABook}. 
Shape Theory, both at the level of configuration space topology and geometry, and at the level of Mechanics, then serves as a model arena 
for many of these foundationally interesting aspects of GR \c{Kuchar92, KieferBook, FileR, APoT2, ABeables, ABeables2, ABeables3, Mercati, ABook, Project-1}.

\m 

\n{\bf Subject Motivation 6} Indeed, Background Independence is realized in most Quantum Gravity programs 
-- Geometrodynamics, Loop Quantum Gravity, canonical Supergravity, M-Theory... but not perturbative covariant quantization or perturbative String Theory.  
Indeed, Background Independence is often touted as a -- or even the -- main ingredient in such programs. 
Such programs often encounter conceptual difficulties due to mismatches in time and related concepts between different branches of Physics, 
most notably between GR and QM, which go by the collective name of `the Problem of Time'. 
This is a substantial and recognized obstacle, which has been widely studied 
\c{DeWitt67, Battelle, Kuchar81, Kuchar91, Kuchar92, I93, Kuchar99, KieferBook, RovelliBook, APoT1, APoT2, ABook}. 
The Problem of Time can moreover be traced back \c{APoT2, BI, ABook} to mismatches between background dependent and Background Independent branches of Physics.
Shape Theory for points in a carrier space is moreover a useful model arena for parts of the study of GR's manifestation of Background Independence. 
For Shape Theory provides examples with known and simple configuration space topology and geometry, upon which computible models of shape mechanics  
and then shape QM can be built. 
Then many Background Independence and Problem of Time calculations, which are too hard to work out for GR itself, can be solved for these shape-theoretic examples. 

\m 

\n Indeed, the above connection with Quantum Gravity and Quantum Cosmology in Motivation 6  
also considerably increases the value of quantizing shape(-and-scale). 
Not only as a test of quantum formalism and technique and as regards increasing the scope of ordinary quantum modelling, 
but also as modelling Background Independent features, whole universe features of interest in Quantum Gravity.
Tension between Background Dependence and Background Independent paradigms in Physics has been argued to lie at the root of 
what manifests itself as as the notorious and foundational Problem of Time in Quantum Gravity \c{APoT3, ABook}.

\m 

\n The study of Background Independence moreover places considerable interest on the effect of varying the type of geometry assumed \cite{AMech, ABook} 
(or yet further levels of mathematical structure \cite{I89, IKR, I91, I03, ASoS, ABook, Project-1}). 
Does Background Independence substantially differ if one includes 
metric geometry, 
affine structure, 
conformal structure, 
supersymmetric structure 
and combinations thereof?\footnote{Some modelling variants do not appear to change much, 
e.g.\ conformal superspace has very similar known theorems to superspace's \c{FM96}, 
whereas others do lead to substantial qualitative difference in the nature of Backgound Independence: \c{ABook} shows this to be the case for Supergravity.}
%
The geometries of most usual interest to Theoretical Physics modelling are affine, conformal and supersymmetric, alongside combinations thereof.
Whereas in flat space affine and conformal are incompatible geometries to simultaneously possess, supersymmetric affine and superconformal are already possible here, 
and all combinations of the three are possible in the broader arena of Differential Geometry.  
This goes toward addressing questions such as whether GR's exhibition of Background Independence (and the consequent Problem of Time) typical, 
or whether altering the geometrical content of the theory substantially changes these features?  
In this regard, I have noted substantial qualitative differences in passing from GR to its supersymmetric counterpart, Supergravity \c{ABeables, ABook}.  
This is significant in determining the extent to which detailed knowledge of GR's Background Independence 
is likely to be useful in developing Background Independent theories in general.  

\m 

\n Let us end by noting Shape Statistics' own interplay with the Problem of Time, through providing solid mathematical foundations 
(for now at the classical level \c{AKendall, ABook}) for the elsewise rather speculative notion of a timeless Records Theory \c{PW83, Page1, H99}.

\subsection{Motivation for the current treatise itself}

Our first main point -- Treatise Motivation 1) below -- is that smaller shape models play a significant role within larger shape models, 
where larger can mean whichever combination of larger $d$, larger $N$ or quotienting out by a larger $G$.   
This implies necessity for a careful and deep treatment of the simplest cases. 
The current treatise moreover brings in three innovations (Treatise Motivations 2 to 4) which are significant even for the smallest nontrivial shape spaces.  

\m 

\n{\bf Treatise Motivation 0} Let us however zerothly\footnote{This is `zeroth' as per Wheeler's asking for zeroth principles to underly first principles.
But also zeroth in the sense that the minimal relationally nontrivial unit already motivated its own article \c{AMech} (expanded upon in \c{Project-1}). 
So the current series of articles is more about how Paper Motivations 1) to 4) build upon Paper Motivation 0)'s minimal relationally nontrivial unit concept.}
point to the {\it minimal relationally nontrivial unit} concept; See Sec \r{Triviality-Criteria} for various strengths of ths concept. 
For Euclidean and similarity geometries in 2-$d$ flat space, 
this is the triangle of the simplest 2-$d$ relational mechanics, and also why Kendall samples specifically in threes therein.  
In this way, the triangle in 2-$d$ constitutes the smallest nontrivial whole-universe model as regards closed-universe modelling and Quantum Cosmological applications.    

\m 

\n Now, the bigger $G$ is, the more particles are needed to have a minimal relationally nontrivial unit.  
This leaves us needing 4 points not as an extension but as a bare necessity, with other $G$, $d$ combinations requiring even 5 and 6 points.  
Then the way in which 3 points in 1-d controls 4 points in 1-$d$ and 3 in 2-$d$, which then significantly control 4 in 2-$d$ recurs.  
So it is best to exposit how this control works for these smallest 4 models here prior to expositing how 5 and 6 point cases work.  
\n Quadrilateralland is the smallest of the nontrivially complex projective shape spaces \c{Kendall84, FORD, FileR, AConfig, ABook, QuadI}, 
and the smallest affine shape space \c{GT09} and M\"{o}bius shape space \c{AMech}, which is a projective, and in some senses also conformal, model.  

\m 

\n{\bf Treatise Motivation 1: Shape-Theoretic Aufbau Principle.}
Shape Theory models are very interconnected, so that one can only build it up well by having cumulative knowledge of 'simpler' models. 
This is because the simpler models are substantially recurrent in the subsequent models. 
`Simpler' and `subsequent' are meant in all four of the following senses. 

\m 

\n 1) Spatially lower-$d$ comes before spatially higher-$d$.
This is because, as we show in this treatise, higher-$d$ exhibits at least all the conceptual features of lower-d, and moreover recurrs withing higher-$d$ as 'lower strata'. 
For instance, the set of planar figures contains the set of collinear figures. 

\m 

\n 2) Lower point or particle number $N$ comes before higher $N$. 
This is because, as this treatise illustrates, higher-$N$ exhibits at least all the conceptual features of lower-$N$. 
This moreover recurrs within higher-$d$ as 'point coincidence' alias 'particle collision' configurations forming submanifolds (or other subspaces) 
within the higher-$d$ cases' configuration spaces.    

\m 

\n 3) Larger automorphism groups held to be physically irrelevant usually need to be treated after the effects of smaller such subgroups are known.
This ordering follows from quotienting being involved: a source of complication \c{Project-2, A-Generic}.
As a simple model, quotienting by a group $G_2$ which extends a smaller $G_1$ could amount to looking at submanifolds within $G_1$'s larger quotient space manifold.
The more quotienting is performed, moreover, the more likely one is dealing with a stratified version of all this.
There is one partial exception to this, for shape theories with shape-and-scale theory partners, such as subgroups of the affine group,  
-- which include the Euclidean and similarity groups of the current treatise. 
Here the shape-and-scale space is the cone (in the sense of Sec \r{MLS-R}) over the shape space, 
by which the more reduced shape theory is in many ways the simpler of the two (by not possessing a conical apex point). 
On the other hand, conformal Shape Theory and its descendants do not have such pairings or conical structure.  

\m 

\n All in all, Shape Theory has a strong flavour of cumulativeness.
Namely, study of the $N_2$, $d_2$ and $G_2$ case with 
\be
N_2 \geqs N_1 \m  ,
\ee 
\be 
d_2 \geqs d_1 \mma \mbox{and}
\ee
\be 
G_2 \geqs G_1
\ee 
(meaning that $G_1$ is a subgroup of $G_2$) uses many results about the smaller $N_1$, $d_1$ and $G_1$ cases).  
The current treatise is thus needed to complete underlying studies prior to bringing in the bigger groups, dimensions and particle numbers; 
see Fig \r{Aufbau} for the first few targets to be attempted.  
%
{            \begin{figure}[!ht]
\centering
\includegraphics[width=0.45\textwidth]{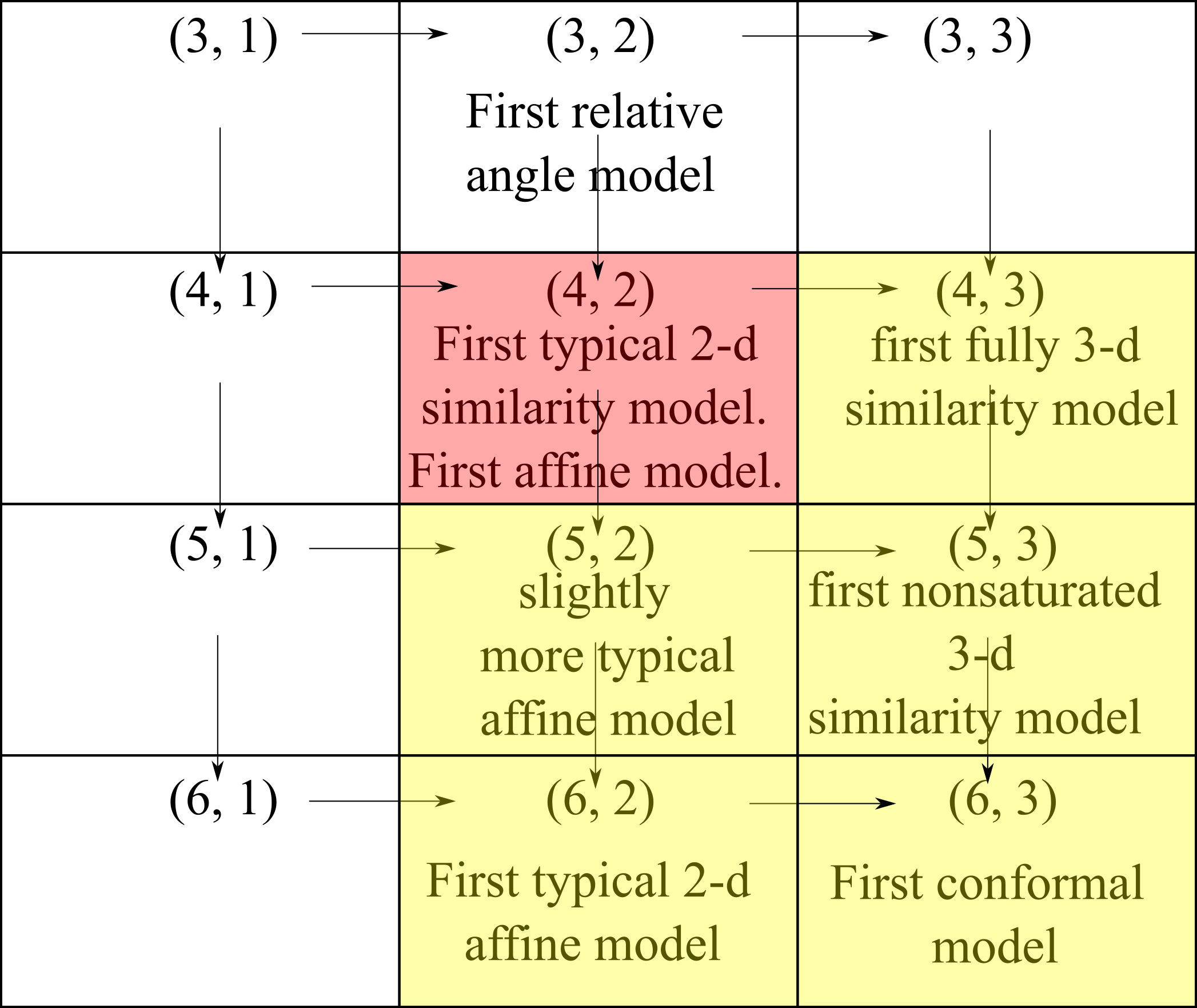}
\caption[Text der im Bilderverzeichnis auftaucht]{\footnotesize{The Shape-Theoretic Aufbau principle involves accessing target problems along all combinations of 
right and down arrows leading there.
The current treatise involves reaching the 4 particles in 2-$d$  -- (4, 2) -- model marked in red. 
This suffices for 2-$d$ similarity typicality and for the smallest affine model. 
Further interesting targets are marked in yellow; these are determined by \c{AMech}'s consideration of minimal relationally nontrivial units, 
and by going past atypically geometrically simple minimal models as per the current treatise and \c{Affine-Shape-1}.} }
\l{Aufbau} \end{figure}          }

\n Whereas Similarity and Euclidean theories for small $d$ and $N$ have been well-studied to date,  
these studies however no longer meet all of the requirements brought about by wanting to study affine, conformal and supersymmetric variants.  
As such, we give here a sharper, 
extended and where necessary corrected version of previous work with the Euclidean and similarity groups in dimensions 1 and 2 going up to 4 particles. 
Note that 4 particles is a bare minimum for affine Shape Theory though some features there require 5 or 6 particles.  
Another application is further understanding the tetrahaedron case (4-body problem in 3-$d$).
Conformal shapes also require higher particle number; see Fig \r{Aufbau} for some minimal-sized targets.  
There are also new comparisons to be made with newer work with points on manifolds other than $\mathbb{R}^d$; see Sec 2.1 for a summary and references.  

\m 

\n{\bf Treatise Motivation 2} Our second innovative feature is systematic study of topological content of shapes-in-space.  
We not only document the topologically distinct shapes for a given $N$ and $d$, 
but also show how the corresponding {\sl topological shape spaces are graph-theoretic in nature}, providing examples and some general solutions to this matter.
This gives a novel combinatorial flavour to topological Shape Theory.  
These graphs furthermore capture some of the shape space structure and, are less hampered than metric manifold tessellation representations in higher-$d$ models

\m 

\n{\bf Treatise Motivation 3} Our third innovative feature is systematic study of indistinguishable point and mirror image identified shapes and their corresponding shape spaces.
\n As discussed in Sec \r{Mirr-Dist-Leibniz}, we use `Leibniz space' to mean the smallest shape space among these options: mirror images identified 
and indistinguishable particles. 
This corresponds to a kind of `unit cell' topologically and geometrically.  
Kendall's spherical blackboard of Fig \r{Leib(3, 2)-Intro}, which is well known in the Shape Statistics literature, is a Leibniz space, 
for the minimal relationally nontrivial case of 3 points in 2-$d$.  
%
{            \begin{figure}[!ht]
\centering
\includegraphics[width=0.45\textwidth]{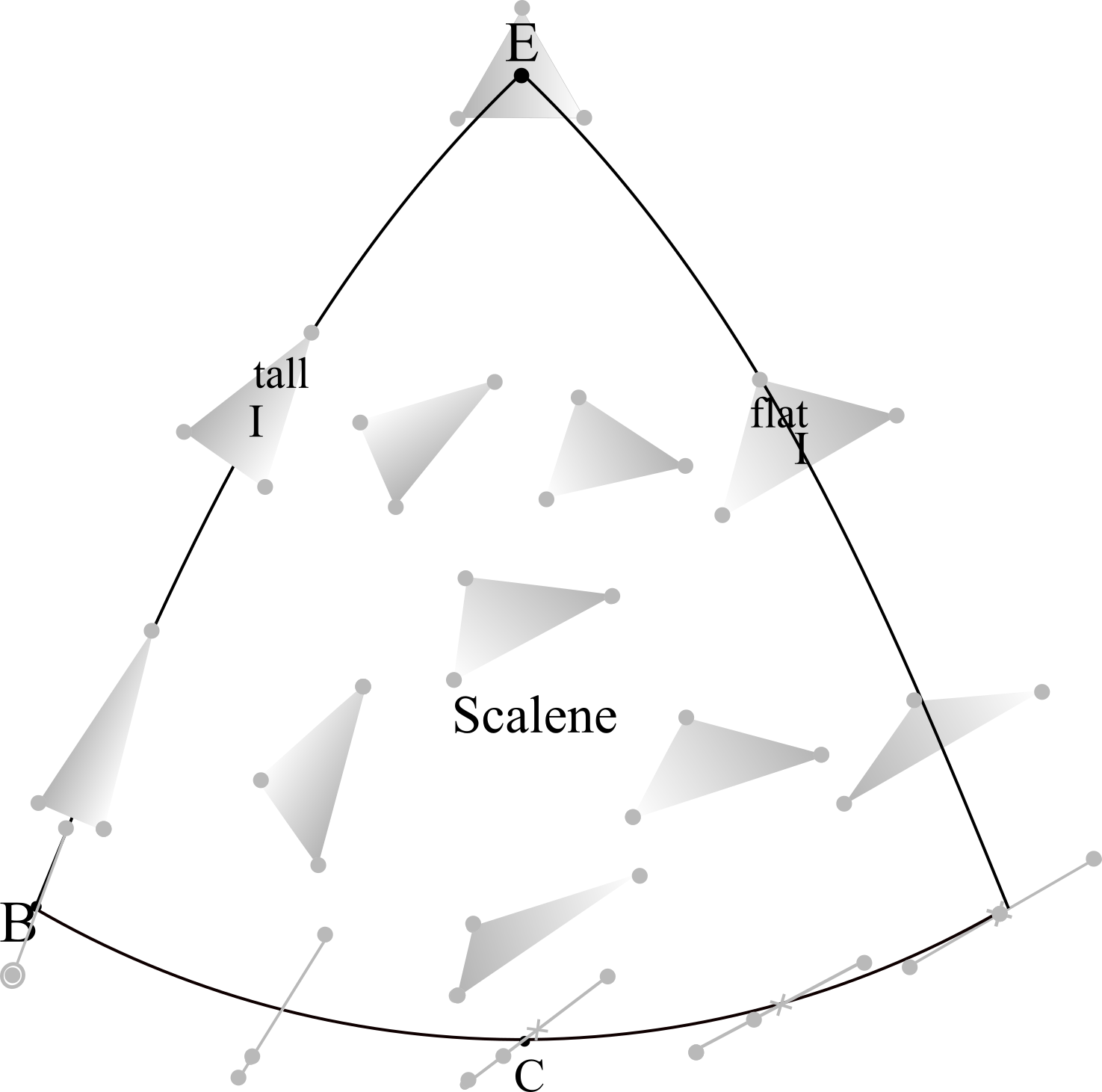}
\caption[Text der im Bilderverzeichnis auftaucht]{\footnotesize{We here motivate Leibniz spaces by presenting a reminder of Kendall's spherical blackboard.  
This is an interesting mathematical object, and in this treatise we also detail its simpler 1-$d$ counterpart, 
the comparably complex 4-particle 1-$d$ counterpart and the 4 particles in 2-$d$ generalization of both of the preceding.  } }
\l{Leib(3, 2)-Intro} \end{figure}          }
 
\m 

\n Leibniz spaces have various natural features.

\m 

\n 1) These are the type of shape space which maximizes the number of aspects held to be unphysical.
There is just one copy of every type of shape here, 
indeed embodying Leibniz's {\it Identity of Indiscernibles} \c{L2} that if two things are exactly alike, they are one and the same.    
Let us note here that the current treatise makes a number of sharp observations about prices to pay for taking 
Leibniz' Identity of Indiscernibles to its logical conclusion for $N = 3$ or $4$ models.
Aside from Secs \r{MII-Indis}, \r{Rel-Sp} and \r{Top-Shape}, discussion of Leibniz's Identity of Indiscernibles is chiefly concentrated in each Part's Conclusion.

\m 

\n 2) Furthermore, it is indistinguishability of this kind which plays a major role in QM.  
In this way, studying shapes at the quantum level further motivates study of the Leibniz spaces in particular. 

\m 

\n The current treatise also gives careful consideration to the topological counterpart of these interesting blackboard spaces.

\m 

\n{\bf Treatise Motivation 4} This treatise's fourth innovative feature is substantial application to Shape Theory of 
Differential Geometry Methods familiar from Theoretical Physics.  
The current treatise's instalment of this considers systematic study of which shape(-and-scale) space Killing vectors survive the quotienting process.
These form isometry groups: a type of automorphism group as laid out in Appendix B. 
In particular, given the current treatise's Leinbiz spaces being manifolds with boundary, 
the current treatise reminds us how to handle Killing Vectors for these (Appendix C).
This is a nice piece of 1950s-1960s Applied Geometry \c{Yano-Integral},   
known in pure geometry, PDE theory and GR communities but which as far as I am aware has not yet been considered in the Shape Statistics literature.  
We include also the corresponding treatment of similar Killing vectors of shape(-and-scale) spaces.  
This extended notion forms the similarity group: an in general larger type of automorphism group.   

\m 

\n It moreover also makes sense to discuss automorphisms for the graphs which are topological shape spaces as well.  
All in all, we find the automorphism groups for the various graph-theoretic and differential-geometric blackboard spaces of indistinguishable-point shapes.  

\m 

\n Which shape space (similarity) Killing vectors survive the quotienting procedures of particle indistinguishability and mirror image identification 
is in particular crucial for quantization.  

\m 

\n A second instalment of Differential Geometry Methods for Shape Theory consists of carefully picking out shape theoretically meaningful notions 
as (pieces of) submanifolds of shape space.
We have already mentioned this as part of how the Shape-Theoretic Aufbau Principle works.  
There is furthermore a strong tie between part of this and the first instalment.
Namely, it is also (similarity) Killing vector preserving submanifolds 
which play a particularly distinguished Shape-Theoretic Aufbau Principle role as regards the quotienting of $G$'s with further generators.  

\m 

\n The third instalment of Differential Geometry Methods for Shape Theory consists of use of fibre bundle methods, where applicable, for $d \geqs 2$ models.
In a nutshell, we will show that such as the first two Figures of the current treatise are better thought of as {\it sections of representative shapes}.

\m 

\n{\bf Treatise Motivation 5} Our fifth innovative feature is substantial application to Shape Theory of 
Differential Geometry Methods beyond those familiar from Theoretical Physics.  
The fourth instalment of Differentiable Geometry concerns {\it stratified manifolds} arising in Shape Theory.
This both from a) the more general and global form that the second instalment comes to take, 
           and b) from gauge orbit considerations by which whole shape spaces themselves are stratified manifolds for $d \geqs 3$ \c{Kendall} 
                  (or even 2 for affine or projective shapes \c{GT09, PE16, KKH16}).  
While only a) is new, the fifth instalment is that the sections evoked in instalment 3 must in general transcend Fibre Bundle Theory. 
They pertain rather to General Bundle Theory \c{Husemoller} or, with more local and global tools available, Sheaf Theory \c{Sheaves1, Sheaves}.   

\m 

\n All in all, the whole package of topological and metric shapes
-- quartered by whether or not to mirror image identify and/or consider particle indistinguishability, with corresponding shape spaces and their automorphisms -- 
amounts to a substantial application of each of Combinatorics, Topology, Differential Geometry and Group Theory.  
In conjunction with the holistic nature of the Shape-Theoretic Aufbau Principle, 
this justifies a substantial treatise with upward of 140 colour figures to cleavly discern between the many layers of structure involved.     
Subsequent works will involve quotienting by larger $G$, for which attaining at least 4 point models in 2-$d$ -- the current treatise's pinnacle -- is a prerequisite.  
As such, the current treatise's broader foundations of what Shape Theory is, 
and demarcation of how even the smallest examples of shape theories illustrate plenty of Shape Theory's qualities and features, paves the way to myriad further research projects.

\subsection{Outline of Part I}

We first consider point positions in Sec 2 and relative separation configurations -- in both Lagrange and Jacobi form -- in Sec 3. 
In each case, we consider the corresponding configuration spaces -- constellation space and relative space -- which are flat.
We introduce in mirror image identification and indistinguishable particle modelling features in Sec 4; both of these are discrete quotients of the preceding configuration spaces.

\m 

\n In Sec 5, we next outline continuous quotients by one or both of dilations and rotations. 
This gives Kendall's \cite{Kendall84, Kendall89, Kendall} preshape space and shape space notions 
alongside the relational space most natural to Celestial Mechanics \cite{Marchal, LR97}, Molecular Physics \cite{LR97} 
and the classic Absolute versus Relational motion debate \cite{BB82, FileR, ABook}.  
We identify each model's invariants, and count out the degrees of freedom so as to identify the minimal relationally nontrivial units in each case.  
We proceed to give an invariants-grounded account of collisions, clumping, uniformity, inhomogeneity and centre of mass mergers in Sec 6. 
This is the first mostly new section in this treatise, 
though some of the intervening sections are more general or structurally distinguished than any previous works in the literature. 
We then compare constellation, perimeter, all-separations and other perspectives in Sec 7.

\m 

\n In Secs 8, 9 and 10, we review the topology and geometry of preshape spaces, shape spaces and relational spaces.  
Sec 11 is a partly-new extension of this to include mirror image identification and indistinguishable particle modelling features, thus including the aforementioned Leibniz spaces.  
Secs 8 and 9 give spheres and complex projective spaces as 1- and 2-$d$ shape spaces, which have $SO(p)$ and $SU(p)$ isometry groups.
Sec 10 gives cones thereover as relational spaces, and 
Sec 11 points to various portions and topological identifications of the previously mentioned spaces, small cases of which are carefully worked out in Papers II to IV. 
Sec 12 concludes the first half of this paper by outlining both the analogies with GR's configuration spaces and the Background Independence 
content shared between simple shape models and GR (including various of its cosmological simplifications).  

\m 

\n In the second half of Part I, we concentrate on the (3, 1) example, which is minimal for many of the shape properties introduced so far to be defined.
The topological content of (3, 1) shapes is discussed in Sec 13, with the corresponding shape spaces -- which are graphs --  in Sec 14; 
the collision structure belongs at this level. 
The metric content of (3, 2) shapes is considered in Sec 15, with further details at the Lagrangian level in Sec 16 and at the Jacobian level in Sec 17.  
Both of these levels have notions of uniformity, whereas centre of mass mergers belong at the Jacobian level. 
Counts of numbers of qualitative types are provided in each case in Sec 18.   
For (3, 1), the relational space is just the relative space again, 
but as we have not yet presented the collision, uniformity and merger structures for this, we provide them in Sec 19. 
Finally Sec 20 provides automorphisms for the topological graphs and the metric-level configuration spaces, 
including under mirror image identification and indistinguishable particle modelling features.

\m 

\n We subsequently consider the (4, 1)                              model in Part II, 
                            the (3, 2) and (3, 3) triangles               in Part III and \c{A-Pillow, 2-Herons, Ineq, Max-Angle-Flow, A-Monopoles, A-Perimeter}, and 
                            the (4, 2)            quadrilaterals          in Part IV, and \c{A-Sylvester, A-Coolidge, A-Quad-Ineq}
in each case in a similar structural vein to the second half of the current treatise. 
The current treatise's shape concepts become very complicated very quickly, so this treatise's range of examples already covers almost all of it very well.    
The current treatise also includes Appendices on each of Graph Theory, automorphisms, and (plain and similarity) Killing vectors on manifolds without boundary; 
these Appendices usefully support all four of the current series' Parts. 

\vspace{10in}

\section{Carrier spaces, points, constellations and configuration space}\l{Preamble}

\subsection{Carrier, absolute, and location data sample spaces}\l{Abs}

\n We adopt an indirect approach to Shape Theory. 
This begins with an incipient notion of {\it carrier space}, $\Space(d)$, of dimension $d$ 
whose features are gradually removed as the formulation in terms of shapes is built up.   
This is the space in which the shapes under consideration are realized in.  

\m 

\n{\bf Structure 1} In the current treatise, we entertain the most usual flat space choice,     
\be
\Space(d) = \mathbb{R}^{d} \m  , 
\ee
equipped with the standard Euclidean inner product, or equivalently, metric. 
We denote this inner product by $(\m  , \m )$, 
      the corresponding norm by $|| \m  ||$ 
  and the corresponding norm by $\delta_{ab}$.\f{The 
$a$ and $b$ index is here a vector index, in the sense of a vector in the flat carrier space;  
more generally lower-case Latin indices from the start of the alphabet denote spatial indices in this treatise. 
The current treatise also uses underlining to denote spatial vectors.}

\m 

\n{\bf Subcase 1} In the case in which $\Space(d)$ is interpreted as physical space's flat-space model, 
it is the {\it absolute space} of Newtonian Mechanics \c{Newton} (in dimension $d$), $\Abs(d)$.  

\m 

\n [While Geometry was originally conceived of \c{Euclid} as occurring in physical space or objects embedded therein (parchments, the surface of the Earth...), 
we consider the Geometry version of our problem in terms of the abstract carrier space rather than its absolute space interpretation subcase.]   

\m 

\n{\bf Subcase 2} $\Space(d)$ can also be interpreted as a {\it sample space} in the context of Probability and Statistics.

\m 

\n{\bf Aside 1} See \c{ASphe} and \c{ATorus} respectively for the outcome of taking $\Abs(d)$ to be each of the following.

\m  

\n i) A sphere $\mathbb{S}^k$. 

\m  

\n ii) A torus $\mathbb{T}^k$. 

\m 

\n These serve to investigate what happens if flat open absolute space is replaced by two of the simplest possible alternatives: closed space without loss in symmetry 
and flat space closed by topological identification respectively.  
Moreover $\mathbb{S}^2$ models the observed {\it celestial sphere} \c{Kendall}, 
whereas  $\mathbb{S}^3$ and $\mathbb{T}^3$ are also the most commonly used closed space model manifold topologies in General Relativity (GR).

\m 

\n iii) Finally, use of projective shapes in Image Analysis is implemented \c{PE16} by considering $\mathbb{RP}^{d - 1}$ as carrier space.

\subsection{Primary points: vertices, data points and material particles}

\n{\bf Definition 1} By a {\it primary point}, we mean a point in carrier space which is ascribed particular significance in one's modelling.
It is represented as a marked point (which is denoted by a heavier grey dot in this treatise's figures) which may additionally be labelled 
(with a likewise grey label in this treatise).

\m 

\n{\bf Remark 1} What its particular primary significance is depends on interpretation. 

\m 

\n 1) In Geometry, it will play the role of a {\it vertex} of one or more figures within the abstract Euclidean space.    

\m 

\n 2) In Probability and Statistics, it is a locational data point within a sample of such in the flat sample space

\m 

\n 3) In Dynamics and Physics, it has {\it material significance} -- further material properties -- by which it is a {\it material point} alias 
{\it particle}.

\m 

\n Common examples of material properties are an associated mass, charge, or further quantum-level attributes such as spin and colour.  
In the current treatise, however, we limit ourselves to neutral classical particles of equal nonzero mass, for reasons of simplicity, 
maximal use of symmetry and the Introduction's reasons for interest in indistinguishable particles.
See \c{FileR} for comments on non-equal mass counterparts.  
We also largely work in the context of point-particle models as opposed to modelling bodies of finite extent by particles, 
thus precluding also a range of properties supported by finite bodies such as rigidity or angular momentum corresponding to rotation about an axis of the body.  
%

\m 

\n{\bf Structure 1} Relative to our carrier space's choice of point of origin, our primary point's location is at $\u{q}$: the {\it position vector}.
At this level in the modelling, evoking a carrier space also entails choice of a set of Cartesian axes and of a scale.  
In this treatise, position vectors are depicted as in Fig \r{Constell}.b). 

\m 

\n{\bf Remark 2} In modelling Dynamics or Physics, these are an absolute origin, absolute axes and absolute scale. 
The first two of these in particular have been regarded as physically meaningless features that Physics should be freed from, 
by the relationalist side of the Absolute versus Relational Motion Debate.   

\m 

\n{\bf Remark 3} In modelling Dynamics or Physics, one permits $\u{q}$ to be time-dependent, denoted by $\u{q}(t)$.
In this sense, the Dynamics or Physics case exceeds the abstract Geometry setting, in which a vertex position $\u{q}$ has a fixed meaning.  
This is marked by particles having further material properties such as velocity, momentum and acceleration, 
alongside laws of motion by which some of these are not independent. 
In conventional Dynamics and Physics, the laws are second-order, by which position at a given value of time, alongside one of velocity or momentum at that time, 
suffice as data for evolution over time.  
%
{            \begin{figure}[!ht]
\centering
\includegraphics[width=1.0\textwidth]{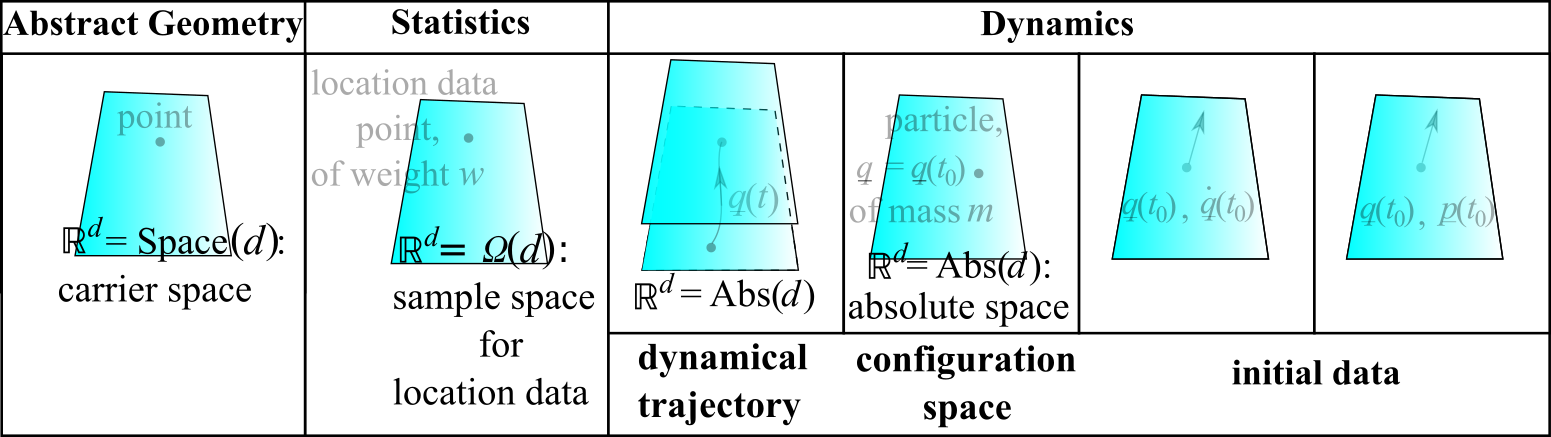}
\caption[Text der im Bilderverzeichnis auftaucht]{        \footnotesize{Geometric point, data point, dynamical point evolution, dynamical configuration point 
and dynamical data points.} }
\l{Model-Multiplicity}\end{figure}            }

\m 

\n{\bf Remark 4} Partly because of the above excess content, 
              and partly because of some relevant concepts being more familiar or developed therein 
			  (or comparably familiar but to different audiences using different nomenclature), 
			  we quite often use joint names from Geometry and Dynamics or Physics. 
Let our first term for a such -- point-or-particle -- illustrate the hyphenation convention by which such terminology can be immediately spotted and understood 
in the current treatise.  
In this particular case, this is short for `primary point in the abstract Geometry context or material point-particle in the Dynamics or Physics context'.   
[A location data point in Probability and Statistics remains a subcase of the abstract Geometry side of this portemanteau concept.]

\subsection{Collections of points-or-particles as constellations}\l{Constellations}
%
{            \begin{figure}[!ht]
\centering
\includegraphics[width=0.42\textwidth]{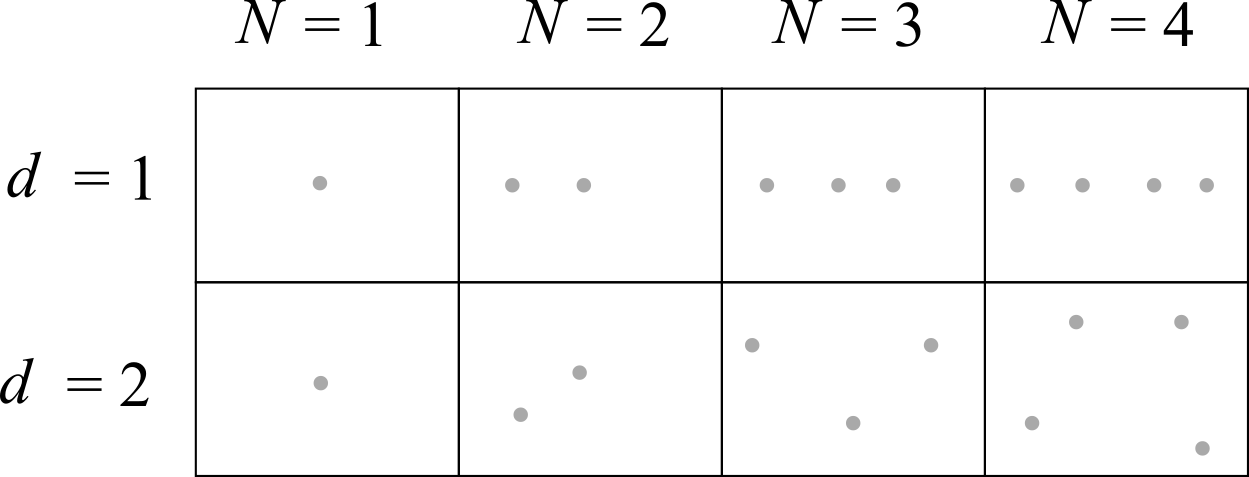}
\caption[Text der im Bilderverzeichnis auftaucht]{        \footnotesize{a) The main range of models in the current treatise: 
(1, 1) though to (4, 2) constellations; their points-or-particles are each marked with a larger grey dot.  } }
\l{Constell}\end{figure}            }

\n{\bf Definition 1} In the Classical Physics or Dynamics context, a {\it configuration} is the instantaneous state of the whole system under consideration. 
In the context of abstract Geometry, the same definition continues to make sense, 
excepting that the word `instantaneous' is now redundant since the object of one's modelling is now fixed.  

\m 

\n{\bf Definition 2} A `{\it constellation}' is a collection of $N$ (possibly superposed) points-or-particles.  

\m 

\n{\bf Remark 1} Constellations can be defined for any geometric model of carrier space $\Space(d)$, 
though in the current treatise we restrict attention to $\Space = \mathbb{R}^d$. 
In this case, labelling out points-or-particles with index $I$ running from 1 to $N$, their Cartesian coordinates of their position vectors in carrier space 
are denoted by $q^{aI}$, and particles' masses by $m_I$.  
Constellations are indeed $N$-point-or-particle models' notion of configuration. 
For now, our precise incipent notion of configuration consists of $N$ points-or-particles in carrier space.
We shall moreover show below that the constellation concept retains `intrinsic' content as the carrier space (in particular absolute space) is stripped away; 
while (some part of) the carrier space structure remains in place, constellations possess some other `extrinsic' features relative to this.  
Indeed, scale-normalized constellations turn out to be the shapes that the Shape Theory in this treatise alludes to 
(more precisely, this is {\sl similarity} Shape Theory).  

\m 

\n{\bf Remark 2} In Shape Theory we focus on constellations, which are configurations and thus instantaneous in the Dynamics or Physics context. 
Let us comment more generally that this instanteneousness wipes out many of the differences between this and the abstract Geometry context, 
by which the Dynamics or Physics and abstract Geometry portemanteau is rather more cohesive than one might originally expect.

\m 

\n{\bf Definition 3} The above-mentioned inclusion of superposition is realized moreover by {\it point coincidences} in abstract Geometry and 
{\it particle collisions} in Dynamics or Physics, referred to jointly as {\it coincidences-or-collisions}.  
 
\m 

\n{\bf Remark 3} In the abstract Geometry context itself, these are {\it vertex coincidences} -- which we shall see are one way of having degenerate figures -- 
whereas in its Probability and Statistics application, these are {\it data point coincidences}, corresponding to identical-location data points in one's sample.  
																								
\m 
																								
\n{\bf Definition 4} The {\it maximal coincidence-or-collision} O refers to the maximal coincidence -- of all points -- 
                                                                          or the maximal collision   -- of all particles -- in one's constellation. 
									 
\m 

\n{\bf Remark 4} We shall see that this often plays a particularly distinguished role, 
whether as the most dominant point or as a point which must be excluded from certain structures and conceptualizations.  

\m 

\n{\bf Definition 5} A {\it partial coincidence-or-collision} is one which is not maximal.

\m  

\n{\bf Remark 5} These relatively seldom require excision; Fig \r{Constell} shows that there are a variety of such, 
alongside this treatise's pictorial representation of these.

\m 

\n{\bf Definition 6} Let us also introduce the notation ($N$, $\Space(d)$) for the N-point-or-particle model on $\Space(d)$. 
For the current treatise series' $\Space(d) = \mathbb{R}^d$ case, we simplify this notation to ($N$, $d$). 
This brevifies the means of keeping track of which examples are being referred to.  
The Dynamics and Physics names for these examples are {\it d-dimensional N-body problem}.

\m 

\n{\bf Remark 6} These are often-studied problems which nonetheless rapidly become intractably complicated as $d$ and $N$ increase, 
even for just moderate values of $N$ such as 3, 4, 5 or 6; Shape Theory is starting to offer a new perspective on this.   
%
{            \begin{figure}[!ht]
\centering
\includegraphics[width=1.0\textwidth]{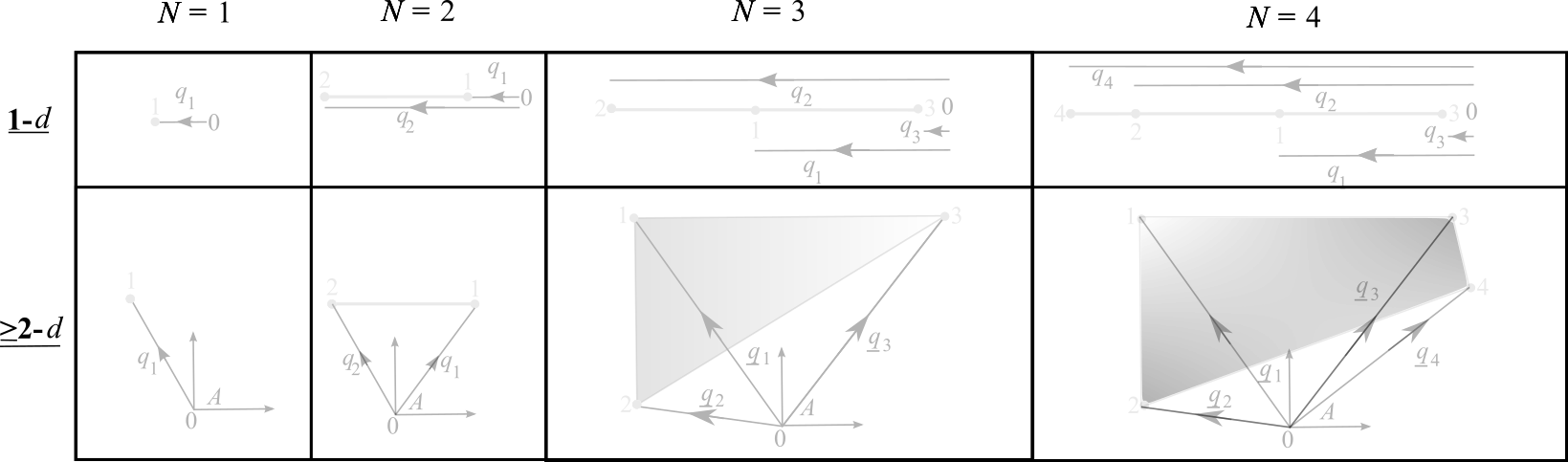}
\caption[Text der im Bilderverzeichnis auftaucht]{        \footnotesize{Position coordinates relative to a fixed absolute scale $S$, 
                                                                                                                absolute origin $0$, 
																										      and absolute axes $A$ (in $d \geqs 2$).} }
\l{Position-Coordinates}\end{figure}            }

\subsection{Configuration spaces}

\n{\bf Definition 1} A {\it configuration space} $\FrQ$ is the set of all possible configurations a given system can exhibit.   

\m 

\n{\bf Remark 1} N.B.\ that each configuration corresponds to a {\it single point} in configuration space $\FrQ$. 
With configurations being sets of points, we distinguish between points in space and points in configuration space by colouring the former in grey and the latter in black.

\m 

\n{\bf Structure 1} Configuration space is usually taken to be furthermore an {\sl equipped} set: with a topology, 
                             and on some occasions, with differential structure, 
							                            Riemannian structure, 
											       or a metric in the `Analysis sense' of a distance function.

\m  

\n{\bf Definition 2} {\it Incipient constellation space} $\FrQ(N, \Space(d))$ 
is the configuration space of constellations in the incipient extrinsic sense of collections of points in $\Space(d)$.  
For the current treatise series' $\Space(d) = \mathbb{R}^d$ case, we simplify this notation to $\FrQ(N, d)$ 
(but the full notation is needed once $\Space(d)$ attains further geometrical diversity \c{Kendall, ASphe, ATorus, PE16}). 

\m 

\n{\bf Example 1} Among point-or-particle models, the 1-body problem makes no distinction between configuration space and carrier space,
\be
\FrQ(1, \Space(d)) = \Space(d) \m  , 
\ee
but they are clearly distinct from the 2-body problem upward: 
\be
\FrQ(N, \Space(d))      \es      \bigtimes_{I = 1}^N \Space(d)  
                   \m  \neq \m   \Space(d)                      
				   \mbox{ for } 
				   N \geqs 2 \m  , 
\ee
where $\bigtimes$ denotes Cartesian product.  
This is an important distinction to emphasize on the physical side, in particular for the following reason. 
Many Theoretical Physics teaching programs jump straight from 1-body Quantum Theory to infinite-body Quantum Theory. 
This leaves a widespread feeling that `Quantum Theory unfolds in space', 
whereas even just the 2-body problem illustrates that Quantum Mechanics actually unfolds on configuration space (or any other `half-polarization' of phase space).

\m 

\n{\bf Example 2} The current treatise's $\Space(d) = \mathbb{R}^d$ case moreover simplifies further:  
\be
\FrQ(N, d) = \bigtimes_{I = 1}^N \mathbb{R}^d 
           = \mathbb{R}^{Nd}, 
\ee
which is itself a flat space but now of dimension 
\be
\mbox{dim}(\FrQ(N, d)) = Nd  = \mbox{(particle position vector components)}  \m  .
\l{dim-Q}
\ee  
{\bf Structure 2} This is equipped with the corresponding Euclidean inner product or, equivalently, metric.   
In the case in which masses are attributed to material point particles, the inner product comes with mass weightings, 
\be
(\bq, \bq^{\prime})_{\mbox{\scriptsize\boldmath$m$}} := \sum_{I = 1}^N \sum_{a = 1}^d  m_I q^{aI} q^{\prime aI} 
                                                      =     \delta_{ab}\delta_{IJ }    m_I q^{aI} q^{{\prime}bJ} \mbox{ (no sum) } . 
\ee
where we use boldface for vectors in configuration space (now regarding $aI$ as a multi-index).  
The norm version of this defines the scalar {\it total moment of inertia}, or {\it inertia quadric} 
\be
I(\bq) := ||\bq||_{\mbox{\scriptsize\boldmath$m$}}\mbox{}^2 
   = \sum_{I = 1}^N \sum_{a = 1}^d  m_I q^{aI} q^{aI} 
   = \sum_{I = 1}^N m_I||\u{q}^I||^2
   = \delta_{ab}\delta_{IJ} m_I q^{aI} q^{bJ} \mbox{ (no sum) } , 
\l{In-Quad}
\ee
The kinetic energy is then the version of this built out of velocities rather than positions and divided by 2.  
The above objects can furthermore be viewed as built out of the {\it configuration space metric},  
\be
M_{aIbJ} = m_I \delta_{ab}\delta_{IJ} \m  .  
\ee
\n {\bf Structure 3} One can moreover pass to {\it mass-weighted position coordinates},\footnote{We use Greek letters for mass-weighted quantities, 
though $\alpha, \beta, \gamma, \theta, \phi, \varphi, \Theta$ and $\Phi$ remain reserved for angles.} 
\be
\chi^{aI} = \sqrt{m_I}q^{aI} \m  .
\ee
The corresponding inner product is now with respect to the unit metric 
\be 
U_{aIbJ} = \delta_{ab}\delta_{IJ} \m  .  
\l{Unit-Metric}
\ee
\be
I(\chi) = \sum_{I = 1}^N || \chi_I ||^2 \m .  
\l{Chi-Quad}
\ee 
One also uses (\r{Chi-Quad}) {\it ab initio} in applications in which no concept of mass is associated (points rather than particles).  

\m 

\n The rest of this series of articles gives many further examples of configurations and configuration spaces.  
The main case of Remark 1's correspondence in the current treatise is the {\it `shape-in-space to position-in-shape-space' correspondence},  
shape spaces being a subcase of configuration spaces.

\section{Relative configurations and their configuration space}
				
\subsection{Relative Lagrange coordinates}\l{Rel-Lag}

\n{\bf Definition 1} We use {\it relative} to refer to all concepts which are free from the notion of an absolute (or elsewise preferred) origin $0$. 

\m 

\n{\bf Remark 1} This is attained by taking differences of the position vectors $\u{q}^I$ so as to cancel out any reference to $0$. 

\m 

\n{\bf Definition 2} {\it Relative Lagrange coordinates} are the differences between each pair of position vectors, 
\be
\uir^{IJ} = \uiq^J - \uiq^I \m  . 
\ee 
\n{\bf Example 1} 
\be
N \geqs 2 \mbox{ is required for relative Lagrange coordinates to be defined } ; 
\ee 
see Fig \r{Relative-Coordinates}.a) for these in (2, 1) through to (4, 2) models.   
%
{            \begin{figure}[!ht]
\centering
\includegraphics[width=1.0\textwidth]{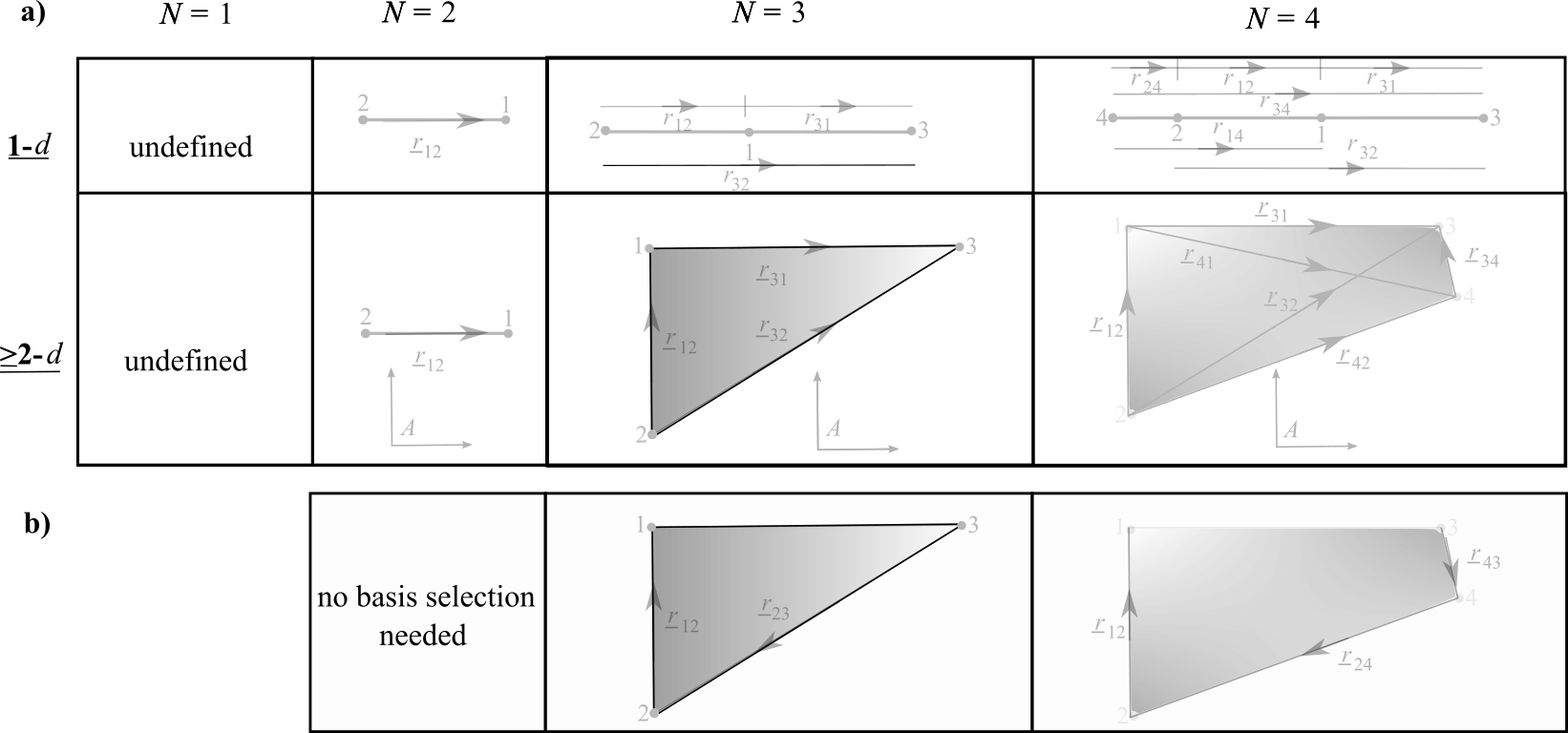}
\caption[Text der im Bilderverzeichnis auftaucht]{        \footnotesize{a) Relative separation vectors; 
Note that these are still defined with respect to a fixed absolute scale $S$, and absolute axes $A$ (in $d \geqs 2$).  

\m 

\n b) Some basis choices amongst these.} }
\l{Relative-Coordinates}\end{figure}            }

\m  

\n{\bf Example 1} On the one hand, $N = 2$ has just one relative separation vector, $\ur_{12}$.
On the other hand, 
\be 
N \geqs 3 \m  \mbox{ has multiple $\uir^{IJ}$ which are moreover not all linearly independent } ; 
\ee
for instance 
\be
\uir_{13} = \uir_{12} + \uir_{23} \m  . 
\ee
One can get around this by picking a basis of $n := N - 1$ of the $\uir^{IJ}$ to work with.  
One would then express all of one's relative separation quantities in terms of this choice of basis. 

\m 

\n{\bf Remark 1} While this is more mathematically convenient in some ways than using the entirety of the relative separation vectors, it also 

\m 

\n a) entails a choice.  

\m 

\n b) It can cause some properties to be less directly manifest.  

\m 

\n c) It complicates some expressions.
In more detail a) is that a basis choice is an `undemocratic privileging of some separations over others'.
In considering quantities which are symmetric in all of the relative separations, 
a basis choice has the effect b) that these quantities cease to be expressed in a {\sl manifestly} symmetric manner.    
Examples of complications c) stem from 
\be
\mbox{the configuration space metric becomes nondiagonal when expressed in terms of a basis of $\uir_{IJ}$ for $N \geqs 3$ } . 
\ee
Consequently, quantities built from this, such as the total moment of inertia or the kinetic energy cease to be diagonal.   

\m 

\n{\bf Remark 2} $n$ is conceptually $\#(\mbox{independent relative separation vectors})$.

\subsection{Centring, translations and their invariants, and relative space}

\n{\bf Definition 1} The {\it centring map} involves passing to centre of mass coordinates alongside a basis of $n$ relative coordinates.  

\m 

\n{\bf Remark 1} `Passing to cente of mass coordinates' can be thought of as placing the carrier space origin $0$ at the total centre of mass of the system $\fO$: 
a point that is now not arbitrarily placed but rather determined by the distribution of the points-or-particles.  
Note that {\it Barycentre} is an alias of centre of mass, as in e.g. barycentric coordinates or barycentric frames.  
The Statistics analogue of this passage is `standardizing the position of the mean'. 

\m 

\n{\bf Remark 2} Applying the centring map followed by discarding the centre of mass position vector itself can moreover be viewed as 
quotienting out the corresponding $d$-dimensional translations $Tr(d)$.
This procedure serves the following two purposes. 

\m 

\n{\bf Purpose 1} It identifies the aforementioned differences of position vectors $\ur^{IJ} = \uq^J - \uq^I$ as {\sl invariants} of the translation group $Tr(d)$; 
moreover the {\sl general} such invariants are of the form 
\be
f(\u{r}_{IJ} \mbox{ alone} ) = f(\mbox{differences } \uq^I - \uq^J \mbox{ alone}) \m  .  
\ee
\n{\bf Purpose 2} It provides the configuration space corresponding to the relative coordinates, as follows.  

\m 

\n {\bf Definition 2} {\it Relative space} is the quotient  
\beq
\lFrr(N, d)  \es  \frac{\FrQ(N, d)}{Tr(d)} \m  .
\eeq 
\n{\bf Remark 3} $Tr(d)$ can moreover readily be identified as 
\be
Tr(d) = \mathbb{R}^d \m  .
\l{Tr(d)}
\ee
Thus 
\be
\mbox{dim}(Tr(d)) = \mbox{dim}(\mathbb{R}^d) 
                  = d                                                     \m  , \l{dim-Tr}
\ee
By this and elementary consideration of Euclidean spaces, we obtain a general explicit geometrical form for relative space,  
\beq
\lFrr(N, d)  \es  \frac{\mathbb{R}^{N d}}{\mathbb{R}^d} 
              =   \mathbb{R}^{n d}                        \m  .  
\eeq
I.e.\ relative space is itself a Euclidean space, now of dimension   
\be
\mbox{dim}(\lFrr(N, d)) = n  d 
                        = \#( \mbox{ independent separation vectors' components } ) \m  .  
\l{dim-r}
\ee
\n{\bf Definition 3} In this treatise, we say a configuration space is {\it reduced} if it is obtained by quotienting out a continuous group. 

\m 

\n{\bf Remark 4} $\lFrr(N, d)$ is thus a first example of reduced configuration space. 
It is however an atypically simple example since it is still topologically nontrivial and flat; all the other instances of reduced configuration spaces in this treatise are curved.
%

\m 

\n{\bf Remark 5} Because of the large degree of geometrical similarness between $\FrQ(N, d) = \mathbb{R}^{Nd}$   and 
                                                                                 $\lFrr(N, d) = \mathbb{R}^{nd}$, 
there is little difference \c{FileR} between using one or the other as one's starting point in the study of shapes.
Regarding $\lFrr(N, d)$ as an incipent configuration space, moreover, is more convenient in setting up Sec \r{Gdyn}'s analogy with GR.  

\m 

\n{\bf Remark 6} The effect on the inertia quadric (\r{In-Quad}) is as follows.  
Translating the origin by some arbitrary amount $\u{a}$, 
\be
I(\u{q}_I, \u{a}) \es  \sum_{I = 1}^N \mm_I||\u{q}_I - \u{a}||^2             \m  .
\l{Ia}
\ee 
Furthermore, extremizing with respect to $\u{a}$,   
\be
\u{a} \es  \frac{1}{\mM} \sum_{I = 1}^N \mm_I q_I =: \u{a}_{\sC\so\sM}       \m , 
\ee 
where `CoM' stands for `centre of mass', and where 
\be
\mM \m  := \m  \sum_{I = 1}^N \mm_I \m  
\ee 
is the {\it total mass}.  
So introducing an arbitrary $\u{a}$ and extremizing thereover picks out the centre of mass position. 
Substituting for this back in (\r{Ia}) gives the {\it relative Lagrangian version of the inertia quadric}, 
\be
\bigiota(\u{r}_{IJ}) \es  \frac{1}{\mM}\s{    \mbox{$\sum_{I = 1}^N\sum_{J = 1}^N$}    }{    \mbox{\scriptsize$I < J$}    } \, m_I m_J r_{IJ}\mbox{}^2 \m  . 
\l{ILag}
\ee 
\n{\bf Remark 7} For $N \geqs 3$, the inertia quadric (\r{ILag}) has technical disadvantages stemming from not all the $\u{r}_{IJ}$ being independent.
Non-diagonality ensues.

\m 

\n{\bf Remark 8} (\r{ILag}) can be rewritten in Linear Algebra form as 
\be
\bigiota(\u{r}_{IJ}) = L_{IJ} q_I q_J
\ee
for `relative Lagrange matrix' $\u{\u{L}}$.  
For example, narrowing down consideration to $N = 3$, 
\be
\frac{1}{3}(r_{12}\mbox{}^2 + r_{13}\mbox{}^2 +  r_{23}\mbox{}^2) \m  , 
\ee 
for which 
\be
\u{\u{L}} \es  \frac{1}{3} \left( \s{  \s{  \mbox{$\m  \m  1  \m  \m                   -1 \m  -1$}  }  
                                         {  \mbox{$\m          -1  \m  \m  \m  \m  1 \m  -1$}  }  }
										 {  \mbox{$\m          -1  \m                            -1 \m  \m  \m  \m  1$}  }  \right) \m  .  
\ee
\n{\bf Remark 9} Another linear algebra formulation is in terms of the chosen basis $\ur_i$ of the $\ur_{IJ}$:\f{$i, j, k$ are used as indices for these; 
these are relative space indices and take values from 1 to $n$.}
 \be
\bigiota(\ur_i)  = L^{\prime}_{ij}r_ir_j
\ee
For example, narrowing down consideration to $N = 3$, 
\be
\u{\u{L}} \es  \frac{1}{3} \left(  \s{  \mbox{$\m  \m  2  \m  \m 1$}  }  
                                     {  \mbox{$\m  \m  1  \m  \m 2$}  }  \right) \m  .  
\ee

\subsection{Relative Jacobi coordinates}

\n{\bf Remark 1} Since (\r{ILag}) is symmetric, it can moreover be diagonalized.

\m 

\n{\bf Definition 1} This gives the {\it relative Jacobi coordinates} $\u{R}_i$. 

\m 

\n{\bf Remark 2} On the one hand, by diagonality these are more mathematically convenient to work with:  
they are widely used in e.g.\ Celestial Mechanics \c{Marchal} and in Molecular Physics \c{LR97}.

\m 

\n On the other hand, these are also more involved to formulate:
now in general not just relative point-or-particle separations are required but also separations between different cluster (subsystem) centres of mass.  
Indeed, the subsystem centres of mass (CoM) which enter consideration in this way play a substantial enough role in the current treatise to warrant their own notation. 

\m 

\n{\bf Definition 2} Let us introduce $\mbox{CoM}(ab...z)$ to denote the centre of mass of points-or-particles $a$ to $z$.
This notation permits inclusion of single particles as a special case -- $\mbox{CoM}(a) = a$ -- when convenient; 
this simplifies Remark 2 and a number of subsequent discussions.  
Let us also use $\fX$, $\fT$ and $\fQ$ for 2-, 3- and 4-point-or-particle CoMs (we stop at 4 because the current treatise's specific examples go no further). 
This also explains the font choice already made for a model's total CoM, $\fO$. 
Some figures include distinctions and/or superpositions of two $\fX$, in which case we use rather ${\bf +}$ for one of them.  
Let us finally use ${\cal P}$,  ${\cal X}$, ${\cal T}$ and ${\cal Q}$ for a configuration's sets of $\fP$, $\fX$, $\fT$ and $\fQ$ respectively, 
alongside ${\cal A}$ and ${\cal B}$ as symbols running over all of a model's types of CoM.   

\m  

\n{\bf Definition 3} A {\it clustering} is a recursive build-up of a system from its constituent subsystems as follows.   
One starts with the set of all points-or-particles, and picks two, say $a$, $b$, to make a first cluster. 
One then replaces $a$, $b$ by $\mbox{CoM}(ab)$ in the set of eligible objects from which the next two objects are picked. 
One continues until one ends up with $\fO$.  

\m 

\n {\bf Examples 1 and 2} Since  $N = 2$ has just one $\u{r}_{IJ}$ and thus is already diagonal, 
                              and $N = 1$ does not even have any  $\u{r}_{IJ}$ defined in the first place, 
\be
N \geqs 3 \mbox{ is required for relative Jacobi to Lagrange coordinate distinction } \m  . 
\ee
\n{\bf Example 3}	The minimal $N = 3$ case has moreover 3 choices of clustering, 
corresponding to {\it labelling ambiguity} in picking a first relative separation $\uir_{IJ}$ in forming it. 
We refer to this as the {\it base pair} of the clustering and to the remaining point-or-particle $K$ as the {\it apex}.  
The diagonality condition then requires the other relative Jacobi vector to be from $\mbox{CoM}(IJ)$ to $K$, so we have
\beq
\uiR_1 = \uiq_3 - \uiq_2                                   \m  \mbox{ and } \m  
\uiR_2 = \uiq_1 - \frac{m_2\uiq_2 + m_3\uiq_3}{m_2 + m_3}                        \m  
\ee
(see Fig \r{Relative-Jacobi-Coordinates}.a-b).
The clustering's corresponding reduced masses are 
\beq
\mu_1 \es  \frac{m_2m_3}{m_2 + m_3}                          \m  \mbox{ and } \m  
\mu_2 \es  \frac{m_1\{m_2 + m_3\}}{m_1 + m_2 + m_3}                                \m  . 
\eeq
For the equal mass case used in the current treatise, the above equations simplify to 
\beq
\uiR_1  =  \uiq_3 - \uiq_2                                   \m  \mbox{ and } \m  
\uiR_2 \es \uiq_1 - \frac{\uiq_2 + \uiq_3}{2}                                      \m  .
\l{3-Jacs}
\ee
\beq
\mu_1 \es  \mbox{$\frac{1}{2}$}                                                \m  \mbox{ and } \m  
\mu_2 \es  \mbox{$\frac{2}{3}$}                                                                      \m  . 
\eeq
We call the (3, $d$) model's first and second relative Jacobi vectors the {\it base} and {\it median} vectors.
Note that for the equal-mass case, $\mbox{CoM(base)}$ is at the centre of the base chosen by the clustering. 
The second Jacobi vector then runs from this to the opposing vertex, 
by which `median' is indeed an appropriate name for it in the $d \geqs 2$ case's triangular configuration. 
%
{            \begin{figure}[!ht]
\centering
\includegraphics[width=0.75\textwidth]{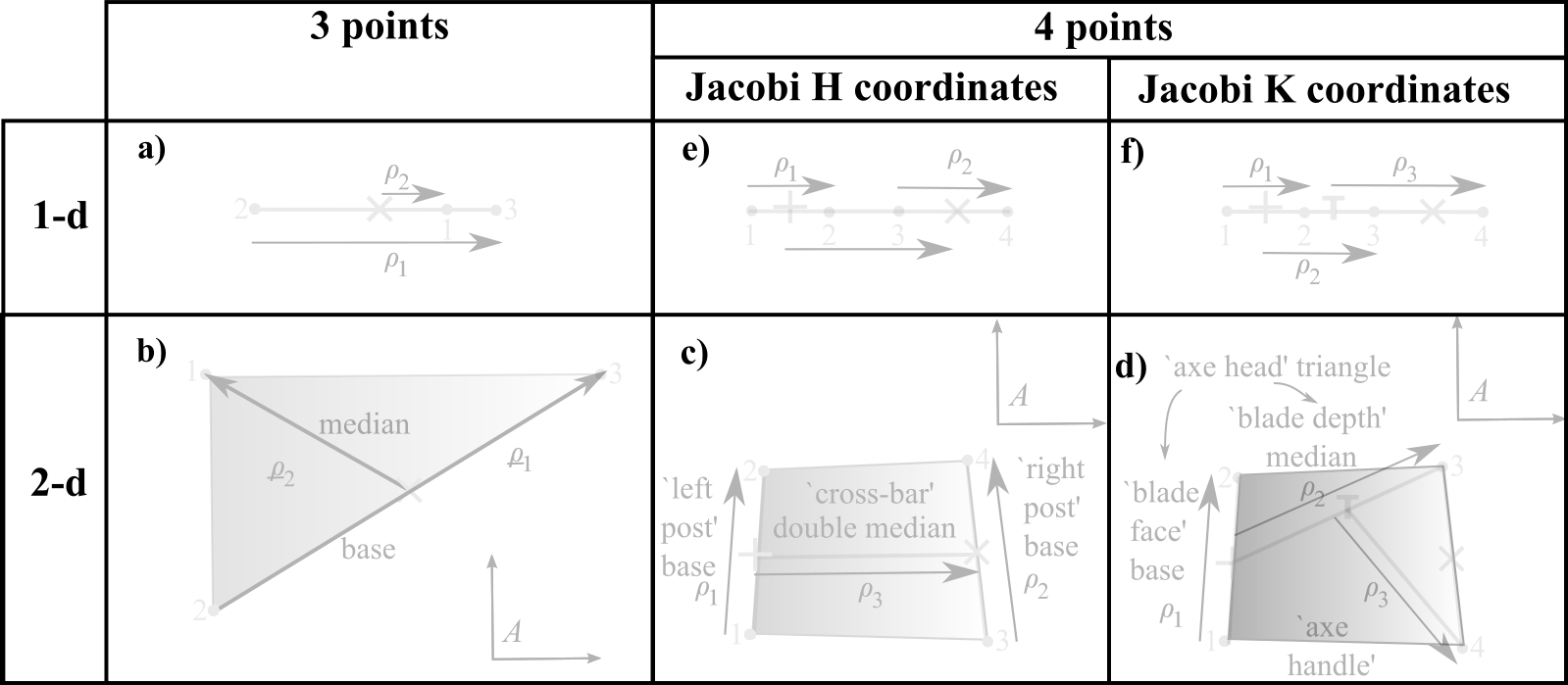}
\caption[Text der im Bilderverzeichnis auftaucht]{        \footnotesize{Specific examples of relative Jacobi coordinates used in this article;  

\m 

\n a) depicts these for (3, 1) and b) depicts these for (3, 2). 

\m 
 
\n c) and d) depict the H- and  K-clusterings for (4, 2); the H and K names are clearer for $d \geqs 2$ than for the 1-$d$ case depicted in e) and f).} }
\l{Relative-Jacobi-Coordinates}\end{figure}            }

\m 

\n{\bf Remark 3} Setting  
\be
0 = \mbox{det} \left( \Lmat - \lambda \, \Imat \right) \m , 
\ee
the eigenvalues are 
$\lambda = 0$ with multiplicity 1, and 
$\lambda = 3$ with multiplicity 2.

\m 

\n The corresponding orthonormal eigenvectors are, respectively, 
\be 
\frac{1}{\sqrt{3}}\left( \s{  \s{ \mbox{1}  } 
                                              { \mbox{1}  }  }
										      {  1 } \right)                   \mma\m 
\frac{1}{\sqrt{2}}\left( \s{  \s{  \m   \m  \mbox{0} }  
                                              {                     \mbox{$-1$} }  }
										      {  \m  \m   1 } \right)\mma\m 
\frac{1}{\sqrt{6}}\left( \s{  \s{  \m   \m  \mbox{2} }  
                                              {                    \mbox{$-1$} }  }
										      {                    \mbox{$-1$} } \right) \m  . 
\l{L-evectors}
\ee
\n{\bf Remark 3} The first of these is just the centre of mass coordinate. 
This occurs regardless of what $N$ is, and contributes nothing to the diagonalized relative Jacobi form of the inertia quadric.
Thereby, this can be considered to involve 1 coordinate vector less: $n = N - 1$ coordinate vectors, so we write it as
\be
\bigiota(\u{\w{R}}_a) = \w{Y}_{aa}\w{R}_a\mbox{}^2 \m  . 
\ee 
The $\w{R}_i$ here are proportional to the conventional relative Jacobi coordinates. 
We tilde everything for now so as to reserve the untilded version for the conventionally used proportions themselves. 

\m  

\n{\bf Remark 4} For $N = 3$, 
\be
\u{\u{\w{Y}}} = \mbox{diag}(1, \, 1) \m  , 
\ee  
so we arrive at
\be
\bigiota(\u{\w{R}}_i) = \w{R}_1\mbox{}^2 + \w{R}_2\mbox{}^2  \m  .
\ee 
The conventional scaling is moreover as given in (\r{3-Jacs})
\be 
\uR_1  := \uq_3 - \uq_2\mma\m  \uR_2 \m  
       :=  \m  \uq_1 - \frac{\uq_2 + \uq_3}{2}               \m  ,
\l{Jac-Defs}
\ee 
which, as promised, is recognizable as consisting entirely of cluster separation vectors. 
The first is a fortiori an interparticle separation vector, whereas the second involves a 2-particle centre of mass (see Fig \r{Jac-Med-Ineq-Fig-2}.c).  
If these are used, the diagonal relative Jacobi separation matrix $\u{\u{Y}}$ moreover consists of the reduced masses of the clusters in question, 
\be 
\u{\u{Y}} \es  \mbox{diag}\left(\frac{1}{2}, \, \frac{2}{3}\right) \m  . 
\ee 
This indeed makes use of the standard definition of {\it reduced mass}, i.e.\ conceptually 
\be
\frac{1}{\mu} \es  \frac{1}{\mm_1} + \frac{1}{\mm_2} \m  , 
\ee 
which rearranges to the more computationally immediate form
\be
\mu \es  \frac{\mm_1\mm_2}{\mm_1 + \mm_2} \m  . 
\ee
Thus the relative Jacobi separation matrix can be allotted a further, now conceptual, name -- {\it reduced mass matrix} -- 
with reference to the cluster subsystems picked out in the allocation of the particular Jacobi coordinates in hand.
We mark this be replacing the $\u{\u{Y}}$ notation with $\u{\u{M}}$, 
which we take to be a capital $\mu$ standing for both `mass' and `diagonal' (in the manner that $\Lambda$ is probably the most common notation for a diagonal matrix).  
So we end up with a relative Jacobi inertia quadric of the form 
\be 
\bigiota(\u{R}_i) = M_{ij} R_i R_j \m  . 
\ee 
\n{\bf Remark 4} For $N = 3$, the sole ambiguity in picking out cluster subsystems in forming Jacobi coordinates is which two particles to start with. 
So there are 3 possible clustering choices, corresponding to the second orthonormal eigenvector above being free to have its zero in whichever component.\footnote{For $N \geqs 4$, 
there are further ambiguities, which can be shown to result from $N \geqs 4$ points supporting multiple shapes of tree graph (see e.g.\ Appendix A). 
Jacobi coordinates are widely used for instance in Celestial Mechanics \c{Marchal} and in Molecular Physics \c{LR97}.}
%
We denote the above choice by $\u{R}^{(1)}$ alias $\u{R}^{(a)}$, 
and the clusters with $\u{r}_{31}$ and $\u{r}_{12}$ as their first relative Jacobi coordinate 
               by $\u{R}^{(2)}$ alias $\u{R}^{(b)}$ 
			  and $\u{R}^{(3)}$ alias $\u{R}^{(c)}$ respectively.

\m 

\n{\bf Example 3} $N = 4$ exhibits {\it clustering hierarchy ambiguity} that goes beyond labelling ambiguity.   
On the one hand, a set of H-coordinates picks out a 2 + 2 partition into subsystem clusters: two binary systems. 
On the other hand,          a set of K-coordinates picks out a 3 + 1 partition into subsystem clusters: a ternary system and a single particle  
(or, more accurately, a \{2 + 1\} + 1 hierachical partition, by which furthermore a particular binary is picked out within the ternary). 
See Fig \r{Relative-Jacobi-Coordinates}.c-d) for this distinction. 
  	 
\m 

\n This ambiguity can be accounted for as an underlying topological-level distinction, 
which can moreover be mapped to a rather more well-known mathematical problem. 
Namely, H and K correspond to the non-isomorphic tree graphs \c{ACG86} on 4 vertices: the claw alias 3-star graph $S_3$ and the 4-path $P_4$ 
[see Fig \r{Jacobi-Trees} for how and Part II for ($N$, $d$) use of this observation].  
%
{            \begin{figure}[!ht]
\centering
\includegraphics[width=0.75\textwidth]{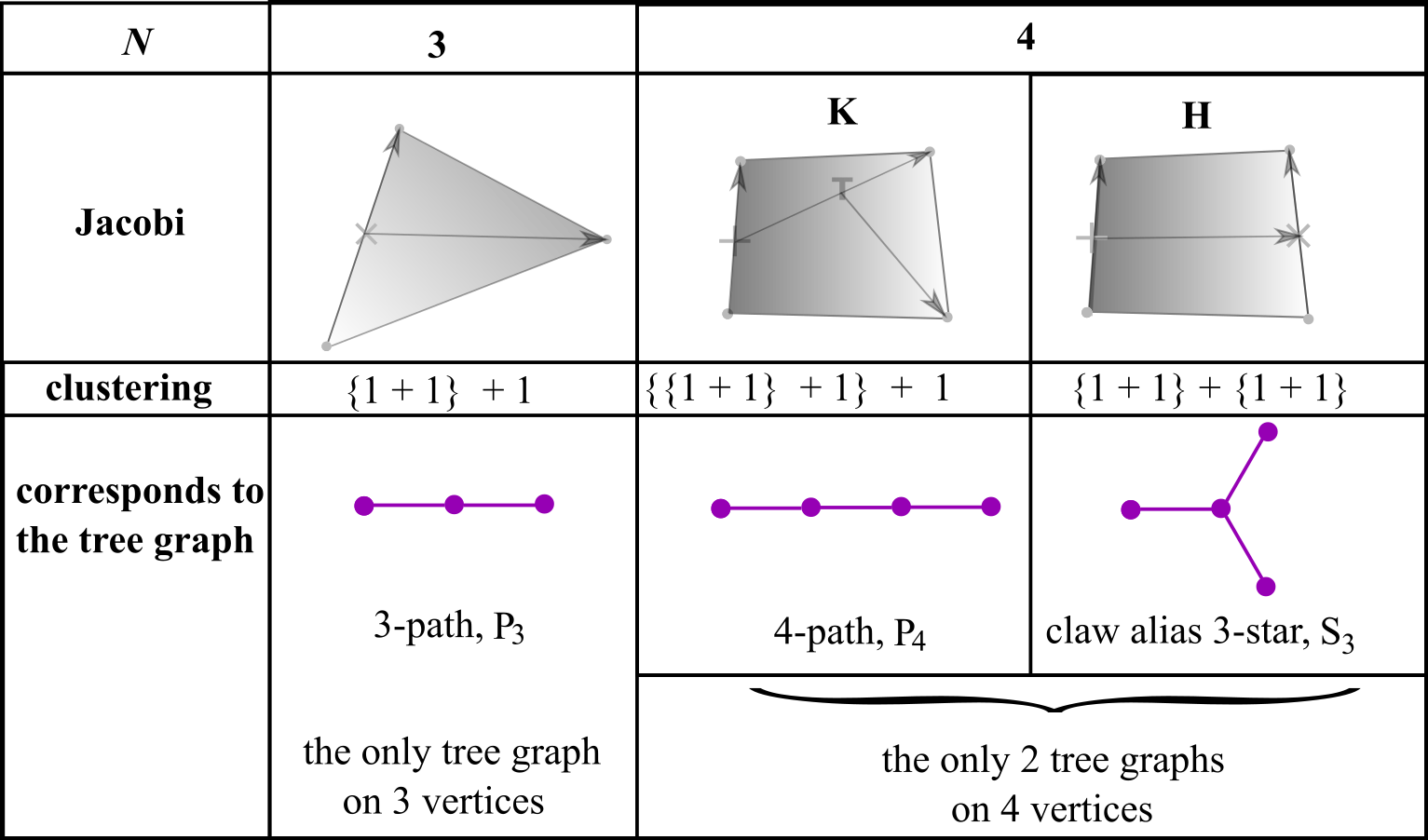}
\caption[Text der im Bilderverzeichnis auftaucht]{        \footnotesize{Correspondence between Jacobi coordinate clusterings and tree graphs.  
} }
\l{Jacobi-Trees}\end{figure}            }

\m 

\n Thus, in summary,  
\be
N = 4 \m  \mbox{ is minimal for exhibiting topological cluster hierarchy ambiguity }
\ee
since 
\be
N = 4 \m  \mbox{ vertices is minimal for tree graph nonuniqueness } \m  . 
\ee
\n There are moreover 3 and 6 label choices for H and K clusters respectively.  

\m 

\n Papers II and IV of this series use the following choice of labelling of $H$ clustering: 
\beq
\uiR_1   =   \uiq_2 - \uiq_1                                                                    \mma
\uiR_2   =   \uiq_4 - \uiq_3                                                                    \mma
\uiR_3  \es  \frac{m_3\uiq_3 + m_4\uiq_4}{m_3 + m_4} - \frac{m_1\uiq_1 + m_2\uiq_2}{m_1 + m_2}   \m  . 
\eeq
The corresponding cluster masses are 
\beq
\mu_1 \es  \frac{m_1 m_2}{m_1 + m_2}                                              \mma
\mu_2 \es  \frac{m_3m_4}{m_3 + m_4}                                                \m    \mbox{ and } \m  
\mu_3 \es  \frac{\{m_1 + m_2\}\{m_3 + m_4\}}{m_1 + m_2 + m_3 + m_4}                \m  .  
\eeq
For the case of equal masses used in the current series, the above simplify to 
\beq
\uiR_1  =   \uiq_2 - \uiq_1                                                       \mma
\uiR_2  =   \uiq_4 - \uiq_3                                                       \mma
\uiR_3 \es  \mbox{$\frac{1}{2}$} \{\uiq_3 + \uiq_4 - \uiq_1 - \uiq_2\}                           \m  , 
\eeq
\beq
\mu_1  \es  \mbox{$\frac{1}{2}$}                                                   \mma
\mu_2  \es  \mbox{$\frac{1}{2}$}                                                    \m  , \m 
\mu_3   =    1                                                                       \m  .  
\eeq
\n Papers II and IV also use the following choice of labelling of K clustering: 
\beq
\uiR_1  =   \uiq_2 - \uiq_1                                                       \mma
\uiR_2 \es  \uiq_3 - \frac{m_1\uiq_1 + m_2\uiq_2}{m_1 + m_2}                      \mma
\uiR_3 \es  \uiq_4 - \frac{m_1\uiq_1 + m_2\uiq_2 + m_3\uiq_3}{m_1 + m_2 + m_3}     \m  , 
\eeq
with corresponding cluster masses  
\beq
\mu_1 \es \frac{m_1 m_2}{m_1 + m_2}                                             \mma
\mu_2 \es  \frac{m_3\{m_1 + m_2\}}{m_1 + m_2 + m_3}                               \m  \mbox{ and } \m 
\mu_3 \es \frac{m_4\{m_1 + m_2 + m_3\}}{m_1 + m_2 + m_3 + m_4}                   \m  .  
\eeq
For the case of equal masses used in the current series, these simplify to 
\beq
\uiR_1  =   \uiq_2 - \uiq_1                                                      \mma
\uiR_2 \es  \uiq_3 - \mbox{$\frac{1}{2}$} \{\uiq_1 + \uiq_2\}                    \mma
\uiR_3 \es  \uiq_4 - \mbox{$\frac{1}{3}$}\{\uiq_1 + \uiq_2 + \uiq_3\}             \m  ,  
\eeq
\beq
\mu_1 \es  \mbox{$\frac{1}{2}$}                                                                  \mma
\mu_2 \es  \mbox{$\frac{2}{3}$}                                                                   \m    \mbox{ and } \m 
\mu_3 \es  \mbox{$\frac{3}{4}$}                                                                   \m  .  
\eeq
{\bf Exercise 1} Derive the previous 8 equations by diagonalization.   

\m

\n{\bf Remark 5} In the relative Jacobi coordinatization of relative space, the configuration space metric indeed takes a diagonal form,  
\beq
m_{iajb} := \mu_i\delta_{ij}\delta_{ab} \m  . 
\eeq
The moment of inertia $\bigiota$ and the kinetic energy are thereby diagonal in relative Jacobi coordinates.  
We observe moreover that these quantities take the same form as in position coordinates, but with one vector's worth of coordinates less.  

\m 

\n{\bf Definition 4} Let us term this form-preserving passage -- 
from objects in position coordinates to their counterparts in relative Jacobi coordinates -- the {\it Jacobi map}. 

\m 

\n{\bf Remark 6} This corresponds to the removal of the $d$ degrees of freedom in the total CoM being a mapping between Euclidean spaces $\FrQ(N, d)$ 
and $\lFrr(N, d)$. 
Coordinatiazions of the latter which match the former's original-given diagonal form then exist as per Jacobi.  

\m 

\n{\bf Remark 7} All in all, for the metric quadric 
\be 
\bigiota :=  \sum_{I = 1}^N   m_I q^{I \, 2} 
          =  \sum_{i = 1}^n \mu_i R^{i \, 2}
\ee
where the first equality uses the Jacobi map and the second uses Definition 4. 

\m 

\n{\bf Remark 8} For all that relative Jacobi coordinates are more mathematically convenient than relative Lagrange coordinates,  
they remain defined relative to absolute axes $A$. 
This reflects that both of these are just different coordinatizations of the same relative space $\lFrr(N, d)$.  

\m 

\n{\bf Definition 5} We subsequently make considerable use of {\it mass-weighted relative Jacobi coordinates} 
\beq
\rho^{ia} := \sqrt{\mu_i}R^{i a} \m  .  
\l{rho-R}
\eeq      
\n{\bf Remark 9} The $\rho_i$ can furthermore be interpreted as
\be
\rho_i = \sqrt{I_i} \m  ,  
\l{rhoi-Ii}
\ee
where the $I_i$ are {\it partial moments of inertia}, i.e.\ quantities summing up to form 
\be
\sum_{i = 1}^n I_i  = I \m  : 
\ee 
the total moment of inertia. 

\m 

\n{\bf Remark 10} Using the $\rho_i$ further simplifies the configuration space metric to a unit array with components 
\beq
\delta_{iajb} = \delta_{ij}\delta_{ab} \m  .
\eeq  
Thus also 
\be
\bigiota = \sum_{i = 1}^n {\rho_i}^{2} \m  . 
\ee
\n{\bf Definition 6} Let us use the term {\it mass-weighted relative space} to mean the mass-weighted relative Jacobi coordinate space, 
i.e.\ the space spanned by the $\rho_i$.  
[This gains significant distinction from relative space once {\sl length ratios} enter consideration in Sec 5.1.]

\subsection{Democracy, and the Lagrangian--Jacobian distinction}

\n{\bf Definition 1} The existence of multiple choices of clustering furthemore motivates considering linear transformations between different clusterings,  
termed `{\it democracy transformations}' in the Molecular Physics literature \c{Zick-1, Zick-2, Zick-3, ACG86, LR95}.  

\m  

\n{\bf Remark 2} N.B.\ that this refers to  `democratic treatment' of all possible clustering choices, 
as opposed to Sec \r{Rel-Lag}'s mention of `democratic treatment', which amounts to considering all possible Lagrange basis choices.    

\m 

\n{\bf Structure 1} One can usually add, or average, to form a democratic version of an a priori undemocratic quantity; 
the exception is when the democratic version just returns zero, as occurs in Sec III.15.14.  

\m 

\n{\bf Remark 3} Let us finally distinguish between 
{\it Lagrangian level notions}, which are built from Lagrange particle separations $\u{r}_{IJ}$, and 
{\it Jacobian level   notions}, which involve relative Jacobi inter-particle cluster separations $\u{\rho}_i$ other than just their Lagrangian subset.
As a first application, we can reassess the above two uses of `democratic' as {\it Lagrangian democracy} of relative separations 
                                                                           to {\it Jacobian democracy}   free from clustering choices.

\section{Mirror image identification and indistinguishability}\l{MII-Indist}
													 
\subsection{Configurations}\l{MII-Indist-Config}
%
{            \begin{figure}[!ht]
\centering
\includegraphics[width=0.6\textwidth]{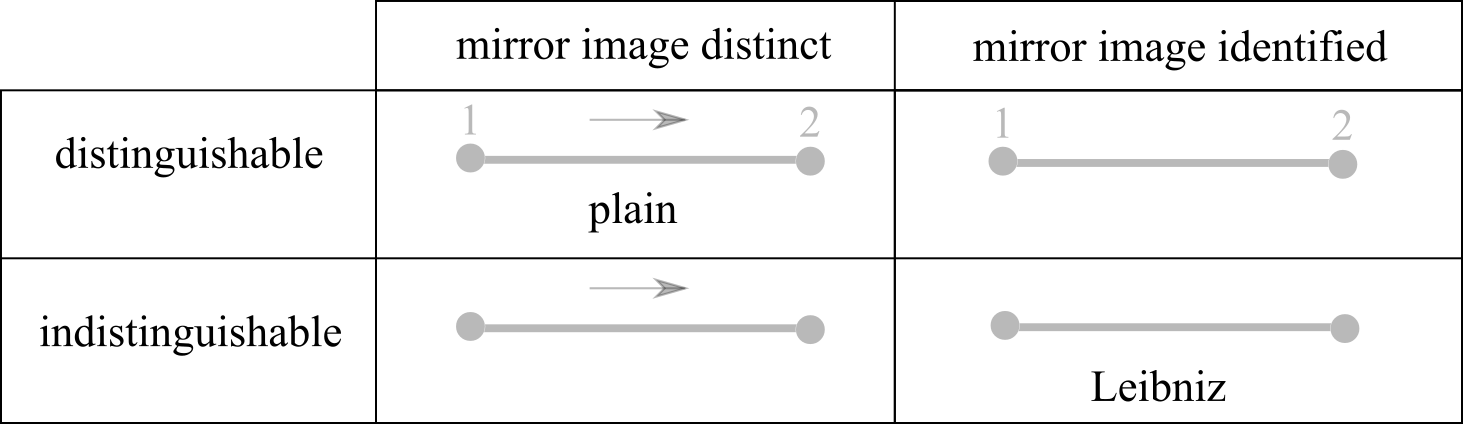}
\caption[Text der im Bilderverzeichnis auftaucht]{        \footnotesize{Mirror image identified and label distinguishability notation for configurations.} }
\l{PMIL} \end{figure}          }

\n We next incorporate two modelling ambiguities of note in considering constellations.      

\m 

\n{\bf Modelling Ambiguity 1)} is whether mirror image configurations are to be held to be distinct. 

\m 

\n{\bf Modelling Ambiguity 2)} is that a constellation's constituent points-or-particles can furthermore be equipped with distinguishable labels. 
In the absense of such, the points are indistinguishable. 
If all the labels are distinct, one has total distinguishability.
Partial distinguishability is also possible, though the current treatise does not consider this beyond the current section; see \cite{A-Monopoles} for further development.    

\m  

\n{\bf Remark 1} See Figure \r{PMIL} for the depictional convention used each of the ensuing conceptual types of configuration.

\m 

\n{\bf Remark 2} Moreover, on the one hand, 
\be
N < d \m \mbox{ is required for there to exist the option to treat mirror images as distinct }  
\l{N = d}
\ee 
of Ambiguity 1.
For if this does not hold, the configuration is contained in a codimension-1 hyperplane, which can be rotated through the remaining dimension into its mirror image.  

\m 

\n{\bf Remark 3} On the other hand, Realizing Ambiguity 2) requires 
\be 
N \geqs 2 \mbox{ is required for distinguishable labels to be meaningful } , 
\ee
as per Ambiguity 2). 
This is since for single-particle models whether or not one labels it does not change the nature of the configuration.  
Finally, 
\be
N \geqs 3 \mbox{ is required for partially distinguishable labels to be meaningful } , 
\ee
and
\be
\mbox{(number of types of distinguishability supported by $N$)} \es  (  \m  \mbox{partition number, $p(N)$} \m  ) \m  .  
\ee
\n These observations combine to give the counts of numbers of disinctly realized concepts of configuration in Fig \r{q-types}.
%
{            \begin{figure}[!ht]
\centering
\includegraphics[width=0.75\textwidth]{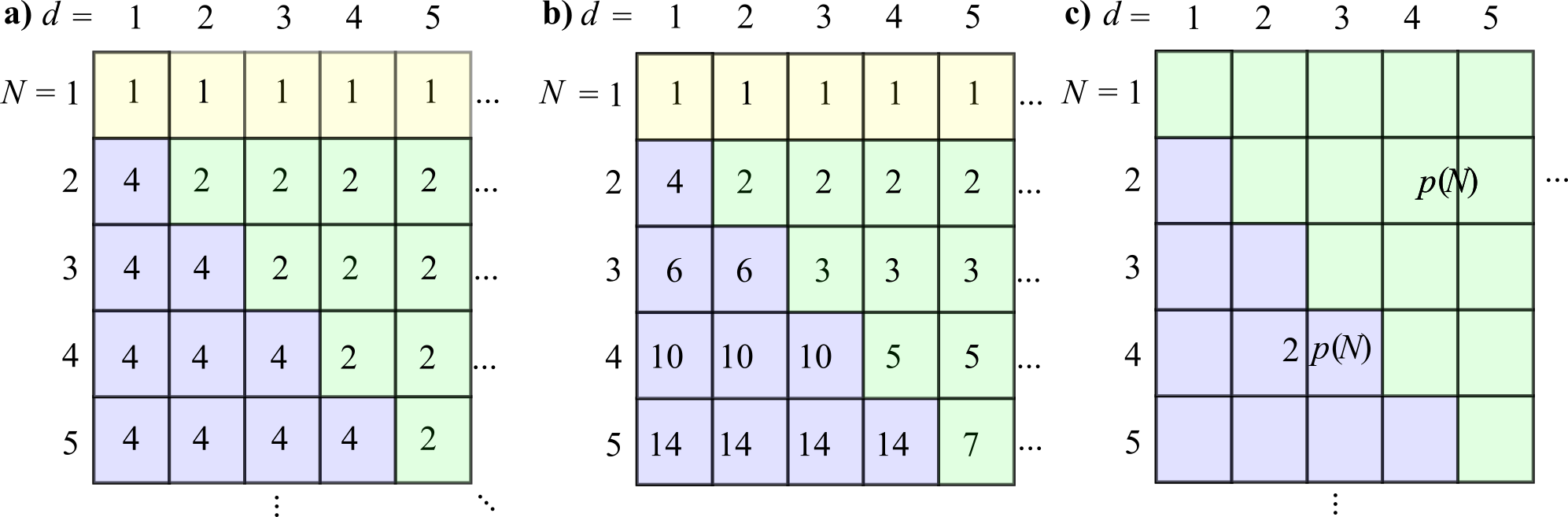}
\caption[Text der im Bilderverzeichnis auftaucht]{\footnotesize{Number of concepts of configuration model by model, 
taking into account mirror image identification alongside  
a) indistinguishability versus total distinguishability.  
b) Including partial distinguishability as well.
Finally c) establishes a reduction of b) to a more standard mathematical problem.} }
\l{q-types} \end{figure}          }
						
\m 

\n{\bf Remark 4} Distinguishable particles cover the case of generic masses or, via some further unspecified mysterious classical labels, the case of equal masses. 
Particle distinguishability of the kind discussed here also plays a major role in Quantum Theory. 
Within each particle type, at the quantum level it is a {\it principle} that they are all the same; e.g.\ all electrons are the same. 
This is a postulation that {\sl meaningless} distinguishing labels -- ennumerations `painted onto' particles -- 
are a fortiori {\sl nonexistent} at the quantum level. 
This is clearly another -- now explicitly quantum -- realization of Leibniz's Identity of Indiscernibles.\footnote{Meaninglessness 
of gauges          in Particle Physics and 
of diffeomorphisms in GR                   are two further realizations. 
So Leibniz's Indentity of Indiscernibles can be considered to underpin quite a number of key aspects of modern Theoretical Physics!}  
of nonexistence of meaningless distinguishing labels at the quantum level.  
This further motivates the study of indistinguishable particle configuration spaces.

\subsection{The corresponding configuration spaces}\l{Mirr-Dist-Leibniz}
%
{            \begin{figure}[!ht]
\centering
\includegraphics[width=0.35\textwidth]{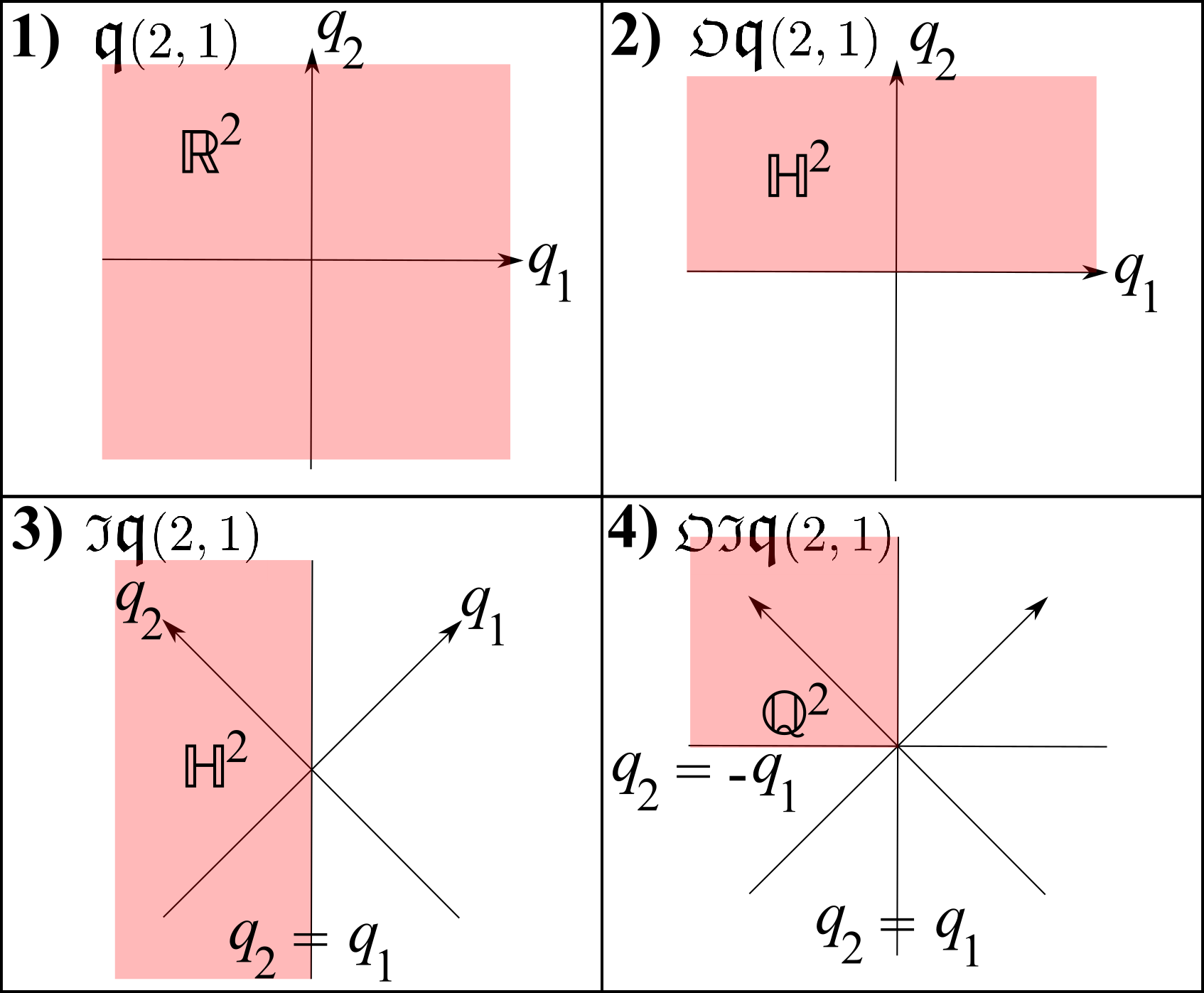}
\caption[Text der im Bilderverzeichnis auftaucht]{        \footnotesize{The (2, 1) model's four configuration spaces} }
\l{4-Q} \end{figure}          }

\m 

\n {\bf Definition 1} Let us denote the configuration space of mirror image identified   distinguishable particle configurations by $\FrO\FrQ(N, d)$, 
											        that of mirror image distinct   indistinguishable particle configurations by $\FrI\FrQ(N, d)$, 
											    and that of mirror image identified indistinguishable particle configurations by $\FrO\FrI\FrQ(N, d)$.
We also have occasion to use $\FrG\FrQ$ to denote the general such configuration space, taken in this series of paper to mean whichever of the previous four.  
We shall use this prefix notation with more reduced configuration spaces in the role of $\FrQ$ as well. 

\m 

\n{\bf Remark 1} 
\be 
\FrO\FrQ(N, d)     \es  \frac{\FrQ}{\mathbb{Z}_{2}} \m  ,
\ee
\be 
\FrI\FrQ(N, d)     \es  \frac{\FrQ}{S_{N}} \m  , 
\ee
\be
\FrO\FrI\FrQ(N, d) \es  \frac{\FrQ}{\mathbb{Z}_{2} \times S_{N}} \m  .
\ee 
The       $\mathbb{Z}_{2}$    group of order 2    involved has reflective action on space,                thus encoding mirror image identification, 
while the $S_{N}$ permutation group of order $N!$          has the obvious action on the particle labels, thus encoding indistinguishability.

\m 

\n{\bf Remark 2} Thus straightforwardly 
\be 
\FrO\FrQ(N, d)     \es  \frac{\mathbb{R}^{Nd}}{\mathbb{Z}_{2}} \es  \mathbb{H}^{Nd}         \m    
\ee
-- half space -- whereas 
\be 
\FrI\FrQ(N, d)     \es   \frac{1}{N \, !}\mbox{-}\left(  \mbox{wedge of } \right) \m \m \mathbb{R}^{Nd}  \m  , 
\ee
\be
\FrO\FrI\FrQ(N, d) \es   \frac{1}{2 \times N \, !}\mbox{-}\left(  \mbox{wedge of } \right) \m \m \mathbb{R}^{Nd}  \m  .
\ee 
Note that the first of these is a complete characterization, whereas the other two benefit from the following characterization with simple examples.  

\m 

\n Example 1) For (2, 1), the four possibilities are realized as in Fig \r{4-Q}. 
This indicates that  
\be 
\FrQ(2, 1) \cong \FrI\FrQ(2, 1)
\ee
are half-spaces $\mathbb{H}^2$ but with unaligned realizations (c.f.\ where the axes point), 
so that the cumularive effect of the corresponding quotients is that $\FrO\FrI\FrQ(2, 1)$ is a quadrant $\mathbb{Q}^2$. 

\m 

\n{\bf Example 2} $\FrI\FrQ(3, 1)$ is determined by three planes at $2\pi/3$ to each other which coincide solely at the origin 0.  

\m 

\n{\bf Remark 3} In the current treatise, 
we adopt the convention of mirror images being vertically on top of each other in mirror image distinct configuration spaces, 
with mirror image identified configuration spaces represented as upper half spaces.

\m 

\n{\bf Remark 4} Observe also the presence of a horizontal hyperplane of reflection-symmetric shapes; in the above convention, this is the horizontal hyperplane.  
These configurations are already present just once in the mirror image distinct case.

\subsection{Relative configuration space versions}
%
{            \begin{figure}[!ht]
\centering
\includegraphics[width=0.75\textwidth]{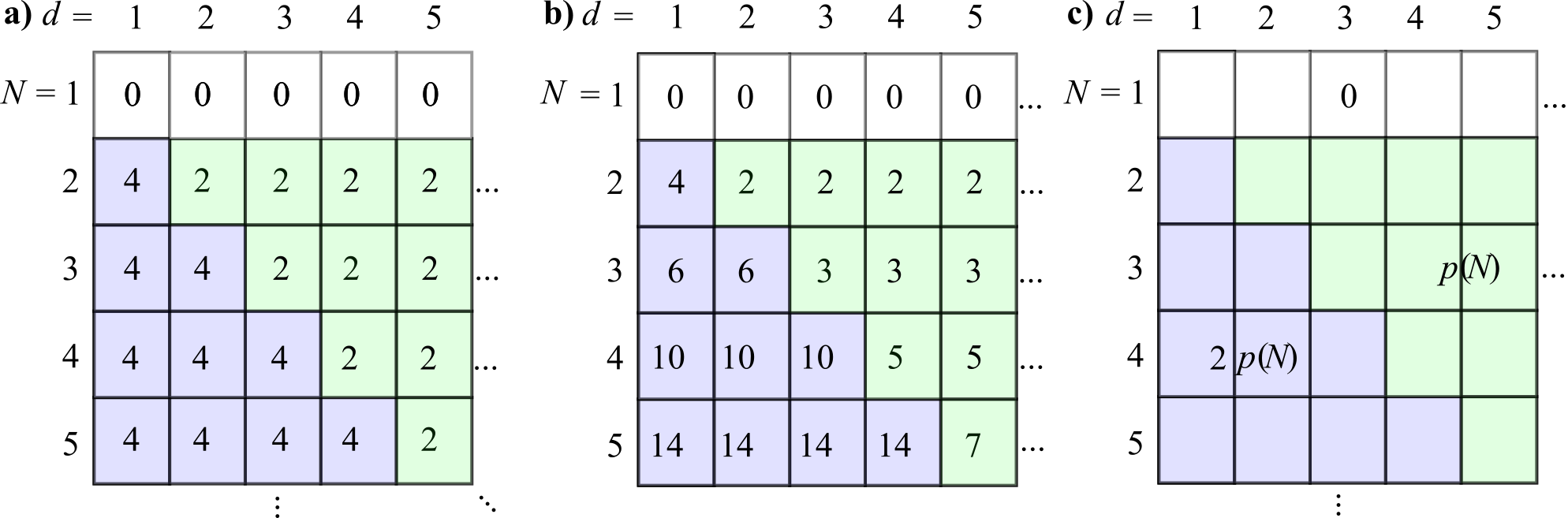}
\caption[Text der im Bilderverzeichnis auftaucht]{\footnotesize{Relative space counterpart of Fig \r{q-types}.  } }
\l{r-types} \end{figure}          }
%
We now also need $N \geqs 2$ for nontriviality, so the number of conceptual types of configuration is amended as per Fig \r{r-types}.  
The condition 
\be 
n = d \mbox{  } \mbox{ for linear independence saturation of relative $d$-vectors } , 
\l{LI-Sat}
\ee
is realized diagonally within the lattice of ($N$, $d$) values. 
So on the one hand, for the diagonal and lower-triangle parts of this array,  
\be
n \leqs d \m  , 
\ee
mirror image identification is optional; on the other hand, for the upper-triangle part, 
\be
n \ls d \m  , 
\ee
it is obligatory. 

\m 

\n{\bf Remark 1} Now
\be
\Frr(N, d) = \mathbb{R}^{Nd} \m  , 
\ee
\be 
\FrO\Frr(N, d)         \es  \frac{\Frr(N, d)}{\mathbb{Z}_{2}} 
                       \es  \frac{\mathbb{R}^{nd}}{\mathbb{Z}_{2}} 
					   \es  \mathbb{H}^{nd}                        \m  ,
\ee
\be 
\FrI\Frr(N, d) \es  \frac{\Frr(N, d)}{S_{N}}  
               \es  \frac{\mathbb{R}^{nd}}{S_{N}} 
               \es  \frac{1}{N \, !}\mbox{-}\left(  \mbox{wedge of } \right) \m \m  \mathbb{R}^{nd} \m  , 
\ee
\be
\FrO\FrI\Frr(N, d) \es  \frac{\Frr(N, d)}{\mathbb{Z}_{2} \times S_{N}} 
                   \es  \frac{\mathbb{R}^{nd}}{\mathbb{Z}_{2} \times S_{N}}
                   \es  \frac{1}{2 \times N \, !}\mbox{-}\left(  \mbox{wedge of } \right) \m \m \mathbb{R}^{Nd}  \m  .
\ee

\section{Introducing some furtherly reduced configuration spaces}\l{Preamble-2}

\subsection{The further groups involved, with corresponding types of invariants}

The only reduction we have considered so far involved quotienting out the translations $Tr(d) = \mathbb{R}^d$.
The corresponding invariants are functions of differences $\uq^{J} - \uq^{I} = \ur^{IJ}$, or, under basis change, of $\urho^i$.
This case's quotient is moreover just a simple map between flat Euclidean configuration spaces, for which Linear Algebra suffices to handle all matters.  

\m 

\n{\bf Remark 1} Consider instead just the dilations $Dil$.
Since these transformations just cover the diversity of possible (global) scale factors, we identify this group to be    
\be
Dil = \mathbb{R}_+ \m  , 
\ee 
which is indeed independent of the dimension $d$.
The corresponding invariants are functions of ratios, 
\be
f
\left(
\frac{q^{Ia}}{q^{Jb}}
\right)                            \m  , 
\ee
where the numerator and denominator are individual components. 

\m 

\n{\bf Remark 2} Consider instead just the rotations $Rot(d)$, whose mathematical nature as a group is 
\be
Rot(d) = SO(d) \m  : 
\ee
the {\it special orthogonal group}. 
The corresponding invariants are functions of dot products, 
\be
f(\uq^I \cdot \uq^J) \m  .  
\ee
Dot products moreover include both squared norms, giving the squares of what are now {\it separation lengths}  
\be 
||\uq^I|| \m  , 
\ee
and {\it angle} information via 
\be
(\uq^I \cdot \uq^J) = ||\uq^I|| \, ||\uq^J|| \, \mbox{cos} \, \omega \m  , 
\ee
which inverts to 
\be
\omega \es  \mbox{arccos}
\left(
\frac{(\uq^I \cdot\uq^J)}{||\uq^I|| \, ||\uq^J||} 
\right)                                            \m  . 
\ee
\n{\bf Remark 3} We are furthermore interested in quotienting out {\it all combinations} of the above three types of transformation. 
These compose into the following transformation groups.  

\m 

\n 1) The rotations and dilations combine in the particularly simple {\it direct product} manner, 
\be
Rot(d) \times Dil = SO(d) \times \mathbb{R}_+ \m  .   
\ee
\n 2) The {\it dilatations} consist of the translations alongside the dilations, 
composing in a slightly more complicated mathematical form as the {\it dilatational group} 
\be
Dilatat(d) = Tr(d) \rtimes Dil 
           = \mathbb{R}^d \rtimes \mathbb{R}_+ \m  , 
\ee 
where $\rtimes$ denotes semidirect product \c{Cohn}. 

\m 

\n 3) The (continuous) {\it Euclidean transformations} consist of the translations alongside the rotations, composing as the {\it Euclidean group}
\be
Eucl(d) = Tr(d) \rtimes Rot(d) = \mathbb{R}^d \rtimes SO(d) \m  .  
\ee 
We know that these are precisely the right continuous transformations to consider for Euclidean geometry, 
not only by common experience of flat-space isometries but also because they are the solutions of Appendix B's Killing equation in the case of flat space.
\n 4) Finally putting all of these together, the (continuous) {\it similarity transformations} consist of the translations, rotations and dilations, 
composing as the {\it silimarity group} 
\be
Sim(d) = Tr(d) \rtimes \{ Rot(d) \times Dil \} 
       = \mathbb{R}^d \rtimes \{ SO(d) \times \mathbb{R}_+ \} \m  .  
\ee
Again, we know that these are precisely the right continuous transformations to consider for similarity geometry, 
not only by common experience of flat-space similarites but also because they are the solutions of Appendix B's 
similarity Killing equation in the case of flat space.
%

\m 

\n{\bf Proposition 1} The corresponding types of invariants turn out to {\it functionally compose}.
I.e.\ 

\m  

\n 1) $Rot(d) \times Dil$ invariants are functions of ratios of dots, 
\be
f
\left(
\frac{(\uq^I \cdot \uq^J)}{(\uq^K \cdot \uq^L)}
\right)                                             \m  , 
\ee
\n 2) Dilatational invariants are functions of ratios of differences, 
\be
f
\left(
\frac{\uq^J - \uq^I}{\uq^L - \uq^K}
\right)                                                              \m  , 
\ee
which can be re-expressed as functions of ratios of linear combinations of `separation components $r^{IJa}$ and self-component differences $q^{Ja} - q^{Jb}$', 
                                                        or equivalently of `$\rho^{ia}$ and $q^{Ja} - q^{Jb}$'.

\m 														
														
\n 3) $Eucl(d)$ invariants are functions of dots of differences, 
\be
f
\left( \{\uq^J - \uq^I\} \cdot \{\uq^L - \uq^K \}  
\right)                                             \es   
f
\left(
\ur^{IJ} \cdot \ur^{KL}
\right)                                             \es                   
f
\left(
\urho^i \cdot \urho^j
\right)                                                  \m  .
\ee
\n{\bf Example 1} An important special case among these are the {\it relative Lagrange separations}, 
\be
||\ur_{IJ}||  \m  .  
\ee
\n 4) Finally, $Sim(d)$ invariants are functions of ratios of dots of differences, 
\be
f
\left( 
\frac{    \{\uq^J - \uq^I\} \cdot \{\uq^L - \uq^K \}    }{    \{\uq^N - \uq^M\} \cdot \{\uq^P - \uq^O \}    }  
\right)                                                                                                       \es 
f
\left(
\frac{\ur^{IJ} \cdot \ur^{KL}}{\ur^{MN} \cdot \ur^{OP}}                                                  
\right)                                                                                                       \es 
f
\left(
\frac{\urho^i \cdot \urho^j}{\urho^k \cdot \urho^l}                                                                   
\right)                                                                                                             \m  .  
\ee
\Proof Much as Remark 3 is underpinned by the (generalized) Killing equation, this result is undepinned by the {\it configurational observables equation} 
(although this is less well-known, so we give a reference: \c{ABeables2}). $\Box$

\m 

\n{\bf Example 2} A special case which plays a substantial role in this treatise is the {\it ratio of relative separations}, 
\be
\frac{||\ur_{IJ}||}{||\ur_{KL}||} \m  . 
\ee

\m 

\n{\bf Remark 3} These invariants serve    moreover as each corresponding Relational Particle Mechanics' (RPM) potential functional dependence \c{FileR}, 
                                                                                              observables \c{ABeables, ABeables2}, 
																				 and end up being its quantum wavefunction dependencies as well \c{FileR}. 

\m 

\n{\bf Remark 4} While the above give a total of eight groups to consider, the just $Dil$               invariant, 
                                                                                just $Rot(d)$            invariant,  
															                and just $Rot(d) \times Dil$ invariant cases are of somewhat less interest. 
This is due to absolute origin -- and origins of other interpretations of carrier spaces -- quite often being meaningless, 
and due to quotienting out translations not affecting the geometry within the `subgroups of the affine group' family of geometries.\f{Statistical location data 
may moreover have a privileged origin; see \c{AMech, Project-1} for a start on the theory of these other cases.}
%
Thus we just consider quotienting out $Tr(d)$ first, followed by quotienting out $Rot(d)$ and $Dil$ in either order.
This gives the subgroup lattice of   Fig \r{Group-Inv-Config}.a) 
with the corresponding invariants in Fig \r{Group-Inv-Config}.b) 
and the configuration space in       Fig \r{Group-Inv-Config}.c).
We already dedicated Sec 2 and Sec 3 to the first two of these; the rest of the current section serves to introduce the other three.  
%
{            \begin{figure}[!ht]
\centering
\includegraphics[width=0.8\textwidth]{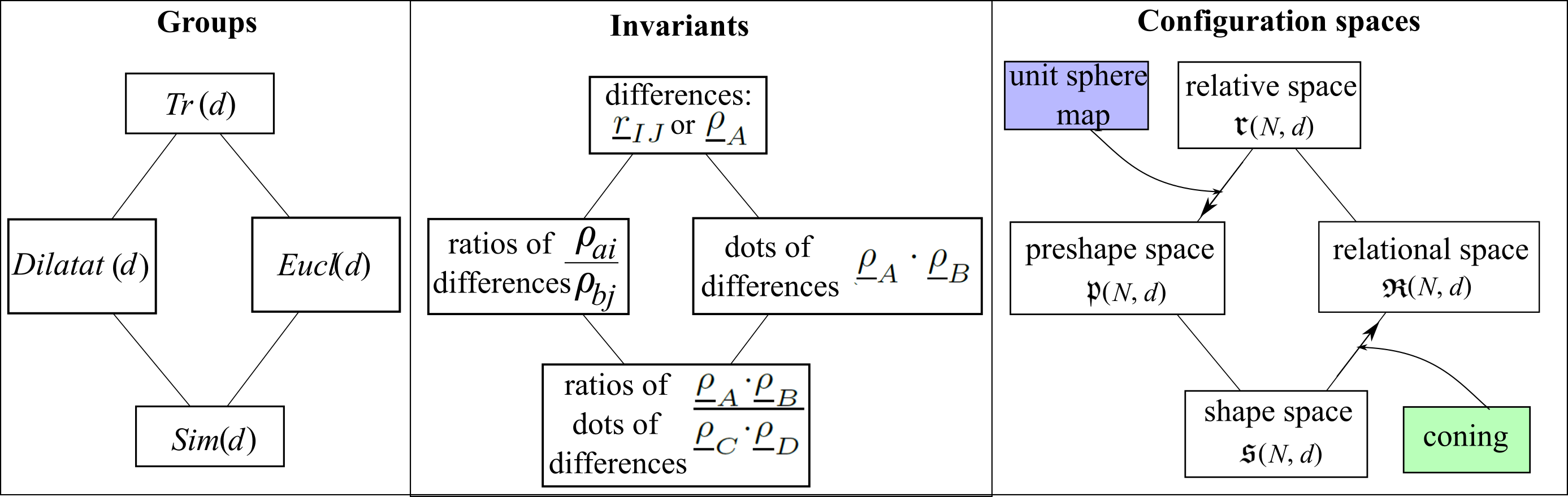}
\caption[Text der im Bilderverzeichnis auftaucht]{        \footnotesize{a) A lattice of four subgroups $\nFrg$ of the similarity group $Sim(d)$, 
                                                                   with b) corresponding invariant quantities 
																    and c) reduced configuration space names and symbols.} }
\l{Group-Inv-Config} \end{figure}          }

\subsection{Preshape, shape and relational spaces}\l{Preamble-1}

{\bf Definition 1} Firstly suppose that carrier space scale $S$ is to join the carrier space origin $0$ in having no meaning.
This corresponds to quotienting the incipient configuration space $\FrQ(N, d)$ by the dilatational group $Dilatat(d)$, thus forming Kendall's \c{Kendall}
\be
(\mbox{\it preshape space}) \mma  \FrP(N, d) \:=  \frac{  \FrQ(N, d)  }{  Tr(d) \times Dil  }  
                                             \es  \frac{  \lFrr(N, d)  }{  Dil  } 
                                             \es  \frac{  \mathbb{R}^{nd}  }{  \mathbb{R}_+  } \m  .
\l{Shape-Space}
\ee
\n{\bf Definition 2} Next suppose instead that the carrier space axes $A$ is to join the carrier space origin $0$ in having no meaning.
This corresponds to quotienting $\FrQ(N, d)$ by the Euclidean group $Eucl(d)$, thus forming the 
\be
(\mbox{\it relational space}) \mma    \m  {\cal R}(N, d)  \:=  \frac{\FrQ(N, d)}{Eucl(d)} 
                                                             \es  \frac{\lFrr(N, d)}{Rot(d)} 
	    			                                         \es  \frac{\mathbb{R}^{nd}}{SO(d)} \m  .
\ee
of the relational side of the noted absolute versus relational debate. 

\m 

\n{\bf Definition 3} Finally, suppose that all three of $S$, $A$ and $0$ are to have no meaning.
This corresponds to quotienting $\FrQ(N, d)$ by the similarity group $Sim(d)$, thus forming Kendall's \c{Kendall84, Kendall} 
\be
(\mbox{\it shape space}) \mma    \m  \FrS(N, d) \:=  \frac{\FrQ(N, d)}{Sim(d)} 
                                                    \es  \frac{\lFrr(N, d)}{Rot(d) \times Dil} 
 			                                        \es  \frac{\FrP(N, d)}{Rot(d)} 
 			                                        \es  \frac{\FrP(N, d)}{SO(d)}               \m  .
\ee
Given subsequent formulations of affine          \c{PE16}, 
                                 conformal       \c{AMech}, 
								 projective      \c{PE16, KKH16} 
							 and supersymmmetric \c{AMech} shape theories 
warrants further qualifying this name as {\it metric shape space} alias {\it similarity shape space}.  
 
\m  
 
\n{\bf Remark 1} Sec \r{Preshape}'s topological and geometrical considerations demonstrate that 
preshape space is a useful halfway house en route to Kendall's shape space.  
Kendall's shape space is in turn a useful intermediary en route to relational space.  

\m 

\n{\bf Remark 2} In the traditional setting of the aforementioned debate, one is to furthermore bring in velocities, momenta and force laws, 
and then also formulate everything without reference to Newtonian absolute time either \c{BB82, FileR, ABook}.  
The traditional premises of this debate can be addressed by modelling precisely such a space additionally in the absense of Newtonian absolute time.

\subsection{Configuration space dimension counting}\l{Dimension}

\n{\bf Remark 1}
\be
\mbox{dim}(Rot(d)) = \mbox{dim}(SO(d)) 
                   \es  \frac{d\{d - 1\}}{2}                                 \m  \mbox{ and }
\ee
\be
\mbox{dim}(Dil) = \mbox{dim}(\mathbb{R}_+) 
                = 1                                                       \m  .
\ee
Thus, using also (\r{dim-Tr}),  
\be
\mbox{dim}(Eucl(d)) = \mbox{dim}(Tr(d) \rtimes Rot(d)) 
                    \es  d + \frac{d\{d - 1\}}{2} \es  \frac{d\{d +  1\}}{2}                    \m  \mbox{ and }
\ee
\be
\mbox{dim}(Sim(d))   =   \mbox{dim}(Tr(d) \rtimes \{ Rot(d) \times Dil \}) 
                    \es  d + \frac{d\{d - 1\}}{2} + 1 
					\es  \frac{d\{d +  1\}}{2} + 1                                   \m  .  
\ee
\n{\bf Remark 2} 
\be
\mbox{dim}\left(\frac{A}{B}\right) = \mbox{dim}(A) - \mbox{dim}(B) \m  , 
\ee
counts the dimension of a quotient; moreover 
\be
\frac{A}{B} \m  \mbox{ need not be of a single dimension, even if $A$ and $B$ are of a single dimension } \m  . 
\ee
so this may at most refer just to the top stratum. 
See the Conclusions of Papers I, II, III, and IV for more on caveats concerning stratification. 

\m  

\n{\bf Remark 3} Consequently
\be
\mbox{dim}(\FrP(N, d)) \es  \mbox{dim}\left(  \frac{    \lFrr(N, d)    }{     Dil       }  \right) 
                       \es  \mbox{dim}\left(  \frac{ \mathbb{R}^{nd}  }{ \mathbb{R}_+  }  \right) 
				       \es  \mbox{dim}(  \mathbb{R}^{nd}  ) - \mbox{dim}(  \mathbb{R}_+  )
                       \es  nd - 1                                                              \m  , 
\ee
\be
\mbox{dim}(\FrS(N, d)) \es  \mbox{dim}\left(  \frac{\FrQ(N, d)}{Sim(d)}  \right) 
                       \es  \mbox{dim}(\mathbb{R}^{Nd})   - \mbox{dim}(Sim(d)) 
			           \es  N d - \frac{d\{d + 1\}}{2}    - 1 
		               \es  \frac{d\{2 \, n + 1 - d\}}{2} - 1                                              \mma\mbox{ and }
\ee
\be
\mbox{dim}({\cal R}(N, d)) \es  \mbox{dim}\left(  \frac{\FrQ(N, d)}{Eucl(d)} \right) 
                           \es  \frac{\lFrr(N, d)}{Rot(d)} 
						   \es  \mbox{dim}(\mathbb{R}^{Nd}) - \mbox{dim}(Eucl(d)) 
					       \es  Nd - \frac{d\{d + 1\}}{2} 
						   \es  \frac{d\{2 \, n + 1 - d\}}{2}                                                 \m  .						   
\ee
Note that none of these dimension counts have a `simple product' form, unlike for $\FrQ(N, d)$ or $\lFrr(N, d)$.
Also note in particular that, in dimensions 1, 2 and 3, $\mbox{dim}({\cal R}(N, d)) = N - 1$, $2 \, N - 3$ and $3 \, N - 6$ respectively, 
                                                  while $\mbox{dim}(\FrS(N, d))     = N - 2$, $2 \, N - 4$ and $3 \, N - 7$ respectively.

\m 

\n {\bf Remark 4} Finally, quotienting out discrete transformations such as -- reflections or permutations --  
does not affect the configuration space dimension count (these amount to taking a same-dimensional portion).
Thus, because the current section only considers counts, we postpone introduction of these discrete elements of diversity until Sec \r{MII-Indis}.  

\m

{            \begin{figure}[!ht]
\centering
\includegraphics[width=1.0\textwidth]{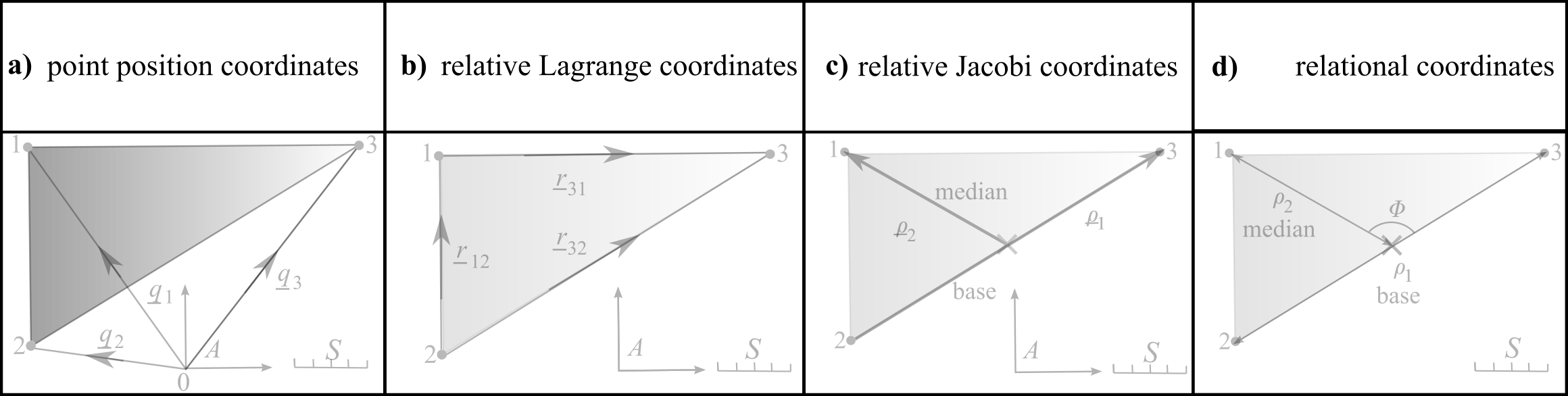}
\caption[Text der im Bilderverzeichnis auftaucht]{        \footnotesize{a) Point particle position coordinates.
b) Relative Lagrange separation vectors.   
c) Relative Jacobi separation vectors; the cross denotes the centre of mass of particles 2 and 3. 
d) Coordinates not depending on the absolute axes either: Jacobi magnitudes and the angle between them. 
To finally not depend on the scale, take the ratio $\rho_2/\rho_1$ of the Jacobi magnitudes alongside this angle. 
On the shape sphere, moreover, $\Phi$ plays the role of polar angle and the arctan of this ratio plays the role of azimuthal angle.} }
\l{Jac-Med-Ineq-Fig-2}\end{figure}            }

\n{\bf Structure 1} Let us now consider our triangle's vertices to be equal-mass particles with position vectors $\uq_I$ ($I = 1$ to $3$) 
relative to an absolute origin 0 and axes $A$ (Fig \r{Jac-Med-Ineq-Fig-2}.a).

\subsection{Triviality criteria}\l{Triviality-Criteria}

\n Some triviality criteria which are crucial in selecting smallest nontrivial models are as follows.

\m 

\n{\bf Definition 1} {\it Dynamical triviality} is the (pre-)Newtonian criterion that at least one degree of freedom is required; 
this evolves in a given or absolute alias background time.

\m  

\n{\bf Definition 2} {\it Relational triviality} is that at least two degrees of freedom are required so that one evolves with respect to the other 
(in the absense of an absolute or background time: a Temporal Relationalism criterion, as explained in Sec \r{Gdyn}).  

\m  

\n Let us next regard a group $G$ of spatial (or more generally configurational) transformations as physically irrelevant: 
a Configurational Relationalism criterion.  
Common instances of this are regarding the Euclidean group of translations and rotations as physically irrelevant, 
and internal gauge groups such as $U(1)$ for electromagnetic gauge theories. 

\m  

\n{\bf Definition 3} {\it No full group action}. 
$G$ does not act fully on a) small enough particle number configurations or b) some nongeneric configurations.

\subsection{Smallest (Relationally) Nontrivial Examples}\l{Smallest-Nontrivial}
%
{            \begin{figure}[!ht]
\centering
\includegraphics[width=1.0\textwidth]{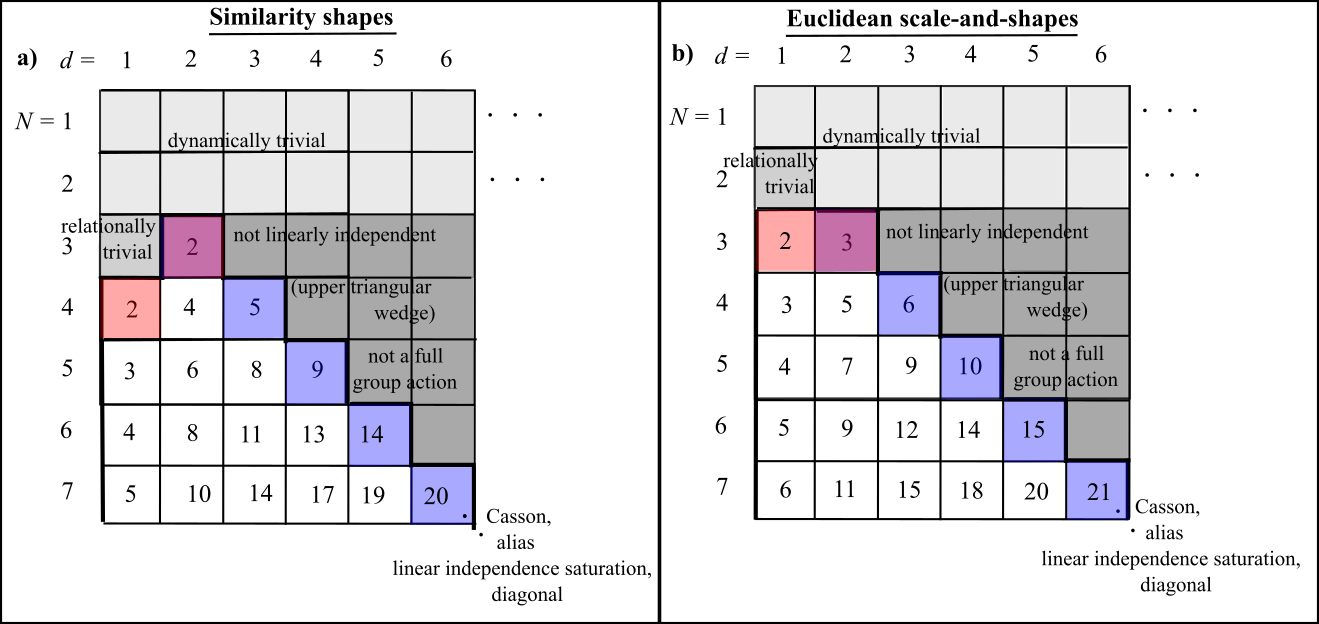}
\caption[Text der im Bilderverzeichnis auftaucht]{        \footnotesize{Overview of a) pure-shape and b) Metric Shape and Scale RPMs configuration space dimensions.
We mark in blue the hitherto so-called Casson diagonal \c{Kendall}, 
for which we now provide the conceptual name of {\it linear independence saturation diagonal} as per (\r{LI-Sat}).
Beyond this, in the upper triangular wedge, the counting entries on each row of the upper triangular block are equal to the corresponding diagonal entry.
We also mark particularly interesting minimal models in red; 
the (3, 2) case -- triangleland -- is doubly minimal through also lying on the diagonal, 
and thereby ends up marked in purple (= blue + red)).
This is an exceptionally simple model, though one needs to look beneath it in the figures to find the minimal 
{\it typical} 2-$d$ similarity and Euclidean model: the (4, 2) case of quadrilateralland.}}
\l{Shape-Triviality-Count}\end{figure}            }

\n Fig \r{Shape-Triviality-Count}.a)-b) tabulates configuration space dimension $k$, so as to display inconsistency, 
                                                                                                         triviality, 
																								     and relational triviality by shading.
See \c{AMech, ABook, Affine-Shape-1} for the corresponding counts -- and minimal relationally nontrivial units -- for a wider range of $G$.  
It is figures of this kind which first provide Fig \r{Aufbau}'s targets for the Shape-Theoretic Aufbau Principle is to build toward.   
For the moment, (3, 2) and (4, 3) lie on the critical linear independence saturation diagonal, 
and (4, 2) and (5, 3) are maximal in the lower triangular region preceding this diagonal.

\section{Notions of inhomogeneity and uniformity}\l{Inhom-Uni}																		   

This Section contains some new conceptual-level research, leading to quite a number of technical results for Part I to IV's four models.  

\m 

\n Let us begin by giving a general principle. 

\m 

\n{\bf Principle 1} Notions of inhomogeneity and uniformity are of the general conceptual forms
\be
(\mbox{Invariant}) = 0
\ee
and 
\be
(\mbox{Invariant}) = (\mbox{Invariant}^{\prime}) \m  ,  
\ee
[The Author abstracted this from consideration of the separation magnitude (ratio) cases in the next two sections. 
This principle moreover applies to many cases beyond the scope of the current treatise: see \c{Affine-Shape-2, Project-1} for further instalments.]

\subsection{The coincidence-or-collision equations}\l{Coll}

\n{\bf Definition 1} The {\it coincidence-or-collision equations} are 
\be 
||\ur_{IJ}|| = 0 \m  .  
\l{Collision-Eqs}
\ee
These are to be interpreted in the following manner.   
Any configuration solving at least one such equation is a coincidence-or-collision. 
Configurations solving more than one of these equations simultaneously pick out furtherly special types of coincidence-or-collision. 
This further information being significant, 
coincidence-or-collision equations are to be interpreted by choosing any nonempty subset thereof as a system of simultaneous equations.  
The totality of the solutions of these systems of simultaneous equations picks out what is usually called the {\it collision set} in Dynamics. 
We however consider this is a context which jointly embraces geometical modelling, whereby `coincidence-or-collision' enters our concept and naming. 
We are also in the business of treating configuration spaces as spaces -- {\it equipped} sets -- which extends to significant subsets being equipped 
and thus also spaces.
Thus we pass to the {\it coincidence-or-collision structure} conceptualization, which we denote by $\Co(\FrG\Frr)$.  

\m 

\n{\bf Motivation 1} In particular, we place additional emphasis on confluent coincidences-or-collisions having properties in excess of their elementary parts, 
due to which our conceptualization of $\Co(\FrG\Frr)$ is specifically layered. 
This amounts to viewing this not as {\sl a} subspace of $\FrG\FrS$ but as a {\it collection of subspaces}, 
which typically form a partly-joined network of spaces of different dimensions. 
We will have more to say about the mathematical nature of such networks once Papers II and III have generated sufficient variety of examples.  

\m 

\n{\bf Motivation 2} This line of thought moreover covers a multitude of parallel situations, 
providing a unified picture for much hitherto merely piecemeal modelling and conceptualization.
The next of this section provides the first inkling of the modelling diversity thus entailed; 
see the end summary Fig \r{10-Comparer} to keep track of what these look like.

\subsection{Two-tailed notions of inhomogeneity}\l{2-Tail}  

\n{\bf Remark 1} Coincidences-or-collisions are configurations with `elements of (exact) clumpiness', with the maximal coincidence-or-collision as its extremum. 
In fact, the next Section begin to develop some reasons why this usually is, or needs to be, excluded, 
leaving the coincidence-or-collision of all but one of the particles as the extremum of clumpiness.

\m 

\n{\bf Remark 2} Many concepts of inhomogeneity moreover have two extremes, the other being the most uniform configuration. 
Because of this, we next pass to considerations of uniformity, itself both a foundational and a cosmological topic.

\subsection{The perpendicularity and parallelism equations}  

\n Dimension $d \geqs 2$ has further relative-angle counterparts -- inner product rather than norm -- of clumping as follows. 

\m 

\n{\bf Definition 1} The {\it perpendicularity equations} are 
\be
(    \ur_{IJ} \cdot \ur_{KL}    ) = 0 \m  . 
\l{Perp-Eqs-1}
\ee
Again, these are to be interpreted as any shape solving at least one such equation contains a perpendicularity alias right angle.    
Moreover shapes solving more than one of these equations simultaneously pick out furtherly special types of configuration with multiple right angles.  
Thus one chooses any nonempty subset of the perpendicularity equations as a system of simultaneous equations.  
The totality of the solutions of these systems of simultaneous equations defines the 
{\it perpendicularity structure} $\FrP\mbox{erp}(\FrG\Frr)$ of the model system.  

\m 

\n{\bf Remark 2} To keep this distinct from coincidence-or-collision detection, we furthermore impose the restriction that $IJ \neq KL$. 
Thereby $||\ur_{IJ}||, \m  \ur_{KL} \neq 0$, so we can divide by both of these and so rewrite \r{Perp-Eqs-1} in terms of unit vectors,
\be
(    \hat{\ur}_{IJ} \cdot \hat{\ur}_{KL}    ) = 0 \m  . 
\l{Perp-Eqs-2}
\ee
A third form for these equations is then the manifestly-angular 
\be
\mbox{arccos}(     \hat{\ur}_{IJ} \cdot \hat{\ur}_{KL}) = \pi/2 \m  \mbox{ or } \m  3 \, \pi/2 \m  . 
\l{Perp-Eqs-3}
\ee 
{\bf Definition 2} The {\it parallelism equations} are 
\be
(    \hat{\ur}_{IJ} \cdot \hat{\ur}_{KL}    ) = 1 \m \mbox{ or }
\l{Para-Eqs-1}
\ee
\be 
\mbox{arccos}(    \hat{\ur}_{IJ} \cdot \hat{\ur}_{KL}    ) = 0 \mbox{ or } \pi \m  . 
\l{Para-Eqs-2}
\ee 
\n{\bf Remark 3} The nonzero right hand side in the first form for these means that unit vectors are now ab initio required so as to have a ratio expression.

\m 

\n{\bf Remark 4} These are once again subject to $IJ \neq KL$, now so as not to just be the identity $1 = 1$, 
and to be interpreted as a set of equations from which subsets of simltaneous equations are to be drawn from.  
The totality of the solutions of these systems of simultaneous equations defines the {\it parallel structure} $\FrP\mbox{ara}(\FrG\Frr)$ of the model system.  

\m 

\n{\bf Remark 5} In both of the above definitions, partial overlap, such as $K = J$ is allowed; 
in this case, $\ur_{IJ}$ and  $\ur_{JL}$ directly define an angle.
Another case in which direct definition occurs is if $K \neq J$ but $\uq_{K}$ is coincident-or-colliding with $\uq_{J}$.  
Elsewise, the angle in question is defined indirectly by translating one of the disjoint separation intervals on top of the other. 
Note that for the current treatise's model, $N = 3$ renders partial overlap inevitable, so only directly defined angles feature.

\m  

\n{\bf Remark 6} In the case of perpendicularity, the partially overlapping case corresponds to an angle between separations {\sl at} a point-or-particle.
Part III already realizes this, but more trivially than in Part IV's example.  

\m 

\n{\bf Remark 7} In the case of parallelism, the partially-overlapping index case is a fortiori an example of {\it collinearity}.  
The non-overlapping case can realize either of collinear or non-collinear parallelism, 
depending on whether extending each relative separation's interval to a line realizes the same line or two different lines.
We also note that if $\ur_{IJ}$ and $\ur_{KL}$ have no overlapping indexes but are collinear, 
then there also exist partially-overlapping index collinearities spanning between them, such as $\ur_{IJ}$, $\ur_{JK}$. 
In this way, one can represent all collinearity information with partially overlapping index cases, with 
and parallel but not collinear cases corresponding to those which do not admit such a representation.  
In this way, we can identify a subset of the parallelism set as the {\it collinearity structure}, $\FrC\mbox{ollin}(\FrG\Frr)$.  

\m 

\n{\bf Remark 8} Upon passing to the Lagrangian angles by taking the arccos, moreover, the accordance of zero status is reversed between the perpendicularity 
and parallelism equations.
[The accompanying doubling of values to diametrically-opposite pairs 
is a reflection of the two possible choices of relative orientation of the two Lagrangian separation vectors involved.] 
In the $N = 3$ context, this means that it is collinearity which is clumpy, with perpendicularity offering an opposite extreme.

\m  

\n{\bf Remark 9} The words `{\it linear correlation}' (Statistics), `{\it filament}' and `{\it cigar-shaped}' (which are both used in Astrophysics and Cosmology) 
give some indication of the conceptual class of collinearity as a type of clumpiness.
Collinearity corresponds to some lines containing more substantial distributions of points-or-particles than others.
Upon placing observers in the physical space, these become lines-of sight along which more points-or-particles are distributed than in other directions, 
from which points of view the space's matter distribution has preferred directions: a manifestation of {\it anisotropy}.  
It is moreover instructive to emphasize the difference between {\it collinearity in threes} (and other {\sl subsystem} collinearities) and 
the much more familiar concept of linear correlation for one's entire data set.  
Sampling for collinearity in threes was introduced by Kendall in connection with archaeological landsite analysis \c{Kendall84, Kendall}.
Use of threes corresponds to 3 point-or-particle subsystems being of the minimal size with which one can probe relative angle information.
One can think of this as following either from the definition of collinearity 
-- we need at least 2 lines to compare, which between them require a minimum of three points to define -- 
or from the more general consideration that the triangle of points is the minimal relationally nontrivial unit for $d \geqs 2$ Euclidean and similarity geometries.

\m 

\n{\bf Remark 10} Parallelism's conceptual class as a type of clumpiness is somewhat more subtle and structured.
`{\it Parallel filaments}', `{\it self alignment}' and `{\it nematic phase}' give some indication of the conceptual class of clumpiness entailed 
(words from the Topologial Defects literature, more from liquid crystal applications than cosmological ones).  
Also note that while collinearity filaments are of minimal point-or-particle number three, parallelism fragments can be as short as two, 
but on the other hand require a minimum of two non-overlapping such unit filaments for parallelism to be realized. 
Thus the minimal relationally nontrivial unit for parallelism to be realized is 
\be
N = 4 \m \mbox{ points-or-particles in dimension } \m d = 2  \m . 
\ee 
{\sl So sampling for parallelism requires probing in fours}. 
This minimality gives further motivation for detailed study of the space of quadrilaterals. 
[This 4 moreover bears relation to 4 being \c{AMech, Affine-Shape-1} the minimal relationally nontrivial unit for Affine Shape Theory \c{PE16}, 
via parallelism being the central subject matter of Affine Geometry].  

\m 

\n{\bf Remark 11} Collinearity and parallelism are moreover {\sl cumulative}, meaning that any number of separations can be mutually collinear or parallel.
On the other hand, mutual perpendicularity is much more restrictive: in the absense of parallelism, 
only 2 separation vectors can enjoy this in 2-$d$, or 3 in 3-$d$ (`tangent, normal and binormal') 
Because of this, perpendicularity does not offer a strong opposite to parallelism for systems in which $N$ is at all substantially larger than $d$.  
Moreover (and just using 2-$d$ as the simplest example) even if non-mutual perpendicularity were heavily represented, 
one would just have a different more subtle structuring rather than homogeneity.  
Up to a certain threshold, a random distribution of right-angled triples is possible, 
though beyond this the only way for the proportion of right angles to increase is for the right angled triples to start to form parallel structures.
In this manner, a point-or-particle distribution manifesting a large enough proportion of right angles is a subcase of, not an opposite to, 
a point-or-particle distribution manifesting a large proportion of parallelism.

\subsection{The uniformity equations}\l{Lag-Uni}

\n{\bf Definition 1} We define {\it incipient uniformity equations} by  
\be 
||\ur_{IJ}|| = ||\ur_{KJ}||\mma\mbox{ or in manifest-ratio form }   \m  
\frac{||\ur_{IJ}||}{||\ur_{KJ}||} = 1                                                \m  ,
\l{Lag-Incipient-Uniformity-Eqs}
\ee
also to be interpreted as a set of equations from which any subset can be chosen to hold as simultaneous equations.  
The totality of their solutions constitute the {\it incipient uniformity structure} of the model, $\Uni^{\mbox{\scriptsize incipent}}(\FrG\Frr)$. 

\m 

\n{\bf Remark 1} Concrete notions of uniformity require further qualification.  
First note that 
\be
||\ur_{IJ}|| = 0                        \mma    
||\ur_{KL}|| = 0                        \Rightarrow   
||\ur_{IJ}|| - ||\ur_{KL}|| = 0 - 0 
                            = 0         \m ,   
\ee
by which a configuration containing a coincidence-or-collision suffices for it to register as containing `elements of (the corresponding incipient notion of) uniformity'.  
This however runs against intuitions of uniformity. 
What is happening is that the incipient definition is picking out {\sl both} tails of inhomogeneity. 
This is remedied by sculpting the incipient uniformity equations by means of the following excisions.

\m 

\n{\bf Excision i)} We require the relative Lagrange coordinates entering the zero differences themselves to be nonzero.
(This builds on us already having a specific characterization of this other extreme of inhomogeneity as per two subsections back.)      
Once this is applied, (\r{Lag-Incipient-Uniformity-Eqs})'s ratio form is indeed always defined and equivalent to its other given form.  

\m 

\n{\bf Excision ii)} Coincidences-or-collisions moreover cause another problem by automatically implying some nonzero differences: 
those in which the separations have coincidences-or-collisions at both ends. 
Thus we also decree that coincidence-or-collision positions count {\sl only once} in determining the more refined notion of uniformity structure.

\m 

\n{\bf Remark 2} With these considerations in mind, two particular refinements of notions of uniformity are as follows. 

\m 

\n{\bf Definition 2} The {\it entire Lagrangian uniformity equations} are (\r{Lag-Incipient-Uniformity-Eqs}), 
to be interpreted as a set of equations from which any subset of equations are treated as simultaneous equations, but now within Excisions i) and ii).
The solutions of this generally smaller complex of equations give the {\it entire Lagrangian uniformity structure}, $\Uni^{\mbox{\scriptsize entire}}(\FrG\Frr)$.

\m 

\n{\bf Definition 3} The {\it adjacent Lagrangian uniformity equations} are 
\be 
||\ur_{IJ}|| = ||\ur_{JK}|| \m  , 
\l{Adjacent-Uniformity-Eqs}
\ee
now for all particles occurring in the order $\uq_I \leq \uq_J \leq \uq_K$ along the line in 1-$d$, 
or along a complete `joining of the dots' circuit (a choice of perimeter).  
These are once again be interpreted as a set of equations from which any subset of equations are treated as simultaneous equations, subject to Excision i).  
[Excision ii) is rendered unnecessary by the ordering stipulation of the adjacent configurations.]
The solutions constitute the {\it adjacent Lagrangian uniformity structure}, $\Uni^{\mbox{\scriptsize adjacent}}(\FrG\Frr)$.

\m 

\n{\bf Remark 3} On the one hand, entire uniformity exemplifies Sec \r{Rel-Lag}'s notion of Lagrangian democracy.
On the other hand, adjacent uniformity picks a preferred subset, resting upon a particular choice of joining the dots (Sec \r{Constellations}).  
While either of these notions turns out suffice to characterize uniformity in (3, 1), 
entire uniformity already turns out to be a sharper quantifier (in the sense of Appendix II.A) for (4, 1). 

\m 

\n{\bf Definition 4} The minimal elements of uniformity are pairs of non-superposed separations of equal length. 

\m 

\n{\bf Remark 4} All $\FrG\Frr$ points in $\Uni^{\mbox{\scriptsize entire}}(\FrG\Frr)$ contain at least one element of uniformity.
On the other hand, those in $\Uni^{\mbox{\scriptsize adjacent}}(\FrG\Frr)$ contain at least one {\sl along the line} (1-$d$) or {\sl designated perimeter}.

\subsection{The Lagrangian angle-uniformity equations}\l{Lag-Angle-Uni}

These are the `norm to inner product' and `vector to unit vector' counterparts of the preceding subsection, 
which are now meaningfully more sharply characterized as {\it Lagrangian separation-uniformity equations}.

\m 

\n{\bf Definition 1} We define {\it incipient Lagrangian angle-uniformity equations} by  
\be
(    \hat{\ur}_{IJ} \cdot \hat{\ur}_{KL}) = \pm (    \hat{\ur}_{MN} \cdot \hat{\ur}_{OP}    ) \m  .  
\ee
\be
\frac{           \mbox{arccos}(\hat{\ur}_{IJ} \cdot \hat{\ur}_{KL})    }
     {    \pm \, \mbox{arccos}(\hat{\ur}_{MN} \cdot \hat{\ur}_{OP})    } = 1 \m  .  
\ee
is an alternative form for this equation.  
\n{\mbox Remark 2} These require various restrictions so as to pick out Lagrangian angle-uniformity in isolation.
Firstly, none of the separations involved are to be zero -- i.e.\ not coincidences-or-collisions -- 
so that the unit vectors and corresponding angles are well-defined. 
Secondly, these equations are to compare different angles, so no solutions by 1 = 1 identity are admissible.  
Thirdly, if we wish to diagnose equiangularlty separately from parallelism, we forbid parallel separation pair inputs. 
For if say the right-hand side contained such, then it is 1, turning the equiangularity equation into a parallelism equation on the left-hand-side inputs.  

\m 

\n{\bf Remark 3} Subject to these restrictions, 
one is to treat the Lagrangian angle-uniformity equations as a set from which subsets of simultaneous equations are extracted, 
whose combined solutions form the {\it Lagrangian angle-uniformity structure}, $\FrA\mbox{ngle-}\FrU\mbox{ni}(\FrG\Frr)$

\m 

\n{\bf Remark 4} So far, this subsection has considered just Lagrangian notions; these moreover have clear Jacobian counterparts to which we now pass.

\subsection{Preliminary position space analogue}\l{Q-Prelim}

\n We now regress our general invariants form to the case with absolute origin as well. 

\m 

\n The {\it origin coincidence equations} are 
\be 
||\uq_I|| = 0   \m  .  
\l{O-C-Eqs}
\ee
The equal distance from the origin equations are 
\be 
||\uq_I|| = ||\uq_J||  \m  .  
\l{E-D-O-Eqs}
\ee
The {\it perpendicular relative to the origin equations} are 
\be
(\uq_I \cdot \uq_J) = 0 \mma   \mbox{ or }
\mbox{arccos}(\hat{\uq}_I \cdot \hat{\uq}_J) = \pi/2 \m \mbox{ or } \m 3\,\pi/2 \m  . 
\l{O-Perp-Eqs-1}
\ee 
The {\it parallelism relative to the origin equations} are 
\be
(\hat{\uq}_I \cdot \hat{\uq}_J) = 1                \mma \mbox{ or }     \m 
\mbox{arccos}(\hat{\uq}_I \cdot \hat{\uq}_J) = 0        \m    \mbox{ or } \m \pi \m  ,  
\l{O-Para-Eqs-1}
\ee 
which include collinearity with the origin as a subcase.
The {\it uniformity relative to the origin} equations are 
\be
||\uq_I|| = ||\uq_J||\mma\mbox{ or in manifest-ratio form } \m  
\frac{    ||\uq_I||    }{    ||\uq_J||    } = 1                                              \m  .
\l{O-Uniformity-Eqs}
\ee
Finally, the {\it angle uniformity relative to the origin equations} are 
\be
     (\hat{\uq}_I \cdot \hat{\uq}_J) = 
\pm  (\hat{\uq}_K \cdot \hat{\uq}_L)    \m  .  
\ee
\be
\frac{       \mbox{arccos}(    \hat{\uq}_I \cdot \hat{\uq}_J    )    }
     {  \pm  \mbox{arccos}(    \hat{\uq}_K \cdot \hat{\uq}_L    )    } = 1   \m  .  
\ee

\subsection{Mergers: the Jacobian-level counterpart of coincidences}\l{Merger}

We now start to go in the opposite direction, to Jacobi counterparts of the Lagrangian invariants.  

\m 

\n{\bf Definition 1} The {\it general merger equations} are 
\be 
||\urho_i|| = 0 \m  .  
\l{General-Merger-Eqs}
\ee
Yet again, these are to be interpreted as a set of equations from which any subset can be taken and interpreted as a system of simultaneous equations.  
These correspond to coincidences between CoMs (including of particles, through viewing these as their own 1-particle CoMs).  
Any configuration solving at least one such equation is a general merger.  
The totality of such solutions defines the {\it general merger structure} $\Merger(\FrG\Frr)$ of the model system.  
These equations moreover run over all cluster choices, labelling and topological.

\m 

\n{\bf Remark 1} The notion of general merger is thus the Jacobian generalization of the Lagrangian notion of coincidence.  
It corresponds to the coincidences of the centres of mass of any two distinct clusters of any size. 
In the case in which both clusters are distinct single particles, the Lagrangian notion of primary-point coincidence or material-particle collision is recovered.  

\m   

\n{\bf Definition 2} We moreoveruse {\it merger} to mean a nontrivially Jacobian general merger, i.e.\ one which is not of the minimal form $\fP\fP$. 
%

\m 

\n{\bf Remark 2} On the one hand, for many applications, Jacobian modelling of interparticle clusters supercedes Lagrangian modelling of individual particles. 
Because of this, splitting of general mergers into collisions and mergers is far from always desirable 
(this is to be contrasted with the refinement of notions of uniformity to exclude collisions).  
On the other hand, the collision--merger split is between already topologically picked out features -- collisions -- and purely metric information: mergers. 

\m 

\n{\bf Excision i)} If we wish to consider {\it purely Jacobian mergers}, we exclude all the $\fP\fP$. 

\m 

\n{\bf Definition 3} We also have occasion to consider in isolation mergers of each other $\fA\fB$ hierarchical type in turn.  
We term $\fA\fA$ mergers {\it self}-$\fA$ {\it mergers} and $\fA\fB$ mergers {\it Mutual}-$\fA\fB$ {\it mergers}.  

\m  

\n{\bf Definition 4} {\it Coarse-grained Lagrangian coincidences} 
consist of coincidences between centres of mass in the specific case of the pieces of a partition of a given configuration.

\m 

\n{\bf Remark 3} This amounts to firstly applying a coarse-graining 
in which each piece P of the partition is to be approximately modelled as a single particle at CoM(P), 
with the pieces themselves then admitting a purely Lagrangian treatment. 

\m 

\n{\bf Remark 4} The Jacobian framework's capacity to incorporate coarse-grained modelling 
is another pointer to its conceptual superiority over the Lagrangian framework. 

\m 

\n{\bf Definition 5} With general mergers consisting of the purely-Lagrangian collisions and the mergers which are new to the Jacobian level of the analysis, 
{\it collision--merger equations, elements and structures} become better final names. 

\m 

\n{\bf Remark 5} Notions of merger are moreover constrained by some rigidity inter-relations. 
See Sec \r{(3,1)-Mergers} for how (3, 1) is `doublly rigid': allowing just one notion of merger, which is realized by just one shape, 
which, in any case, is an already-known shape at that stage of the exposition. 
Part II, however, already exhibits the more general situation of nontrivially partial rigidity.

\subsection{Jacobian, alias centre of mass, uniformity}\l{Jac-Uni}

\n{\bf Definition 1} The {\it incipient Jacobian uniformity equations} are 
\be 
||\urho_i|| = ||\urho_j|| \m  , \mbox{ or in manifest-ratio form } \m  
\frac{||\urho_i||}{||\urho_j||} = 1 
\m  ,
\l{Incipient-Jacobi-Uniformity-Eqs}
\ee
yet again to be interpreted as a set of equations from which any subset can be chosen to hold as simultaneous equations.  
These are for equal separations between whichever centres of mass, so one can substitute `Jacobi' for `CoM' in all of this section's names.  
The totality of their solutions constitute the {\it incipient Jacobian uniformity structure} of the model, $\CoM$-$\Uni^{\mbox{\scriptsize incipient}}(\FrG\Frr$.  
As per the previous subsection these equations run over all cluster choices (labelling or topological).

\m 

\n{\bf Definition 2} The {\it entire Jacobian uniformity equations} then i) count CoM coincidence positions once only each, and ii) preclude 
solutions based on pairs of $\rho_i = 0$ solutions to the general merger equations.  
These form the {\it entire Jacobian uniformity structure} $\CoM$-$\Uni^{\mbox{\scriptsize entire}}(\FrG\Frr)$.   

\m 

\n{\bf Definition 3} The {\it adjacent Jacobian uniformity equations} are based on i) 
and on ordering the centres of mass along the line (1-$d$) or a perimeter ($\geqs 1$-$d$) and considering only differences of adjacent such.
These form the {\it entire Jacobian uniformity structure} $\CoM$-$\Uni^{\mbox{\scriptsize adjacent}}(\FrG\Frr)$.    

\m 

\n{\bf Remark 1} Many applications of uniformity, however, concern directly material uniformity, as quantified in solely Lagrangian terms. 
On the other hand, even if direct material uniformity is sought, when coarse-graining replacing some particle clusters by their CoMs is performed, 
Jacobian modelling does enter even these most conventional notions of uniformity.  

\m 

\n{\bf Remark 2} The previous subsection's hierarchical splitting and rigidity remarks also impinge upon Jacobi uniformity. 
See Sec \r{UJ} for how (3, 1)'s Jacobi uniformity is also `doubly rigid' and yet does yield a new special shape in this case.

\subsection{Jacobian-level counterparts of angular notions}\l{Jac-Angles}

{\bf Definition 1} The {\it Jacobi perpendicularity equations} are 
\be
(    \urho_i \cdot \urho_j    ) = 0 \mbox{  } , 
\ee
or, in terms of unit Jacobi coordinates, 
\be
(    \hat{\urho}_i \cdot \hat{\urho}_j    ) = 0 \m  ,  
 \l{Jac-Perp-Eqs-1}
\ee
or, in terms of angles, 
\be
\mbox{arccos}(    \hat{\urho}_i \cdot \hat{\urho}_{j}) = \pi/2 \m  \mbox{ or } \m  3 \, \pi/2 \m  . 
\l{Jac-Perp-Eqs-2}
\ee
{\bf Remark 1} On some occasions, we preclude input pairs of relative Jacobi separations which are both already relative Lagrange separations.
        If we do not, we have {\it entire} alias {\it Lagrange--Jacobi perpendicularity equations}, 
whereas if we do,     we have {\it pure-Jacobi perpendicularity equations}.  

\m 
			  
\n{\bf Remark 2} Another source of diversity is perpendicularity within one topological and labelling clustering choice 
(cluster self-perpendicularity) versus mutual perpendicularity between clusterings.  

\m 
		
\n{\bf Definition 2} The {\it Jacobi parallellism equations} are 
\be
(    \hat{\urho}_i \cdot \hat{\urho}_j    ) = 1 \m  . 
 \l{Jac-Para-Eqs-1}
\ee
or, in terms of angles, 
\be
\mbox{arccos}(    \hat{\urho}_i \cdot \hat{\urho}_{j}    ) = 0 \m  \m \mbox{ or } \m \m  \pi \m  . 
\l{Jac-Para-Eqs-2}
\ee
The totality of the solutions of these systems of simultaneous equations defines the {\it Jacobi parallelism set} of the model system.  

\m 

\n{\bf Remark 3} Lagrange--Jacobi versus pure-Jacobi and self versus mutual are also sources of variety here.  

\m 

\n{\bf Definition 3} We define {\it incipient Jacobian angle-uniformity equations} by  
\be
(    \n_i \cdot \hat{\urho}_j    ) = \pm (    \hat{\urho}_i \cdot \hat{\urho}_j    ) \mma \mbox{ or } \m  
\ee
\be
\frac{       \mbox{arccos}(    \hat{\urho}_i \cdot \hat{\urho}_j    )    }
     {  \pm  \mbox{arccos}(    \hat{\urho}_k \cdot \hat{\urho}_l    )    } = 1 \m  .  
\ee
\n{\bf Remark 4} These require various restrictions so as to pick out Lagrangian angle-uniformity in isolation.
Firstly, none of the cluster separations involved are to be zero -- i.e.\ not mergers -- 
so that the unit vectors and corresponding angles are well-defined. 
Secondly, these equations are to compare different relative Jacobi angles, so no solutions by 1 = 1 identity are admissible.  
Thirdly, if we wish to diagnose Jacobi equiangularity separately from Jacobi parallelism, we forbid Jacobi-parallel separation pair inputs. 
For if say the right-hand side contained such, then it is 1, 
turning the Jacobi equiangularity equation into a Jacobi-parallelism equation on the left-hand-side inputs.  

\m 

\n{\bf Remark 3} Subject to these restrictions, 
the Jacobian angle-uniformity equations are to be considered as a set from which subsets of simultaneous equations are extracted, 
whose combined solutions form the {\it Jacobian angle-uniformity structure}, $\CoM$-$\FrA\mbox{ngle-}\Uni^{\mbox{\scriptsize adjacent}}(\FrG\Frr)$.

\m 

\n{\bf Remark 4} This also comes in {\it figurative} and {\it entire} forms.
%

\m 

\n{\bf Remark 5} Hierarchical layering applies to Jacobi angle concepts as well.  
Finally, notions of merger are also constrained by some rigidity inter-relations.

\subsection{Approximate notions of clumpiness and of uniformity}\l{Approx-Cl-Uni}

\n Let us end Section 2 by considering approximate notions of {\it highly uniform} and {\it highly clumped} states.   

\m 

\n{\bf Definition 1} The $\epsilon$-{\it tolerant collision-or-coincidence inequalities}
\be 
\frac{||\ur_{IJ}||}{||\ur_{KL}||} \m  \leq  \m  \epsilon  \m  .  
\l{Epsi-Collision-Eqs}
\ee
replace the collision-or-coincidence equations (\r{Collision-Eqs}), where here and below $\epsilon > 0$ is a small tolerance parameter.  
When interpreted along the lines of Sec \r{Coll} subsections, the solutions of these form the $\epsilon$-{\it approximate coincidence-or-collision structure}.  

\m 

\n{\bf Definition 2} The $\epsilon$-{\it tolerant perpendicularity inequalities} are 
\be
\pi/2 - \epsilon  \m  \leq  \m  \mbox{arccos}(\hat{\ur}_{IJ} \cdot \hat{\ur}_{KL})  \m  \leq  \m  \pi/2 + \epsilon 
\ee
\n{\bf Definition 3} The $\epsilon$-{\it tolerant parallelism inequalities} are 
\be
\mbox{arccos}(\hat{\ur}_{IJ} \cdot \hat{\ur}_{KL})  \m  \leq  \m        \epsilon        \mma \m 
\mbox{arccos}(\hat{\ur}_{IJ} \cdot \hat{\ur}_{KL})  \m  \geqs  \m  \pi - \epsilon 
\ee
\n{\bf Definition 4} The $\epsilon$-{\it tolerant Lagrangian uniformity inequalities}
\be 
1 - \epsilon  \m  \leq  \m  \frac{||\ur_{IJ}||}{||\ur_{KL}||} \m  \leq \m  1 + \epsilon  \m  .  
\l{Epsi-Lag-Uniformity-Eqs}
\ee
replace the Lagrangian uniformity equations subset of (\r{Lag-Incipient-Uniformity-Eqs}). 
When interpreted along the lines of Sec \r{Lag-Uni}, the solutions of these form the $\epsilon$-{\it approximate Lagrangian uniformity structure}.  

\m  

\n{\bf Definition 5} The $\epsilon$-{\it tolerant angle-uniformity inequalities} are 
\be
\left|     \mbox{arccos}(    \hat{\ur}_{IJ} \cdot \hat{\ur}_{KL}    ) - 
           \mbox{arccos}(    \hat{\ur}_{MN} \cdot \hat{\ur}_{OP}    )   \right| \m  \leq \m  \epsilon  \m . 
\ee
\n{\bf Definition 6} The $\epsilon$-{\it tolerant merger inequalities} are 
\be 
\frac{||\urho_i||}{||\urho_j||} \m  \leq \m  \epsilon  \m  .  
\l{Epsi-General-Merger-Eqs}
\ee
replace the merger equations subset of (\r{General-Merger-Eqs}).    
When interpreted along the lines of Sec \r{Merger}, the solutions of these form the $\epsilon$-{\it approximate merger structure}.  

\m 

\n{\bf Definition 7} The $\epsilon$-{\it tolerant Jacobi perpendicularity inequalities} are 
\be
\pi/2 - \epsilon \m  \leq \m  \mbox{arccos}(    \hat{\urho}_i \cdot \hat{\urho}_j) \m  \leq \m  \pi/2 + \epsilon 
\ee
\n{\bf Definition 8} The $\epsilon$-{\it tolerant Jacobi parallelism inequalities} are 
\be
\pi - \epsilon \m  \leq  \m  \mbox{arccos}(    \hat{\urho}_i \cdot \hat{\urho}_j    ) \m  \leq \m  \pi/2 + \epsilon 
\ee
\be
\mbox{arccos}(    \hat{\urho}_i \cdot \hat{\urho}_j    ) \m  \leq \m  \epsilon 
\m  , \m  \m 
\mbox{arccos}(    \hat{\urho}_i \cdot \hat{\urho}_j    ) \m  \geqs \m  \pi - \epsilon 
\ee
\n{\bf Definition 9} The $\epsilon$-{\it tolerant Jacobian uniformity inequalities}
\be 
1 - \epsilon \m  \leq \m  \frac{    ||\urho_i||    }{    ||\urho_j||    }  \m  \leq  \m  1 + \epsilon  \m  .  
\l{Epsi-Jacobi-Uniformity-Eqs}
\ee
replace the Jacobian uniformity equations subset of (\r{Incipient-Jacobi-Uniformity-Eqs}). 
When interpreted along the lines of Sec \r{Jac-Uni}, the solutions of these form the $\epsilon$-{\it approximate Jacobian uniformity structure}.  

\m 

\n{\bf Definition 10} The $\epsilon$-{\it tolerant Jacobian uniformity inequalities}
\be
\mbox{arccos}(    \hat{\urho}_i \cdot \hat{\urho}_j    ) \leqs  \epsilon 
\m  , \m  \m 
\mbox{arccos}(    \hat{\urho}_k \cdot \hat{\urho}_l    ) \geqs  \pi - \epsilon 
\ee
\n{\bf Remark 1} Note that the collision-or-coincidence equations, and the merger equations, have $\epsilon = 0$ limits which cease to require ratios to quantify.  
On the other hand, {\sl all} the approximate $\epsilon$-tolerant versions of the above inequalities are ratio equations.  

\m 

\n{\bf Remark 2} Once further information about the possible geometrical structures of spaces of ratios become apparent in the next Section, 
the further caveat that some uses of $\epsilon$ may correspond to pieces of balls, rather than whole balls, 
will come into play as a further restriction on the valid solution sets of the above problems.

\subsection{Discussion}

\n See Fig \r{10-Comparer} for a summary of this section's fifteen notions with simple examples which make for useful comparison.
%
{            \begin{figure}[!ht]
\centering
\includegraphics[width=0.8\textwidth]{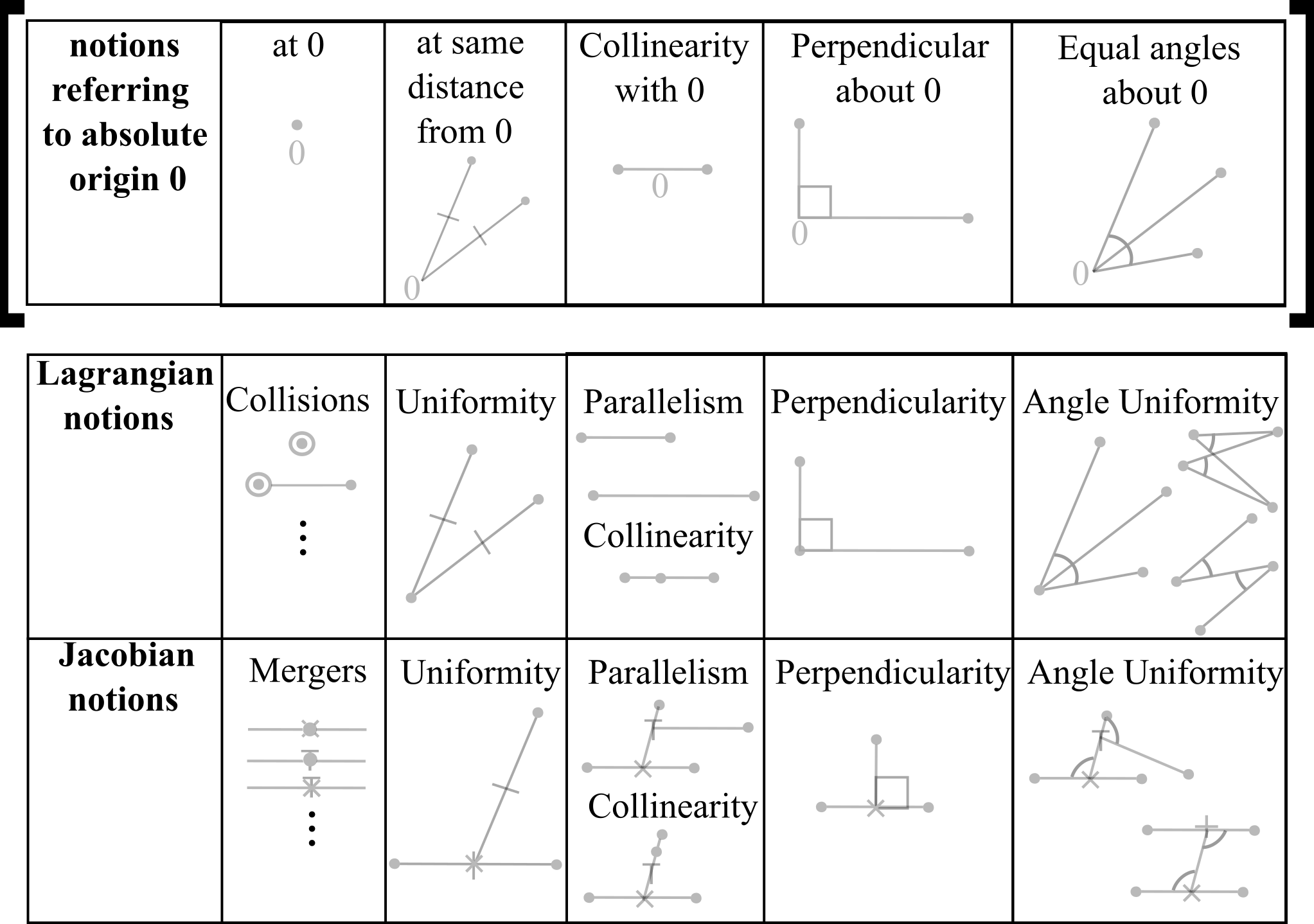}
\caption[Text der im Bilderverzeichnis auftaucht]{        \footnotesize{This section's Lagrangian and Jacobian notions 
(and less interesting origin-dependent notions) compared with simple examples.}}
\l{10-Comparer}\end{figure}            }

\m 

\n Most of this Section's notions are furthermore based on {\sl length ratios}, which are metric-level structures.    
It is coincidences-or-collisions  -- the Lagrangian elements of exact clumpiness -- which are the exception; these have a purely topological character.   
The two-tailed notion of inhomogeneity is thus a metric-level conceptualization, 
which breaks down to a one-tailed notion at the more structurally sparse topological level.  
While Lagrangian uniformities are {\sl obviously} based on ratios, 
those notions which are Jacobian but not Lagrangian moreover have no topologically significant status either.  
In the case of mergers, this is via the prerequisite CoMs being a metric notion.  
Overall, this indicates that (material) point coincidences or particle collisions can be studied in isolation before the other three types of features.  

\m 

\n Moreover, Lagrangian uniformity's status as the opposite tail to coincidence or collision clumpiness, 
alongside the further subtlety and complexity of the Jacobian notions, makes Lagrangian-level metric structure the most natural second port of call.  
The current treatise follows this pattern of development in turn: 

\m 

\n A) topological-level coincidence-or-collision structure of the primary points or material particles.

\m 

\n B) Its metric-level extension to Lagrangian-level structure thus including also Lagrangian uniformity. 

\m 

\n C) Its further metric-level extension to Jacobi-level structure: mergers and Jacobi uniformity.

\section{Constellational, separational, figurative and other perspectives}\l{Perspectives}

\n{\bf Modelling Ambiguity 1}  The following are a non-exhaustive subset of the possible positions one might take depending on the application. 

\m 

\n a) The {\it constellational perspective} of Sec 2 involves giving the points-or-particles primary status.  

\m 

\n b) The {\it separation perspective} involves instead giving relative separations primary status.  

\m 

\n In each case, this can amount to giving primary status i) to the named objects or ii) to the associated vectors. 

\m 

\n c) The {\it angle perspective} is to give primary significance to the relative angles; this requires $d \geqs 1$.   

\m 

\n d) The {\it relative invariants perspective} is to ascribe primary significance even-handedly to relative separations and relative angles.    

\m 

\n e) The {\it joint perspective} is to give all three co-equal status.  

\m 

\n f) The {\it perimeter perspective} involves giving the most primary status to a 
cycle of separations going through all points; the latter is termed a {\it perimeter} = `joining of the dots'.  
Note that this in general entails a partition by conceptual role of the set of relative separations into those used in `joining the dots' and those omitted.
For 1-$d$ this first shows up for $N = 3$, and for 2-$d$ for $N = 4$: the quadrilaterals.
The latter is known as the {\it edge-diagonal} distinction for $N$-a-gons. 

\m 

\n g) The {\it path perspective} in 1-$d$ involves giving the most primary status to a path of separations going through all points; 
this includes one edge less than the perimeter perspective.
%

\m 

\n h) The {\it minimal perspectives} are to give primary status to a minimal set of objects that specify the figure. 
This can be done at various different levels, e.g.\ choosing $n$ relative vectors, 
but also choosing one of the familiar congruence or similarity conditions for a triangle at the level of the corresponding invariants. 
Some of the latter are moreover mixed separation and angle criteria, such as two sides and the angle between them.   
Of course, various ab initio different such conditions end up interchangeable by theorems.  

\m 

\n{\bf Remark 1} The constellational position is physically (and in some senses relationally) natural, 
by ascribing {\it material primality}: primary status to the materially significant particles themselves.  
An order in which to `join up the dots' to form a `perimeter' or `laminar figure' plays at most a secondary role here, 
viewed as practitioner- or particle-label-dependent -- rather than intrinsic -- in its content.   

\m 

\n{\bf Remark 2} On the other hand, the separation, perimeter and minimal perspectives are natural and commonplace in basic flat geometry; 
these form a broader {\it figurative perspective}.  
Especially in the absense of a meaningful carrier space origin, one does not specify geometrical figures by where their vertices are, 
but rather in terms of what side-lengths and relative angles they have, especially edge lengths and angles between edges.  

\m 

\n{\bf Remark 3} On the one hand, perspectives a) to e) are in one sense superior through considering all of the objects in some given class, to 
path, perimeter and minimal entailing choices.
On the other hand, not considering redundant information points in the opposite direction, in particular to the minimal choice.  

\m 

\n{\bf Remark 4} Moreover, not all models have f) to h) meaningfully exist as distinct options. 
One requires 
\be
N \geqs 2 \mbox{ to have a bona fide path } , 
\ee
\be
N \geqs 3 \mbox{ to have a bona fide perimeter } ,  
\ee
and 
\be 
N \geqs 4 \mbox{ for the perimeter to be distinct from the totality of relative separations }: 
\l{edge-diagonal}
\ee 
i.e.\ quadrilaterals are the smallest figures to have {\it diagonals} as well as edges (Fig \r{Join-the-Dots}.h).   

\m 

\n{\bf Remark 5} Also, not all models' configurations are uniquely specified by b), c), f) or g).  
For instance, path only uniquely determines a figure in 1-$d$, whereas in 2-$d$, perimeter is only a unique specification for $N = 3$'s triangles.  
Also, whereas specifying three side lengths fully determines a triangle (if no length exceeds the sum of the other two),
giving four side lengths at most only specifies a continuum family of quadrilaterals.

\m 

\n{\bf Remark 6} The path perspective supports two directions of path, 
whereas the perimeter perspective distinguishes between clockwise and anticlockwise `joining of the dots' (Fig \r{Join-the-Dots}.e-f) 
Including this further ambiguity, there are $N!$ path orders in 1-$d$ (permutations) and $\{N - 1\}!$ perimeter orders (cyclic permutations).
Thus if no direction is picked, the 
\be
\mbox{(number of paths)}  \es  \frac{N!}{2} \m 
\ee 
and if no orientation is picked, the 
\be
\mbox{(number of perimeters)}  \es  \frac{\{N - 1\}!}{2} \m  .
\ee

\m 

\n{\bf Remark 7} (\r{edge-diagonal}) has the consequence that quadrilaterals are minimal for the re-entrant, convex and crossed figure distinction: 
Fig \r{Join-the-Dots}.d).  
%
{            \begin{figure}[!ht]
\centering
\includegraphics[width=0.75\textwidth]{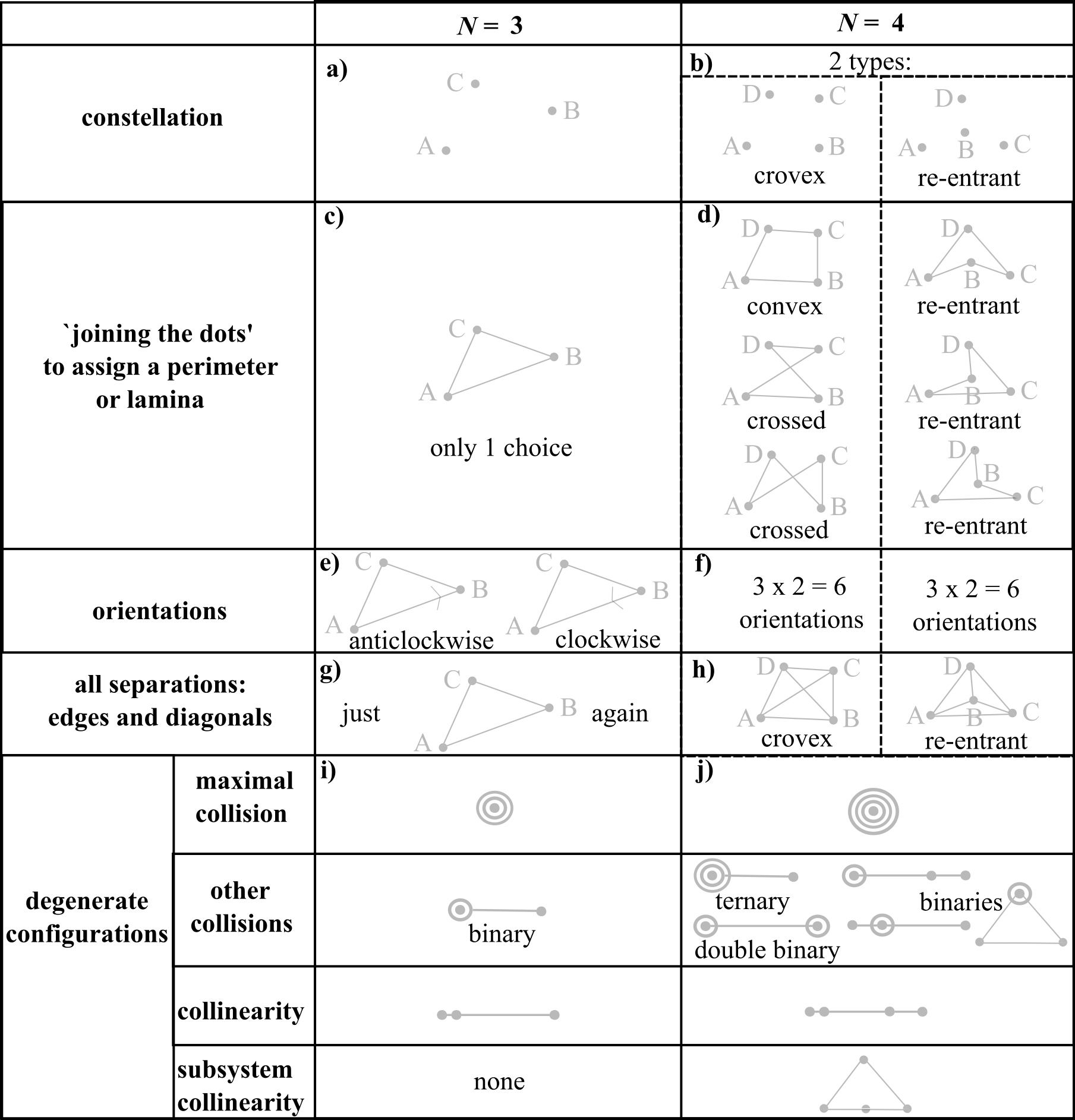}
\caption[Text der im Bilderverzeichnis auftaucht]{        \footnotesize{For each of (3, 2) and (4, 2) in turn, we depict 
nondegenerate constellations a)-b), 
joining the dots to make a perimeter or lamina in c)-d), 
a count of orientations in e)-f), 
inclusion of all separations -- diagonals as well as edges -- in g)-h), 
and demonstration of the variety of types of degenerate configurations in i)-j). 
Some further features requiring at least (4, 2) are as follows. 

\m 

\n 1) The distinction between crovex and re-entrant constellations and total separation diagrams, 
and the finer distinction between convex, crossed and re-entrant quadrilaterals.

\m 

\n 2) The existence of multiple types of partial (non-maximal) coincidences-or-collisions.  

\m 

\n 3) The existence of non-collinear coincidences-or-collisions. 

\m  

\n 4) The existence of non-collinear configurations with collinear subsystems, which are moreover intermediate between crovex and re-entrant configurations.
} }
\l{Join-the-Dots}\end{figure}            }

\m 

\n{\bf Remark 8} This distinction is moreover partially retained at the constellational level, in the `crovex versus re-entrant' form. 
Here {\it crovex} is the well-defined class of of those constellations (Fig \r{Join-the-Dots}.b) 
for which, upon ascribing a perimeter produce crossing or convex quadrilateral figures; this class has substantial applications in Part IV.  

\m 

\n{\bf Remark 9} b).i)'s consideration of all separations is moreover graph-theoretically dual to a)'s consideration of none of them, 
by which ascription of primality becomes less meaningful in hindsight.
In this way, joint primality is preferred 
(and also it is the crovex and re-entrant classes which are meaningful once again upon consideration of all separations: Fig \r{Join-the-Dots}.h).
Moreover, if one considers all separations together, one can always work in triangle subsystems to determine all relative angles, 
so the content of this joint primality ends up coinciding with perspective c)'s.  

\m 

\n{\bf Remark 10} Going full circle back to Remark 1's comment, however, Physics' material reasons for particle primality break this graph-theoretic duality. 
None the less, even here, it is often advantageous for the theoretician to `draw in' separation lines (so as to compute and clearly present relative angles). 
This practicality does not make distinction between edges and diagonals, so one actually works in practise with c) even if one's conceptions are in terms of a).

\section{Ratios, and their preshape space's topology and geometry}\l{Preshape}

Let us first consider various types of dimensionless ratio variables for use here; a subset of these moreover transcend to shape space as well.

\subsection{Simple ratio coordinates}

\n We first give a more precise consideration to the ratio variables already introduced in Sec \r{Preamble-2}.

\m 

\n{\bf Definition 1} In 1-$d$, the signed 
\be
{\cal S}_{ij}  \:=  \frac{\rho^i}{\rho^j}    \mma i,  \m  j = 1  \mbox{ to } n  \mma    \m  i \neq j \m   
\ee
and the unsigned  
\be
{\cal R}_{ij}  \:=  \frac{|\rho^i|}{|\rho^j|}\mma i,  \m  j = 1  \mbox{ to } n  \mma    \m  i \neq j \m   
\ee
{\it simple ratio coordinates} are available.  
For $d \geqs 2$, these extend to the more cumbersome {\it ratios of individual components}, e.g.\ in the signed case 
\be
{\cal S}_{i a j b}  \:=  \frac{\rho^{ia}}{\rho^{jb}}\mma ia,  \m  jb = 1 \m \mbox{ to } \m nd  \mma    \m  ia \neq jb \m  ,  
\ee 
and also to the more limited number of {\it magnitude ratio coordinates} (necessarily unsigned), 
\be
{\cal R}^{\prime}_{ij} \:=  \frac{||\u{\rho}^{i}||}{||\u{\rho}^{j}||}\mma 
i,  \m  j = 1  \mbox{ to } n  \mma    \m  i \neq j \m   .  
\ee 
Note also the {\it unit relative Jacobi vectors} 
\be
\u{\hat{\rho}^i} :=  \frac{\u{\rho}^i}{||\u{\rho}^{i}||} 
                          =  \frac{\sqrt{\mu_i}\u{R}^i}{||\sqrt{\mu_i}\u{R}^{i}||} 
						  =  \frac{\sqrt{\mu_i}}{\sqrt{\mu_i}}  \frac{\u{R}^i}{||\u{R}^i||} 
                          =: \frac{\u{R}^i}{||\u{R}^i||}                                     \m  .  
\ee
(so these are the same irrespective of mass-weighting) since these feature in the expressions for relative angles in space, 
\be
\theta_{ij} = \mbox{arccos}
\left(
\u{\hat{\rho}}^i \cdot \u{\hat{\rho}}^j
\right) \m  
\ee
(which are therefore also independent of whether one mass-weights).  
These generalize the {\it unit relative Lagrangian vectors} 
\be
\u{\hat{r}_{IJ}} :=  \frac{\u{r}_{IJ}}{||\u{r}_{IJ}||} 
\ee 
familiar from basic force law expressions.  

\m 

\n{\bf Remark 1} Both of the above are unit vectors {\sl in space}.  

\m 

\n{\bf Remark 2} The ${\cal R}_{ij}$ -- length ratios -- {\sl are} in general affected by mass weighting: 
\be
\frac{||\u{\rho}^i||}{||\u{\rho}^j||}   =   \frac{||\sqrt{\mu_i}\u{R}^i||}{||\sqrt{\mu_j}\u{R}^j||}
                                        =   \sqrt{\frac{\mu_i}{\mu_j}}  \frac{||\u{R}^i||}{||\u{R}^j||}  
									  \neq  \frac{||\u{R}^i||}{||\u{R}^j||}                                   \m , 
\l{s-mws}
\ee
apart from on the relatively infrequent occasions for which $\mu_i = \mu_j$.  
This ratio ambiguity means that we need to pay careful attention to whether one's concepts are in space or mass weighted space. 
As we shall see in Sec \r{Z(3,1)}, this can incurr ambiguities or multiplicities of concept.

\m 

\n{\bf Remark 3} Our collective name for these coordinates is {\it simple ratio coordinates}.

\subsection{Normalized ratio coordinates}

\n{\bf Definition 1} Let us next introduce  
\be
\rho   :=  ||\brho||_{\Frr}  
      \:=  \sqrt{\sum_{i  = 1}^n\rho^{i \, 2}} 
	  \es  \sqrt{\sum_{i  = 1}^n I_i} 
	    =  \sqrt{\bigiota}                        \m  . 
\ee
The second expression here indicates relative space norm, where the bold $\brho$ is a relative space vector and thus carries $ia$ indices. 
The third just encapsulates that $\Frr$ carries the $nd$-dimensional Euclidean norm.
The fourth sums over (\r{rhoi-Ii}), and the fifth just expresses that the total moment of inertia is the sum of partial moments of inertia.  
This provides a natural normalization for the $\rho_i$, as follows.  

\m 

\n{\bf Definition 2} The {\it normalized mass-weighted relative Jacobi coordinates} are  
\beq
\nu^{i a}  \:=  \frac{\rho^{i a}}{\rho} \m  .  
\l{n-def}
\eeq
This is the relational space, rather than space, version of a unit vector, and represents another natural way of forming dimensionless ratio variables out of the $\rho_i$. 

\m 

\n{\bf Remark 1} $\rho$ has been named {\it hyperradius} in the Molecular Physics and Mathematical Physics literatures \c{MFII}. 
The `hyper' part of this name is not however very descriptive, 
so I refer to this by the more memorable and conceptually meaningful name of {\it configuration space radius}.  

\m 

\n{\bf Motivation 1} A first reason for $\rho$ meriting this name is that when the configuration space is relative space: 
$\rho$ is a {\it relative space radius} since it has the `square-root of a sum of squares' form that radial variables take in ($k$-dimensional) polar coordinates.  

\m 

\n{\bf Motivation 2} A second reason is that if we work in $\u{\rho}_i$ coordinates, preshape space is realized by a radius-$\rho$ sphere 
by the defining part of Remark 1.  
Thus $\rho$ is also {\it preshape space radius}, and we have derived the following. 

\m

\n{\bf Remark 2} We now follow up Subsection \r{Dimension}-\r{Smallest-Nontrivial}'s dimension counts by identifying tractable topological manifolds 
and metric geometries as per Fig \r{Shape-Triviality-Count}.c)-d) \c{FORD}. 

\m 

\n {\bf Lemma 1} Preshape spaces are topologically spheres,  
\beq
\FrP(N, d) \m   \s{t}{=} \m  \mathbb{S}^{n \, d - 1} \m  .
\l{1-d-spheres}
\eeq 
Here $\s{\st}{=}$ denotes `equals as a topological manifold'. 
See also Fig \r{Preamble-Support}.a).

\m 

\n{\bf Remark 3} On the other hand, if we work in $\u{\nu}^i$ coordinates, 
we have passed to a standardized unit-radius realization of the preshape sphere, with
\be
\nu^{i a \, 2} = 1
\ee
as its on-sphere condition.

\subsection{Simple versus normalized ratio coordinates}

\n{\bf Remark 1} Let us now begin a comparison of the previous two subsections' dimensionless coordinates. 
One advantage of the $\nu^{i a}$ is that they are coordinate-choice-democratic; 
one uses all of these alongside the above on-sphere condition, which is itself isotropic in the $\nu^{i a}$ and thus itself a democratic condition.
On the other hand, Definition 1's ratios require various undemocratic choices: which way up the simple ratios are defined, 
                                                                               as well as which basis set of simple ratios to take.  

\m 

\n{\bf Remark 2} The simple ratios moreover tie in well with preshape space's projective structure, whereas the existence of the $\nu^{i a}$ 
reflect that the preshape space is (part of) a sphere.  
Of couse, spheres admit real-projective coordinates, accounting for the presence of both kinds of useful coordinate systems.  
In the projective context, simple ratios are known furthermore by the geometrical and algebraic name of {\it inhomogeneous coordinates} \cite{Nakahara}.

\m 

\n{\bf Remark 3} If one expresses the $\nu^{i a}$ as unit Cartesian coordinates in hyperspherical polar coordinates, 
these spherical polar coordinates are the $\theta_k$ featuring in metric (\r{HS}).  
On the other hand, if one uses a basis choice among the simple ratio coordinates, 
one obtains this metric in the slightly more involved -- but recognizable -- `Beltrami coordinates' form \c{FORD, FileR}.  

\m 

\n{\bf Remark 4} A further reason for considering the simple ratio coordinates is their featuring in Sec \r{Coll} to \r{Approx-Cl-Uni}'s expressions.

\subsection{Preshape space's topology and geometry}

\n{\bf Proposition 1} $\FrS(N, 1) = \mathbb{S}^{n - 1}$ and $\FrP(N, d) = \mathbb{S}^{nd - 1}$ 
admit a natural realization of the standard {\it (hyper)spherical metric} 
\beq
\d s^2_{\sss\sp\sh\se} = \sumpn \prod\mbox{}_{\mbox{}_{\mbox{\scriptsize $m$ = 1}}}^{p - 1} \mbox{sin}^2\theta_m \d\theta_p^2 \m  .
\l{HS}
\eeq

\section{Shape space topology and geometry}\l{MLS-S}

\n{\bf Lemma 1}     
\beq
\FrS(N, d) \m   \s{\sm}{=} \m  \mathbb{S}^{nd - 1} \m  ,  
\eeq
both topologically and metrically. 

\m 

\Proof
${\FrS}(N, 1) = \FrP(N, 1)$ since there are no rotations in 1-$d$, and then use subcases of (\r{1-d-spheres}) and (\r{HS}). $\Box$

\m 

\n {\bf Corollary 1} The 1-$d$ shape spaces' isometry groups are 
\be
Isom(\FrS(N, 1) = SO(n) \m  . 
\ee

\m 

\n {\bf Theorem 1} Shape space $\FrS(N, d)$ is compact, connected and path-connected \c{Kendall}.

\m 

\Proof Preshape space $\FrP(N, d) \, \s{\sm}{=} \, \mathbb{S}^{nd - 1}$ elementarily has these properties, 
which are moreover preserved by quotients and thus in particular by $\FrP(N, d)$.

\m 

\n This subsection's quotient spaces are taken to be not just sets but also normed spaces.   
               
\m    

\n {\bf Remark 3} Quotienting out a 2-$d$ rotation is moreover straightforward, due to one of the well-known generalizations of the Hopf map \c{Hopf} and fibre bundle.  

\m  

\n {\bf Definition 2} The {\it Hopf fibration} itself is 
\beq
\frac{\mathbb{S}^3}{\mathbb{S}^1} \m   \s{\sm}{=} \m  \mathbb{S}^2 \m  .
\eeq 
Or, in short exact sequence \c{Hatcher} form, 
\be
1 \rightarrow \mathbb{S}^1 \rightarrow \mathbb{S}^3 \rightarrow \mathbb{S}^2 \rightarrow 1 \m  .  
\ee
See also Fig \r{Preamble-Support}.d).
In the context of Shape Theory, this gives the following.

\m 

\n {\bf Proposition 1}
\beq
\FrS(3, 2)  \es  \frac{\FrP(3, 2)}{SO(2)}  
            \es  \frac{\mathbb{S}^{3}}{\mathbb{S}^1} \m   
			\s{\sm}{=} \m  \mathbb{S}^2                  \m  .
\eeq
\n {\bf Definition 3} One generalization of the Hopf fibration is 
\beq
\frac{\mathbb{S}^{2 \, p + 1}}{\mathbb{S}^1} \m   \s{\sm}{=} \m  \mathbb{CP}^p \m  .
\eeq
\n{\bf Remark 4} The `accidental relation' 
\beq
\mathbb{S}^2 \m   \s{\sm}{=} \m  \mathbb{CP}^1 \m  ,
\l{S2=CP1}
\eeq 
turns out moreover to lead to the following generalization.

\m 

\n {\bf Definition 3} The {\it generalized Hopf fibration} is 
\beq
\frac{\mathbb{S}^{2p + 1}}{\mathbb{S}^1} \m   \s{\sm}{=} \m  \mathbb{CP}^p \m  ,   
\eeq 
or, in short exact sequence form, 
\be
1 \rightarrow \mathbb{S}^1 \rightarrow \mathbb{S}^{2p + 1} \rightarrow \mathbb{CP}^{p} \rightarrow 1 \m  .  
\ee
\n In the context of Shape Theory, this gives the following.

\m 

\n {\bf Proposition 2}
\beq
\FrS(N, 2)  \es  \frac{\FrP(N, 2)}{SO(2)}  \es  \frac{\mathbb{S}^{2n - 1}}{\mathbb{S}^1} \m   \s{\sm}{=} \m  \mathbb{CP}^{n - 1} \m  .
\eeq
\n{\bf Remark 4} The `accidental relation' (\r{S2=CP1}) means that triangleland is atypically simple at the topological level 
                                                                                          (and subsequently at the metric level). 
Thus many of the generic features of $N$-a-gonland only start to show up upon consideration of the quadrilateralland $\mathbb{CP}^2$.
This is the first motivation for the current series of articles to document not just triangleland but quadrilateralland as well. 

\m 

\n{\bf Proposition 3} (\r{HS}) is moreover also the 1-$d$ shape space metric. 

\m 

\n{\bf Remark 5} $Sim(d)$'s invariants are ratios of scalar products. 
Thus we look among the previous subsection's ratio coordinates for those which are built solely from scalar products. 
These are the length ratios ${\cal R}_{ij}$ and the relative angles ${\theta}_{ij}$, 
but not the unit vectors $\u{\hat{\rho}^i}$ or the component ratios ${\cal S}_{i a j b}$.  

\m 

\n{\bf Proposition 4} The $\FrS(N, 2) = \mathbb{CP}^{n - 1}$ admit a natural realization of the standard {\it Fubini--Study metric} \c{Kendall}, 
\beq
\d s^2_{\sF\sS} = 
\frac{\{1 + ||\mbox{\boldmath$Z$}||_{\sC}^2\} ||\d\mbox{\boldmath$Z$}||_{\sC}^2 -  |(\mbox{\boldmath$Z$} \cdot \d \mbox{\boldmath$Z$})_{\sC}|^2} 
{\{1 + ||\mbox{\boldmath$Z$}||_{\sC}^2\}^2} \m  .
\l{FS-2}
\eeq
The $\sC$ suffix here denotes the $\mathbb{C}^{n - 1}$ version of inner product and norm, with $Z_{\bar{p}}$'s indices running over $n - 1$ copies of $\mathbb{C}$.
\be
Z_{i} = {\cal R}_{i} \mbox{exp}(i\slPhi_{i}) \m  :
\l{Zpolar}
\ee 
the `multiple copies of $\mathbb{C}$ plane-polar coordinates' version of ratios of the $\u{\rho}_i$.
Therein, the $\slPhi_{i}$ are an independent set of $n$ relative angles $\theta_{ij}$ between $\u{\rho}_i$'s, 
whereas the ${\cal R}_{i}$ are the corresponding set of $n$ ratios of magnitudes $||\u{\rho}_i||$ \c{FileR}.
Thus in 2-$d$ relative angles and ratios of magnitudes occur in 1 : 1 proportion as modulus--phase pairs.  

\m 

\n{\bf Remark 6} These Fubini--Study coordinates are furthermore also projective 
(now complex-projective, and indeed a close analogue \c{FileR} of the above-mentioned Beltrami coordinates; 
both the ${\cal R}_{ij}$ and the $\mbox{\boldmath$Z$}$ are examples of the inhomogeneous coordinates in common use in Projective Geometry).     
 
\m  

\n{\bf Remark 7} Each of the above metrics is presented in {\sl very} standard coordinates for the corresponding {\sl shape space} geometries: 
hyperspherical angles and inhomogeneous coordinates respectively. 
Yet in the current Shape Theory setting, these coordinates have {\it additional spatial geometry meaning} in terms of the spatial coordinates describing the 
distribution of points or particles within space itself: see \c{FileR}.  
This means that {\sl very standard and highly developed mathematics of spheres and complex-projective spaces} can be used to solve 
{\sl a vast amount of new problems about inhomogeneity, planar shapes, and associated Probability, Statistics, Dynamics and Quantum Theory questions} 
\c{Kendall, FileR, ABook}.  

\m 

\n{\bf Remark 8} The Fubini--Study metric is of constant curvature.

\m 

\n{\bf Corollary 2} 
\be
Isom(\FrS(N, 2))  \es  \frac{SU(n)}{\mathbb{Z}_n} \m  , 
\ee
the quadrilateral case of which is mathematically the same as \c{MacFarlane, QuadI} the colour group in the theory of the strong force. 

\m 

\n{\bf Corollary 3} For triangleland -- the smallest model with nontrivial relative angle information -- the Fubini--Study metric furthermore simplifies to 
\beq
\d s^2  \es  \frac{\d Z^2}{\{1 + |Z|^2\}^2} 
         =    \d \slTheta^2 + \mbox{sin}^2\slTheta \, \d\slPhi^2 \m  ;
\l{Tri-Sphericals}
\eeq
The last of these expressions uses the following coordinates. 
\be
\slPhi = \mbox{arccos}
\left( 
\u{\hat{\rho}}_1  \cdot  \u{\hat{\rho}}_2 
\right) \m  , 
\l{Swiss}
\ee
which is further explained in Fig \r{Relational-Coordinates}, plays the role of polar angle $\slPhi \in [0, 2\pi$).   

\m 

\n The venerable substitution 
\be 
{\cal R} = \mbox{tan} \, \frac{\Theta}{2} \m   
\l{RTheta}
\ee 
then serves to convert the stereographic radius to the standard azimuthal coordinate $\Theta$ (about the U-axis).  
Inverting,  
\be 
\slTheta := 2\,\mbox{arctan} \, {\cal R} \m : 
\l{tan}
\ee
an azimuthal angle $\slTheta \in (0, \pi)$.
This casts the shape sphere metric into the standard spherical metric form,  
\be 
\d s^2 = \d {\Theta}^2 + \mbox{sin}^2 \Theta \, \d\Phi^2 \m  . 
\ee
Finally note the complex presentation 
\be
Z = {\cal R} \, \mbox{exp}(i \, \slPhi) \m  .  
\ee 
so as to make contact with eq. (\ref{Tri-Sphericals}).  
%
{           \begin{figure}[!ht]
\centering
\includegraphics[width=0.65\textwidth]{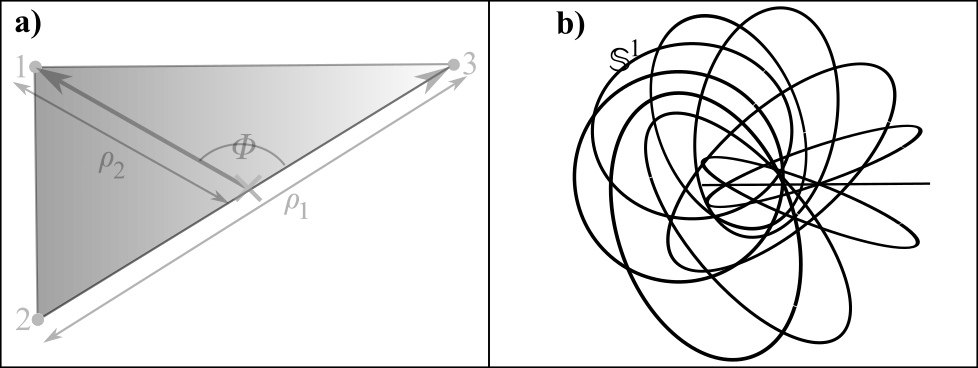}
\caption[Text der im Bilderverzeichnis auftaucht]{  \footnotesize{a) For 3 points in 2-$d$, use as coordinates the magnitudes of the two relative Jacobi coordinates. 
$\slPhi$ is the `Swiss-army-knife' relative angle between the two relative Jacobi vectors, given by formula (\r{Swiss}).  
%
%
These three coordinates {\sl do not} make reference to absolute axes A.  
Pure-shape coordinates are then the relative angle $\slPhi = \theta_2 - \theta_1$ and some function of the ratio quantity ${\cal R} = \rho_2/\rho_1$, 
among which formula (\r{tan})'s is geometrically clearest.    

\m 

\n b) Sketch of some of the $\mathbb{S}^1$ fibres in the Hopf bundle.
}   }
\l{Relational-Coordinates} \end{figure}         } 

\section{Relational space's topology and geometry}\l{MLS-R}
%
%
{            \begin{figure}[!ht]
\centering
\includegraphics[width=0.87\textwidth]{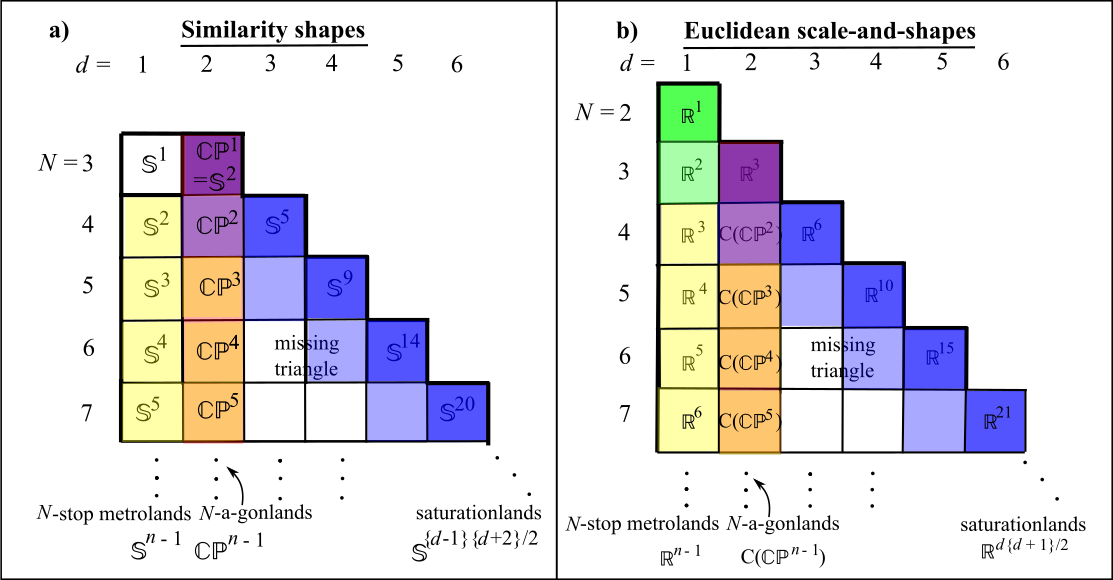}
\caption[Text der im Bilderverzeichnis auftaucht]{        \footnotesize{a) and b) are pure-shape and Shape and Scale  topological manifolds 
(\c{FileR} summarizes further topological results about shape and relational configuration spaces).  
This gives 3 tractable series per table.  

\m 

\n 1) The 1-$d$ such: a yellow column which we term 
{\it N-stop metroland} models since their configurations look like underground train lines, which are topologically and metrically spheres. 

\m 

\n 2) The 2-$d$ ones: an orange column which we term 
{\it N-a-gonland} models since their configurations are planar $N$-sided polygons, which are complex-projective spaces.

\m 

\n 3) Fig \r{r-types}'s linear independence saturation diagonal, alias Casson diagonal, in deep blue 
{\it saturationland} models, which Casson furthermore showed to also be spheres at the topological level \c{Kendall}.

\m 

\n Moreover, only 1) and 2) transcend to tractable series of metric geometries.    

\m 

\n Dynamical and relational nontriviality already covered why pure-shape (3, 1) (white) and (4, 1) (top yellow) models are special minimal cases, 
and partly likewise for scaled (2, 1) and (3, 1) models.
The scaled (2, 1) model (deep green) are furtherly special through being the sole intersection of the metroland and the satuarationland series.  

\m 

\n (3, 2) --   {\it triangleland}  -- (deep purple) is mathematically very special through being the sole nontrivial intersection of the $N$-a-gonland and satuationland, 
which is possible by virtue of (\r{S2=CP1}).
One must however then turn to (4, 2) -- {\it quadrilateralland}  \c{QuadI} -- to have the first mathematically-generic $N$-a-gonland: 
$\mathbb{CP}^k$ mathematics that is not equivalent to spherical mathematics.  

\m 

\n (4, 3) -- {\it tetrahaedronland} -- (pale purple) is also rather mathematically special through 
being the first nontrivial saturationland which is not also an $N$-a-gonland.

\m 

\n But to have a 3-$d$ model with the generic feature of not being saturated -- and consequently of not having a merely spherical shape space -- 
we must turn to (5, 3) (top pale blue).  
This offers a shape-theoretic explanation -- a {\sl partial} explanation -- for many aspects of the 3-$d$ 5-body problem \c{Roberts, LS09} 
being vastly more complex than the 3-$d$ 4-body problem \c{Albouy}.\f{To these references, 
let us add the extra observations that almost all Molecular Physics papers working with a specific $N$ are restricted to $N \leq 4$, 
whereas quite a few Celestial Mechanics papers about 5-bodies in fact restrict themselves to $d = 2$.
On the other hand, the Painlev\'{e} conjecture was established by Jeff Xia \c{Xia} for $N \geqs 5$, whereas it remains an open question for $N = 4$; 
this is however both about a {\sl complicating feature} and a {\it dynamical matter} including reference to a {\sl particular choice of potentials}: gravitational. 
The most salient reason for partialness of explanations in terms of the nature of the underlying shape space is that some matters additionally require 
potential function inputs, most usually of inverse powers of the separations: gravitational in Celestial Mechanics or Coulomb in Molecular Physics.}
}   }
\l{Shape-Triviality-Top} \end{figure}         } 

\n {\bf Definition 1} A generalized notion of {\it cone} over some topological manifold $\FrM$ is
\beq
\mC(\FrM) = \FrM \times [0, \infty)/\m  \w{\m } \m  . 
\eeq
$\w{\m }$ \m  here means that all points of the form \{p $\in$ \FrM, 0 $\in [0, \infty)$\} 
are `squashed' or identified to a single point termed the {\it cone point}, 0 (Fig \r{Preamble-Support}.b-c).
%
{            \begin{figure}[!ht]
\centering
\includegraphics[width=0.54\textwidth]{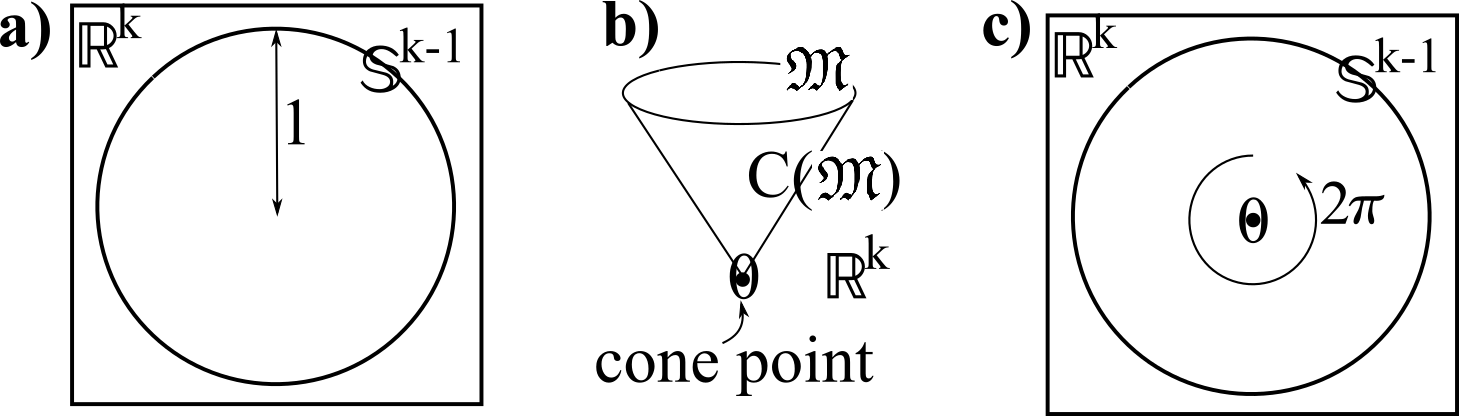}
\caption[Text der im Bilderverzeichnis auftaucht]{        \footnotesize{a) Preshape sphere from quotienting out scale. 

\m 

\n b) The general cone; this gives back c)'s $\mathbb{R}^k$ when over $\mathbb{S}^{k - 1}$ with the all-space angle $2^k\pi$ 
rather than some deficit angle at its cone point.} }
\l{Preamble-Support}\end{figure}            }
	
\m 	

\n{\bf Proposition 1} The Euclidean shape-and-scale spaces are topological cones \c{Kendall, FileR} over the corresponding shape spaces, 
\beq
{\cal R}(N, d) = \mC(\FrS(N, d)) \m  . 
\l{relational-space-cones}
\eeq
\n {\bf Corollary 1} 
\beq
{\cal R}(N, 1) = \mC(\FrS(N, 1)) = \mC(\mathbb{S}^{N - 2}) = \mathbb{R}^{N - 1} \m  .
\l{1-d-relational-spaces}
\eeq
\n {\bf Corollary 2}
\beq
\lFrr(N, d) = \mC(\FrP(N, d)) = \mC(\mathbb{S}^{nd - 1}) = \mathbb{R}^{N - 1} \m  .  
\eeq
\n {\bf Remark 1} This is the last result for general $(N, d)$.
I.e.\ $\FrQ(N, d)$, $\Frr(N, d)$ and $\FrP(N, d)$ all have explicit and tractable general forms, 
whereas ${\FrS}(N, d)$ and ${\cal R}(N, d)$ do not. 
In this way, these last two benefit considerable from indirect formulations as quotients of more tractable spaces.

\m 

\n {\bf Corollary 3} 
\beq
{\cal R}(N, 2)             =      \mC({\FrS}(N, 2)) 
              \m   \s{\sm}{=} \m  \mC(\mathbb{CP}^{n - 1}) \m  :
\eeq 
a tractable albeit hitherto somewhat less studied series. 

\m 

\n{\bf Remark 2} Among these, 
\beq
{\cal R}(3, 2) = \mC({\FrS}(3, 2)) = \mC(\mathbb{S}^{2}) = \mathbb{R}^3 \m  
\eeq 
is an atypically simple case.  

\m 

\n{\bf Remark 3} See Fig \r{Shape-Triviality-Top} for an overview of the tractable series of shape(-and-scale) spaces.

\m 

\n{\bf Definition 2} The {\it cone over a metric manifold} $\langle\FrM, \bm \rangle$  has a natural line element of the form 
\beq
\d s^2_{\scc\so\sn\se} := \d \sigma^2 + \sigma^2 \d s_{\sbm}\mbox{}^2\mma \mbox{ (for } \m  
\sigma \in \mathbb{R}_0 =  [0, \infty) \m  \mbox{ a `radial' coordinate) }         \m  , 
\l{CSR}
\eeq
where $\d \sigma$ is the line element corresponding to $\bm$ itself.  

\m 

\n{\bf Proposition 2} Relational space ${\cal R}(N, d)$ has the cone metric over the corresponding shape space metric \c{FileR}, 
with the $\rho$ of equation (\r{CSR}) playing the role of radius.  

\m 

\n{\bf Motivation 3} $\rho$'s conceptual name `configuration space radius' is thereby further justified.
For now with relational space playing the role of configuration space, $\rho$ plays the role of {\it relational space radius} 
radial coordinate thereupon -- {\it relational space radius} --  and a fortiori of {\it cone radius} 
for the cone structure underpinning the geometrical nature of relational space.
Cone radius moreover ceases to be merely a matter of coordinate choice whenever the cone in question carries a conical singularity, 
for then $\rho = 0$ pinpoints the location not of a coordinate artifact but of a geometrical phenomenon.  

\m  

\n{\bf Motivation 4} $\rho$ is finally the {\it shape space radius}, 
in the sense of the radial value taken throughout a given shape space representative within relational space.  

\m  

\n{\bf Remark 4} Let us finally account for $\rho$ playing a joint role as radius in all of relative space, preshape space, relational space and shape space 
(this is a new result).
These form two (space, cone over space) pairs, within each of which radius is shared due to its privileged role in the metric-level cone construction. 
The two are moreover linked because of $Rot(d) \times Dil(d)$'s direct product form.
By this, relative space's {\it scale--preshape split} -- which is geometrically obvious due to its geometrical nature of an embedding of a sphere 
into a real space with one dimension more -- projects down to relational space's {\it scale--shape split}.  
For quotienting out $Rot(d)$ has no effect whatsoever on its direct product cofactor Dil.

\m 

\n{\bf Remark 5} Proposition 2's coning construct is moreover independent of which shapes it is being adjoined to.  
This gives a sense in which scale is but a spurious appendage at the {\sl mathematical} level 
(a position for which distinct {\sl conceptual} level arguments were given by Barbour in e.g.\ \c{B03}).  
This sense applies for instance to all notions of shape corresponding to sugroups of the affine group, 
indeed including the current treatise's similarity shapes.

\m 

\n{\bf Corollary 4}
\be
\mC({\FrS}(N, 1)) = {\cal R}(N, 1)  \mbox{ is just } \mathbb{R}^{n} \m  , 
\ee
admits a cone metric 
\be
\d s^2 = \d \rho^2 + \rho^2 \d_{\sss\sp\sh\se}^2 \m  , 
\ee
which is moreover the flat metric.

\m 

\n{\bf Corollary 5}
\be
{\cal R}(N, 2) = \mC(\mathbb{CP}^{n - 1})
\ee 
admits the cone metric 
\beq
\d s^2 = \d \rho^2 + \rho^2 \d_{\sF\sS}^2 \m  . 
\eeq 
\n{\bf Corollary 6} The scaled version of triangleland is the cone over the pure-shape case's configuration space,    
\beq
\d s^2 = \d \rho ^2 + \frac{\rho^2}{4}\left\{\d \slTheta^2 + \mbox{sin}^2\slTheta_2 \d \slPhi^2\right\} 
        \es  \frac{1}{4\,\bigiota}\left\{\d \bigiota^2 + \bigiota^2\{\d \slTheta^2 + \mbox{sin}^2\slTheta \d \slPhi_2^2\}\right\} \m  . 
\l{Tri-Scale}
\eeq
{\bf Remark 6} While $\mC({\FrS}(3, 2))$ is topologically $\mathbb{R}^3$, it is not equipped with the flat metric.  
Its metric is however conformally flat \c{FileR}: just apply the conformal factor $4 \, \bigiota$ to the second form of (\r{Tri-Scale}).
%
{            \begin{figure}[!ht]
\centering
\includegraphics[width=0.7\textwidth]{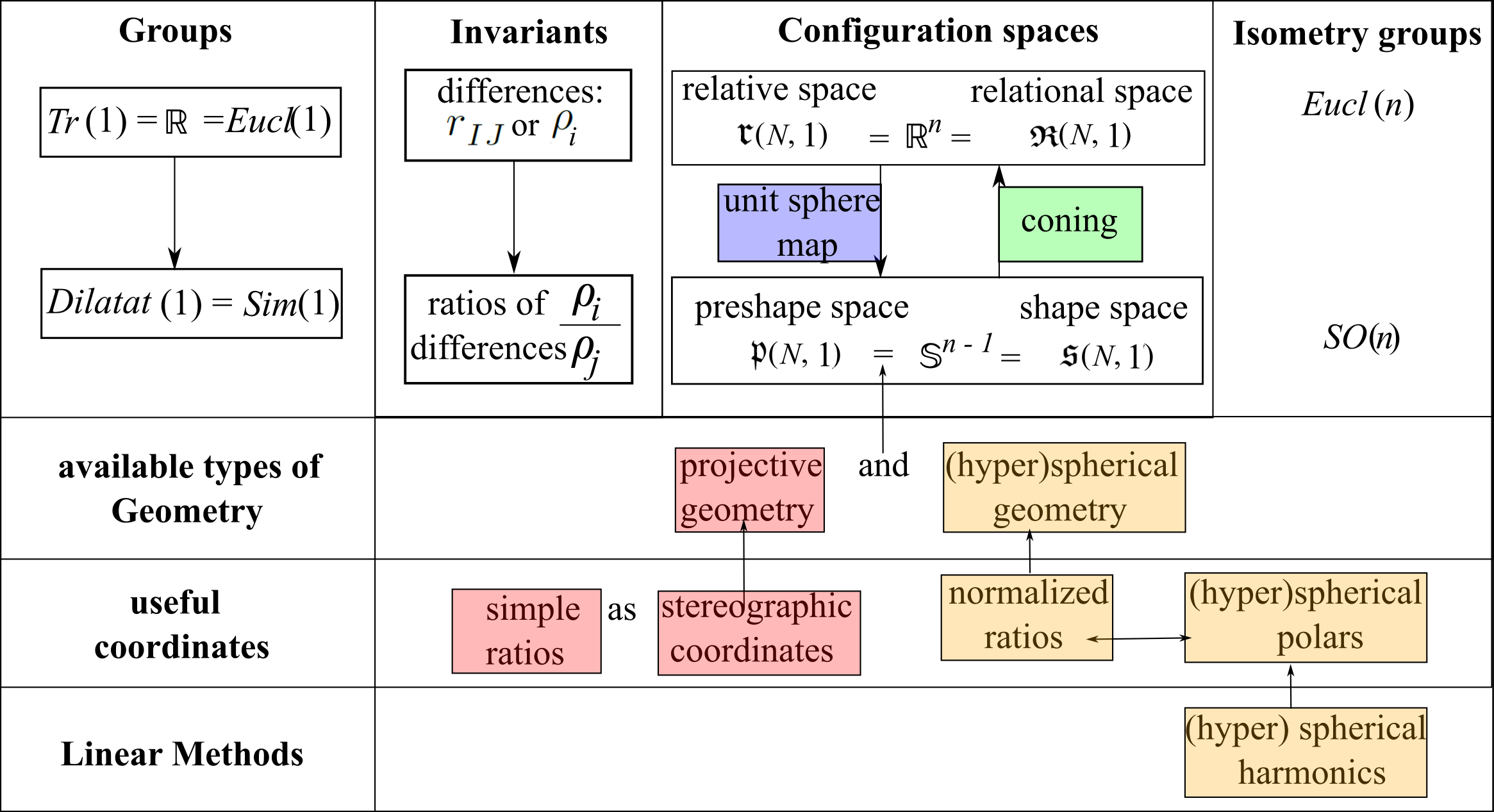}
\caption[Text der im Bilderverzeichnis auftaucht]{        \footnotesize{Simplified version of Fig \r{Group-Inv-Config} for 1-$d$, 
followed by an outline of how $N$-stop metrolands are very straightforward to work with at the metric level.
Moreover, hyperspherical becomes just spherical for $N = 4$ and circular for $N = 3$, so these two cases are particularly straightforward for work with.} }
\l{Group-Inv-Config-1d} \end{figure}          }
%
{            \begin{figure}[!ht]
\centering
\includegraphics[width=0.7\textwidth]{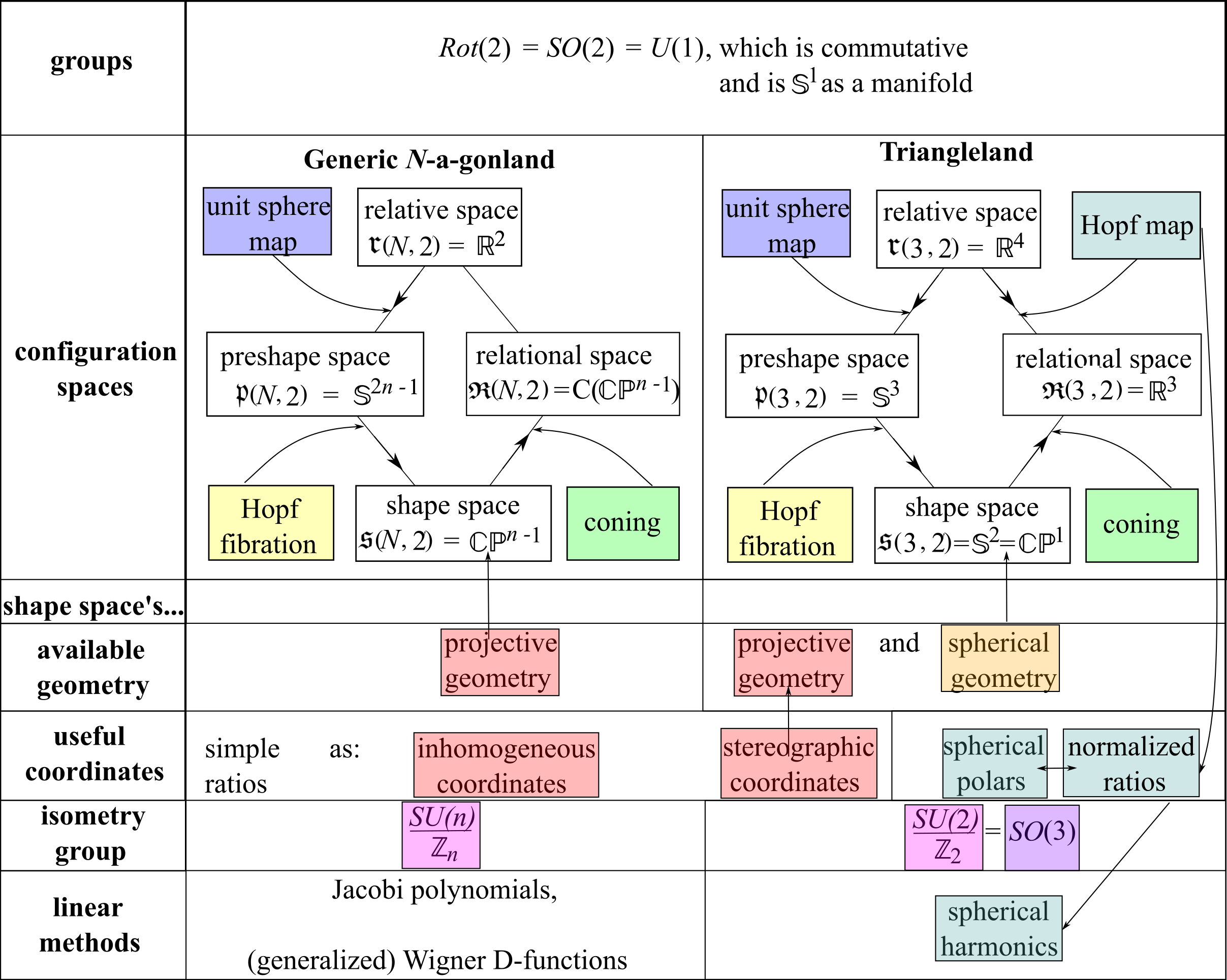}
\caption[Text der im Bilderverzeichnis auftaucht]{        \footnotesize{Summary of ways in which $N$-a-gonland models are special and triangleland models are extra special, 
From the third row onward, we outline how $N$-a-gonland is {\sl quite} straightforward to work with at the metric level, 
whereas triangleland is {\it very} straightforward.
See Part III for more about the Hopf map and \c{QuadII} for more about Jacobi Polynomials and (generalized) Wigner D-functions in the quadrilateralland context.
The general ($N$, $d$) model Fig \r{Group-Inv-Config} makes for good comparison for the second row, 
and the $N$-stop metroland Fig \r{Group-Inv-Config-1d} for all rows.} }
\l{Group-Inv--Tri} \end{figure}          }

\m 
	
\m 
	
\m 
	
\m 
	
\m 
																			   
\vspace{10in}
	
\section{Mirror image identified and indistinguishable reduced configuration spaces}\l{MII-Indis}

\subsection{Preshape spaces}

Now the previously encountered (mirror images distinct, distinguishable point-or-particle) preshape space $\FrP(N, d)$ is accompanied by the following.  

\m 

\n {\bf Definition 1}  The {\it mirror image identified preshape space} is  
\be
\FrO\FrP(N, d)  \es  \frac{\FrP(N, d)}{\mathbb{Z}_2}                                              \m  ,
\ee 
the {\it mirror image distinct indistinguishable preshape space} is 
\be
\FrI\FrP(N, d)  \es  \frac{\FrP(N, d)}{S_N}                                                       \m  ,
\ee
and the {\it mirror image identified indistinguishable preshape space} is 
\be
\FrO\FrI\FrP(N, d)  \es  \frac{\FrP(N, d)}{S_N \times \mathbb{Z}_2}                                \m  .  
\ee
\n{\bf Proposition 1} These four spaces are  
\be
\FrP(N, d) = \mathbb{S}^{nd - 1} \m  , 
\ee
\be
\FrO\FrP(N, d)  \es  \frac{\FrP(N, d)}{\mathbb{Z}_2} = \mathbb{RP}^{nd - 1} \m  ,
\ee 
\be
\FrI\FrP(N, d)  \es  \frac{\FrP(N, d)}{S_N}  \es  \frac{1}{N!}\mbox{-(sector of) } \m \mathbb{S}^{nd - 1} \m  ,
\ee
and 
\be
\FrO\FrI\FrP(N, d)  \es  \frac{\FrP(N, d)}{S_N \times \mathbb{Z}_2}  \es  \frac{1}{2 \times N!}\mbox{-(sector of) } \m \mathbb{S}^{nd - 1} \m  . 
\ee
The first two of these are topological and metric level equations.
The last two are topologically $\{nd - 1\}$-balls and geometrically as given. 
All four naturally admit the standard spherical metric.  

\m 

\n{\bf Remark 1} The third and fourth are of {\it spherical prime} form
\be
\frac{\mathbb{S}^p}{\Gamma} \m  \mbox{ for $\Gamma$ a finite group } \m  : 
\l{Spherical-Primes}
\ee 
a notion studied elsewhere in Mathematics and Physics (group actions on manifolds \c{AMP, Kobayashi}, orbifolds \c{Thurston}...)

\subsection{Shape spaces}

\n{\bf Remark 1} This is a useful point at which to introduce and comment on the Shape Statistics literature's notation \c{GT09, PE16}. 
Now the mirror image identified distinguishable shape space is considered primary, and is thus reassigned the notation $\FrS(N, d)$. 
The mirror image distinct distinguishable shape space already discussed 
is now denoted rather by the topologically-standard double-cover notation $\w{\FrS}(N, d)$. 
Then 
\be 
\FrS(N, d)  \es  \frac{\w{\FrS}(N, d)}{\mathbb{Z}_2} \m  .
\ee
\n{\bf Remark 2} Indistinguishable particle spaces have hitherto been studied rather less; 
again we place primary focus the mirror image identified case first, 
calling it Leibniz space $\Leib_{\sFrS}(N, d)$ as per the Introduction's argument.  
\be
\Leib_{\sFrS}(N, d)  \es  \frac{\w{\FrS}(N, d)}{S_N \times \mathbb{Z}_2} \m  .  
\ee
\n{\bf Remark 3} We retain the notation $\FrI\FrS(N, d)$ for the mirror image distinct counterpart, 
since double-covering a Leibniz space goes back on what is special about Leibniz spaces.  
\be
\FrI\FrS(N, d)  \es  \frac{\w{\FrS}(N, d)}{S_N}   \m  .  
\ee                                                 

\m 

\n{\bf Corollary 1} In 1-$d$, 
\be
\w{\FrS}(N, 1)   =   \FrP(N, 1)   
                 =   \mathbb{S}^{n - 1}   \m  , 
\ee
\be
\FrS(N, 1)       =   \FrO\FrP(N, 1)  
                 =   \FrP(N, 1)   
				 =   \mathbb{RP}^{n - 1}   \m  ,
\ee 
\be
\FrI\FrS(N, 1)   =   \FrI\FrP(N, 1)  
                \es  \frac{1}{N!}\mbox{--(sector of) } \m \mathbb{S}^{nd - 1}   \m  ,
\ee
and 
\be
\Leib_{\sFrS}(N, 1)   =   \FrO\FrI\FrP(N, d)  
                     \es  \frac{1}{2 \times N!}\mbox{--(sector of) }   \m   \mathbb{S}^{nd - 1}   \m  . 
\ee
\n The first of these is trivial; the second is in \c{Kendall}.
Each has the same topological and metric status as in the previous subsection's Proposition.  

\m 

\n{\bf Remark 4} See \cite{A-Monopoles} for a notation capable of accounting for a more extensive range of models including partial distinguishability.  

\m 

\n{\bf Proposition 1} In 2-$d$,
\be
\w{\FrS}(N, 2)   =   \mathbb{CP}^{n - 1}                        \m  , 
\ee
\be
\FrS(N, 2)      \es  \frac{\mathbb{CP}^{n - 1}}{\mathbb{Z}_2}   \m  ,
\ee 
\be
\FrI\FrS(N, 2)  \es  \frac{1}{N!}\mbox{-(`sector' of) }   \m   \mathbb{CP}^{n - 1}   \m  ,
\ee
and 
\be
\Leib_{\sFrS}(N, 2)   =   \FrO\FrI\FrP(N, d)  
                     \es  \frac{1}{2 \times N!}\mbox{-(`sector' of )} \mathbb{CP}^{n - 1}   \m  . 
\ee 
The first of these was already in Sec \r{MLS-S}; the second is a simple example of \c{FileR} {\it weighted projective space} \c{WCP}.

\m 

\n{\bf Remark 5} The third and fourth are of the form 
\beq
\frac{\mathbb{CP}^p}{\Gamma} \m  \mbox{for $\Gamma$ a finite group }   \m  :  
\l{CP-Primes} 
\eeq 
a `complex projective primes' counterpart of the spherical primes, which, as far as the Author is aware, remains unstudied.  
We postpone further discussion of $\mathbb{CP}^n$ $n \geqs 2$ `sectors' to Part IV.  

\m 

\n{\bf Remark 6} In the upper triangular block beyond the saturation alias Casson diagonal, 
the now obligatorily  mirror image identified shape spaces repeat along each row ad infinitum.  
This further cements the upper triangular block's row-by-row indiscernibility while hilighting the reason for the diagonal itself to differ from the upper triangular block.
Namely, that if the dimension $\geqs N$, availability of a higher dimensional-rotation renders identification of mirror images inevitable. 
Whereas if the dimension is $n = N - 1$, this identification is a matter of choice, 
which is then not made for the $\w{\FrS}$ shape space currently under discussion but is made for the thus-distinct shape space $\FrS$.

\subsection{Relational spaces}\l{Rel-Sp}

\n{\bf Definition 1} Mirror image identified Euclidean shape-and-scale space is the cone 
\beq
{\cal R}(N, d) = \mC(\FrS(N, d)) = \mC
\left(
\frac{\FrS(N, d)}{\mathbb{Z}_2}
\right)                                     \m  .  
\eeq
\n{\bf Proposition 1} At the topological level
\beq
{\cal R}(N, 1) = \mC(\mathbb{RP}^{N - 2})   \m  , 
\eeq
and 
\beq
{\cal R}(N, 2) = \mC
\left(
\frac{\mathbb{CP}^{N - 2}}{\mathbb{Z}_2}
\right)                                     \m  : 
\eeq
another tractable albeit hitherto somewhat less studied series of spaces. 

\m  

\n Note that identification does not affect the local form of metric that these spaces carry (it can however affect which Killing vectors remain globally valid).  

\m 

\n{\bf Proposition 2} 
\beq
\FrI{\cal R}(N, 1) = \mC(\FrI\FrS(N, 1)) 
                   = \mC
\left(
\frac{\mathbb{S}^{n - 1}}{S_N}
\right)                                          \m  , 
\eeq
and  
\beq
\FrI{\cal R}(N, 2) = \mC(\FrI\FrS(N, 2))
                   = \mC
\left(
\frac{\mathbb{CP}^{N - 2}}{S_{N}}
\right)                                          \m  : 
\eeq 
another tractable albeit hitherto somewhat less studied series. 

\m 

\n{\bf Remark 1} In symmetry arguments in the rest of this treatise, equal masses or no notion of mass at all (position data points) are assumed.
This is a major and usefully simplifying feature. 

\m 

\n{\bf Proposition 3} 
\beq
\Leib_{\cal R}(N, 1) = \mC(\Leib_{\sFrS}(N, 1)) 
                     = \mC
\left(
\frac{\mathbb{S}^{N - 2}}{S_N \times \mathbb{Z}_2}
\right)                                                                  \m  , 
\eeq
and  
\beq
\Leib_{\cal R}(N, 2) = \mC(\Leib_{\sFrS}(N, 1))
                     = \mC
\left(
\frac{\mathbb{CP}^{N - 2}}{S_{N} \times \mathbb{Z}_2}
\right)                                                                  \m  : 
\eeq 
yet another tractable albeit hitherto somewhat less studied series. 

\m  

\n{\bf Observation 1} Note as a general feature that Leibniz spaces' labels are always all distinct; this indeed implements Leibniz's Identity of Indiscernibles.

\subsection{Discussion}\l{Approx-Discussion}

Approximate clumping, uniformity and merger notions correspond to regions in configuration space 
(see Sec \r{Approx-(3, 1)} for a simple concrete example).  
In many cases, this is a curved geometric region.
The correspondence in question is of course the `shape-in-space to position-in shape-space' correspondence. 
This correspondence moreover tells one how the small $\epsilon$ tolerance parameter in space maps to another small tolerance parameter $\eta$ in shape space.  

\m 
 
\n The main issue to mention at this point, moreover, is that it is by now clear that many instances of $\FrG\FrG$ have boundaries. 
By this, approximations to exact shape which lie on a shape space's boundary involve not whole small balls, but merely those pieces of them which are admissible. 
This needs to be taken into account in determining the approximate coincidence-or-collision, merger, and uniformity structures in those $\FrG\FrG$ with boundaries.

\section{Some General Relativity counterparts}\l{Gdyn}

\subsection{Analogy between shapes and GR configuration spaces}\l{Gdyn-Q}
%
{            \begin{figure}[!ht]
\centering
\includegraphics[width=1.0\textwidth]{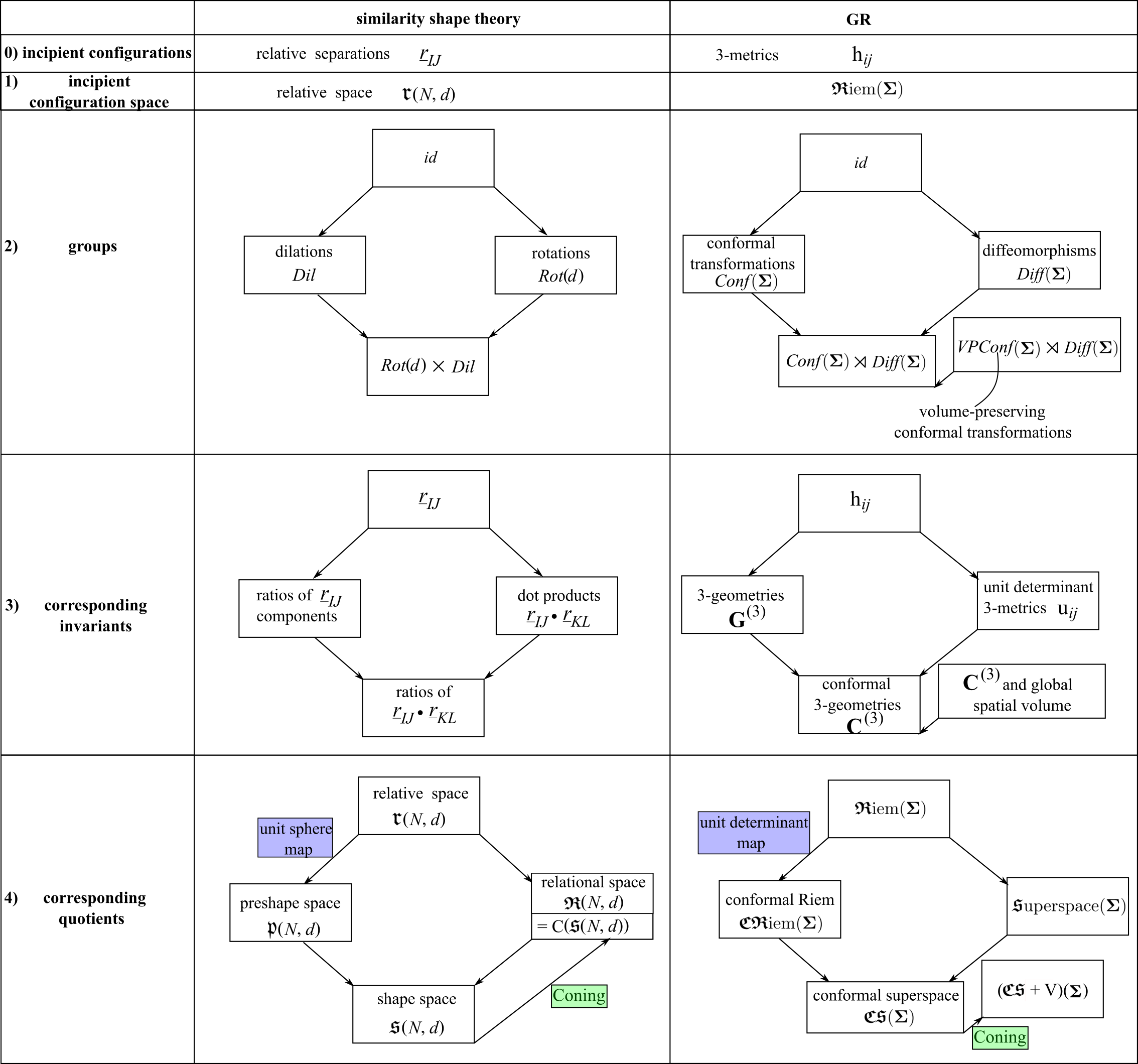}
\caption[Text der im Bilderverzeichnis auftaucht]{        \footnotesize{Similarity Shape Theory to General Relativity initial value problem analogy. 
Row 0) gives incipient notions of configuration, 
Row 1) the corresponding configuration spaces, 
row 2) analogous groups, 
row 3) the corresponding invariants and 
row 4) the corresponding reduced configuaration spaces. 
Note that the GR analogue of coning does not now lead back to an already-formulated configuration space, because there is now a local--global distinction.} }
\l{Shape-Gdyn} \end{figure}          }
																																			 
\n{\bf Analogy 0} GR's incipient configurations are Riemannian 3-metrics $\bh$ with components $\mh_{ab}(x^c)$ 
                   on a fixed 3-topology $\bupSigma$ interpreted as a spatial slice of spacetime 
				   (itself a semi-Riemannian 4-metric $\bg$ with components $\mg_{\mu\nu}$ on a 4-topology $\FrM$). 
As per Fig \r{Shape-Gdyn}, we take this to be most closely analogous to relative configurations $\{\rho^{ia}, i = 1 \mbox{ to } n\}$.  

\m 

\n{\bf Analogy 1} The totality of GR's $\bh$ on a fixed $\bupSigma$ constitute GR's incipient configuration space $\Riem(\bupSigma)$.  
Then on the one hand, the totality of the relative configurations form 
an incipient configuration space $\lFrr(N, d) = \mathbb{R}^{nd}$   
with a finite number of degrees of freedom, 
and which consists of a finite product of $n$ copies of an absolute space $\mathbb{R}^d$. 
On the other hand, $\Riem(\bupSigma)$ has an infinite number of degrees of freedom due to its dependence on the spatial point $x^c$, 
by which it consists of an infinite number of $3 \times 3$ symmetric matrices, one for each point in space.  
Each of these has 6 degrees of freedom. 
These are moreover not in general copies of each other because of the general 3-metric's spatial dependence (alias inhomogeneity).  
While relative space is equipped with the flat metric (\r{Unit-Metric}), $\Riem(\bupSigma)$ 
is provided with its own metric by GR's action in split space-time form \c{ADM}: the inverse DeWitt metric \c{DeWitt67} $\bM$ 
(after Quantum Gravity pioneer Bryce DeWitt; we do not require this object's precise form for the current treatise).  

\m  

\n{\bf Analogy 2} The spatial diffeomorphisms $Diff(\bupSigma)$ are GR's analogue of the rotations $Rot(d)$ in Mechanics 
as the group acting on the incipient configurations which it is most directly desirable to regard as physically meaningless.
$Diff(\bupSigma)$ is moreover far harder to handle at the technical level (see Chapter 35 ov \c{ABook} for a summary table of how).

\m 

\n{\bf Analogy 3} The $Diff(\bupSigma)$-invariant content of a 3-metric $\bh$ is then GR's analogue of the rotationally-invariant dot products. 
This content has 6 - 3 = 3 degrees of freedom, since a $3 \times 3$ metric has a 3-vector's worth of mere coordinate infomation. 
This remaining content is termed the {\it 3-geometry} $\bG^{(3)}$.

\m 

\n{\bf Analogy 4} In each case (subject to $d \geqs 3$ in Mechanics), this most desirable group is moreover a hard one to quotient out, causing stratification. 
The corresponding reduced configuration spaces are the absolute versus relational motion debate's relational space $\FrR(N, d)$ 
versus Wheeler's superspace, $\Superspace{(\bupSigma)}$ [already mentioned in the Introduction: eq. (\r{Intro-Superspace})].  
For GR, it is $\bh$ with nonzero Killing vectors -- spatial metrics with symmetries -- which form the nontrivial stratification.  
DeWitt gave a conceptual account of this in \c{DeWitt70}, 
with mathematical physicist Arthur Fischer concurrently providing a superbly detailed technical account \c{Fischer70}.  

\m 

\n{\bf Analogy 2$^{\prime}$} Both shape(-and-scale) theory and GR moreover have a second group which is more straightforward to quotient out.
In each case, this moreover bears some relation to scaling: the global scaling $Dil$ 
versus the local scaling $Conf(\bupSigma)$ of spatial conformal transformations.  

\m 

\n{\bf Analogy 3$^{\prime}$} The corresponding invariants are ratios of relative separations versus unit-determinant metrics 
\be
\buu  \es  \frac{\bh}{^3\sqrt{\mh}}
\ee
for $\mh := det \, \bh$.  

\m 

\n{\bf Analogy 4$^{\prime}$} Quotienting out instead $Dil$ and $Conf(\bupSigma)$ give, respectively, Kendall's preshape space $\FrP(N, d)$ 
versus \c{DeWitt67, FM96} conformal Riem $\CRiem(\bupSigma)$ (see also Chapter 21 of \c{ABook} for a conceptual discussion).  
Moreover, $\Riem(\bupSigma)$ and $\CRiem(\bupSigma)$ are mathematically simpler \c{DeWitt67, DeWitt70} than $\Superspace(\bupSigma)$ or $\CS(\bupSigma)$, 
with some parallels to how \c{Kendall} and the current treatise argue $\Frr(N, d)$ and $\FrP(N, d)$ to be simpler than ${\cal R}(N, d)$ and $\FrS(N, d)$.  

\m  

\n{\bf Analogy 2$^{\prime\prime}$} Both pairs of easy-and-hard groups moreover combine in the reasonably straightforward semidirect product form at the 
bottom of row 2 in Fig \r{Shape-Gdyn}. 

\m 

\n{\bf Analogy 3$^{\prime\prime}$} The corresponding invariants are now ratios of dots of differences versus conformal 3-geometries $\bC^{(3)}$ \c{York73}.  

\m 

\n{\bf Analogy 4$^{\prime\prime}$} One can at least formally guotient each pair out at once, 
giving Kendall's shape space $\FrS(N, d)$ versus mathematical physicist Jimmy York's \c{York72, York74} {\it conformal superspace} $\CS(\bupSigma)$.

\m 

\n{\bf Remark 1} It is worth commenting that while $Diff(\bupSigma)$ is associated with GR's momentum constraint, 
$Conf(\bupSigma)$ is merely associated with the maximal slicing equation $\mp = 0$ for $\mp$ the trace of GR's momentum.  
This does bear {\sl some} relation to GR's other constraint -- the Hamiltonian constraint -- in that on a maximal slice, this and the momentum constraint decouple; 
this was mathematical physicist Andr\'{e} Lichnerowicz's starting point \c{Lich} for considering GR's initial-value problem. 

\m  

\n{\bf Remark 2} For closed universes, however, $p = 0$ is frozen rather than maintainable by the evolution.  
York remedied this by showing that constraint decoupling persists for constant mean curvature slices 
\be 
\mK \m \propto \m \frac{\mp}{\sqrt{\mh}} = const
\ee
(where $\mK$ is the extrinsic curvature of the slice within spacetime).  

\m 

\n{\bf Analogies 2$^{\prime\prime\prime}$, 3$^{\prime\prime\prime}$ and 4$^{\prime\prime\prime}$} The corresponding group, invariants and quotient are, 
respectively, the (global spatial) {\it volume-preserving conformal transformations} $VPConf(\bupSigma)$, 
              the conformal geometries with solitary global spatial volume adjoined, 
			  and {\it conformal superspace plus volume} \c{YorkTime1, York73, ABFKO} $\{\CS + V\}(\bupSigma)$.  
For shape(-and-scale) models, adding back in a scale degree of freedom has no local--global distinction and so returns one to relational space. 
But GR enjoys such a distinction, whereby $\Superspace(\bupSigma)$ and  $\{\CS + V\}(\bupSigma)$ are in general not at all the same configuration space.  	  

\m 

\n{\bf Remark 3} The conventional GR initial problem stance is moreover as follows.   

\m 

\n 1) Decouple the constraints on a CMC slice.

\m 

\n 2) This decoupling is such that one can solve the GR momentum constraint first.  

\m 

\n 3) One then substitutes this solution into the scheme's conformally-transformed GR Hamiltonian constraint 
-- the so-called {\it Lichnerowicz--York equation} \c{York72, York73} -- which is solved as an equation for the local (conformal) scale factor. 

\m 

\n Thus while one temporarily makes use of the $\{\CS + V\}(\bupSigma)$ configuration space, the last step -- solving the Hamiltonian constraint -- 
{\sl breaks} the conformal symmetry, so that one ends up having solved GR's Hamiltonian constraint {\sl instead of} the constant mean curvature condition.  

\m  

\n Thereby, there is plenty of motivation to consider quotienting Superspace by the Hamiltonian constraint rather than by the maximal or constant mean curvature 
slicing conditions. 
On the one hand, this corresponds to what GR as a dynamical system requires.
On the other hand, for all that the GR initial value problem involves two other quotients in its first steps, this approach itself reverts to 
a quotient by the Hamiltonian constraint in its last step.  
Quotienting out the Hamiltonian constraint is moreover more complicated, so we postpone discussion of this point to the next subsection.  

\m 

\n{\bf Analogy 5} In spatially open models, large diffeomorphisms play an analogous role to this paper's discrete transformations, 
by which multiple versions of $\Riem(\bupSigma)$, $\CRiem(\bupSigma)$ $\Superspace(\bupSigma)$, $\CS(\bupSigma)$ and $\{\CS + V\}(\bupSigma)$ come into use.  
Taking into account large diffeomorphisms gives what Fischer and mathematical physicist Vincent Moncrief term quantum (conformal) superspace, 
in connection with the expected conceptual and technical relevance of these particular variants in formulating quantum General Relativity.
This further motivate the current treatise's detailed pursuit of these spaces' Leibniz space analogues, 
which are simple enough to admit a precise analytic quantum treatment \c{AF, +Tri, FileR, QuadII, QLS, Quantum-Triangles}.

\subsection{Shape Theory in the study of Background Independence}\l{BI}

\n{\bf Observation 1} A sizeable part of Background Independence concerns quotienting by automorphism groups and consequences thereof.  

\m 

\n{\bf Observation 2} Automorphisms acting on space and thus on configuration space in $N$-particle models corresponds to the usual Shape Theory,
\be
(\mbox{general shape space})  \es  \frac{\times_{I = 1}^N\Space(d)}{Aut\langle\Space(d), \sigma\rangle} \m  .  
\ee
$\sigma$ is here the levels of structure which a given choice of automorphisms preserve.  

\m 

\n{\bf Observation 3} Suppose $Aut\langle\Space(d), \sigma\rangle$ has a further contribution from Temporal Relationalism.
This is the Leibnizian criterion that there is no time at the primary level for the universe as a whole, which is mathematically implemented 
by time reparametrization invariance.  
It is in this context that the previously-mentioned relational nontriviality criterion 
-- that we need to express change in one configuration variable in terms of another -- makes sense, since there is then no independent time variable 
for a single configuration variable to evolve in terms of. 
Moreover, the ensuing reparametrization-invariant actions necessarily imply \c{Dirac} constraints, 
which in the case of relational particle mechanics are equation of time reinterpretations of quadratic energy equations. 
In the case of GR, this is another way of obtaining the quadratic Hamiltonian constraint, 
which is central to GR's dynamics and to many of GR's manifestations of the Problem of Time \c{Kuchar92, I93, APoT2, ABook}. 

\m 

\n Thus Temporal Relationalism incurs additional constraints, which is a first reason for Background Independence outstripping Shape Theory in complexity.  
One now has the extended quotient 
\be
\TrueSpace  \es  \frac{\FrQ}{Aut\langle\FrQ, \sigma\rangle \m  \Box \m  Mon \langle \FrQ, \sigma \rangle} \m  ;   
\ee
$\TrueSpace$ refers here to the theory's space of true degrees of freedom, and $Mon$ to the monotonic reparametrizations afforded by its timefunctions.  

\m 

\n{\bf Analogy 6} For relational particle models, the product $\Box$ is just $\times$, as in the third expression in eq.\ (\r{Shape-Space}). 
$\Box$ is however capable or outstripping even $\rtimes$ in complexity, as the GR example below shows.  
This $\Box$ product complexity is intimately tied to the second reason for Background Independence outstripping Shape Theory in complexity: 
the Configurational and Temporal Relationalism `parts' of this may exhibit nontrivial interplay. 
One no longer has a priori guarantees of algebraic closure of the corresponding generators (which are, dynamically, constraints). 
In this way, Background Independence picks up also a Constraint Closure part.  
See \c{Kuchar92, I93, APoT2, APoT3, ABook, Project-1} for further aspects of Background Independence.

\m 

\n On the other hand, in GR viewed as dynamics, we have 
\be 
\TrueSpace(\bupSigma)  \es  \frac{\Riem(\bupSigma)}{Dirac(\bupSigma)} 
                  \es  \frac{\Riem(\bupSigma)}{RI(\bupSigma) \, \Thomas \,  Diff(\bupSigma)} 
				  \es  \frac{\Superspace(\bupSigma)}{RI(\bupSigma)} \m  ,
\ee 
where $RI(\bupSigma)$ is now the reparametrization invariance group of {\sl pointwise} monotonic reparametrizations associated with the Hamiltonian constraint.  
On the other hand, $Dirac(\bupSigma)$ is the Dirac algebroid formed by GR's momentum and Hamiltonian constraints; 
$\Thomas$ indicates that $Diff(\bupSigma)$'s momentum constraint is an integrability of $RI(\bupSigma)$'s Hamiltonian constraint.    
Much of the rest of the literature on Background Independence in GR concerns $Diff(\bupSigma)$ and $Dirac(\bupSigma)$.    
 
\m 

\n The current treatise does not consider Temporal Relationalism and thus the $RI(\bupSigma)$ transformations.  
Including this in ordinary Shape Theory gives Relational Particle Mechanics \c{BB82, B03, FORD, FileR, QuadI, AMech, ABook}.
Including this in dynamical study of spatial 3-geometries gives back GR and the full Problem of Time \c{Kuchar92, I93, APoT2, APoT3, ABook, Project-1}.  
On the other hand, the current shape space study fuels indistinguishable particle extension of relational particle mechanics, 
further and topologically correct quantizations of relational particle mechanics.

\m 

\n{\bf Analogy 7} GR spacetime moreover affords a more direct analogue of Shape Theory:   
\be
\Superspacetime(\FrM)  \es  \frac{\PRiem(\FrM)}{Diff(\FrM)} \m .  
\ee
Much recognition, and prior study, of Background Independence in GR concerns this. 
This can be seen as a direct GR spacetime analogue of Shape Theory; 
see \c{Project-1} for simpler analogues of Shape Theory in SR spacetime and GR spacetime solutions with symmetries.  

\m  

\n This is an important point to make since, firstly, a substantial potion of the literature considering Background Independence in GR is in this spacetime setting.  

\m 

\n Also, while $Diff(\FrM)$ is already technically hard to handle, $Dirac(\bupSigma)$ is vastly harder still: 
an infinite-$d$ Lie algebroid rather than `merely' and infinite-$d$ Lie algebra.  

\m 

\n{\bf Analogy 8} A further source of vast difficulty is allowing for spatial topology change in GR \c{GH92}. 
One then has `$\GrandRiem$' \c{Fischer70} -- the space of $\Riem(\bupSigma)$'s for any (or some subclass of) $\bupSigma$'s, 
and an entirely formal rather than technically worked-out notion of $\GrandSuperspace$.
This situation has moreover a substantially simpler analogue in Shape(-and-Scale) Theory: Shape Theory with variable particle number $N$. 
Some inkling of what topology change in GR itself means can be gleaned from spatially 2-$d$ GR's much simplified version. 
Here genus $g$ plays an analogous role to $N$ in relational particle models  
(this can be considered with or without a fixed orientability status $o$ as an extra 2-valued parameter: orientable or nonorientable).  
Changing $g$ (and $o$) here amounts to allowing for change of absolute space at the topological manifold level.  
This goes beyond the  metric and differential structure remit of conventional GR's incorporation of Background Independence. 

\m 

\n This consideration is moreover a start on Backgound Independence at an arbitrary level of structure 
rather than just at the metric-and-differential-geometry level. 
Chris Isham wrote some pioneering papers on {\sl quantum} Backgound Independence at the level of metric spaces, topological spaces and categories 
\c{I89, IKR, I91, I03}, 
whereas the Author has recently provided some rather more pedestrian {\sl classical precursors} for the earlier of Isham's works 
(\c{ASoS}, Epilogues II.C and III.C of \c{ABook} and \c{Forthcoming}).  

\m 

\n{\bf Further Analogies} Prior to Analogy 8's maelstrom, many further analogies between fixed (N, d) Shape Theory and fixed spatial topology GR 
can be made at the level of the Principles of Dynamics.
These are detailedly exposited in \c{FileR, ABook} (or see \c{Project-1} for a summary).

\subsection{Minisuperspace and Midisuperspace restrictions of GR}\l{Mini-Midi}

\n 1) In the restriction to homogeneous spaces -- metrics independent of spatial position $\ux$ \c{Misner-70, Magic, Ryan, MacCallum} (global) scale and anisotropy are all.
Thus the corresponding notion of shape is anisotropy (see e.g. \c{Misner-Ani} for a simple case of GR anisotropyspace); 
minsuperspace itself \cite{Magic, Ryan} is the corresponding homogenous GR shape-and-scale space.  

\m 

\n The momentum constraint is moreover trivial in these models, 
since it is based on a covariant derivative which returns zero whan acting on $\ux$-independent metrics. 

\m 

\n{\bf Remark 1} These models parallel 1-$d$, rather than full $d \geqs 2$, shape(-and-scale) models.
Moreover, 1-$d$ shape(-and-scale) models already have notions of localization, clumping, inhomogeneity, structure and thus structure formation.  
Minisuperspace does not, since these and the momentum constraint are interlinked by both here following from spatial derivatives being nontrivial.
Thus 1-$d$ shape(-and-scale) models decouple two aspects which GR cannot study in isolation of each other.  

\m 

\n 2) Midisuperspace is GR's restriction to inhomogeneous models -- $\ux$-dependent -- which moreover have some symmetry rather than the generic metric's 
total lack of symmetry. 
Examples include spherically-symmetric collapse models \c{Kuchar94, Krasinski, MacCallum} such as Lemaitre--Tolman--Bondi's, 
                 Einstein--Rosen cylindrical gravitational waves \c{Kuchar71} 
				 or the Gowdy universes (spatially $\mathbb{S}^3$ or $\mathbb{T}^3$) \c{Gowdy}.  

\m 

\n 3) Perturbative midisuperspace considers small inhomogeneous perturbations about minisuperspace models. 
This is a typical subject area both in classical observational cosmology  \c{Bardeen, Cos} and in Quantum Cosmology \c{HallHaw, ABook}.  
Upon factoring out scale and homogenous modes, one is left with a perturbative notion of shape coupled to inhomogeneous matter modes.  

\m 

\n{\bf Remark 2} $N$-a-gonlands have not only distance-ratio structure but also relative-angle structure.
These also bring in further midisuperspace-like features due to presence of nontrivial linear constraints 
(here zero total angular momentum corr to rots and possibly also zero dilational momentum).  
So $N$-a-gonlands are more complete that $N$-stop metrolands as GR midisuperspace analogue models.  

\vspace{10in}

\section{Topological shapes with examples for 1, 2 and 3 points on a line}\l{Top-Shape}

We now start in earnest with many new results, 
working from shapes to shape space at each of the topological, metric Lagrangian and metric Jacobi levels as argued for at the end of Sec 3.
We dedicate the rest of this paper to the (3, 1) model, the even more interesting (4, 1) and (3, 2) models forming whole papers 
whose large diversities of results then interweave to form the series' final (4, 2) paper.

\m 

\n{\bf Definition 1} The {\it topological notion of shape} used in this series of articles is the topological content of $N$-point constellations in dimension $d$.

\m 

\n{\bf Remark 1} This is presented for (3, 1) in Figs \r{(1, 1)-(2, 1)-Top-Shapes} and \r{(3, 1)-Top-Shapes}

\m 

\n{\bf Notation} I use pastel sky blue in depictions intended to be solely of topological shapes. 
This renders them immediately distinguishable from topological shapes' spaces (bright blue), 
                                                   metric shapes              (grey) 
                                               and metric shape spaces        (black). 

\m 

\n{\bf Definition 2} I term such depictions {\it topological distribution diagrams}.
{\it Coincidence diagrams} (row 2 of Fig \r{Top-Shape-Coincidence}) are a further simplification of the previous 
by discarding all points not participating in coincidences.

\m 

\n{\bf Remark 2} This is a fairly weak notion of shape as many of the properties usually attributed to $N$ point constellations are metric in nature: 
the angles defining an isosceles triangle, the ratios defining a rhombus...
As the below examples show, the topological notion of shape is not however empty.
Topological shapes furthermore provide useful insights as regards the structure of shape spaces of metric-level shapes.

\m 

\n{\bf Example 1} For $N = 1$, there is one topological class: the point itself (Fig \r{(1, 1)-(2, 1)-Top-Shapes}.1).    
O is however not a shape in many senses, and excluding O from the $N = 1$ case leaves one with nothing at all.
On the other hand, for $N = 1$, many of the reasons for excluding O have faded away, so we also entertain the case including O.
Moreover, even including the O point, this model exhibits numerous other trivialities
(a translation-invariant single-particle universe model has no content by Leibniz's Identity of Indiscernibles) so one needs to pass to considering $N \geqs 2$.

\m 

\n{\bf Example 2} For $N = 2$, there are two topological classes of (non-)shape (Fig \r{(1, 1)-(2, 1)-Top-Shapes}.2). 
In this case, excluding O still leaves us with a topological theory.
Moreover, excluding O from $N = 2$ leaves one with no topological class distinction, in which case one needs to consider at least $N = 3$ to exhibit this feature.
Many reasons for excluding O also fade away in this case:  
the binary collision's good behaviour turns out to trump the maximal collision's bad behaviour in this $N = 2$ case for which these two notions coincide.  
However, even including O the model is still metric dimensionally trivial, giving plenty of reasons to pass to $N \geqs 3$.
%
{            \begin{figure}[!ht]
\centering
\includegraphics[width=0.42\textwidth]{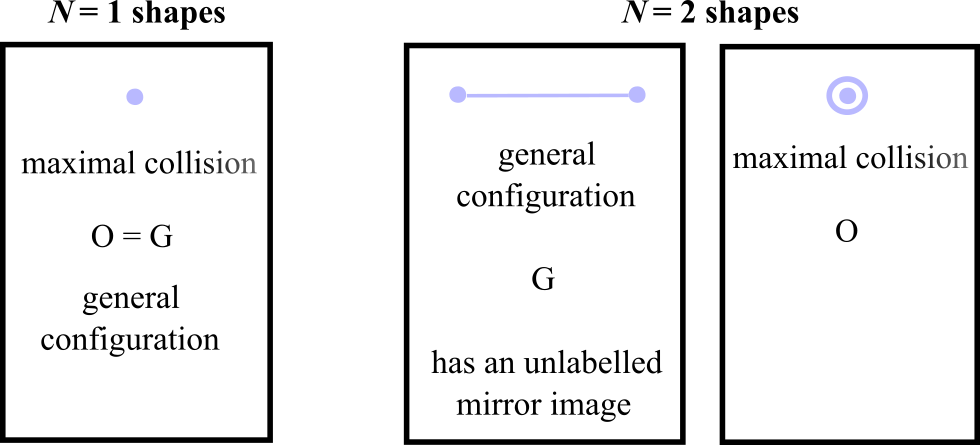}
\caption[Text der im Bilderverzeichnis auftaucht]{        \footnotesize{Topological classes of configurations for 1 and 2 points.  
O denotes the maximal coincidence-or-collision and G denotes the coincidence-or-collision-less generic configuration.  
The 2 point case has distinct classes because, while topology has no notion of length (`rubber sheet'), 
it does distinguish between coincidence and non-coincidence (passing from a coincidence to a non-coincidence involves a tearing and the reverse passage a gluing).}}
\l{(1, 1)-(2, 1)-Top-Shapes} \end{figure}          }

\m 

\n{\bf Example 3} and {\bf Proposition 1}  For $N = 3$, are three topological classes of (non-)shape: Fig \r{(3, 1)-Top-Shapes}.
This is the first metrically dimensionally nontrivial shape space, for all that it is still relationally trivial as a metric shape space (it has 1 degree of freedom).  
The corresponding shape-and-scale relational space, moreover, is not relationally trivial since it has 2 degrees of freedom.  
%
{            \begin{figure}[!ht]
\centering
\includegraphics[width=0.35\textwidth]{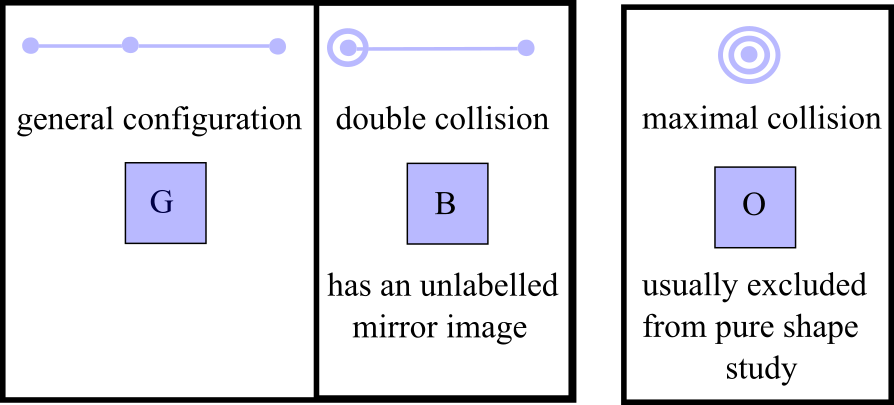}
\caption[Text der im Bilderverzeichnis auftaucht]{        \footnotesize{Topological classes of configurations for 3 points in 1-$d$.  
Here B denotes the binary coincidence-or-collision B, and G denotes the coincidence-or-collision-less generic shape.
In this treatise, blue square labels are used to keep track of such topological equivalence classes in subsequent figures.} }
\l{(3, 1)-Top-Shapes} \end{figure}          }

\m 

\n{\bf Proposition 2} These topological classes of shapes may furthermore be viewed as {\sl equivalence classes} of metric shapes. 
Equivalence classes are always disjoint and exhaustive and so constitute a partition.  
This is this treatise's coarsest-level characterization of topological shape, or, for that matter, of shape.  

\m 

\n{\bf Proposition 3} For $N = 3$, the number and types of topological classes themselves are a dimension $d$-independent characterization.  

\m 

\n{\bf Remark 3} On the other hand, which metric shape representations each class contains is of course dimension-dependent: 
compare Fig \r{(3, 1)-Top-Shapes} with Fig III.1.   

\m 

\n{\bf Proposition 4} If one or both of particle labelling and mirror image distinction are incorporated,  
a finer partition of the topological shapes is being entertained.  

\m  

\n{\bf Returning to Example 1} For $N = 1$, both of these distinctions are meaningless, so there is still just one point.

\m 

\n{\bf Returning to Example 2} For $N = 2$, these distinctions are equivalent, 
in the sense that `left' and `right' assignment has the same labelling content as any other specification of 2 distinct labels. 
In both cases, there are now 2 G's in place of 1, and still just room for 1 realization of $\mB = \mO$.

\m 

\n{\bf Returning to Example 3} For $N = 3$, if the points are labelled and mirror image configurations are held to be distinct, 
\be
\mbox{\#(G)} = \mbox{(label permutations)} = 3 \, ! = 6\mma \mbox{ and }
\ee
\be 
\mbox{\#(B)} = \mbox{(3 ways of leaving a particle out)} \times \mbox{(2 mirror images)} = 6 \m  .  
\ee
If instead the points are labelled but mirror image configurations are held to be identified, 
these values are halved by the mirror image identification to 3 of each.
If moreover the points are not labelled but mirror images are held to be distinct, 2 distinct G's and 2 distinct B's ensue.
Finally, if neither distinction is made, there is just 1 of each.  

\m 

\n{\bf Remark 4} The distinction between `equivalence classes in general give partitions' 
and the unlabelled mirror image identified topological shapes for (3, 1) is in 1 : 1 correspondence with {\sl the} finite partitions themselves 
(row 3 of Fig \r{Top-Shape-Coincidence}).  

\m 

\n Part II shows moreover that this 1 : 1 correspondence is {\sl not} maintained for $N \geqs 4$; instead a somewhat finer partition is realized. 
On the other hand, for $d \geqs 2$ the correspondence is maintained. 
This gives plenty of reason to introduce a second type of depiction which keeps track of {\sl the} finite partition content, as follows, 
for all that for the (3, 1) model the two have equivalent content.   
%
{            \begin{figure}[!ht]
\centering
\includegraphics[width=0.65\textwidth]{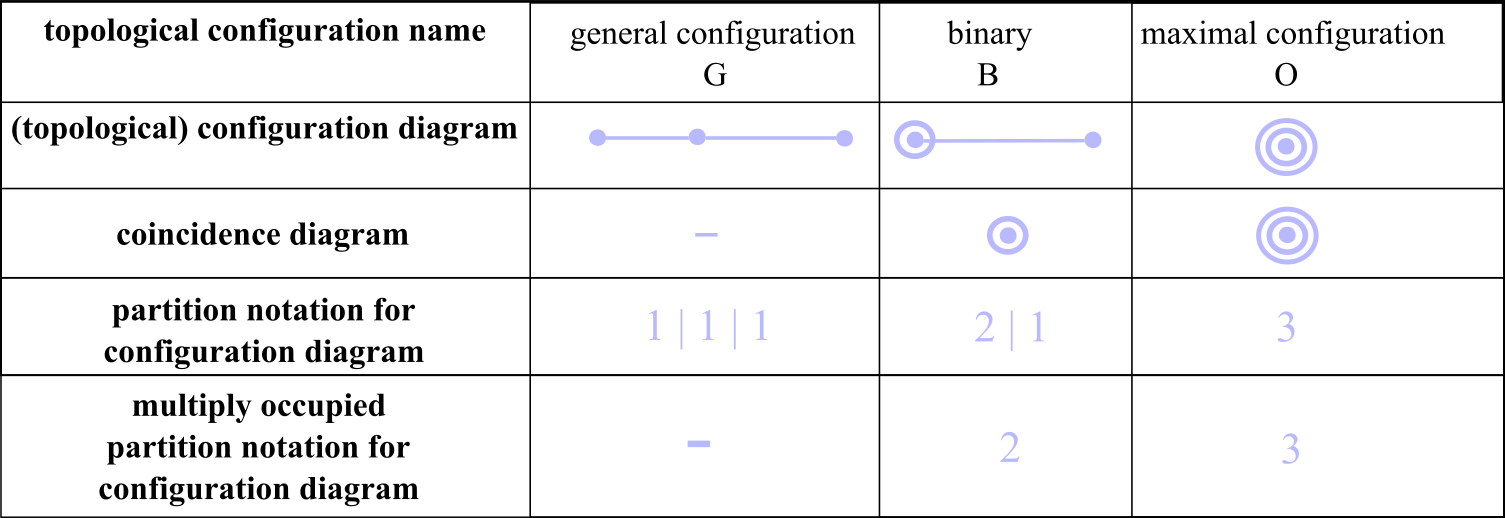}
\caption[Text der im Bilderverzeichnis auftaucht]{        \footnotesize{Topological shape diagrams versus coincidence diagrams, 
with some corresponding partition notation and an inkling of how 1-$d$ shape spaces deviate from the usual conceptual content that gives rise to partitions. } }
\l{Top-Shape-Coincidence} \end{figure}          }

\section{Topological shape spaces}\l{Top-Shape-Spaces}

\n{\bf Definition 1} The {\it topological shape space} is the collection of all of a model's topological shapes.
These shape space graphs are to be distinguished in this treatise in writing by the preface $\FrT$op, 
and graphically by use of bright blue edges and vertices. 

\m 

\n{\bf Proposition 1} For point or particle models, $\FrT\mbox{op}$ are realized by graphs.  

\m  

\n This is the first main justification for this Paper containing a Appendix A on graphs.  

\m   

\n{\bf Proposition 2} For $N = 1$, 
i) excluding O, the only topological shape space is
\be
\Top\mbox{-}\FrS(1, 1)          = 
\Top\mbox{-}\w{\FrS}(1, 1)      = 
\Top\mbox{-}\FrI\FrS(1, 1)      = 
\Top\mbox{-}\Leib_{\sFrS}(1, 1) = \emptyset \m  .  
\ee
ii) Including O (which we indicate by appending an O superscript to the shape space in question) 
\be
\Top\mbox{-}\FrS^{\sO}(1, 1)          = 
\Top\mbox{-}\w{\FrS}\mbox{}^{\sO}(1, 1)      = 
\Top\mbox{-}\FrI\FrS^{\sO}(1, 1)      = 
\Top\mbox{-}\Leib_{\sFrS}^{\sO}(1, 1) = P_1: \mbox{  a single point labelled with } O = G \m  .  
\ee
\n{\bf Proposition 3} For $N = 2$, 
i) excluding O, the topological shape spaces are
\be
\Top\mbox{-}\FrS(2, 1)          = P_1 \, \coprod \, P_1 \mbox{ two points labelled by } G \m  . 
\l{Top-S(2, 1)}
\ee
Also 
\be
\Top\mbox{-}\w{\FrS}(2, 1)      = 
\Top\mbox{-}\FrI\FrS(2, 1)      = 
\Top\mbox{-}\Leib_{\sFrS}(2, 1) = P_1   \mbox{  a single point labelled with } G \m  .  
\ee
ii) Including O, 
\be
\Top\mbox{-}\FrS^{\sO}(2, 1)          = P_3 \m  :
\ee
the 3-vertex {\it 3-path graph} labelled as per Fig \r{S(1, 1)-S(2, 2)-Top}.c.1).  
Also 
\be
\Top\mbox{-}\w{\FrS}^{\sO}(2, 1)      = 
\Top\mbox{-}\FrI\FrS^{\sO}(2, 1)      = 
\Top\mbox{-}\Leib_{\sFrS}^{\sO}(2, 1) = P_2 \m  :  
\ee
the 2-vertex {\it 2-path graph} labelled as per Fig \r{S(1, 1)-S(2, 2)-Top}.c.2).  
%
{            \begin{figure}[!ht]
\centering
\includegraphics[width=0.9\textwidth]{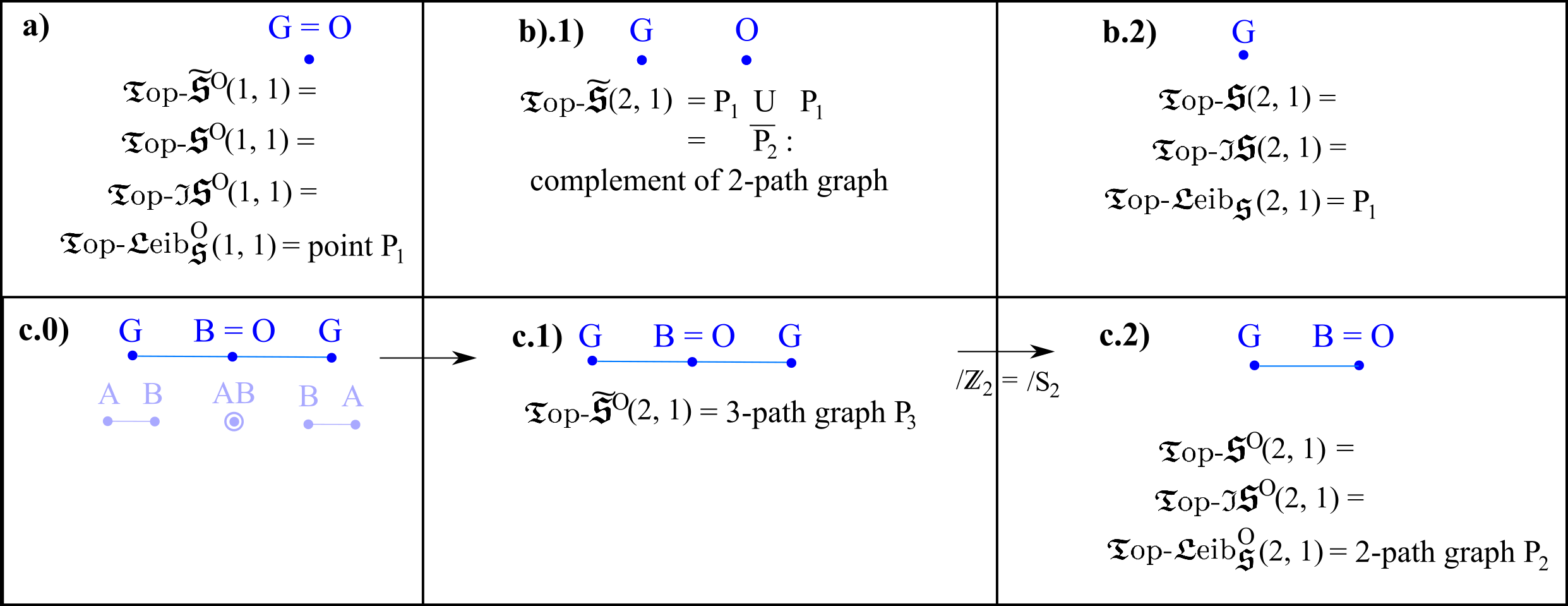}
\caption[Text der im Bilderverzeichnis auftaucht]{        \footnotesize{The smallest topological shape spaces:   
a) $\Top\mbox{-}\sFrS^{\tO}(1, 1)$.
b) $\Top\mbox{-}\sFrS(2, 1)$.

\m 

\n c).0) Continuity method for determining the topology of $\Top\mbox{-}\sFrS(2, 1)$ to be the 3-vertex path graph given in 1).  
2) $\Top\mbox{-}\w{\sFrS}(2, 1) = \Top\mbox{-}\FrI\sFrS(2, 1) = \Top\mbox{-}\Leib_{\tFrS}(3, 1)$'s 2-vertex path graph.} }
\l{S(1, 1)-S(2, 2)-Top} \end{figure}          }

\m 

\n{\bf Remark 1)} Note that in $\Top-\FrG\FrG$ graphs, the shapes are just vertex labels.
On the other hand, the graph edges encode the {\it topological adjacency} relation.   
This is based on topology distinguishing between plain tearing and tearing followed by gluing to another distinguishable object. 
So e.g.\ AB--C is not topologically adjacent to A--BC, since to move between these, one needs to tear B off A and then glue it to C.   

\m 

\n{\bf Remark 2} (\r{Top-S(2, 1)}) is disconnected: there is no edge between the vertices of this graph.  
On the other hand, the structure of $\Top\mbox{-}\FrS^{\sO}(2, 1)$ can be arrived at by a continuity method (Fig \r{S(1, 1)-S(2, 2)-Top}.c.0).   

\m 

\n{\bf Proposition 4} For $N = 3$, the topological shape spaces are as follows.    
\be
\Top\mbox{-}\FrS(3, 1) = \mC_{12} \m  :
\ee
the 12-vertex cycle graph as labelled in Fig \r{S(3, 1)-Top}.1).     
\be 
\Top\mbox{-}\w{\FrS}(3, 1)  \es  \frac{\Top\mbox{-}\FrS(3, 1)}{\mathbb{Z}_2} = \mC_{6} \m  :
\ee
the 6-vertex cycle graph as labelled in Fig \r{S(3, 1)-Top}.d). 
\be
\Top\mbox{-}\FrI\FrS(3, 1)  \es  \frac{\Top\mbox{-}\FrS(3, 1)}{\mathbb{S}_3} = \mP_{3} \m  : 
\ee
the 3-vertex path graph with the symmetric endpoint labelling given in Fig \r{S(3, 1)-Top}.3).  
\be
\Top\mbox{-}\Leib_{\sFrS}(3, 1)  \es  \frac{\Top\mbox{-}\FrS(3, 1)}{S_3 \times \mathbb{Z}_2} = \mP_2 \m  :
\l{Top-Leib(3, 1)}
\ee
the 2-vertex path graph with end-points labelled distinctly as per Fig \r{S(3, 1)-Top}.4). 
%
{            \begin{figure}[!ht]
\centering
\includegraphics[width=0.7\textwidth]{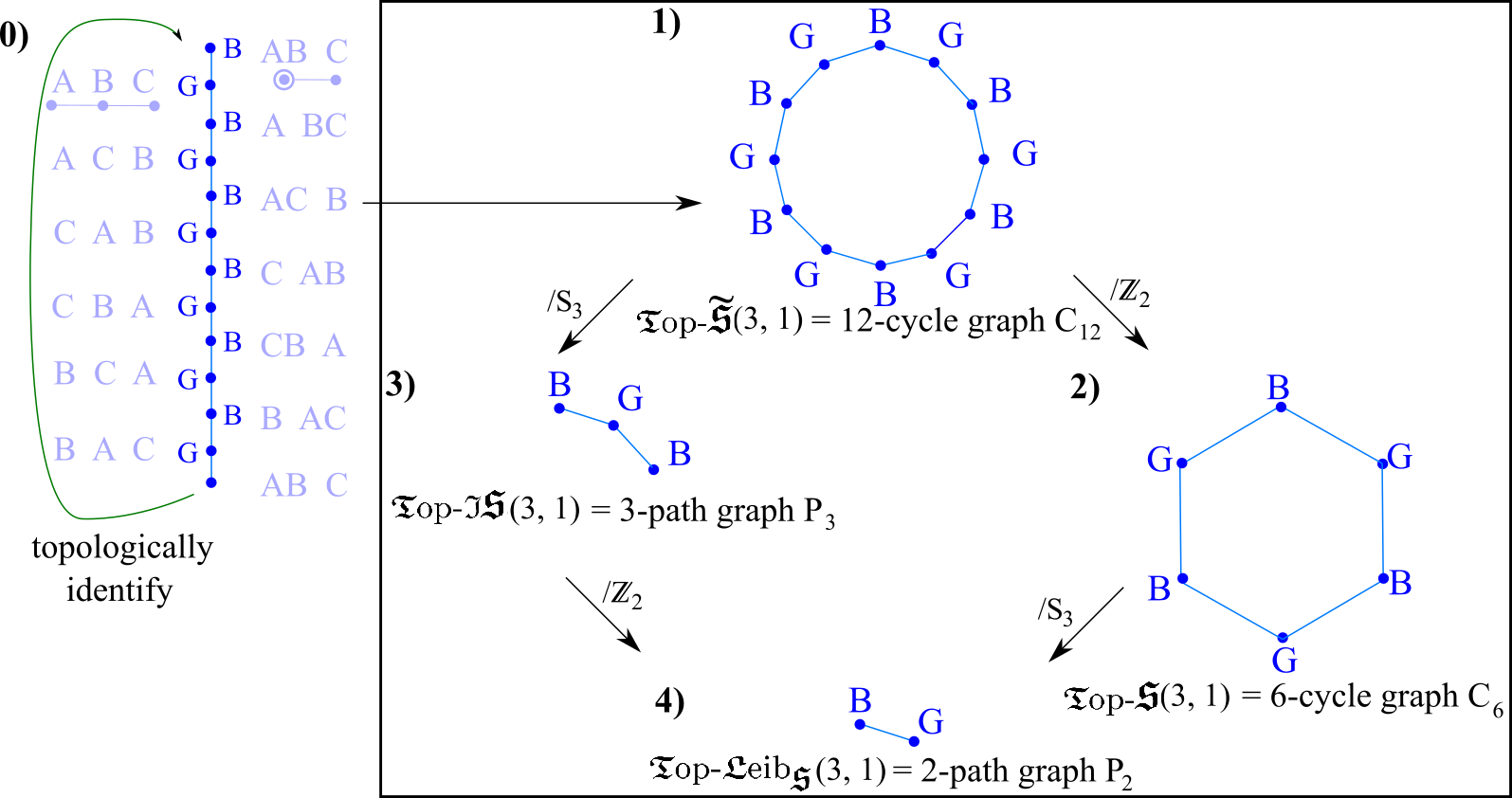}
\caption[Text der im Bilderverzeichnis auftaucht]{        \footnotesize{0) Continuity method for determining the topology of $\Top\mbox{-}\sFrS(3, 1)$ 
to be a loop (i.e.\ a closed path).   
It has 12 vertices thus forming 1)'s dodecagonal perimeter. 
$\Top\mbox{-}\w{\sFrS}(3, 1)$, on the other hand, has 6 vertices forming 2)'s hexagonal perimeter
In both of these cases, the vertices are labelled alternately G and B: G for general configuration and B for binary coincidence-or-collision.    
3) $\Top\mbox{-}\FrI\sFrS(3, 1)$'s 3-vertex path graph with end-points labelled symmetrically with B's.
4) $\Top\mbox{-}\Leib_{\tFrS}(3, 1)$'s    2-vertex path graph; by construction Leibniz spaces' labels are always all distinct.} }
\l{S(3, 1)-Top} \end{figure}          }

\m 

\n {\bf Corollary 1} The B's or G's of $\Top\mbox{-}\Leib_{\sFrS}(3, 1)$ form a hexagon, and  
 \be
S_3 \times \mathbb{Z}_2 \m  \mbox{ acts on this hexagon as } \m  D_6 \mbox{ : } \mbox{ the dihaedral group of order 12 } .
\ee
This permits us to rewrite the labelling and mirror image based definition (\r{Leib(3, 1)}) as 
\be
\Top\mbox{-}\Leib_{\sFrS}(3, 1)  \es  \frac{\Top\mbox{-}\FrS(3, 1)}{D_6}\m  ,
\ee
which is very natural at the shape space level because of the realization of the hexagons therein.

\m 

\n{\bf Structure 1} In $\FrG\FrG$ differential geometries, it is helpful and structurally meaningful to draw the shapes each point represents as follows 
(the start of a standardized and mathematically precise way of drawing Kendall's spherical blackboard and generalizations).  

\m 

\n 1) The centre of mass of the shape is placed upon the point in shape space that represents the shape. 
This takes advantage of a unique and most representative (usually non-material) point on the shape to establish an attachment.  

\m 

\n 2) The shape is depicted of normalized size. 
This way, there is no size ambiguity in the representative shape. 

\m 

\n For $(N, 1)$ similarity shapes, 1) and 2) give a unique prescription for shape representatives; difference in their orientation are purely aesthetic 
(including to avoid overcrowding on the page).
Once rotations are in play, 
we will add to this prescription and identify it as in fact being part of some fairly sophisticated mathematics (see Sec III.17-18).

\section{Metric-level theory}\l{Metric-Shape-Space}

\subsection{Metric shapes}\l{(3,1)-Metric-Shapes}

\n We next consider including the metric level of structure's information in the shapes as well.  
In this case, shapes in space are depicted in grey: the {\it spatial distribution diagram}. 
This is how we have been representing most of the shapes encountered so far.
The previous section's topological distribution diagram 
is a simplification of the preceding made by disregarding the metric-level length ratio information in the preceding, depicted in pastel sky blue. 

\m 

\n{\bf Proposition 1} For 1 and 2 points in 1-$d$, the metric level contains no more shape information than at the topological level.

\m 

\n{\bf Derivation} Consider the configuration space dimensionality of the G's and O's that these models possess. 

\m 

\n Firstly, regardless of dimensionality, $Sim(d)$ collapses to just $Tr(d)$ when acting upon O, so 
\be
\mbox{dim}(O) = \mbox{dim}\left(\frac{\mathbb{R}^d}{Tr(d)}\right) 
              = \mbox{dim}(\mathbb{R}^d) - \mbox{dim}(\mathbb{R}^d)
			  = d - d 
			  = 0       \m  .
\ee
For $N = 1$, this gives \mbox{dim(G)} as well, since $\mG = \mO$.  
For $N = 2$,
\be 
\mbox{dim}(\mG) = \mbox{dim}\left(\frac{\mathbb{R}^2}{Tr(1) \times Dil}\right) 
                = \mbox{dim}(\mathbb{R}^2) - \mbox{dim}(\mathbb{R} \times \mathbb{R}_+) 
				= 2 - 1 - 1 
				= 0                                                                              \m  .  
\ee	
Thus these models just manifest nontrivial metric dimensionality. $\Box$

\m 

\n{\bf Remark 1} For $N = 3$ (and $N \geqs 3$) however, the G's do have a nontrivial configuration space dimensionality in configuration space.  
\be
\mbox{dim}(\mG) = \mbox{dim}\left(\frac{\mathbb{R}^3}{Tr(1) \times Dil}\right) 
                = \mbox{dim}(\mathbb{R}^3) - \mbox{dim}(\mathbb{R}) - \mbox{dim}(\mathbb{R}_+) 
				= 3 - 1 - 1 
				= 1                                                                              \m  .  
\ee
Now additionally 
\be
\mbox{dim}(\mB) = \mbox{dim}\left(\frac{\mathbb{R}^2}{Tr(1) \times Dil}\right) 
                = \mbox{dim}(\mathbb{R}^2) - \mbox{dim}(\mathbb{R}) - \mbox{dim}(\mathbb{R}_+) 
				= 2 - 1 - 1 
				= 0                                                                              \m  . 
\ee
Thus in the metric setting, the $N = 1$ model's G's and B's can be interpreted as 1-$d$ continuum edge curves and point vertices respectively.

\subsection{Model diameter per unit moment of inertia maximizing and minimizing shapes}\l{cal L}
%
{            \begin{figure}[!ht]
\centering
\includegraphics[width=0.5\textwidth]{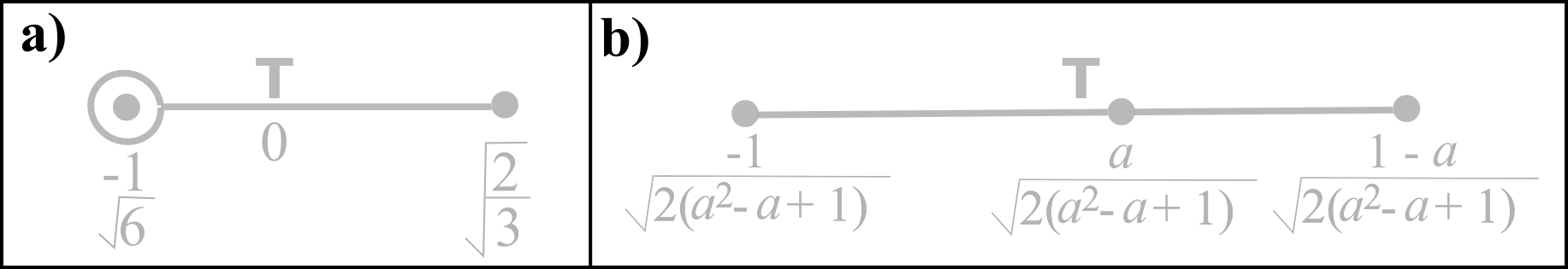}
\caption[Text der im Bilderverzeichnis auftaucht]{        \footnotesize{a) The B configuration at the metric level, 
including normalization by the total moment of inertia which is useful in comparing different shapes.

\m 

\n b) A representation of the general (3, 1) shape; again we normalize by the moment of inertia $\bigiota = 1 + a^2 + \{1 - a\}^2 = 2\{a^2 - a + 1\}$.
Finally note that the maximal coincidence-or-collision configuration cannot be normalized; this is a common underlying reason for excluding it.} }
\l{General-Position-(3,1)} \end{figure}          }
%
This refers more specifically to the shape (and thus dimensionless) quantity 
\be
\diam  \ :=  \frac{\mbox{(mass-weighted diameter)}}{\sqrt{\mbox{(moment of inertia)}}} \m  ;
\ee
It is moreover more convenient to extremize the square of this, 
\be
\Delta := \diam^2 \m  .
\ee
We can evaluate this from the general shape using e.g.\ the representative in Fig \r{General-Position-(3,1)} 
there are three distinct expressions for what the diameter is depending on the value of $a$,  
\be
\Delta  \es  \frac{  \{2 - a\}^2  }{  2\{a^2 - a + 1\}  } \m  -1 < a < \frac{1}{2} \m  , 
\ee
\be
\Delta  \es  \frac{  \{a + 1\}^2  }{  2\{a^2 - a + 1\}  } \m  \frac{1}{2} < a < 2 \m  ,
\ee
\be
\Delta  \es  \frac  {\{1 - 2\, a\}^2  }{  2\{a^2 - a + 1\}  } \m  a < -1 \mbox{ and } \m  a > 2 \m  .
\ee
We have continuity at each interface but not differentiability.
Basic Optimization requires one to check each region's end-point values as well as the interior extrema.
This yields the outcome presented in Figure \r{Max-Min}, by which the B's minimize $\diam$ and a new configuration $\diam_{\sm\sa\sx}$ maximizes it.  
This $\diam_{\sm\sa\sx}$ configuration is sketched in Figure \r{Max-Min}. 
We will repeatedly encounter this shape below as the answer to further interesting questions and thus the realizer of further interesting properties, 
by which it will acquire a number of further names. 
Moreover, as we shall see in Papers II to IV, for $N \geqs 4$ and/or $d \geqs 2$, these interesting questions cease to give coincident answers for the 
next simplest models.    
%
{            \begin{figure}[!ht]
\centering
\includegraphics[width=0.65\textwidth]{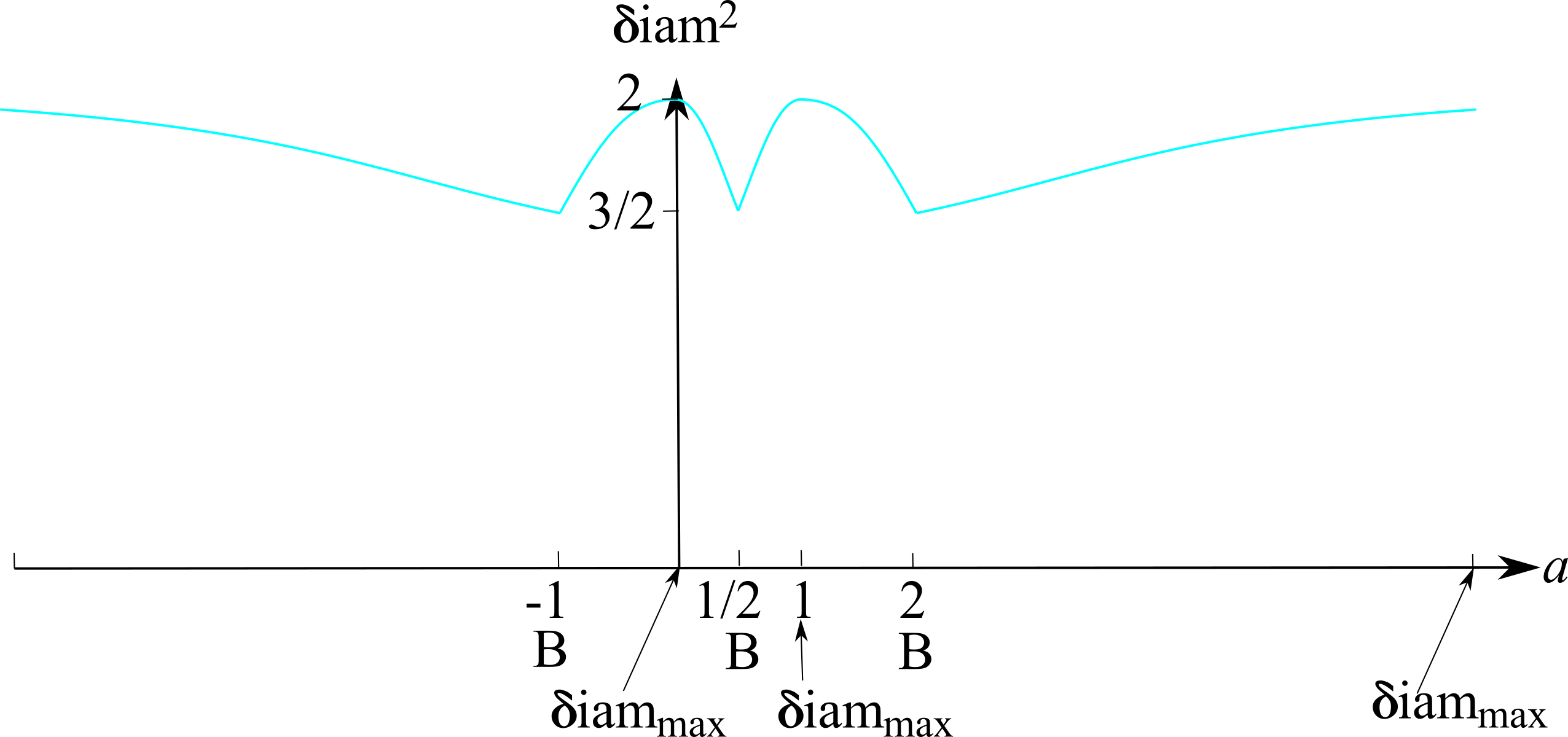}
\caption[Text der im Bilderverzeichnis auftaucht]{        \footnotesize{Sketch of the $\Delta = \diam^2$ measure of size per unit moment of inertia.} }
\l{Max-Min} \end{figure}          }

\subsection{Metric shape spaces}

\n Continuity considerations now show that these fit together in the manner of Fig \r{S(3, 1)-Metric-1}.0) 
to give the shape space $\FrS(3, 1)$ depicted in Fig \r{S(3, 1)-Metric-1}.1).  

\m 

\n{\bf Remark 1} In this continuum context, 
such continuity methods for determining shape space topology rests upon the following underpinning topological features.

\m  

\n{\bf Proposition 1} Metric-level similarity shape spaces a) close up and b) all parts of them can be reached by the continuity method. 

\m 

\n{\bf Derivation} 
\n a) `Closing up' occurs since similarity shape spaces are compact. 

\m  

\n b) One is able to reach the whole of the configuration space in this manner because similarity shape spaces are path-connected. 

\m 

\n Finally, a) and b) occur because preshape space has these properties (and connectedness), 
and all of these properties are additionally preserved by quotients \c{Lee1}. $\Box$ 

\m 

\n{\bf Remark 2} For this treatise's small examples, 
the continuity method's `closing up' occurs furthermore within small finite `operation number time expenditure'. 
This is because numbers of vertices, edges (and eventually higher simplices involved) are finite and small.  
%
{            \begin{figure}[!ht]
\centering
\includegraphics[width=0.72\textwidth]{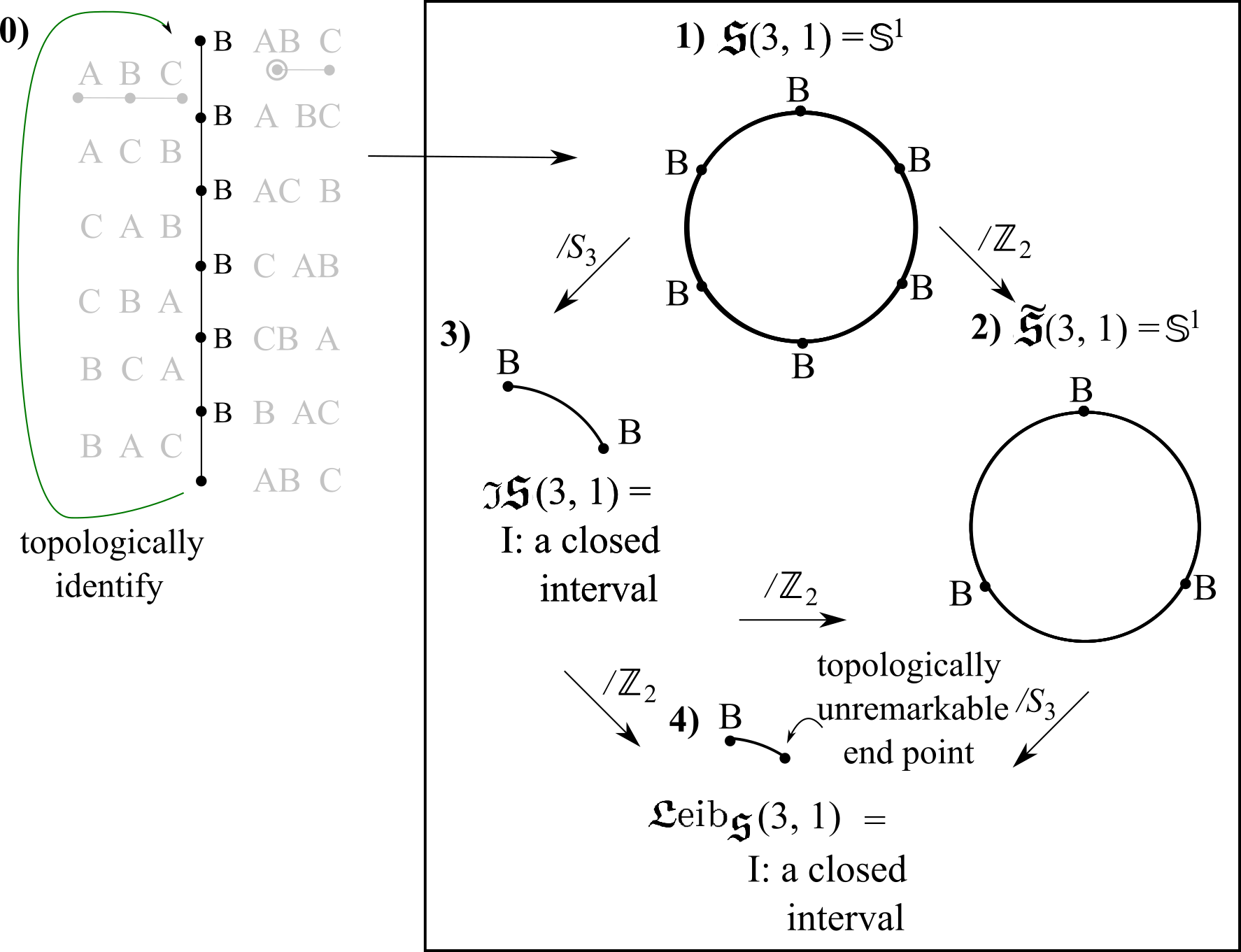}
\caption[Text der im Bilderverzeichnis auftaucht]{        \footnotesize{0) Continuity method for determining the topology of $\sFrS(3, 1)$ to be a circle, 
After symmetry considerations, this is decorated by spatially topologically distinguished points B and spatially metrically distinguished points M as in 1): 
in the manner of a clock-face with even hours marked B and odd hours marked M for `merger' (see Fig \r{(3, 1)-Shapes}).  
2) The $\w{\sFrS}(3, 1)$ circle has only even hours marked at all, alternating between B and M labels.  
3) The $\FrI\sFrS(3, 1)$ 2-hour arc and with end-points symmetrically labelled by B.  
4) The $\Leib_{\tFrS}(3, 1)$    1-hour arc. 
These are all well-known as topological manifolds consequently and at the level of Algebraic Topology.} }
\l{S(3, 1)-Metric-1} \end{figure}          }

\m  

\n{\bf Corollary} of Proposition 1 of Sec \r{(3,1)-Metric-Shapes} 
There is no further metric-level structure for $N = 1$ and $2$ shape spaces; 
the previous section's graphs comprise all of the configuration space structure for these. 

\m 

\n{\bf Proposition 2} For 3 points in 1-$d$, the metric-level shape spaces are as follows. 
\be
\FrS(3, 1) = \mathbb{S}^1 \m  ,
\ee
labelled as per Fig \r{S(3, 1)-Metric-1}.1). 
\be
\w{\FrS}(3, 1)  \es  \frac{\FrS(3, 1)}{\mathbb{Z}_2}  
                \es  \frac{\mathbb{S}^1}{\mathbb{Z}_{2 \, \si\sn\sv}} = \mathbb{RP}^1 = \mathbb{S}^1
\ee
again, though now labelled with half as many distinguished B points, as per Fig \r{S(3, 1)-Metric-1}.2).  
\be
\FrI\FrS(3, 1)  \es  \frac{\FrS(3, 1)}{S_3} \m  
\ee
labelled as per Fig \r{S(3, 1)-Metric-1}.3). 
\be
\Leib_{\sFrS}(3, 1)  \es  \frac{\FrS(3, 1)}{S_3 \times \mathbb{Z}_2}  
                    \es  \frac{\FrS(3, 1)}{D_6}\m  , 
\l{Leib(3, 1)}
\ee
labelled as per Fig \r{S(3, 1)-Metric-1}.4). 

\m 

\n{\bf Remark 3} Once again, the $D_6$ in the last form is very natural in configuration space 
since the B's form a hexagon that $D_6$ acts naturally upon.

\subsection{Shape space coordinates and location of the B's}

The plane polar form for 3-point 1-$d$ relative Jacobi coordinates is 
\be
\rho_1 = \rho \, \mbox{cos} \, \phi   \m  , 
\l{rho-1-(3, 1)}
\ee
\be
\rho_2 = \rho \, \mbox{sin} \, \phi   \m  , 
\l{rho-2-(3, 1)}
\ee
which invert to 
\beq
\phi = \mbox{arctan}
\left(
\frac{\rho_2}{\rho_1}  
\right)                               \m ,
\eeq
\beq
\rho = \sqrt{\rho_1^2 + \rho_2^2}     \m  . 
\eeq
To work on shape space, set $\rho = constant$ and use $\phi$ as an angular coordinate on shape space.  
We shall see that it is most convenient for us to assign $\rho_1$ to be a {\sl vertical} axis, with $\phi$ measured anticlockwise thereabout.  
For $\FrS(3, 1)$ this has the usual polar range $[0, 2\pi)$, so these are {\sl geometrically standard} polar coordinates.
On the other hand, they have acquired an {\sl interpretationally nonstandard} shape-theoretic meaning, as follows.  

\m 

\n 1) $\phi$ is a function of the ratio of the size of a picked out 
binary subsystem $ab$ relative to the separation of CoM($ab$) from the third point, i.e.\ a base to median ratio quantity. 

\m  

\n 2) $\rho$ is a measure of overall size (in a mass-weighted sense): it is the square root of the total moment of inertia. 
There is moreover a $\rho$ with this interpretation in {\sl all} shape-and-scale point models (Sec \r{MLS-R}).  

\m 

\n In the pure-shape context, we can just set $\rho = \mbox{constant}$ so as to be on $\FrS(3, 1)$.  

\m 

\n A first application of these coordinates involves determining where the B points lie on shape space. 
For one clustering choice, 
\be
\rho_1 = 0 \m  \Leftrightarrow \m  R_1 = 0 \m  , 
\l{rho-1}
\ee
corresponding to 
\be
\phi = 0\mma \m  \pi \m  .
\ee
The other two clustering choices give antipodal pairs at $\pm \pi/3$ to this.  

\m 

\n Whereas the B's are topologically distinct configurations in space, there is no configurational space geometric reason to excise them. 
This is firstly because the whole of $G$ acts on these, just as it does on the G's, so these both lie on the same (and hence in this problem only) stratum.  
Secondly, there is no curvature singularity at the G.  

\m 

\n {\bf Aside} Moreover in Dynamics configuration space geometry is not all. 
For potentials, e.g.\ of the form 
\be
\mV \m \propto \m \frac{1}{||\ur_{IJ}||^{\alpha}} \m \mbox{ for positive } \m \alpha   \m  ,
\ee
can provide a second reason to need to excise particle collisions.
In this way, e.g.\ Celestial Mechanics motivates study of further excised configuration spaces.  

\m 

\n Sec \r{cal L}'s $\Delta$ can now be expressed in shape space angle terms, such as 
\be
\Delta  \es  \mbox{$\frac{1}{2}$} \{ \mbox{cos} \,\phi - \sqrt{3} \, \mbox{sin} \, \phi \}  \m  .  
\ee
in one piece.
This gives the polar plot in Fig \r{Cusp-Flower}.1) over $\Leib(3, 1)$, or, extending by symmetry to all of $\FrS(3, 1)$, as in Fig \r{Cusp-Flower}.4).
%
{            \begin{figure}[!ht]
\centering
\includegraphics[width=0.65\textwidth]{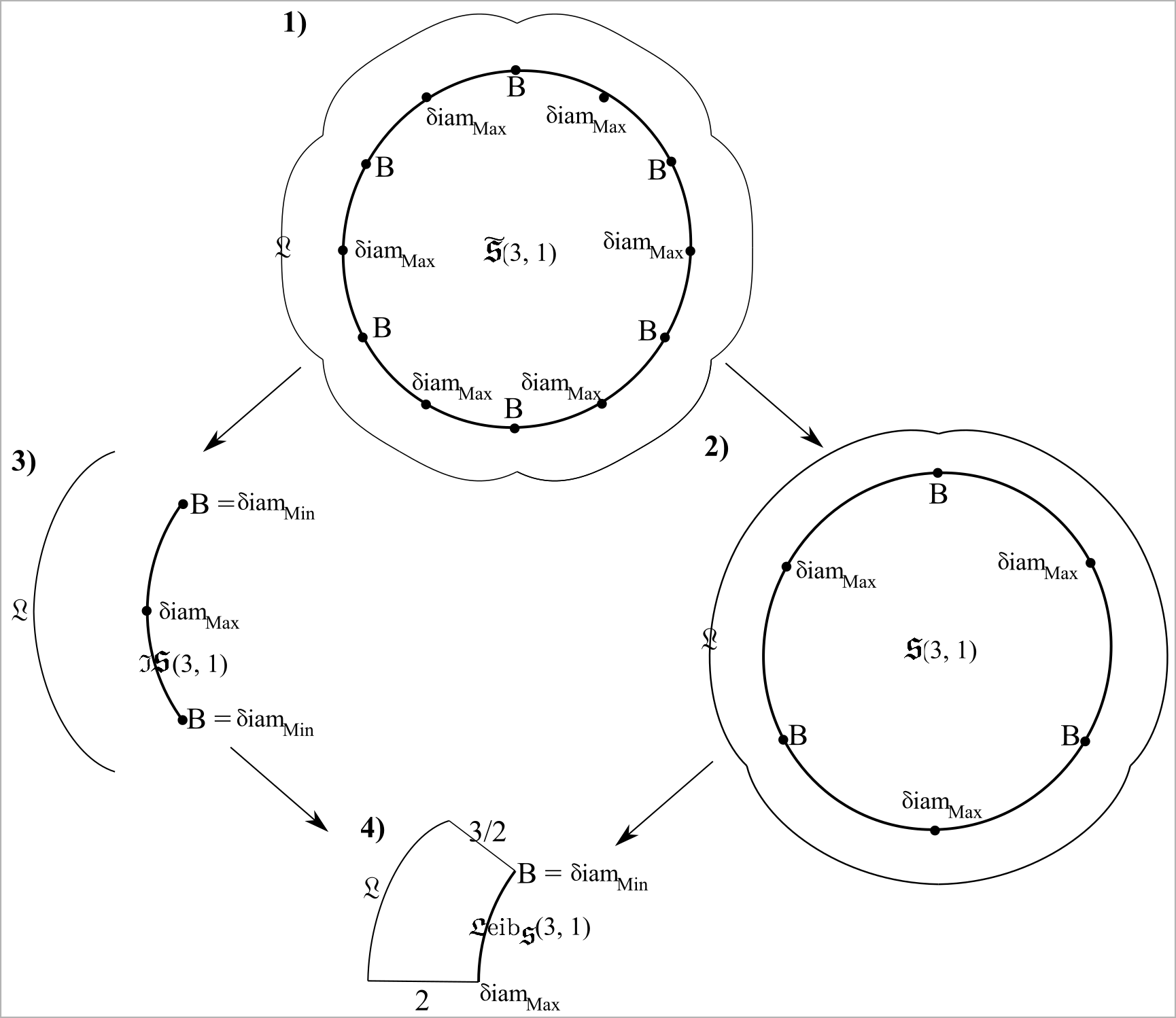}
\caption[Text der im Bilderverzeichnis auftaucht]{        \footnotesize{Diameter per unit moment of inertia is 
4) a differentiable monotone function over $\Leib_{\tFrS}(3, 1)$, 
3) a differentiable symmetric function over $\FrI\sFrS(3, 1)$, 
2) a three-cusped three-petalled flower over $\w{\sFrS}(3, 1)$, and 
1) a six-cusped six-petalled flower over the whole of $\sFrS(3, 1)$.} }
\l{Cusp-Flower} \end{figure}          }

\section{Lagrangian-level metric study}\l{UMS}

\subsection{The (3, 1) coincidence-or-collision equations}

For (3, 1) shapes, the only shape solutions of the coincidence-or-collision equations are the elementary single coincidence-or-collision, B.  
We already know that there are 6, 3, 2 or 1 of these depending on further modelling assumptions. 
So for (3, 1) shapes, the coincidence-or-collision structure $\Co(\FrG\FrS)$ is just a small collection of B points.
If the maximal coincidence-or-collision is allowed as well, this solves all three coincidence-or-collision equations simultaneously, 
leading to the inclusion of one more point in each case.  
This is however too simple an example to appreciate the structure of coincidence-or-collision sets in more general models; see Part II for a somewhat more typical example.

\subsection{(The (3, 1) model's Lagrangian-uniform shapes}\l{Lag-Uni-(3,1)}
%
{            \begin{figure}[!ht]
\centering
\includegraphics[width=0.38\textwidth]{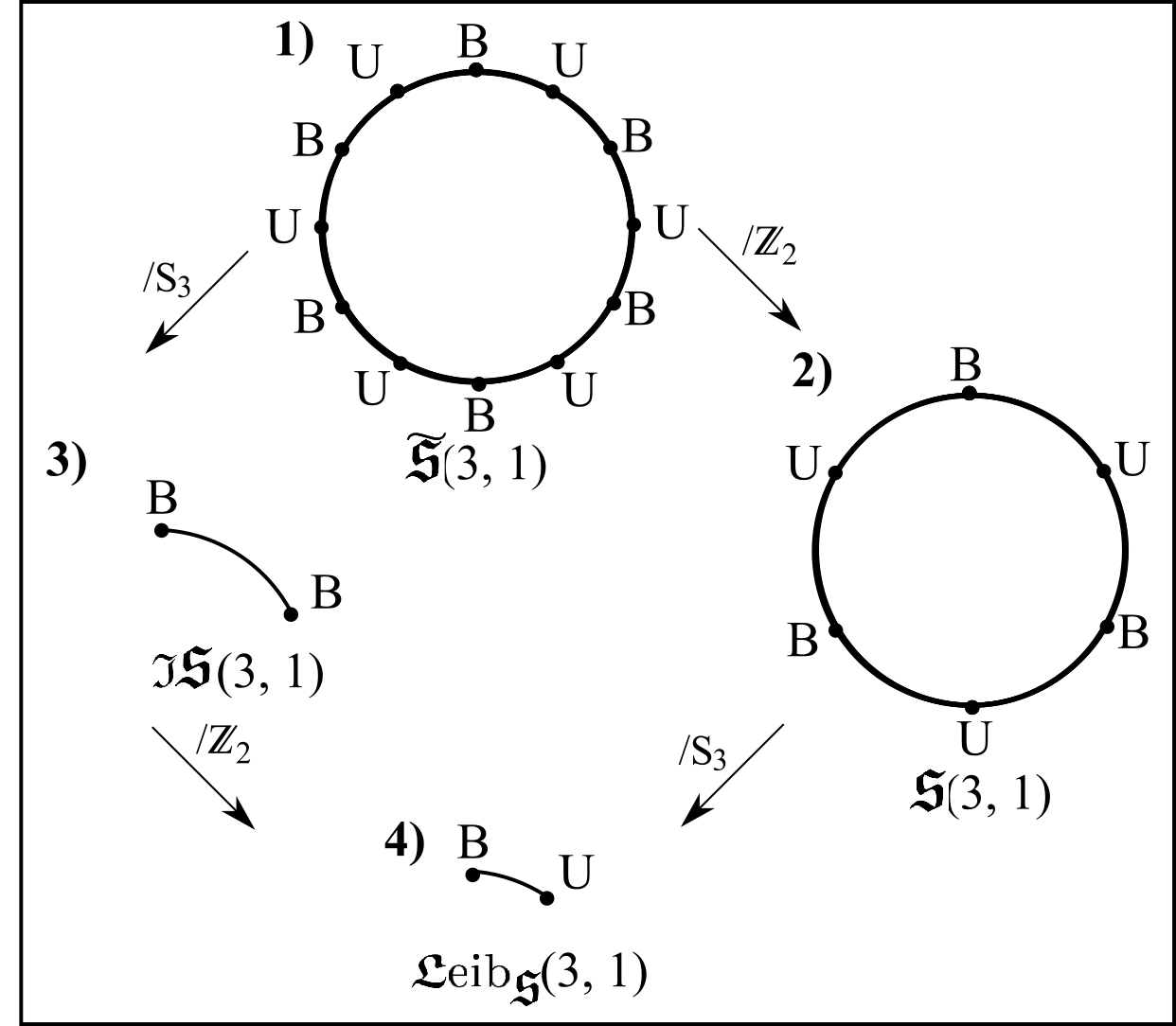}
\caption[Text der im Bilderverzeichnis auftaucht]{        \footnotesize{Shape spaces as further decorated by notions of merger, U.  
1) The $\sFrS(3, 1)$     circle now takes the form of a clock-face with even hours marked B and odd hours marked U.   
2) The $\w{\sFrS}(3, 1)$ circle has only even hours marked at all, once again alternating between B and U labels.  
3) The $\FrI\sFrS(3, 1)$ 2-hour arc and with end-points symmetrically labelled B.  
4) The $\Leib_{\tFrS}(3, 1)$ 1-hour arc with end-points labelled asymmetrically B and U.} }
\l{S(3, 1)-Metric-2} \end{figure}          }

\n{\bf Proposition 1} In $\Leib_{\sFrS}(3, 1)$,  

\m 

\n i) the incipient uniformity equations (\r{Lag-Incipient-Uniformity-Eqs}) pick out precisely just the coincidence-or-collision B and 
the shape U depicted in column 2 of Fig \r{Cluster-Coincidence}. 

\m 

\n ii) Both the entire and adjacent uniformity equations uniquely pick out just one shape U.

\m 

\n{\bf Definition 1} We call this special shape denoted by U -- or U(3, 1) in comparing different $(N, d)$ models -- the {\it uniform shape}. 
The adjective `Lagrangian' is occasionally a useful clarification. 
On the other hand the adjective `maximal' would superfluous for (3, 1), 
as by the above uniqueness U is the {\sl only} such shape with any element of Lagrangian uniformity.

\m 

\n{\bf Remark 1} Thus the uniformity structure $\Uni$ consists of just one point in this case; 
if we work in the other $\FrG\FrS$, $\Uni$ has 2, 3 or 6 distinct U points, depending on modelling assumptions in exactly the same manner as the B's count out. 

\m 

\n{\bf Definition 2} The {\it relative separation diagram} -- displaying all relative separations -- (column 3 of Fig \r{Cluster-Coincidence}) 
is more generally useful in diagnosing and quantifying uniform states, as indicated in its caption.
Elements of uniformity can be marked thereupon on all subsets of equal lengths using the elementary geometrical notation of small perpendicular segments.  

\m 

\n {\bf Proposition 2} In shape space, symmetry places the U points mid-arc between adjacent B points: Figure \r{S(3, 1)-Metric-1}. 

\m 

\n{\bf Remark 2} Finally, for one or both of $N \geqs 4$ and $d \geqs 2$, 
there are multiple points in Leibniz space with elements of Lagrangian uniformity exhibiting moreover various distinct strengths of uniformity.

\subsection{(3, 1) Symmetric shapes}

\n{\bf Observation 1} In the unlabelled case, U is additionally a reflection-symmetric shape (about the midpoint of the shape-in-space).  

\m 

\n{\bf Definition 1} For use as the models become larger, 
define the {\it symmetry structure} $\Sym(\FrG\FrS)$ to be the space of shapes manifesting at least one element of symmetry acting on space.

\m 

\n{\bf Proposition 1} i) The unlabelled U is {\it the only} symmetric state for 3 points in 1-$d$ (alongside O if its inclusion is permissible).  
Thus $\Sym = \Uni$ for (3, 1).  

\m 

\n{\bf Remark 1} Summarizing so far,
\be 
\mU = \mbox{Ref} = \diam_{\sm\sa\sx}  \m  . 
\ee
\be
(\mbox{extremal structure})  =  \{ \diam_{\sm\si\sn}, \, \diam_{\sm\sa\sx} \} 
                             =  \{\mB, \, \mU\}                                 \m  .
\ee
ii) For $N = 1$, there is only one (non-)shape, so 
\be
\mO = \mG 
    = \mU 
	= \mbox{Ref} 
\ee
iii) For $N = 2$, there are just two possible (non-)shapes: 
\be
\mO \neq  \mG 
      =   \mU 
	  =   \mbox{Ref}  \m  . 
\ee
\n{\bf Remark 2} Symmetry in space combines with topology to explain quite a lot of the diversity of shapes.
The current treatise's Lagrangian uniformity notions can moreover be seen as symmetries acting on some set of separations.

\section{Jacobian-level metric study}\l{Jac-UMS}

\subsection{Spatial representations of mergers and cluster hierarchy}

\n{\bf Definition 1} The {\it CoM cluster hierarchy diagram} (row 1 of Fig \r{Cluster-Coincidence})  
gives immediately accessible information concerning the relative positions of clusterings on different echelons of the hierarchy.  
This is useful, firstly for characterizing and spotting notions of merger. 
Secondly for finding further attributes of shapes which are geometrically distinguished 
in one or both of space and shape space, such as characterizations of partial uniformity.    

\m 

\n{\bf Definition 2} The {\it CoM coincidence diagram} (row 2 of Fig \r{Cluster-Coincidence}) 
is a simplification of the preceding which retains the clustering coincidences but discards non-coincident CoM positions. 
This isolates the characterization of merger at the expense of losing sight of other attributes such as characterizations of partial uniformity.

\m 

\n{\bf Remark 1} This treatise also distinguishes between hierarchical conditions which are in fact already implied by point positions 
(or by relatively lower hierarchy conditions), and independent ones.   
The (3, 1) general shape serves as an example of implied hierarchical conditions, according to the following rigidity. 

\m 

\n{\bf Proposition 1} For any (3, 1) shape, the pairwise CoM positions form a back-to front half-sized copy of the point positions themselves (Fig \r{Pre-Euclid}).

\m 

\n{\bf Remark 2} As we shall see in Papers II and IV, however, $N \geqs 4$, ceases to have a counterpart of this total rigidity,  
by which CoM hierarchy positions become interesting as a discerning, rather than just ubiquitous rigidity, feature.  
%
{\begin{figure}[ht]
\centering
\includegraphics[width=0.23\textwidth]{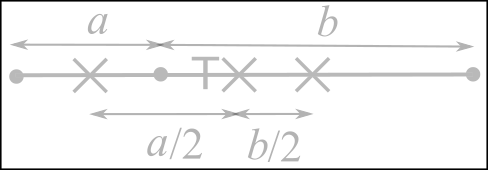}
\caption[Text der im Bilderverzeichnis auftaucht]{\footnotesize{(3, 1) cluster hierarchy rigidity.}} 
\l{Pre-Euclid}\end{figure} } 

\subsection{(3, 1) Mergers}\l{(3,1)-Mergers}

A priori, any of the $\fA$ could coincide with any of the others.
We exclude the self-$\fP$ case as we have already separately analyzed coincidences-or-collisions at the topological level, 
and the self-$\fO$ case as there is only ever one $\fO$.
For (3, 1) normalizable shapes, moreover, the following elements of rigidity occur. 

\m 

\n 1) Mutual-($\fP$, $\fX$) can only occur in the U shape. 

\m 

\n 2) Self-$\fX$ can only occur in the B shape, which is excluded from the pure mergers by being a coincidence-or-collision.   

\m 

\n 3) Mutual-($\fP$, $\fT$) can only occur if the particle in question is at the centre of the U shape, by which it is not a new case. 

\m 

\n 4) Mutual-($\fX$, $\fT$) can only occur if the $\fX$ in question is at the centre of the U shape, by which it is not a new case either. 

\m 

\n These exhaustively conspire to return the `double rigidity' promised in Sec (\r{Merger}). 

\m 

\n{\bf Proposition 1 (Double Rigidity of Uniformity)} i) (3, 1) only supports one type of merger, $\mM^{\sfP\sfX\sfT} =: \mM$. 

\m 

\n ii) In $\Leib_{\sFrS}(3, 1)$, this notion of merger is realized by just one point M, which coincides with the already introduced uniform shape U. 
[So in this case the merger structure $\Merger$ is just a single point.]

\m 

\n{\bf Remark 1} While this is indeed a special shape, it is not a new one at this stage of the analysis, 
having already dropped out in the logically prior Lagrangian-level part of of the analysis.  

\m 

\n{\bf Remark 2} See Column 2 of Fig \r{(3, 1)-Shapes} for M's  cluster hierarchy coincidences.  
%
{            \begin{figure}[!ht]
\centering
\includegraphics[width=0.7\textwidth]{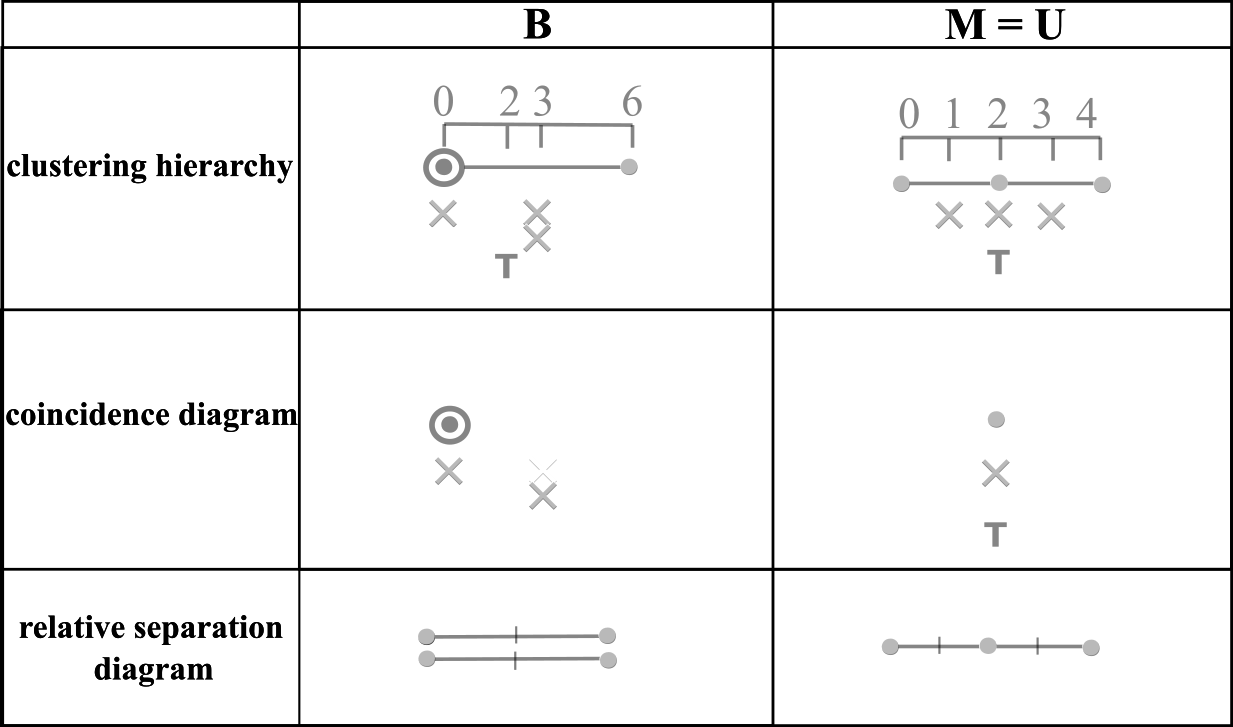}
\caption[Text der im Bilderverzeichnis auftaucht]{        \footnotesize{B and U configurations' hierarchy, coincidence and separation diagrams compared. 
The small perpendicular lines indicate equal lengths.
B's equal lengths are not taken to count toward uniform structure because they are induced by a coincidence, whereas U's do count as uniform structure.} }
\l{Cluster-Coincidence} \end{figure}          }

\m 

\n{\bf Remark 3} The (3, 1) model turns out to be deceptively simple in all three of the following ways (see Part II for a demonstration). 
(3, 1) is moreover minimal as regards supporting normalizable mergers.  

\m 

\n{\bf Remark 4} In summary,  
\be
\mM = \mU 
    = \mbox{Ref} 
	= \diam_{\sm\sa\sx} \m  \mbox{ for (3, 1)} .
\l{MUSym}
\ee 
\n{\bf Remark 5} Part II demonstrates that (4, 1) already has multiple notions of merger, 1-$d$ subspace parts in $\Merger$, $\Co$, $\Uni$ and $\Sym$, 
which are moreover not coincident spaces.   
While the most uniform state U and diameter per unit moment of inertia state $\diam_{\sm\sa\sx}$ remain unique up to labelling, 
they cease to be realized by the same shape. 
Moreover, for (4, 1), $\Uni \subset \Merger$, but on the other hand, for (3, 2) $\Merger \subset \Uni$.

\subsection{(3, 1) Jacobian notions of uniformity}\l{UJ}
%
{            \begin{figure}[!ht]
\centering
\includegraphics[width=0.43\textwidth]{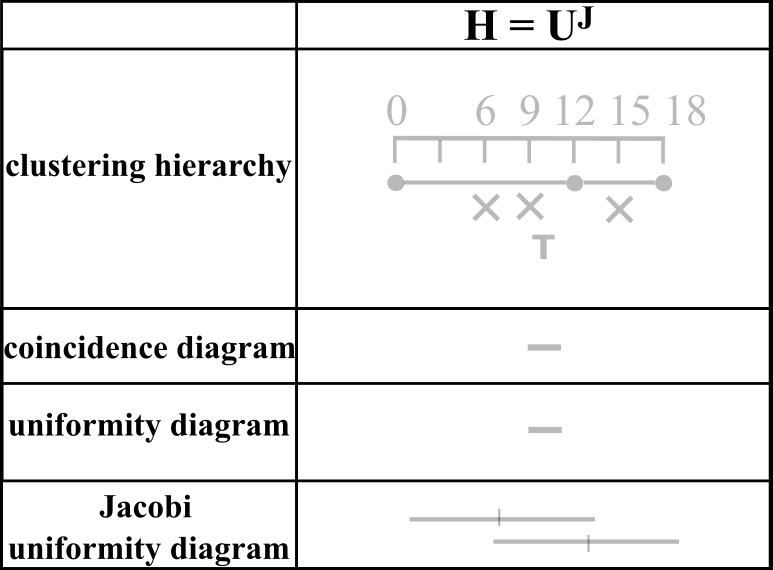}
\caption[Text der im Bilderverzeichnis auftaucht]{        \footnotesize{The Jacobi-uniform alias halfway configuration's clustering, coincidence, and uniformity diagrams.} }
\l{H-UJ} \end{figure}          }

\n{\bf Proposition 1 (Jacobi uniformity's double rigidity} In $\Leib_{\sFrS}(3, 1)$ 

\m 

\n i) There is only one type of Jacobi uniformity.  

\m 

\n ii) The corresponding Jacobi uniformity equations have a unique solution: the shape of Fig \r{H-UJ}.  

\m 

\n{\bf Remark 1} This is a new special shape, call it $\mU^{\sJ}$ where the J superscript emphasizes the Jacobian nature of this notion of uniformity.  
Thus $\CoM$-$\Uni(\Leib_{\sFrS}(3, 1))$ consists of just the point $\mU^{\sJ}$.  

\m 

\n{\bf Remark 2} Again, adding one further point-or-particle to the model suffices to break both of these rigidities; 
moreover, we shall see in papers II and III that H plays a significant Shape-Theoretic Aufbau Principle role in both (4, 1) and (3, 2).

\subsection{Shape space centres}\l{Z(3,1)}

\n{\bf Definition 1} The interesting version of `halfway shape' between $\mB$ and $\mU$  turns out to be the mass-weighted space version, $\mH$.  
This moreover coincides with the Jacobi-uniform shape $\mU^{\sJ}$.  

\m 

\n{\bf Remark 1} By eq.\ (\r{s-mws}), this is distinct from its non-mass-weighted counterpart.

\m 

\n{\bf Structure 1} Let us finally introduce the notion of {\it shape space centre} for each of $\FrI\FrS$ and $\Leib_{\sFrS}$.
For the (3, 1) model, 
\be
Z(\FrI\FrS(3, 1)) = \mU \m  :
\ee
the uniform state in the obvious Lagrangian sense, whereas 
\be 
Z(\Leib_{\sFrS}(3, 1)) = \mU^{\sJ} \m  :
\ee
the uniform state in the more subtle Jacobian sense.  
We shall see furthermore in Parts II to IV that shape space centres {\sl commonly} coincide with notions of maximally uniform shape.

\section{The qualitatively distinct (3, 1) shapes}

\subsection{5 exact qualitative types of shape in (3, 1) Leibniz space}

These exact qualitative types here arise by the breakdown in Fig \r{(3, 1)-Shapes}, 
are laid out over shape space in Fig \r{5}.a) with topological adjacency graph in Fig \r{5}.b) 
each special point's properties collected together in the summary figure Fig \r{5}.c).  
%
{            \begin{figure}[!ht]
\centering
\includegraphics[width=1.0\textwidth]{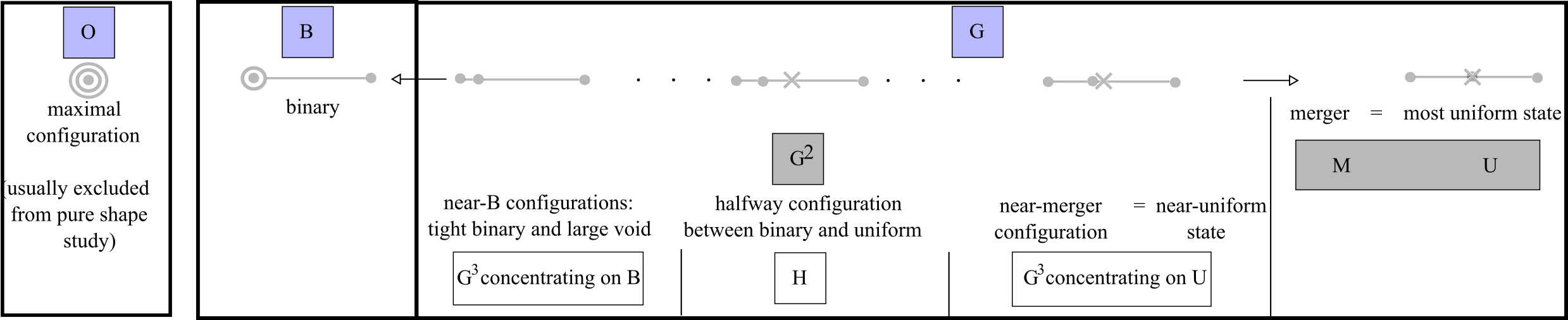}
\caption[Text der im Bilderverzeichnis auftaucht]{        \footnotesize{Upon entertaining metric-level information, 
the 1-$d$ 3-point example's general configuration topology class models extent of clumping-or-uniformity by the position of the interior point.  
This transits from coinciding with an exterior point-or-particle to coinciding with CoM($ab$) for $a$, $b$ the two exterior particles.  
The white arrows are used for now as a rough indication of proximity in configuration space.
The `blue' topological class G splits into the M = U and a disjoint 1-$d$ continuum remainder $\mG^2$ (denoting `doubly generic'). 
This metric class level of structure is labelled by grey squares throughout this treatise.
$\mG^3$ is $\mG^2$ minus its midpoint, which removal further splits $\mG^3$ into two components: 
the tight binaries $\mG^{\sT}$ and the loose binaries $\mG^{\sL}$.} }
\l{(3, 1)-Shapes} \end{figure}          }
%
{            \begin{figure}[!ht]
\centering
\includegraphics[width=0.85\textwidth]{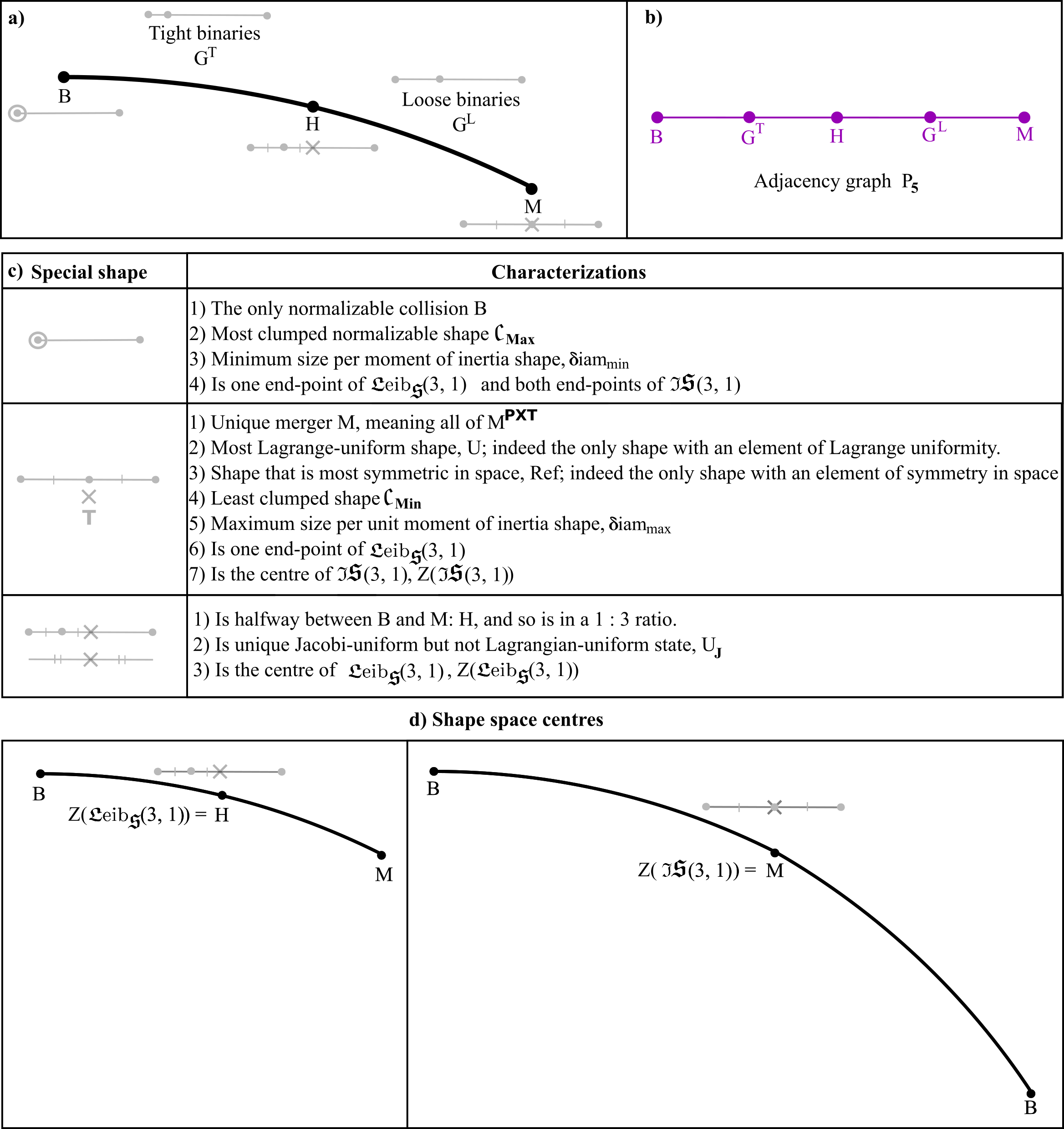}
\caption[Text der im Bilderverzeichnis auftaucht]{        \footnotesize{a) 5 qualitative types of (3, 1) shape. 

\m 

\n b) Representation of these 5 qualitative types as an adjacency graph $\mP_5$, which becomes a useful visual and computational tool for larger shape spaces.

\m 

\n c) Pile-up of conceptual characterizations on the (3, 1) model's 3 special points.  

\m 

\n d) Shape centres of $\FrI\sFrS(3, 1)$ and $\Leib_{\tFrS}(3, 1)$.  
} }
\l{5} \end{figure}          }

\subsection{3 approximate notions of shape in (3, 1) Leibniz space}\l{Approx-(3, 1)}

\n In considering metric-level information about (3, 1) shapes, there are also meaningful notions of shapes which are near one of the coincidence-or-collision B, 
                                                                                                                                   the uniform shape U 
															    													           and the halfway shape H.  
These have the following meanings in analogue model of Astrophysics or Cosmology terminology.  

\m 

\n{\bf Definition 1} {\it Near-B shapes} exhibit a tight binary pair relative to the large void between this and the third point.  

\m 

\n{\bf Remark 1} These are characterized by the $\epsilon$-tolerant approximate coincidence-or-collision inequality (\r{Epsi-Collision-Eqs}), 
which for $\Leib_{\sFrS}(3, 1)$ is solved by just the one small piece of arc $\mB^{\approx}$ in Fig \r{3}. 

\m 

\n {\bf Definition 2} {\it Near}-U {\it shapes} are small perturbations about the most uniform state, which is the principal subject matter in Cosmology.      

\m 

\n{\bf Remark 2} These are characterized by by the $\epsilon$-tolerant approximate uniformity inequality (\r{Epsi-Lag-Uniformity-Eqs}), 
which for $\Leib_{\sFrS}(3, 1)$ is solved by just the one small piece of arc $\mU^{\approx}$ in Fig \r{3}. 

\m 

\n{\bf Remark 3} As regards the relation between the $\epsilon$-tolerance criterion in space to the $\eta$-tolerance criterion in shape space, 
and Sec \r{Approx-Discussion}'s matter of boundary points, e.g.\ for B, 
\be
\phi = \mbox{arctan}(\pm\epsilon) \approx \pm\epsilon
\ee
by Taylor expansion, and we then keep just the positive arc because B is a boundary point so the other arc is inadmissible.  
On the other hand, for U, 
\be
\phi = \mbox{arccot}(\epsilon) = \pi/2 \pm \epsilon \m  ,
\ee
by MacLaurin expansion, which, within the the 0 to $\pi/6$ range Leibniz space, translates to $\pi/6 \pm \epsilon$, 
of which just the negative arc is admissible by U being a boundary point. 
So in each case, the $\epsilon$ to $\eta$ relation is equality (to first order).  

\m 

\n{\bf Remark 4} There is no separate discussion to be had of near-mergers since the only merger M = U in the (3, 1) model.  
On the other hand, there are $\epsilon$-tolerant approximate Jacobi uniform shapes as follows.

\m 

\n{\bf Definition 3} The {\it near-}$\mU^{\sJ} = \mH$ {\it shapes} are small perturbations about the most Jacobi-uniform state. 

\m 

\n{\bf Remark 5} These are characterized by the $\epsilon$-tolerant approximate Jacobi uniformity inequality (\r{Epsi-Jacobi-Uniformity-Eqs}),   
which for $\Leib_{\sFrS}(3, 1)$ is solved by the now two-sided of arc $\mH^{\approx}$ in Fig \r{(3, 1)-Shapes}, since $\mH$ is now an interior point.   
There are moreover two types of near-H shapes, for the H configuration itself splits this two-sided arc into its two sides: 
the shapes starting to tend toward B and the shapes starting to tend toward U.  
See Fig \r{3} for this construct and a notation for it. 
 
\m 

\n{\bf Remark 6} If the points-or-particles are labelled, the aboves notion of closeness are for one specific cluster only.
See the next section for a count of the proliferation of approximate states ensuing from this and/or mirror image distinctness.  
Finally note that one simple way of having label-independent notions of uniformity and clumping is to consider all labellings at once, 
corresponding to a collection of equal small arcs in shape space, about each labelling of whichever of B, U and H.  
%
{            \begin{figure}[!ht]
\centering
\includegraphics[width=0.37\textwidth]{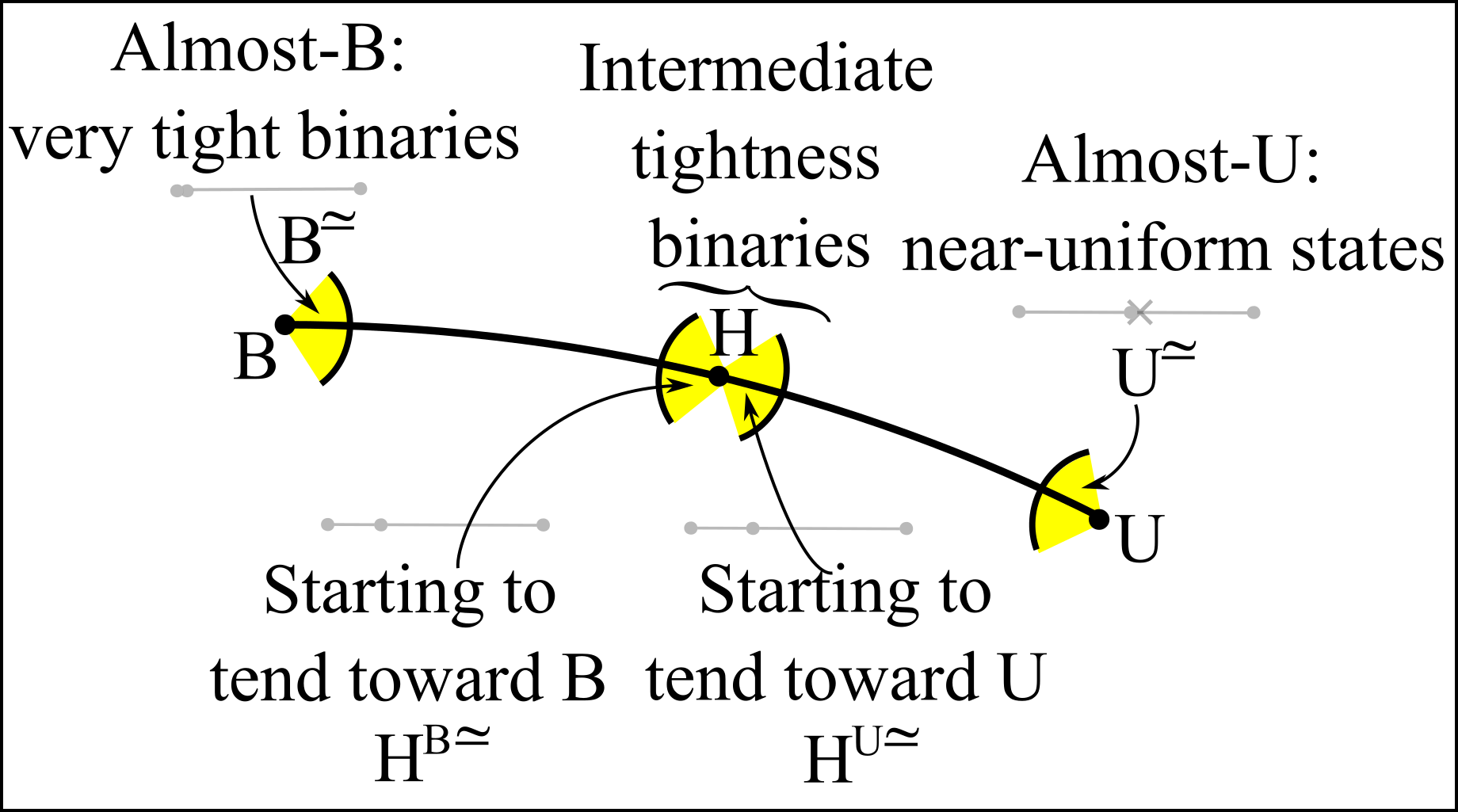}
\caption[Text der im Bilderverzeichnis auftaucht]{        \footnotesize{4 approximate qualitative types of (3, 1) shape.
The general notation for this is $(\mbox{exact qualitative type it is near})^{\approx}$, or 
$(\mbox{exact qualitative type it is near})^{\mbox{\scriptsize in the direction of } \approx}$ when relevant. 
This notation blossoms considerably in Papers II to IV. } }
\l{3} \end{figure}          }

\subsection{Counts of qualitative types for (3, 1) shape spaces more generally}

\n The count of qualitative types is based straightforwardly upon the following.\footnote{See Appendix A for what $E$ and $V$ are, 
and Appendix II.B for the generalization of these counts to the 2-$d$ shape space setting.}

\m 

\n{\bf Definition 1} The number of {\it exact qualitative types} is 
\be 
Q := E + V \m  .
\ee
{\bf Remark 1} This is furthermore 
\be
Q = 2 \, E
\ee 
if the shape space is topologically a cycle. 

\m 

\n{\bf Definition 2} The number of {\it approximate qualitative types} is two around each internal significant point and one for each perimeter significant point.

\m 

\n{\bf Remark 2} This counts out as 
\be 
Q_{\sa\sp\sp\sr\so\sx} = 2 \, V - P \m  , 
\l{Q-Approx}
\ee 
for $P$ the number of endpoints. 

\m 

\n{\bf Remark 3} If the shape space is topologically a cycle, $P = 0$, so (\r{Q-Approx}) simplifies to 
\be
Q_{\sa\sp\sp\sr\so\sx} = 2 \, V                      \m . 
\ee
\n{\bf Definition 3} The {\it total number of qualitative types}
\be
Q_{\st\so\st\sa\sll} := Q + Q_{\sa\sp\sp\sr\so\sx}   \m . 
\ee
\n{\bf Remark 4} This counts out as 
\be
Q_{\st\so\st\sa\sll} = 3 \, V + E - P                \m , 
\ee
which, in the case of a cycle, simpilifies to just 
\be
Q_{\st\so\st\sa\sll} = 4 \, V                        \m .
\ee
\n{\bf Remark 5} The uniformity structure case works out somewhat differently, as follows.
  
\m  

\n{\bf Remark 6} For these shape spaces of top configuration space dimension 1, the number of {\it exact qualitative types of uniformity} counts out as 
\be 
Q^{\sU} = V                                     \m .  
\ee
\n{\bf Definition 4} The number of {\it approximate qualitative types of uniformity} is as per definition 2, 
subject to the caveat now that even if an endpoint is absent from the structure, there is still an approximate regime near where such a deleted endpoint would be.

\m 

\n{\bf Remark 6} Thus this counts out once again as 
\be 
Q^{\sU}_{\sa\sp\sp\sr\so\sx} = 2 \, V - P       \m , 
\l{Q-U-Approx}
\ee
which, in the case of a cycle, simpilifies to just 
\be
Q^{\sU}_{\sa\sp\sp\sr\so\sx} = 2 \, V           \m . 
\ee
\n{\bf Definition 5} Finally the {\it total number of qualitative types of uniformity} is 
\be
Q^{\sU}_{\st\so\st\sa\sll} := Q + Q_{\sa\sp\sp\sr\so\sx}    \m . 
\ee
\n{\bf Remark 7} This counts out as 
\be
Q^{\sU}_{\st\so\st\sa\sll} = 3 \, V - P                     \m , 
\ee
which, in the case of a cycle, simpilifies to just 
\be
Q^{\sU}_{\st\so\st\sa\sll} = 3 \, V                         \m .   
\ee
\n Fig \r{(3, 1)-Q} then tabulates, for each of the 4 shape space, the number of qualitative types at the topological level, 
                                                              at the Lagrangian and Jacobian levels of metric structure, 
															  and of uniform structures. 
%
{            \begin{figure}[!ht]
\centering
\includegraphics[width=0.7\textwidth]{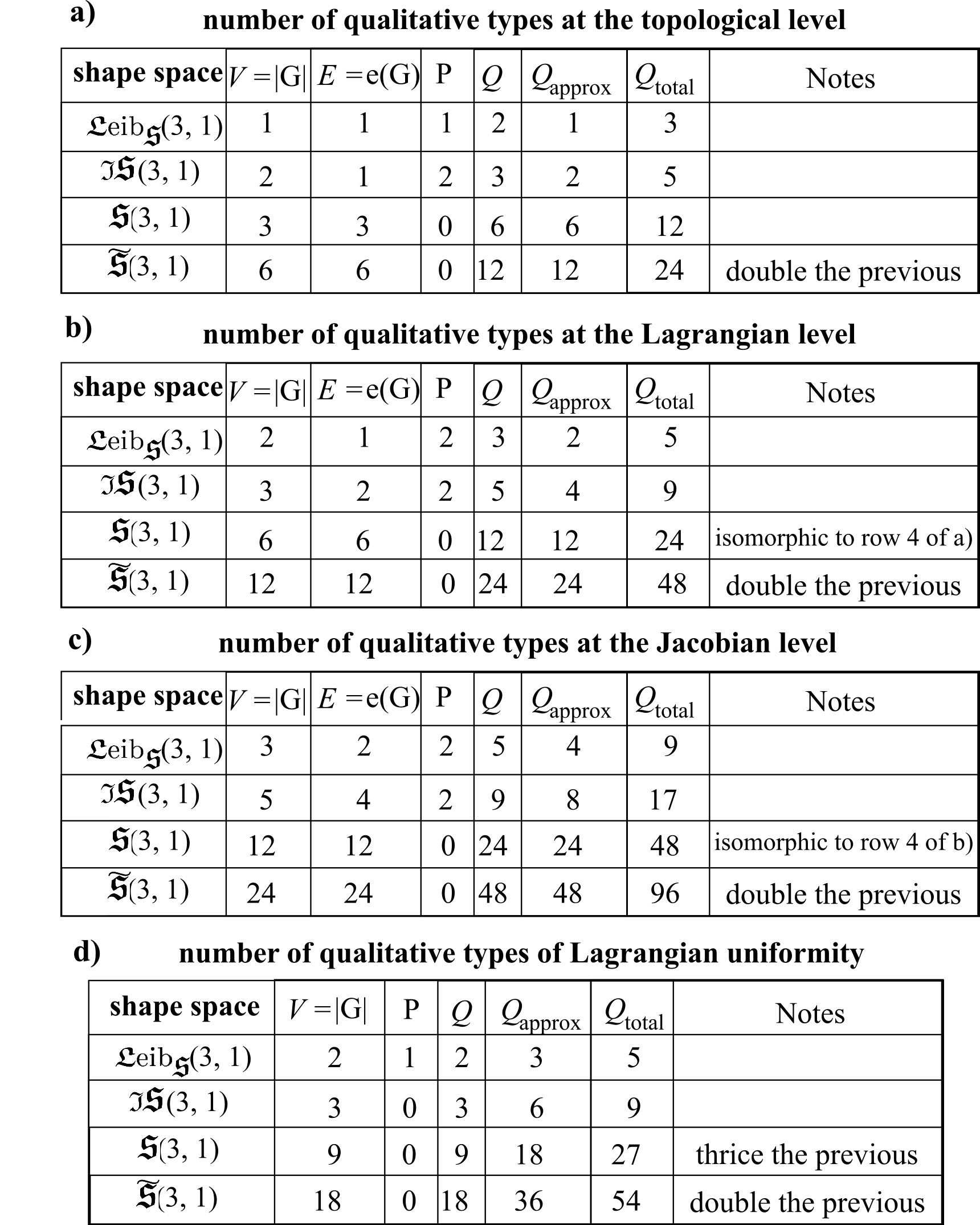}
\caption[Text der im Bilderverzeichnis auftaucht]{        \footnotesize{Count of qualitative types for the topological level, 
                                                              for the Lagrangian and Jacobian levels of metric structure, 
														  and for uniform structures.
															  The isomorphisms are at the level of the underlying graphs.  
															  } }
\l{(3, 1)-Q} \end{figure}          }

\vspace{10in}

\section{Relational spaces and extrinsic structure of shape space}\l{Rel-Ext-(3,1)}
%
{            \begin{figure}[!ht]
\centering
\includegraphics[width=0.7\textwidth]{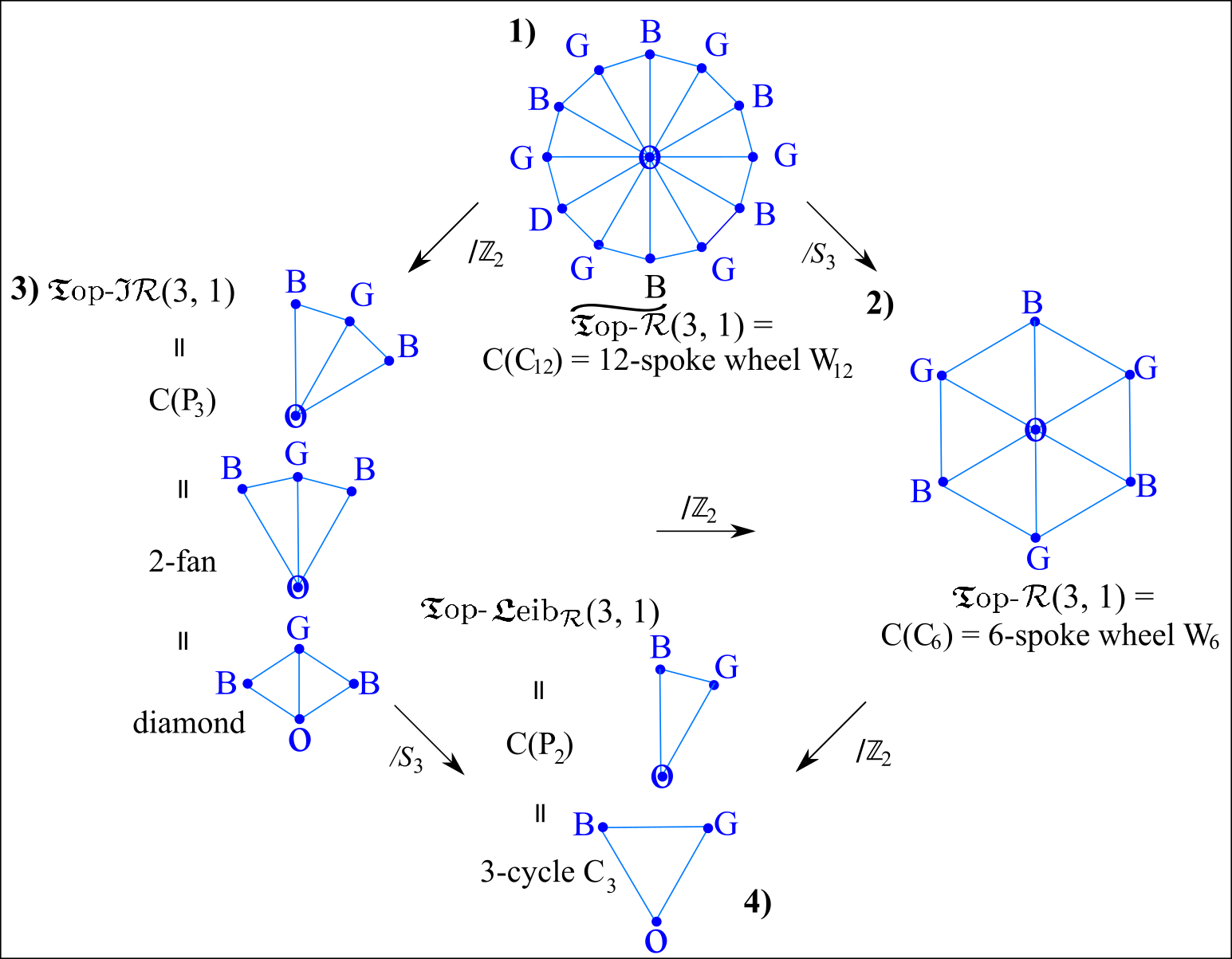}
\caption[Text der im Bilderverzeichnis auftaucht]{        \footnotesize{Relational spaces of topological shapes.} }
\l{R(3, 1)-Top} \end{figure}          }
%
{            \begin{figure}[!ht]
\centering
\includegraphics[width=0.6\textwidth]{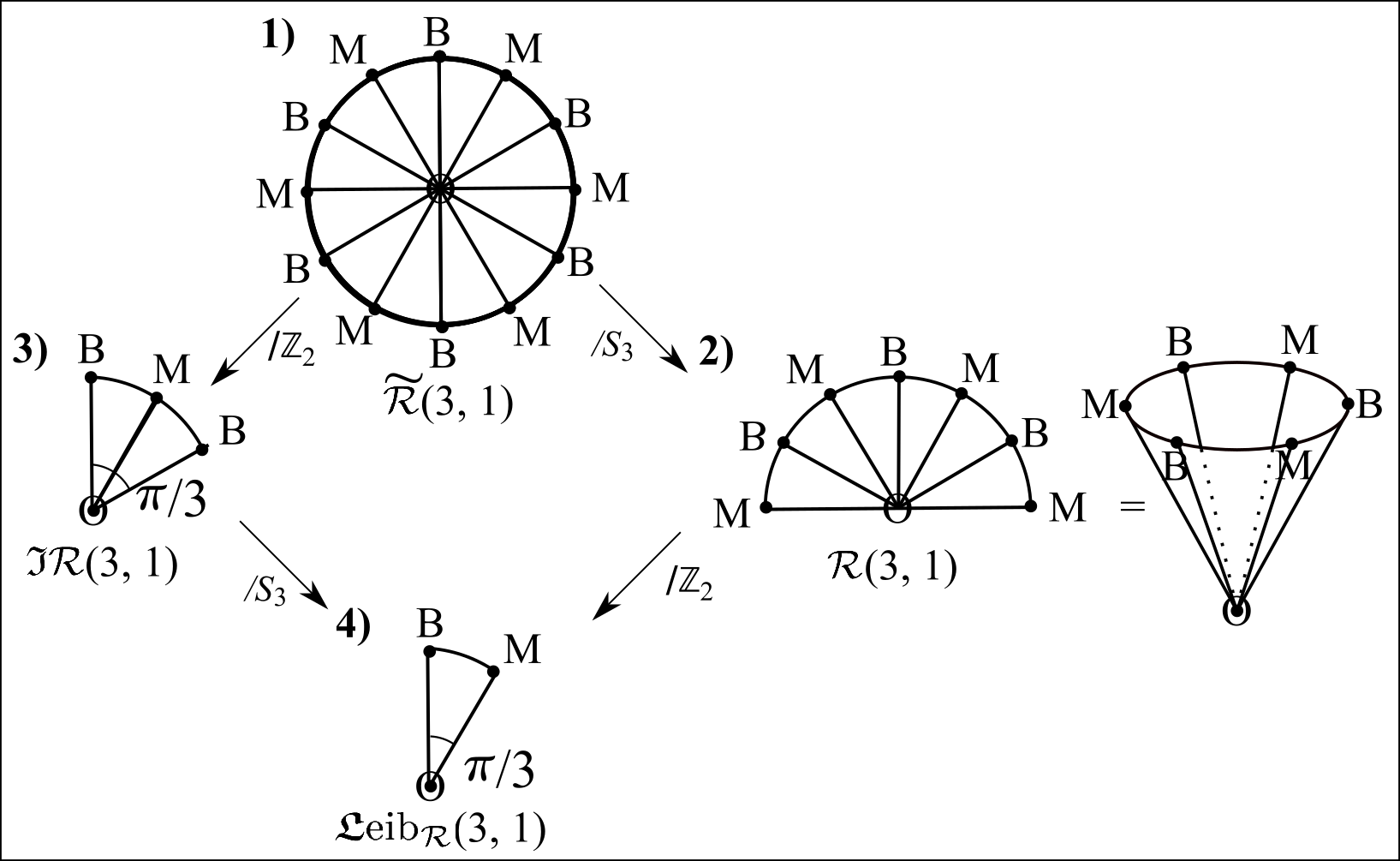}
\caption[Text der im Bilderverzeichnis auftaucht]{        \footnotesize{1) ${\cal R}(3, 1)$ is the cone over the circle, 
which can be viewed as a plane since it has no deficit angle. 
2) At the metric level, $\w{\cal R}(3, 1)$ is best viewed as a cone over the circle with deficit angle $\pi$.  
3) $\FrI{\cal R}(3, 1)$ and 4) $\Leib_{\cal R}(3, 1)$ are cone arcs with edge boundaries.  
These are the outcome of forming the cone over a manifold with boundary, 
in the case of the circle arcs with endpoints of Figs \r{S(3, 1)-Metric-2}.3-4), giving sectors with edges of width $\pi/6$ and $\pi/3$ respectively.
If O is excised, one has punctured planes, apex-less cones and cornerless sectors. 
If the B's are excised as well, e.g.\ due to infinite potentials there, then one has disconnected sectors in the first three cases.} }
\l{R(3, 1)-Metric} \end{figure}          }

\n{\bf Proposition 1} At the topological level, entertaining the relational space is the same as including O, 
so Fig \r{S(1, 1)-S(2, 2)-Top} already covers the outcome of this for 1 and 2 points. 
On the other hand, for 3 points, the underlying topological relational spaces are in Fig \r{R(3, 1)-Top}, 
and their metric decor counterparts are in Fig \r{R(3, 1)-Metric}.  

\m 

\n In the $\FrS(3, 1)$ circle's ambient $\mathbb{R}^2$, 
$\rho_1$ and $\rho_2$ [eqs (\r{rho-1-(3, 1)} and \r{rho-2-(3, 1)})] additionally serve as Cartesian axes (Fig \r{R(3, 1)-Axes}.a).

\m 

\n On the other hand, for $\w{\FrS}(3, 1)$ in circular representation, $\rho_1$ still provides a B--M axis, 
but the axis orthogonal to this is a new H--H axis (Fig \r{R(3, 1)-Axes}.3) with some further subtleties in subfigures 2) and 4).  
%
{            \begin{figure}[!ht]
\centering
\includegraphics[width=0.65\textwidth]{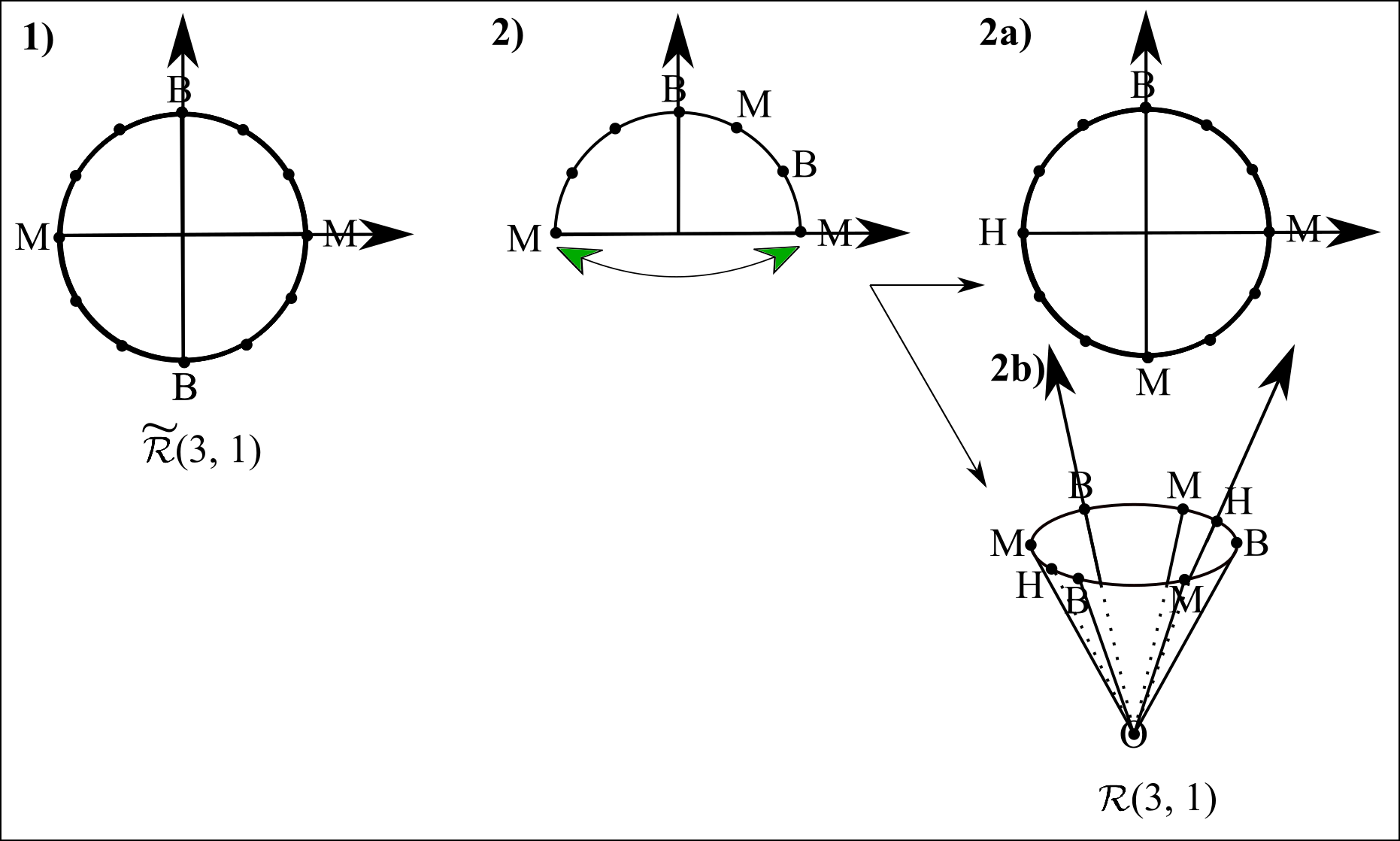}
\caption[Text der im Bilderverzeichnis auftaucht]{        \footnotesize{B--B and M--M directions serve as axes for the mirror images distinct case 1), 
but only a half-axis version of B--B and an identified vesion of M--M survive in the mirror image identified case's half-planar representation. 
2.a) Doubling angles, H--H enters use as a second orthogonal axis for an ambient space for $\w{\cal R}(3, 1)$, 
though in the relational space version itself (2.b), these are cone axes rather than perpendicular Cartesian axes.} }
\l{R(3, 1)-Axes} \end{figure}          }

\section{shape(-and-scale) space automorphisms}\l{Killing-(3,1)}

\subsection{For topological shape spaces}\l{Graph-Aut-Argument}

\n{\bf Proposition 1} 
\be
Aut(\Top\mbox{-}\FrS(3, 1))         = Aut(\mC_{12} \mbox{ as labelled}) 
                             = D_6 
					         = S_3 \times \mathbb{Z}_2         \m  , 
\ee
\be
Aut(\Top\mbox{-}\w{\FrS}(3, 1))     = Aut(\mC_{6} \mbox{ as labelled}) 
                             = D_3 
						     = S_3                             \m  , 
\ee
\be
Aut(\Top\mbox{-}\FrI\FrS(3, 1))     = Aut(\mP_{3} \mbox{ as labelled}) 
                             = \mathbb{Z}_2                    \mma       \mbox{ and }
\ee
\be
Aut(\Top\mbox{-}\Leib_{\sFrS}(3, 1)) = Aut(\mP_{2} \mbox{ as labelled}) 
                             = id                              \m  . 
\ee
Also 
\be 
Aut(\mC(G) \mbox{ with cone point labelled distinctly}) = Aut(G) \m  ,
\ee
so the above automorphism groups are unchanged under $\FrS \longrightarrow {\cal R}$.

\subsection{Shape momenta}

\n{\bf Remark 1} We aim to work in the shape variables $\u{\nu}^i$ of eq. (\r{n-def}) for shape space, 
and in the Cartesian variables $\u{\rho}_i$ for shape-and-scale space.  
We shall require notation for the conjugate momenta of most of the variables in the build-up to these: 
$\u{p}_I$  to the point or particle positions        $\u{q}^I$, 
$\u{P}_i$  to the relative Jacobi coordinates        $\u{R}^i$, 
$\u{\pi}_i$ to the mass-weighted Jacobi coordinates  $\u{\rho}^i$ and finally 
$\u{\upi}_i$  to the                                 $\u{\nu}^i$ themselves. 

\m 

\n{\bf Remark 2} The {\it total dilational momentum} is 
\be
D \:=  \sum_{I = 1}^N \u{q}^I    \cdot \u{p}_I    
  \es  \sum_{i = 1}^n \u{R}^i    \cdot \u{P}_i   
  \es  \sum_{i = 1}^n \u{\rho}^i \cdot \u{\pi}^i 
  \es  \sum_{i = 1}^n \u{\nu}^i    \cdot \u{\pi}^{\nu}_i    \m  . 
\ee
Note that its definition is the dot product analogue of the angular momentum cross product, 
whereas the next equality is by the Jacobi map, the one after that by eq.\ (\r{rho-R}) and the final one by eq.\ (\r{n-def}); 
on shape space $\rho = \mbox{constant}$.  
Similarly, th {\it partial dilational momenta} (c.f.\ the more common term `partial moments of inertia' $\bigiota_i$) are given by  
\be
D_i := \u{\nu}^i    \cdot \u{\pi}^{\nu}_i \mbox{ (no sum) } .
\l{PDM}
\ee
\n{\bf Remark 3} In the case of the (3, 1) model, the {\it relative dilational momentum} is given by 
\beq
\sD\si\sll :=   \pi_{\varphi} 
            =   \rho^1 \pi_2 - \rho^2 \pi_1 
		    =   \nu_1  \upi^{\nu}_2 - \nu_2  \upi^{\nu}_1 
		   \es  D_2 \frac{\nu_1}{\nu_2}   - D_1 \frac{\nu_2}{\nu_1}  
		   \es  D_2 \frac{\rho_1}{\rho_2} - D_1 \frac{\rho_2}{\rho_1}        \m  . 
\eeq
The first equality here is mathematically equivalent to the usual expression for 2-$d$ angular momentum, 
though it now has a different physical interpretation as relative dilational momentum 
(the dilational momentum analogue of the more well-known concept relative angular momentum).
This corresponds to a particular exchange of dilational momentum between the `base' and `median' clusters in the squashed triangle interpretation of (3, 1).
A more correct way therefore of identifying this move's mathematics is to attribute it rather to $SO(2)$'s mathematics 
free from any connotations of physical interpretation.  
That quantities with physical interpretations othere than angular momentum admit the same mathematics has been known at least as far back as the 
work of Felix Smith \c{Smith60}, on which we comment further in Part III after seeing a wider range of such quantities arise from Shape Theory.  

\m 

\n{\bf Remark 4}  The second and third equalities follow from Remarks 1 and 2 applied to 1-$d$.
The final equality manifestly casts this in terms of ratios of relative Jacobi separations by cancelling off two $\rho$ factors in each fraction.

\m 

\n{\bf Remark 5} In this treatise, I use the notation ${\cal D}\mi\ml$ more generally for whichever type of relative dilational momenta.  

\m 

\n{\bf Remark 6} In the shape-and-scale case, one can also work usefully with the shape--scale split variables $\rho$ and $\varphi$.   
There is now also an absolute dilational momentum $\pi_{\rho}$ conjugate to $\rho$, which moreover enters a further expression for $D$, 
\be
\sD = \rho \, \pi^{\rho} \m  .  
\ee

\subsection{Killing vectors and isometry groups}

\n{\bf Proposition 1} The (3, 1) shape-and-scale spaces are pieces of the plane $\w{\cal R}(3, 1) = \mathbb{R}^2$.
This two translational Killing vectors
\beq
\frac{\pa}{\pa\rho_1}  \mma   \frac{\pa}{\pa\rho_2}  \m  , 
\eeq
and one rotational Killing vector 
\beq
\frac{\pa}{\pa\varphi} \es  \rho_1\frac{\pa}{\pa\rho_2} - \rho_2\frac{\pa}{\pa\rho_1}  \m  .
\eeq
These correspond to $\pi_1$, $\pi_2$ and the first form given of $\pi_{\varphi}$.  

\m 

\n{\bf Proposition 2} For mirror image identified shapes, 
the shape space ${\cal R}(3, 1)$ is an identification of the corresponding half-plane (Fig \r{R(3, 1)-Metric}). 
By continuity considerations, this globally breaks both translations but not the rotation.
The same occurs for $\FrI{\cal R}(3, 1)$ and $\Leib_{\cal R}(3, 1)$, the boundaries of both of which are given in Fig (\r{R(3, 1)-Metric}). 
%
{\begin{figure}[ht]
\centering
\includegraphics[width=0.7\textwidth]{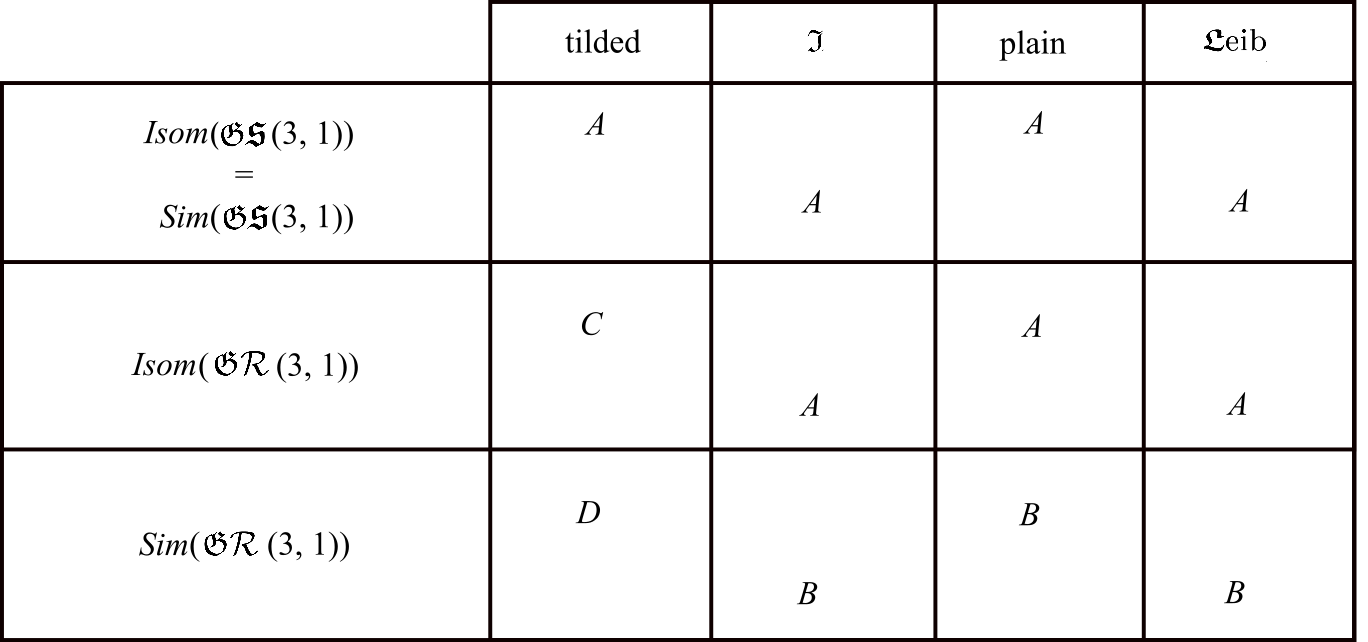}
\caption[Text der im Bilderverzeichnis auftaucht]{\footnotesize{Table of pattern of isometry and similarity groups 
for shape(-and-scale) spaces for 3 particles in 1-$d$.}} 
\l{12-1}\end{figure} } 
 
\m  
 
\n{\bf Structure 1} Fig \r{12-1} makes use of 
\be
A              :=         Rot(2) 
                =         SO(2)
                =         U(1)				
          \m \s{m}{=} \m  \mathbb{S}^1                                             \m  ,
\ee
\be
B              :=         Rot(2) \times Dil 
                =         SO(2) \times \mathbb{R}_+
         \m \s{m}{=} \m   \mathbb{S}^1 \times \mathbb{R}_+
                =         \mC(\mathbb{S}^1) 
				=         \mathbb{R}^2                                     \m  ,
\ee
\be
C              :=         Eucl(2)
                =         Tr(2) \rtimes Rot(2) 
                =         \mathbb{R}^2  \rtimes  SO(2)  
          \m \s{m}{=} \m  \mathbb{R}^2 \rtimes \mathbb{S}^1                      \m  ,
\ee
\be
D              :=         Sim(2)
                =         Tr(2) \rtimes \{ Rot(2) \times Dil\}
                =         \mathbb{R}^2  \rtimes  \{ SO(2) \times \mathbb{R}_+ \}  
          \m \s{m}{=} \m  \mathbb{R}^2 \rtimes \mC(\mathbb{S}^1) 
                =         \mathbb{R}^2 \rtimes \mathbb{R}^2    \m  .  
\ee

\section{Conclusion: (3, 1) similarity shapes as the simplest model of Shape Theory}\l{Conclusion-I}

In Part I, we have reviewed previous results about the shapes made by $N$ points in flat space.
In so doing, we have additionally built in quite a number of principles and of structural variants. 
In particular, 

\m 

\n 1) we have systematically included indistinguishable-point versions, without and with mirror image identification: $\FrI\FrS$ and $\Leib_{\sFrS}$ respectively. 
The latter name is with reference to Leibniz' Identity of Indiscernibles.  

\m 

\n 2) We have also made more use of geometrical invariants than has hitherto been usual in the shape-theoretic literature, 
and used these to sharply pick out notions of inhomogeneity and uniformity. 
This point of view extends to many other shape theories, whether established \c{Kendall, PE16, Affine-Shape-1, Affine-Shape-2} or new \c{Project-1}.  

\m 

\n 3) We additionally identified each `obvious' notion of inhomogenity and uniformity as a condition on invariants based on relative Lagrangian coordinates. 
This analysis moreover points to considering relative Jacobi coordinate versions of each condition, 
due to these being both dynamically useful {\sl and} clean descriptors of (shape-and-scale) space geometry. 
Note that location data in Statistics is open to having its separations, angles... later cluster separations studied in parallel to all of this paper's developments.    

\m 

\n 4) We furthermore emphasized reformulating collision sets as a fortiori differential-geometric structures, 
and follow suit with merger structures and a number of uniformity structures. 
These are a network of layered subspaces of the corresponding shape(-and-scale) space, which shall eventually require description as stratified manifolds. 

\m 

\n 5) We also explained how Shape Theory can be used for all of abstract flat Geometry, location data analysis in Statistics and Dynamics, 
in particular contrasting the different aspects emphasized in each of Geometry and Dynamics and how they none the less contain a common shape-theoretic core.
Papers II to IV shall moreover develop Shape Theory as a {\sl joint} theory of Geometry and of some aspects of Dynamics. 

\m 

\n 6) We also outlined how $N$-point Shape Theory is a partial analogue of General Relativity as a dynamical system, 
and also of General Relativity like notions of Background Independence: a major topic in the Foundations of Physics and in Quantum Gravity
Shape Theory has thereby already yielded many useful clarifications as regards the nature of Background Independence \c{ABook}.  
Moreover only a few of the simplest Shape Theories (small groups, small point numbers, small spatial dimension, flat space) 
have so far been considered in this vein in any detail.  
So we expect many further useful clarifications about Background Independence to ensue from continuing this study, 
for which the current treatise lays out systematic foundations. 

\m 

\n We illustrated this paper's configurational and shape-theoretic notions for now with a simple but none the less pedagogical example: 
the similarity shapes made by 3 points in 1-$d$, and the corresponding shape-and-scale model.  
These examples already serve for the following purposes.  

\m 

\n A) To define a topological notion of shape and to show that the corresponding notions of shape space for this are graphs. 
This (3, 1) similarity model's first limitation is that its topological shapes are identical to partitions, whereby its shape spaces are partition spaces.  
We show however in Part II that for $(N, 1)$ models with $N \geq 4$, this coincidence with partitions no longer occurs.  

\m 

\n B) At the metric level, the (3, 1) shape spaces are various decorations of the circle and of the interval. 
These support very few notions of distinguished shape indeed: just the binary coincidence-or-collision B at the topological level, 
and then also the uniform shape U at the metric level, along with the shape H half-way between these.  
Having very few distinguished configurations means that, while a lot of shape notions and properties can already be defined for (3, 1), 
all of the realizations of these notions pile up on these three special points, as per Fig \r{5}.c).  
This provides a second justification for the (3, 1) model having pedagogically-useful status: it already serves to introduce many notions 
which we then show to be distinctly realized for Part II's (4, 1) model, but which cannot yet be distinctly realized for the (3, 1) model itself.  
For (4, 1), (3, 2) and beyond, there are not only more special points, but also codimensional room for arcs and subsequently submanifold pieces, 
and then yet further freedom (see Part II) in the form of being able to make stratified-manifold 
\c{Whitney46, Thom55, Whitney65, Thom69, Pflaum, Kreck, Sniatycki} networks out of these lower-$d$ subsets.
A great increase in distinctly realized notions of shape ensues.  
Some of the Differential Geometry advances in Shape Theory must however await the realization of such networks in Papers II to IV, 
whereas others require distinct stratified manifold, general bundle and sheaf-theoretic inputs specific to $d \geqs 2$: Papers III, IV and 
\c{Project-2, Affine-Shape-1}.

\m 

\n C) We also noticed that topologically meaningful coincidences-or-collisions do not suffice to understand the periphery of Leibniz space: 
it is B at one end but the only metrically-significant U = M shape at the other end.  
On the other hand, $\FrI\FrS$ has a fully understood periphery: two B points. 
This represents a first price to pay for taking Leibniz's Identity of Indiscernibles to its logical conclusion for the (3, 1) similarity model: 
$\FrI\FrS$'s periphery already admits an adequate topological characterization whereas $\Leib_{\sFrS}$ requires a topological-and-metric level characterization.   
A second price to pay is that $\FrI\FrS$ has a shape theoretic centre which is both a stronger notion of centre and a stronger realization of uniformity 
than $\Leib_{\sFrS}(4, 1)$'s: U versus the somewhat ambiguous H notion (one of which is {\sl Jacobi}-uniform).  
This is in the context that, as we shall further develop in Papers II to IV, maximally uniform states commonly turn out to also serve as shape space centres, 
which is of likely interest to all of the Cosmology, Foundations of Physics and Quantum Gravity communities.
The number of prices to pay, moreover, increases with increase in $N$ and/or $d$.

\m 

\n D) The diagramatic presentations, basic Graph Theory and (similarity) Killing vector considerations for manifolds with boundary 
set up in the current treatise are usefu throughout the current treatise, which successively builds up further such tools. 
In particular, the many of the current treatise's notions of shape receive quantifiers in Part II, alongside a second instalment of Applied Graph Theory 
to be used to count out larger numbers of qualitative types.

\m 

\n E) For the (3, 1) model, the number of qualitative types of merger, symmetry and uniformity can be non-negligible (Fig \r{(3, 1)-Q}), 
but are still up to 2 orders of magnitude smaller than the corresponding counts for (4, 1).
Detailed consideration of quantitative measures for the corresponding properties -- expected to be a major research direction improving the conceptual underpinning of 
notions of uniformity and inhomogeneity in Cosmology and in General Relativity more widely -- is thus best left to (4, 1) and (3, 2) models upward.  

\m 

\n{\bf Acknowledgments} I thank Chris Isham and Don Page for discussions about configuration space topology, geometry, quantization and background independence. 
I also thank Jeremy Butterfield and Christopher Small for encouragement. 
I thank Don, Jeremy, Enrique Alvarez, Reza Tavakol and Malcolm MacCallum for support with my career. 
I also thank Angela Lahee and the typesetting editors for my book manuscript this past summer for prompt and excellent work without which the current treatise could not 
have been completed by this date. 
I finally thank the person who is the main source of strength in my life.  

\vspace{10in}

\begin{appendices}

\section{Some basic Graph Theory}\l{Graphs}

\n{\bf Definition 1} A {\it graph} $\mG$ consists of a set of {\it vertices} $V$ and a set of {\it edges} $E$ interlinking some subset of the vertices.  
The graphs used in the current treatise are all finite, and are `simple graphs' rather than `multigraphs' in the senses of 
       having at most one edge between any two given vertices, 
and of having no loops (edges running between a vertex and itself).
The {\it order} of a graph $\mG$ is its number of vertices, denoted by $|\mG|$, whereas its {\it size} e(G) is its number of edges.  
The {\it degree} or {\it valency} of a vertex $v$ is the number of edges $\md(v)$ emanting from it.
The current treatise's graphs are furthermore {\it labelled}, in the sense that their vertices are decorated with labels.  

\m 

\n{\bf Remark 1} The inter-relation 
\be
\sum_{v \in \sG} \d(v) = 2 \, e(\mG)  
\l{I.A.1}
\ee
is both basic to Graph Theory and indeed often used in Part II.  

\m 

\m 

\n{\bf Definition 2} A graph $\mG$ is {\it connected} if there is a path between any two of its vertices.
$\mG$ is {\it disconnected} if it is not connected.
The {\it components} of $\mG$ are its maximal connected subsets; these partition $\mG$. 
We denote graphs with multiple components by the disjoint union of the names of their components. 
If multiple copies of a given component $\mH$ are present, we use the power notation: $\mH^n = \coprod_{i = 1}^n \mH$.  

\m 

\n{\bf Definition 3} A {\bf totally disconnected graph} is one with no edges at all.
We denote these by $\mD_n$, where $n = |\mG|$.

\m 

\n A {\it path graph} $\mP_n$ is an alternating sequence $v_0 e_1 v_1... e_{n - 1} v_n$, 
where the vertex set is $V = \{v_i: i = 1 \mbox{ to } n\}$ 
and   the edge   set is $E = \{e_j: j = 1 \mbox{ to } n - 1\}$.  
So these are connected graphs in which all but two of the valencies are 2; 
the exceptions to this are the end-point vertices $v_0$ and $v_n$, of valency 1.

\m 

\n A {\it cycle graph} $\mC_n$ is an alternating sequence as above except that $v_n$ is now $v_0$ again: $v_0 e_1 v_1... e_{n - 1} v_0$.  
So these are connected graphs in which all the valencies are 2.  

\m 

\n A {\it complete graph} $\mK_n$ is one in which there is an edge between any two vertices. 
So all vertex valencies are $|\mG| - 1 = n - 1$. 
This is the maximal number of edges which a graph of order $|\mG| = n$ can support.  

\m 

\n See Fig \r{Graph-1} for the first few members of each of the above series, including those examples which are small enough to be multiple of the above, 
and subsequently the first distinct members of each of these series. 
%
{            \begin{figure}[!ht]
\centering
\includegraphics[width=0.9\textwidth]{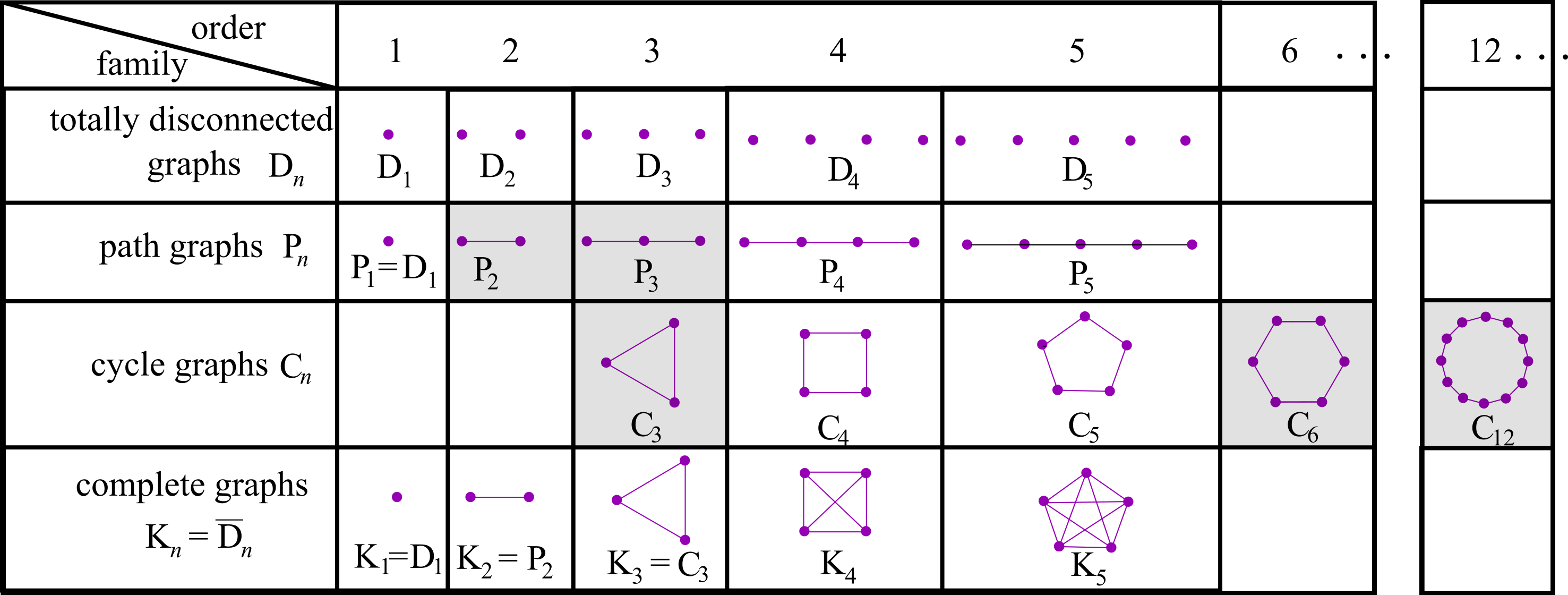}
\caption[Text der im Bilderverzeichnis auftaucht]{        \footnotesize{Totally disconnected, path, cycle and complete graphs. 
Throughout this Appendix's figures, a grey background indicates examples of graphs which are shape-theoretically realized within this treatise.} }
\l{Graph-1} \end{figure}          }

\m 

\n{\bf Definition 4} The {\it complement} $\overline{\mG}$ of a graph $\mG$  
is a graph which has edges between precisely those vertex pairs which have no edges between them in $\mG$.  

\m 

\n For example, the complete graph $\mK_n$ can also be characterized as the complement of the totally disconnected graph $\mD_n$.  

\m 

\n{\bf Remark 2} Whichever of $\mG$ and $\bar{\mG}$ has less edges may be easier to identify. 

\m 

\n{\bf Remark 3} Moreover, $Aut(\mG) = Aut(\overline{\mG})$.  

\m 

\n{\bf Remark 4} A few graphs are self-complementary: $\mG = \overline{\mG}$.  
Small Exercise: show that $\mP_4$ and $\mC_5$ are two of the first nontrivial such graphs 
(these are used in rearranging some of the larger graph's complements in Fig \r{Graph-3}).

\m 

\n{\bf Definition 5} The {\it cone over a graph} $\mG$, $\mC(\mG)$, is a graph of order $|\mG| + 1$ 
whose extra vertex -- the cone vertex -- has edges leading to all of $\mG$'s vertices (i.e.\ its valency is $|\mG|$).   

\m 

\n {\bf Remark 5}  This is a special case of the topological notion of cone (defined in Sec \r{MLS-R}).  

\m 

\n{\bf Example 1} $\mC(\mD_n) = \mS_n$: the $n$-{\it star graph} ($n$-pointed star, of order $n + 1$).  
Fig \r{Graph-2} shows that the first nontrivial star is $\mS_3$, an alias for which is the {\it claw graph}; 
this is realized as the (3, 2) topological shape space.  

\m 

\n{\bf Example 2} $\mC(\mP_n) = \mF_{n - 1}$: the $(n - 1)$-{\it fan graph} (hand fan with $n - 1$ folds, of order $n + 1$).  
Fig \r{Graph-2} also shows that the first nontrivial fan is $\mF_2$, an alias for which is the {\it diamond graph}; 
                                                 an alias for $\mF_3$                     is the {\it gem graph}; 
												 both of these are realized as topological shape(-and-scale) spaces.  

\m 

\n{\bf Example 3} $\mC(\mC_n) = \mW_n$: the $n$-{\it wheel graph} (wheel with $n$ spokes, of order $n + 1$).  
Fig \r{Graph-2} also shows that the first nontrivial wheel is $\mW_4$; 
it is $\mW_6$ and $\mW_{12}$ which are realized as topological shape-and-scale spaces in the current treatise.  

\m 

\n{\bf Example 4} As $\mC(\mK_n)$ is just $\mK_{n + 1}$, none of these are new graphs.  
%
{            \begin{figure}[!ht]
\centering
\includegraphics[width=1.0\textwidth]{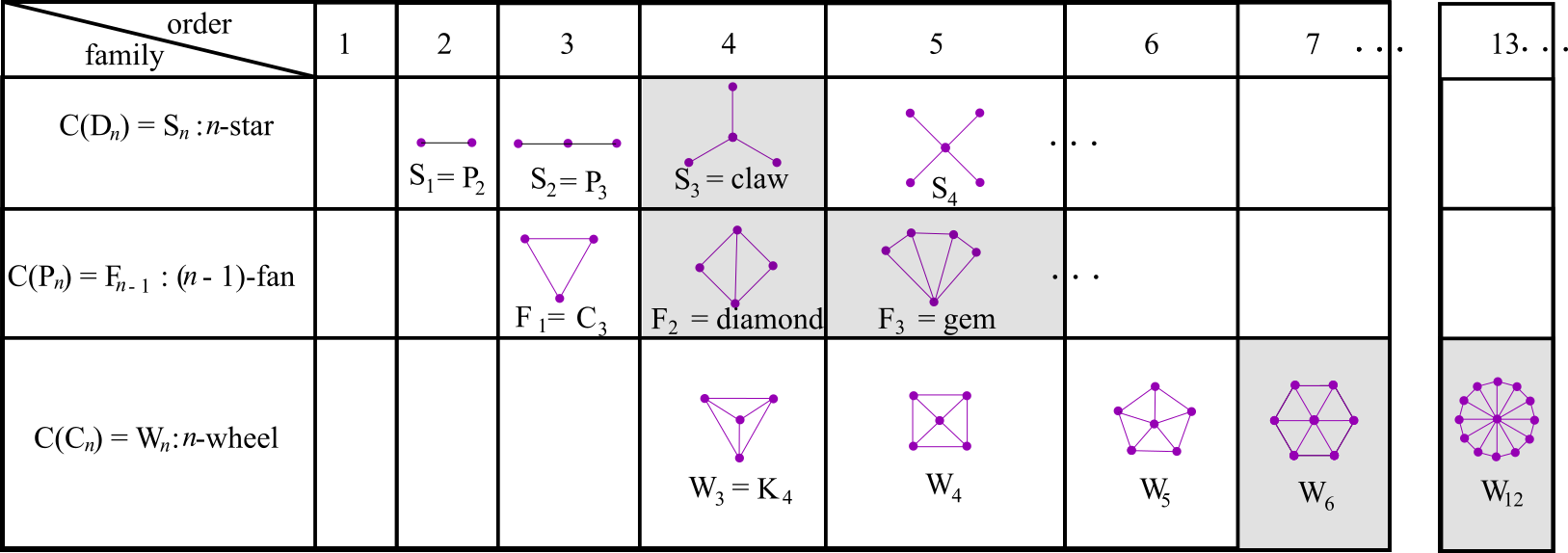}
\caption[Text der im Bilderverzeichnis auftaucht]{        \footnotesize{Cones over totally disconnected, path and cycle graphs: star, fan and wheel graphs.} }
\l{Graph-2} \end{figure}          }

\m 

\n{\bf Definition 6} The {\it suspension over a graph} $\mG$, $\mS(\mG)$, is a graph of order $|\mG| + 2$ 
in which the two extra vertices have edges leading to all of $\mG$'s vertices but not to each other (thus these two vertices have valency $|\mG|$).  

\m 

\n{\bf Remark 6} This is a special case of the topological notion of suspension \c{Hatcher}. 
It is also known as a 2-cone graph, readily generalizing to the notion of $k$-cone graph.  
This refers to its having 2 cone points; take care not to confuse this with the following less commonplace definition, which is however 
the one which is shape-theoretically realized (Fig \r{Graph-3} contrasts the two).  
%
{            \begin{figure}[!ht]
\centering
\includegraphics[width=1.0\textwidth]{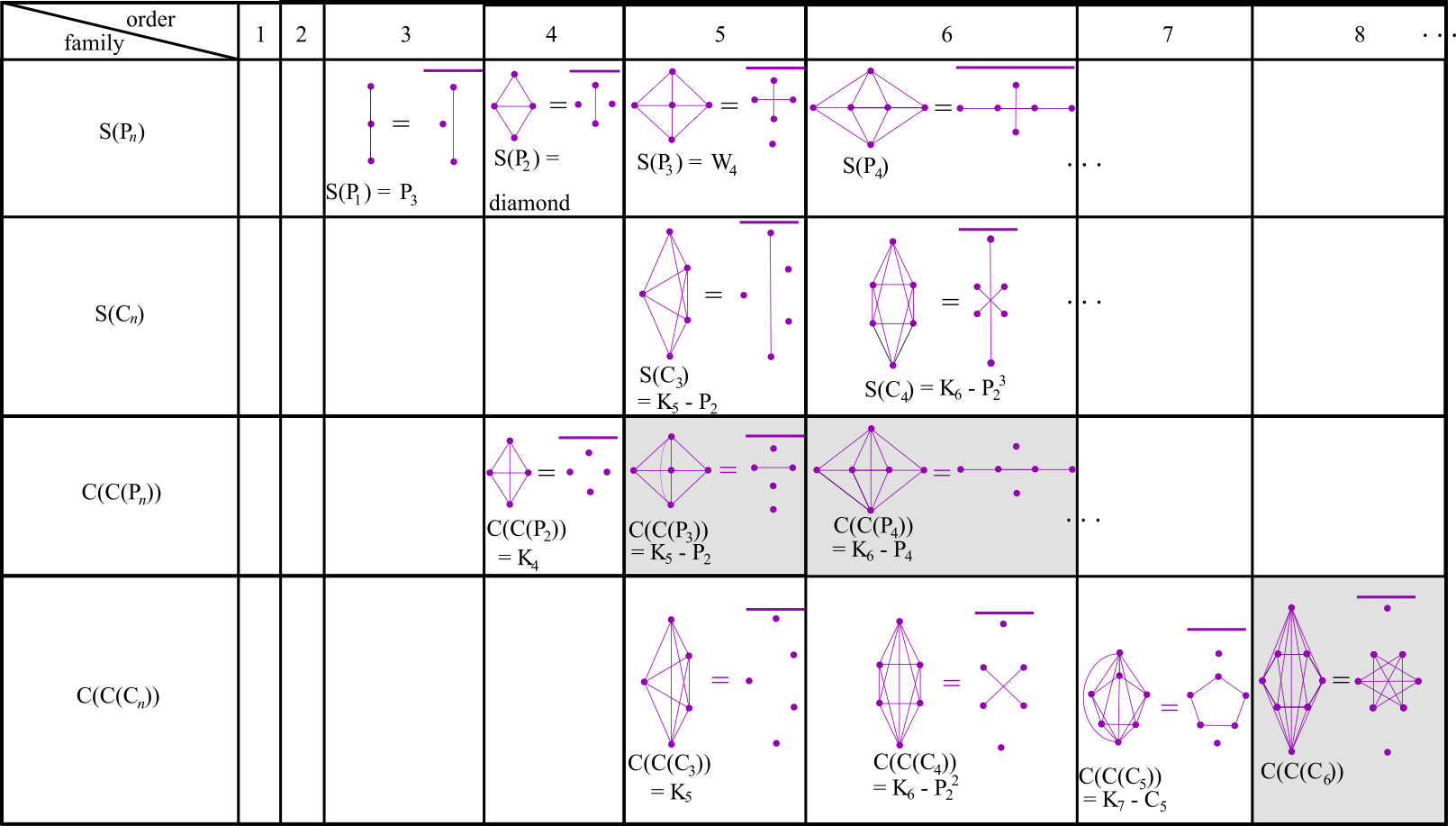}
\caption[Text der im Bilderverzeichnis auftaucht]{        \footnotesize{Suspension graphs versus cones of cones of graphs, 
including their often more readily identifiable complements.} }
\l{Graph-3} \end{figure}          }

\m 

\n{\bf Definition 7} The {\it `cone of a cone graph'},  $\mC(\mC(\mG))$, is a graph of order $|\mG| + 2$ 
in which two vertices have edges leading to all of $\mG$'s vertices {\sl and} to each other (thus these two vertices have valency $|\mG| + 1$).

\m 

\n The current treatise requires $\mC(\mC(\mP_n))$ and $\mC(\mC(\mP_n))$, which are depicted in Fig \r{Graph-3}.  

\m 

\n{\bf Remark 7} The actions on the complement of coning, suspending and coning of a coning are worth pointing out for their simple nature 
and usefulness in classifying small examples such as occur in the current treatise.  

\m 

\n Coning's action on the complement is adjoining a disconnected point, $\coprod \mD_1 = \coprod \mP_1$.  

\m 

\n Suspension's action in the complement is $\coprod \mP_2$ (because its action on the graph has no edge between the two adjoined points).  

\m 

\n Coning of a coning's action on the complement is $\coprod \mD_2 = \coprod \mD_1 \coprod \mD_1$: the action of coning carried out twice.  

\m 

\n To fully appreciate the distinction between these two concepts, note that the generalization of suspension to $k$ cone points 
acts on the complement by $\coprod \mK_k$, by which it is more insightful to replace the above $\mP_2$ by its alias $\mK_2$.  
So the distinction between the more conventional 2-cone and the current treatise's `cone of a cone' 
follows simply from the distinction for $k > 1$ between $\mD_k$ and its complement $\mK_k$.  

\m 

\n{\bf Definition 8} {\it Multipartite graphs} are graphs whose vertices can be partitioned into subsets such that there are no edges within any one given subset.
The simplest nontrivial version of this is for a partition into two subsets: {\it bipartite graphs} (Fig \r{Graph-4}), the best-known of which are complete within this restriction.  
The case which occurs in the current treatises' Shape Theory is, however, both tripartite and not complete (also in Fig \r{Graph-4}).  
%
{            \begin{figure}[!ht]
\centering
\includegraphics[width=0.7\textwidth]{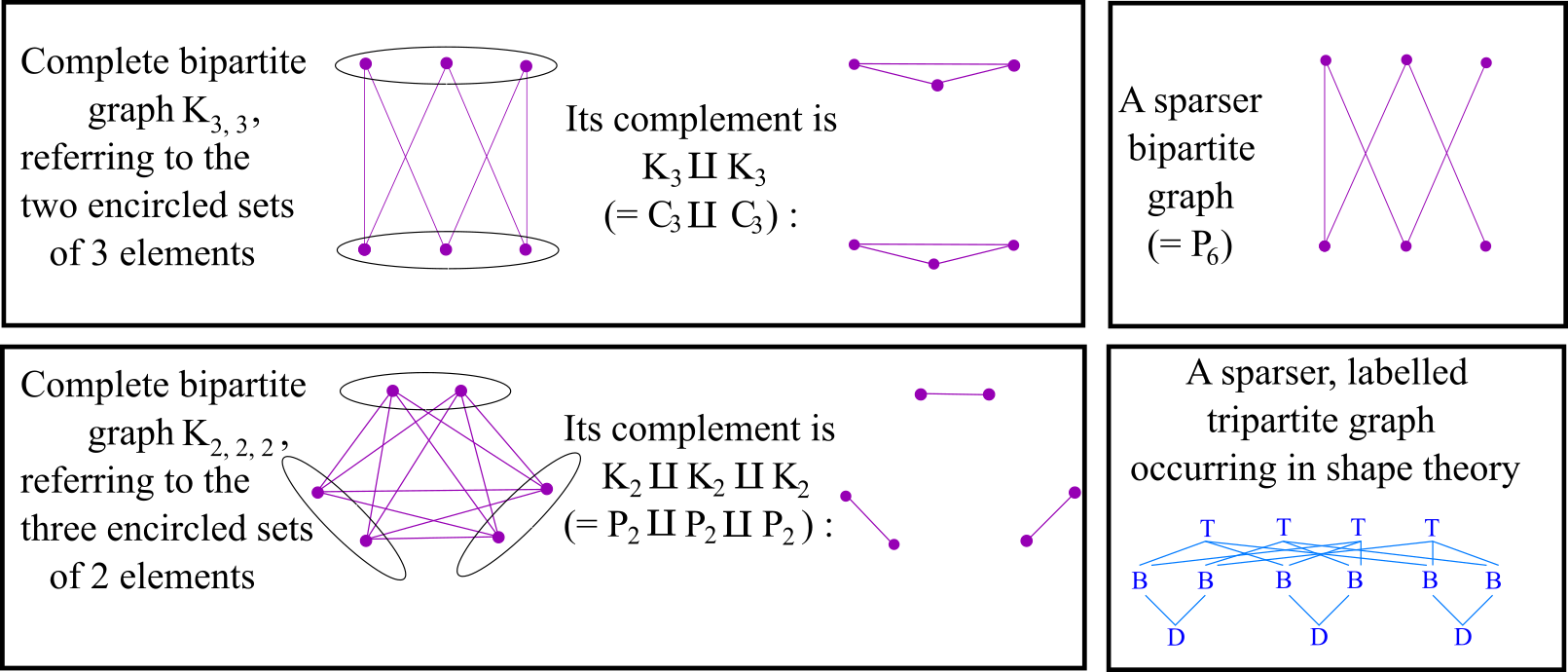}
\caption[Text der im Bilderverzeichnis auftaucht]{        \footnotesize{Bipartite and tripartite graphs, alongside their complements.} }
\l{Graph-4} \end{figure}          }

\m 

\n{\bf Definition 9} A graph is {\it planar} \c{Graphs-2} if it can be embedded in a plane. 
This means that it can be drawn on a piece of paper such that its edges intersect at their vertex end-points alone.  

\m 

\n Fig \r{Graph-5} gives simple examples of planar and non-planar graphs.
Many of the shape-theoretically realized graphs in the current treatise are planar. 
%
{            \begin{figure}[!ht]
\centering
\includegraphics[width=0.7\textwidth]{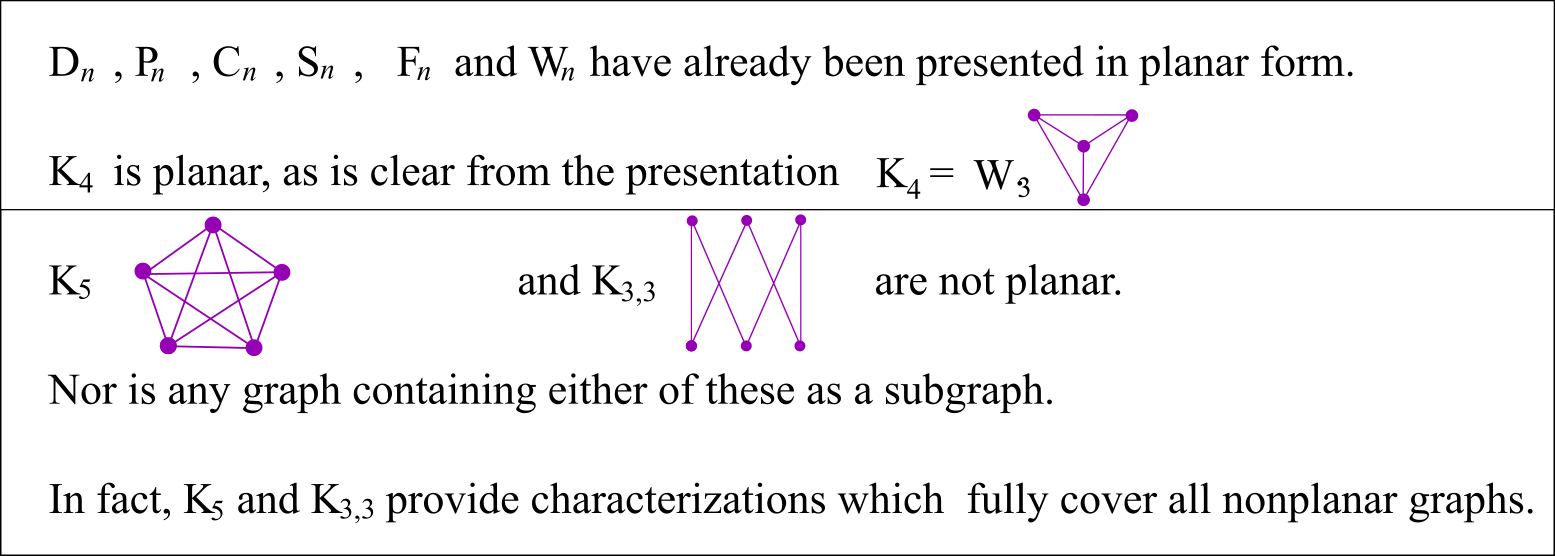}
\caption[Text der im Bilderverzeichnis auftaucht]{        \footnotesize{Simple examples of planar and non-planar graphs.} }
\l{Graph-5} \end{figure}          }

\m 

\n From the graph-theoretic point of view, the current treatise poses the following question. 
How many of the groupings and sequences of graphs arising naturally in Shape Theory ones which have occurred elsewhere in Graph Theory and its applications?  
(``Natural groupings' could e.g.\ refer to a fixed ($N$, $d$)'s quadruplet of plain, mirror image identified, indistinguishable and Leibniz 
topological shape space graphs.  
`Natural series' could refer e.g. to varying one of $N$ or $d$ while keeping the other fixed, separately for each of plain, 
mirror image identified, indistinguishable and Leibniz topological shape space graphs.
We postpone highlighting some specific examples of these to Part IV's Conclusion, as this benefits from first building up more examples in Papers II to IV).

\section{Automorphisms: graphical, isometries and similarities}\l{Aut}

\n{\bf Definition 1} Given a mathematical space $\FrS$, an {\it automorphism} is an endomorphism  -- a map $\FrS \longrightarrow \FrS$ -- 
                                                                    which is also an isomorphism of $\FrS$: 
																	it preserves up to a given level of mathematical structure that $\FrS$ is equipped with.

\m 

\n{\bf Proposition 1} The automorphisms of whichever mathematical structure $\FrS$ form a group, the {\it automorphism} group $Aut(\FrS)$.

\m 

\n The current series of articles makes use of the following examples of automorphisms.

\m 

\n{\bf Example 1} {\it Permutations}, $Aut(\mG) = S_p$, are automorphisms for the finite sets with $p$ elements.

\m 

\n{\bf Example 2} {\it (Labelled) graph automorphisms}, $Aut(\mG)$, are automorphisms for the labelled graphs $\mG$. 
These interchange equally labelled vertices of $\mG$ while maintaining the adjacency relations encoded by $\mG$'s edges.

\m 

\n{\bf Example 3} {\it Isometries} are automorphisms $Aut(\FrM) = Isom(\FrM)$ 
for $\FrM$ a manifold equipped with metric structure (in the current series of articles, a Riemannian metric).
$Isom(\FrM)$ is the group formed by $\FrM$'s Killing vectors.  

\m 

\n{\bf Subexample 3.1} $Isom(\mathbb{R}^p) = Eucl(p)$, which is further detailed in Section \r{Preamble}.   
The constituent Killing vectors are here the translations $Tr(p)$ and the rotations $Rot(p)$.

\m 

\n{\bf Example 4} {\it Similarities} are automorphisms $Aut(\FrM) = Sim(\FrM)$ 
for $\FrM$ a manifold now equipped with metric structure up to rescalings by a positive constant function $k^2$, 
\beq
g_{AB} \longrightarrow g_{AB}^{\prime} = k^2 g_{AB} \m  .  
\eeq
$Sim(\FrM)$ is the group formed by $\FrM$'s similar Killing vectors.  

\m 

\n{\bf Subexample 4.1} $Sim(\mathbb{R}^p) = Sim(p)$, which is also further detailed in Section \r{Preamble}.   
The constituent similar Killing vectors are here the translations $Tr(p)$ and rotations $Rot(p)$ -- which are also Killing vectors -- 
                                                and the dilation $Dil$, which is not.  

\m 

\n Note that the current series of articles makes use of both automorphisms in space -- to formulate shape and shape-and-scale spaces -- 
and of these configuration spaces themselves, toward their further study.  


\section{Some (similar) Killing Lemmas, including with boundary}\l{KV-Bdry}

\n{\bf Definition 1} A {\it Killing vector} $\u{X}$ is a solution of the {\it Killing equation}
\be
0 = \pounds_{\u{X}}\mg_{AB} 
  = \nabla_A X_B + \nabla_B X_A \m  . 
\ee
Here $\mg_{AB}$ denotes a Riemannian metric, 
     $\nabla_A$ the corresponding covariant derivative, 
 and $\pounds_{\u{X}}$ denotes the Lie derivative with respect to the vector $\u{X}$. 
 
\m 

\n{\bf Remark 1} Contracting, one gets a (covariant) divergencelessness condition  
\be
\u{\nabla} \cdot \u{X} = 0 \m  .  
\ee
\n{\bf Killing Lemma} \c{Yano-Integral} On a compact without boundary (CWB) manifold $\FrM$, a vector field $\u{X}$ is Killing iff 
\be
\diamond \u{X} = 0  \mma  \u{\nabla} \cdot \u{X} 
               = 0   \m   \mbox{ on } \m  \FrM .
\l{KL}
\ee 
\n Here, 
\be
\diamond X^A := \triangle X^A - 2 \, {\mR^A}_B X^B \,  \m  ,
\ee
for ${\mR^A}_B$ the Ricci curvature tensor of $\mg_{AB}$. 

\m 

\n{\bf Killing Lemma with boundary} \c{Yano-Integral}
%
%
On a manifold $\FrM$ with a boundary $\pa \FrM$ and normal $\u{N}$ (Fig \r{Cpct-with-Bdry}), 
a vector field $\u{X}$ is Killing iff (\r{KL}) holds and
\be
(\pounds_{\u{X}}\bg)(\u{X}, \u{N}) = 0 \m  \mbox{ on } \m  \pa\FrM \m  .
\ee
\n{\bf Corollary 1} $\u{X} = 0$ on $\pa \FrM$ satisfies this boundary condition.  
%
{            \begin{figure}[!ht]
\centering
\includegraphics[width=0.35\textwidth]{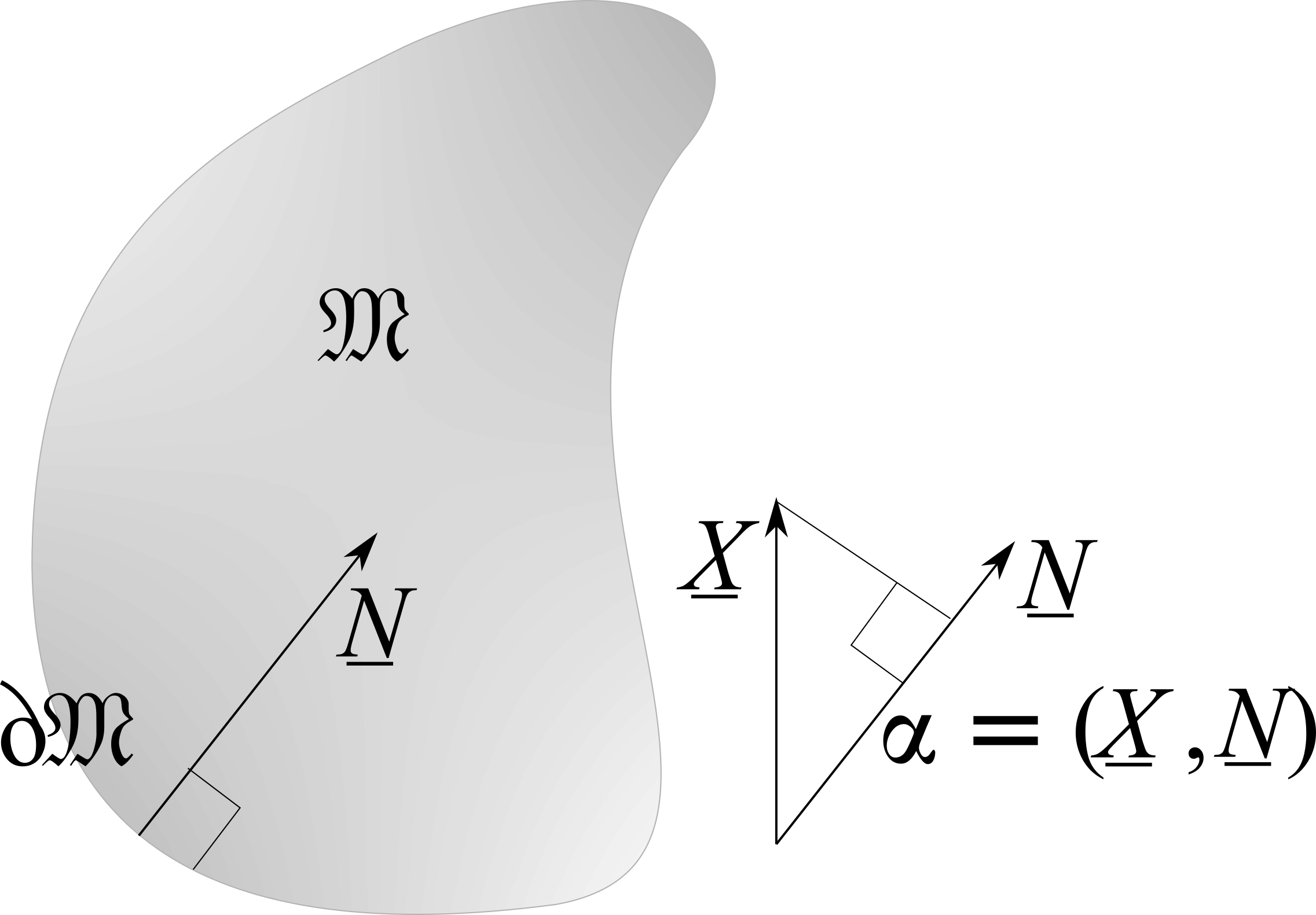}
\caption[Text der im Bilderverzeichnis auftaucht]{        \footnotesize{Notation for a manifold with boundary.} }
\l{Cpct-with-Bdry} \end{figure}          }

\m 

\n{\bf Remark 2} A useful computational form \c{Yano-Integral} for this boundary condition is 
\be
(\nabla \cdot \xi)(\u{N}, \u{X})    + 
\mK(\u{X}^{\prime}, \u{X}^{\prime}) + 
\alpha^2 \, \mK                                     + 
2 \, \alpha(\delta^{\prime}\xi^{\prime})            - 
\delta^{\prime}(\alpha \xi^{\prime})                = 0 \m  \mbox{ on } \m  \pa\FrM \m  . 
\l{bigger-case}
\ee
Here the prime is used to denote a vector on the boundary, and $\xi$ is the 1-form associated with $\u{X}$ as per 
\be 
\xi_A = \mg_{AB} X^B \m  .
\ee
Also 
\be 
\alpha := (\u{X}, \u{N}) \m  , 
\ee 
$\delta$ denotes the codifferential, 
and $\mK$ is the {\it mean extrinsic curvature} of $\pa\FrM$ relative to $\FrM$.   

\m 

\n{\bf Remark 3} If $\u{X}$ is normal to $\pa\FrM$, (\r{bigger-case}) simplifies to 
\be
\alpha \, \mK = 0  \m  \mbox{ on } \m  \pa\FrM \m  .
\l{normal-bc}
\ee
\n{\bf Remark 4} On the other hand, if $\u{X}$ is tangential to $\pa\FrM$, (\r{bigger-case}) simplifies to 
\be
\mK(\u{X}^{\prime}, \u{X}^{\prime}) + 
\alpha^2 \, \mK                                     = 0  \m  \mbox{ on } \m  \pa\FrM \m  .
\l{tangential-bc}
\ee
\n{\bf Remark 5} 
\be
\mK = 0 \m  \mbox{ (minimal/maximal boundary surface) }
\ee
suffices to satisfy the normal case (\r{normal-bc}).

\m 

\n{\bf Remark 6} For $\mK_{AB}$ the {\it extrinsic curvature tensor} of $\pa \FrM$ relative to $\FrM$,  
\be
\mK_{AB} = 0 \m  \mbox{ (symmetric boundary surface) } \mbox{ alongside } \m  \u{\xi} \mbox{ divergenceless }: \m  \m  
\u{\nabla} \cdot \u{\xi} = 0 
\ee
suffices to satisfy the tangential case (\r{tangential-bc}).

\m 

\n{\bf Remark 7}  Due to the nature of great circles and $\mathbb{CP}^k$ geodesics as intersections with planes, $\mK_{AB} = 0$ holds for these. 
Thus $\mK = 0$, so Killing vectors normal to these survive the quotienting. 
These are $\sD$               for 3 points in 1-$d$, 
          $\sD_3$             for 4 points in 1-$d$ (defined in Part II), 
		  $\sJ = \sS_3$  for 3 points in 2-$d$ (defined in Part III), 
      and $\sI_3$, $\sY$ for 4 points in 2-$d$ (defined in Part IV). 

\m 

\n{\bf Remark 8} No other Killing vectors survive, including the indistinguishable particle cases of relational space not having any translations survive.  

\m 

\n{\bf Remark 9} Arbitrary-$N$ counterparts of these results in 1- and 2-$d$ are given in Part IV's Conclusion.  

\m 

\n{\bf Definition 2} A {\it similar Killing vector} \c{Yano-Integral} $\u{X}$ is a solution of the {\it Killing equation}
\be
2 \, k \, \mg_{AB} = \pounds_{\u{X}} g_{AB} 
                   = \nabla_A X_B + \nabla_B X_A \m  . 
\ee
\n{\bf Remark 10} Contracting, 
\be
\u{\nabla} \cdot \u{X} = 2 \, p \ k \m  , \mbox{constant} \m  ,
\ee
where $p$ is the dimension of $\FrM$.  

\m 

\n{\bf Similarity Killing Lemma} \c{Yano-Integral} On a CWB manifold $\FrM$, a vector field $\u{X}$ is similarity Killing iff 
\be
\diamond \u{X} = 0  \mma   \u{\nabla} \cdot \u{X} 
               = \mbox{constant} \m  \mbox{ on } \m  \FrM \m .
\l{HKL}
\ee 
\n{\bf Similarity Killing Lemma with boundary} \c{Yano-Integral}
On a manifold $\FrM$ with boundary $\pa \FrM$ and normal $\u{N}$, a vector field $\u{X}$ is similarity Killing iff (\r{HKL}) holds and 
\be
(\pounds_{\u{X}} \bg)(\u{X}, \u{N}) = 4 \, k \, \xi(\u{N}) \m  \mbox{ on } \m  \pa\FrM \m  .
\ee
\n{\bf Corollary 2} $\u{X} = 0$ on $\pa \FrM$ satisfies this boundary condition.  

\m 

\n{\bf Remark 11} A useful computational form \c{Yano-Integral} for this boundary condition is 
\be
(\nabla \cdot \xi)(\u{N}, \u{X}) + \mK(\u{X}^{\prime}, \u{X}^{\prime}) + \alpha^2 \, \mK + 2 \, \alpha(\delta^{\prime}\xi^{\prime}) - \delta^{\prime}(\alpha \xi^{\prime})                - 
\frac{p - 2}{p} \alpha (\delta \xi^{\prime})        = 0 \m  \mbox{ on } \m  \pa\FrM \m  . 
\ee
\n{\bf Remark 12} If $\u{X}$ is normal to $\pa\FrM$, this simplifies once again to (\r{normal-bc}).   

\m 

\n{\bf Remark 13} $D = \rho \frac{\pa}{\pa \rho}$ is tangential along all boundaries of cone manifolds with boundary considered in this article.
These boundaries are moreover extrinsically flat by Remark 7, so $D$ survives the corresponding quotientings.

\end{appendices}

\vspace{10in}



\begin{thebibliography}{99}

\footnotesize


\bibitem{Euclid}              Euclid {\it Elements}, approximately 300 B.C; 
                              see e.g. {\it The Thirteen Books of Euclid's Elements} for an English translation and commentary by T.L. Heath (Dover, New York 1956).  

\bibitem{Newton}              I. Newton, {\it Philosophiae Naturalis Principia Mathematica} ({\it Mathematical Principles of Natural Philosophy}) (1686).  
%
                              For an English translation, see e.g.\ I.B. Cohen and A. Whitman (University of California Press, Berkeley, 1999).  
%
							  In particular, see the Scholium on Time, Place, Space and Motion therein.  

\bibitem{L}                   See {\it The Leibnitz--Clark Correspondence}, ed. H.G. Alexander (Manchester 1956), originally dating to 1715 and 1716.   

\bibitem{L2}                  G.W. Leibniz, {\it The Metaphysical Foundations of Mathematics} (University of Chicago Press, Chicago 1956) originally dating to  

\bibitem{M}                   E. Mach, {\it Die Mechanik in ihrer Entwickelung, Historisch-kritisch dargestellt} (J.A. Barth, Leipzig 1883).    
%
                              An English translation is {\it The Science of Mechanics: A Critical and Historical Account of its Development} 
							  Open Court, La  Salle, Ill. 1960).  
			
\bibitem{Einstein1}           A. Einstein, ``Dialogue about Objections to the Theory of Relativity", 
                              {\it Die Naturwissenschaften}" {\bf 6} 697 (1918).
%
                              An English translation is available in Janssen et al (2002). 
			
\bibitem{Hopf}                H. Hopf, ``\"{U}ber die Abbildungen der dreidimensionalen Sph\"{a}re auf die Kugelfl\"{a}che", 
                              (Concerning the Images of $\mathbb{S}^3$ on $\mathbb{S}^2$), Math. Ann. (Berlin) Springer {\bf 104} 637 (1931).  

\bibitem{Lich}                A. Lichnerowicz, ``L'Inegration des \'{E}quations de la Gravitation Relativiste et le Probl$\grave{\me}$me des $N$ Corps" 
                             (``The Integration of the Equations of Relaticitic Gravitation and the $N$-Body Problem"), 
                              J. Math. Pures Appl. {\bf 23} 37 (1944).
							  
\bibitem{Whitney46}           H. Whitney, ``Complexes of Manifolds", Proc. Nat. Acad. Sci. USA {\bf 33} 10 (1946).  
							  
\bibitem{Lanczos}             C. Lanczos, {\it The Variational Principles of Mechanics} (University of Toronto Press, Toronto 1949). 							  

\bibitem{Einstein2}           A. Einstein's Foreword to M. Jammer, 
                             {\it Concepts of Space. The History of Theories of Space in Physics. Third, Enlarged Edition.} (Dover, New York 1993).
							  
\bibitem{MFII}                P.M. Morse and H. Feshbach, {\it Methods of Theoretical Physics. Part II} (McGraw-Hill, New York 1953).  

\bibitem{Thom55}              R. Thom, ``Les Singularit\'{e}s des Applications Diff\'{e}rentiables" (Singularities in Differentiable Maps), Ann. Inst. Fourier (Grenoble) {\bf 6} 43 (1955).
							 
							 
\bibitem{Smith60}             F.T. Smith, ``Generalized Angular Momentum in Many-Body Collisions", Phys. Rev. {\bf 120} 1058 (1960). 
							  
\bibitem{ADM}                 R. Arnowitt, S. Deser and C.W. Misner, ``The Dynamics of General Relativity", 
                              in {\it Gravitation: An Introduction to Current Research} ed. L. Witten (Wiley, New York 1962), arXiv:gr-qc/0405109.  

\bibitem{BSW}                 R.F. Baierlein, D.H. Sharp and J.A. Wheeler, ``Three-Dimensional Geometry as Carrier of Information about Time", 
                              Phys. Rev. {\bf 126} 1864 (1962).
							  
\bibitem{A64}                 J.L. Anderson, ``Relativity Principles and the Role of Coordinates in Physics.", in {\it Gravitation and Relativity} 
                              ed. H-Y. Chiu and W.F. Hoffmann p.\  175 (Benjamin, New York 1964).  

\bibitem{Dirac}               P.A.M. Dirac, {\it Lectures on Quantum Mechanics} (Yeshiva University, New York 1964). 
							  
\bibitem{Whitney65}           H. Whitney, ``Tangents to an Analytic Variety", Ann. Math. {\bf 81} 496 (1965). 
							  
\bibitem{A67}                 J.L. Anderson, {\it Principles of Relativity Physics} (Academic Press, New York 1967).  
  
\bibitem{Zick-1}              W. Zickendraht, ``Configuration-Space Approach to Three-Particle Scattering", Phys. Rev. {\bf 159} 1448 (1967). 

\bibitem{DeWitt67}            B.S. DeWitt, ``Quantum Theory of Gravity. I. The Canonical Theory.", Phys. Rev. {\bf 160} 1113 (1967).

\bibitem{Battelle}            J.A. Wheeler, in {\it Battelle Rencontres: 1967 Lectures in Mathematics and Physics} ed. C. DeWitt and J.A. Wheeler 
                             (Benjamin, New York 1968). 
							 
\bibitem{Misner-Ani}          C.W. Misner, ``The Isotropy of the Universe", Astrophys. J. {\bf 151} 431 (1968).  

\bibitem{Zick-2}              W. Zickendraht, ``Configuration-Space Approach to the Four-Particle Problem", J. Math Phys. {\bf 10} 30 (1969). 
							  
\bibitem{Misner-69}           C.W. Misner, ``Quantum Cosmology. I", Phys. Rev {\bf 186} 1319 (1969).

\bibitem{Thom69}              R. Thom, ``Ensembles et Morphismes Stratifi\'{e}s" (Stratified Spaces and Morphisms), Bull. Amer. Math. Soc. (N.S.) {\bf 75} 240 (1969).

							 
\bibitem{Misner-70}           C.W. Misner, ``Classical and Quantum Dynamics of a Closed Universe",
                              in {\it Relativity (Proceedings of the Relativity Conference in the Midwest, held at Cincinnati, Ohio June 2-6, 1969)} 
                              ed. M. Carmeli, S.I. Fickler and L. Witten (Plenum, New York 1970).

\bibitem{Zick-3}              W. Zickendraht, ``Collective and Single-Particle Coordinates in Nuclear Physics", {\bf 12} 1663 (1970). 

\bibitem{DeWitt70}            B.S. DeWitt, ``Spacetime as a Sheaf of Geodesics in Superspace", in {\it Relativity} (Proceedings of the Relativity Conference in 
                              the Midwest, held at Cincinnati, Ohio June 2-6, 1969), ed. M. Carmeli, S.I. Fickler and L. Witten (Plenum, New York 1970). 

\bibitem{Fischer70}           A.E. Fischer, ``The Theory of Superspace", in {\it Relativity} (Proceedings of the Relativity Conference in the Midwest, 
                              held at Cincinnati, Ohio June 2-6, 1969), ed. M. Carmeli, S.I. Fickler and L. Witten (Plenum, New York 1970). 

\bibitem{Yano-Integral}       K. Yano, {\it Integral Formulas in Riemannian Geometry} (Dekker, New York 1970).

\bibitem{Kuchar71}            K.V. Kucha\v{r}, ``Canonical Quantization of Cylindrical Gravitational Waves",  Phys. Rev. {\bf D4}, 955 (1971).  
							 
\bibitem{Ryan}                M.P. Ryan, {\sl Hamiltonian Cosmology} Lec. Notes Phys. {\bf 13} (Springer, Berlin 1972).

\bibitem{Magic}               C.W. Misner, ``Minisuperspace", in {\it Magic Without Magic: John Archibald Wheeler} ed. J. Klauder (Freeman, San Francisco 1972).
							 
\bibitem{Kobayashi}           S. Kobayashi, ``Transformation Groups in Differential Geometry", (1972).   

\bibitem{York72}              J.W. York Jr., ``Role of Conformal Three-Geometry in the Dynamics of Gravitation", Phys. Rev. Lett. {\bf 28} 1082 (1972).

\bibitem{YorkTime1}           J.W. York Jr., ``Mapping onto Solutions of the Gravitational Initial Value Problem", J. Math. Phys. {\bf 13} 125 (1972).  

\bibitem{York73}              J.W. York Jr., ``Conformally Invariant Orthogonal Decomposition of Symmetric Tensors on Riemannian Manifolds and the 
                              Initial-Value Problem of General Relativity", J. Math. Phys. {\bf 14} 456 (1973).
							 
\bibitem{Magnus}              W. Magnus {\it Noneuclidean Tesselations and their Groups}  (Academic Press, New York 1974).  
						
\bibitem{York74}              J.W. York Jr., ``Covariant Decompositions of Symmetric Tensors in the Theory of Gravitation", Ann. Inst. Henri Poincar\'{e} {\bf 21} 319 (1974).  

\bibitem{YW}                  T.T. Wu and C.N. Yang, ``Dirac Monopole Without Strings: Monopole Harmonics", Nu. Phys. {\bf B107} 365 (1976). 

\bibitem{Arnol'd}             V.I. Arnol'd, {\it Mathematical Methods of Classical Mechanics} (Springer, New York 1978).  

\bibitem{GP78}                G.W. Gibbons and C.N. Pope, ``$\mathbb{CP}^2$ as a Gravitational Instanton", Commun. Math. Phys. {\bf 61} 239 (1978).  

\bibitem{Page-Instanton}      D.N. Page, ``Some Gravitational Instantons" (1978 Moscow Conference presentation), arXiv:0912.4922.  
  
\bibitem{Thurston}            W. Thurston, ``The Geometry and Topology of Three-Manifolds" (Princeton University Lecture Notes 1978-1981).  
  

\bibitem{Bardeen}            J. Bardeen, ``Gauge-Invariant Cosmological Perturbations", Phys. Rev. {\bf D22} 1882 (1980). 
 
\bibitem{Kuchar81}            K.V. Kucha\v{r}, ``Canonical Methods of Quantization", in {\it Quantum Gravity 2: a Second Oxford Symposium} 
                              ed. C.J. Isham, R. Penrose and D.W. Sciama (Clarendon, Oxford 1981).

\bibitem{BB82}                J.B. Barbour and B. Bertotti, ``Mach's Principle and the Structure of Dynamical Theories", Proc. Roy. Soc. Lond. {\bf A382} 295 (1982). 

\bibitem{AMP}                 Y. Choquet-Bruhat, C. DeWitt-Morette and M. Dillard-Bleick, 
                              {\it Analysis, manifolds and physics} vol. 1 (Elsevier, Amsterdam 1982).  

\bibitem{BTBook}              R. Bott and L. Tu, {\it Differential Forms in Algebraic Topology} (Springer, New York 1982).

\bibitem{Watson}              G.S. Watson, {\it Statistics on Spheres} (Wiley, Chichester 1983).  

\bibitem{PW83}                D.N. Page and W.K. Wootters, ``Evolution Without Evolution: Dynamics Described by Stationary Observables", Phys. Rev. {\bf D27}, 2885 (1983).                    

\bibitem{Arnol'd-Cat}         V.I. Arnol'd, {\it Catastrophe Theory} (Springer-Verlag, Berlin 1984).  

\bibitem{Kendall84}           D.G. Kendall, ``Shape Manifolds, Procrustean Metrics and Complex Projective Spaces", Bull. Lond. Math. Soc. {\bf 16} 81 (1984). 

\bibitem{I84}                 C.J. Isham, ``Topological and Global Aspects of Quantum Theory", 
                              in {\it Relativity, Groups and Topology {II}}, ed. B. DeWitt and R. Stora (North-Holland, Amsterdam 1984). 

\bibitem{HallHaw}             J.J. Halliwell and S.W. Hawking, ``Origin of Structure in the Universe", Phys. Rev. {\bf D31}, 1777 (1985).
							  
\bibitem{ACG86}               V. Aquilanti, S. Cavalli and G. Grossi, ``Hyperspherical coordinates for Molecular Dynamics by the Method of Trees 
                              and the Mapping of Potential Energy Surfaces for Triatomic Systems", J. Chem. Phys. {\bf 85} 1362 (1986).  
							  
\bibitem{Fischer86}           A.E. Fischer, ``Resolving the Singularities in the Space of Riemannian Geometries", J. Math. Phys {\bf 27} 718 (1986).  

\bibitem{PP87}                R.T. Pack and G.A. Parker, ``Quantum Reactive Scattering in Three Dimensions using Hyperspherical (APH) Coordinates. Theory", 
                              J. Chem. Phys. {\bf 87} 3888 (1987).  

\bibitem{Iwai87}              T. Iwai, ``A Geometric Setting for Internal Motions of the Quantum Three-Body System", J. Math. Phys. {\bf 28} 1315 (1987).  

\bibitem{Sheaves1}            J.L. Bell, {\it Toposes and Local Set Theories} (Clarendon, Oxford 1988 and Dover, New York 2008).  

\bibitem{DoD}                 J.B. Barbour, {\it Absolute or Relative Motion? Vol 1: The Discovery of Dynamics} (Cambridge University Press, Cambridge 1989).    

\bibitem{Kendall89}           D.G. Kendall, ``A Survey of the Statistical Theory of Shape", Statistical Science {\bf 4} 87 (1989).

\bibitem{I89}                 C.J. Isham, ``Quantum Topology and Quantization on the Lattice of Topologies", Class. Quan. Grav {\bf 6} 1509 (1989);  
%
                              ``Quantization on the Lattice of Topologies, in {\it Florence 1989, Proceedings, Knots, Topology and Quantum Field Theories} 
							  ed. L. Lusanna (World Scientific, Singapore 1989).   

                               ``An Introduction To General Topology And Quantum Topology, unpublished, Lectures given at Banff in 1989 (available on the KEK archive). 			
							   

\bibitem{Marchal}             C. Marchal, {\it Celestial Mechanics} (Elsevier, Tokyo 1990).

\bibitem{Nakahara}            M. Nakahara, {\it Geometry, Topology and Physics} (Institute of Physics Publishing, London 1990).   

\bibitem{IKR}                 C.J. Isham, Y.A. Kubyshin and P. Renteln, ``Quantum Metric Topology", in {\it Moscow 1990, Proceedings, Quantum Gravity} 
                              ed M.A. Markov, V.A. Berezin and V.P. Frolov (World Scientific, Singapore 1991);  
%
                              ``Quantum Norm Theory and the Quantization of Metric Topology", Class. Quant. Grav. {\bf 7} 1053 (1990).  					  
	
\bibitem{Bookstein}           F.L. Bookstein {\it Morphometric Tools for Landmark Data: Geometry and Biology} (Cambridge University Press, Cambridge 1991). 
							  
\bibitem{Kuchar91}            K.V. Kucha\v{r}, ``The Problem of Time in Canonical Quantization", in {\it Conceptual Problems of Quantum Gravity} ed. 
                              A. Ashtekar and J. Stachel (Birkh\"{a}user, Boston, 1991).  
 							  							
\bibitem{I91}                 C.J. Isham, ``Canonical Groups And The Quantization Of Geometry And Topology", in {\it Conceptual Problems of Quantum Gravity} ed. 
                              A. Ashtekar and J. Stachel (Birkh\"{a}user, Boston, 1991).  

\bibitem{Xia}                 Z. Xia, {\it The Existence of Noncollision Singularities in Newtonian Systems}, Ann. Math. {\bf 135} 411 (1992).  
							  
\bibitem{GH92}                G.W. Gibbons and S.W. Hawking, ``Selection Rules for Topology Change", Commun. Math. Phys. {\bf 148} 345 (1992); 
%
                              G.W. Gibbons, ``Topology Change in Classical and Quantum Gravity", arXiv:1110.0611 covers further details of this.  
																				  
\bibitem{Kuchar92}            K.V. Kucha\v{r}, ``Time and Interpretations of Quantum Gravity", 
                              in {\it Proceedings of the 4th Canadian Conference on General Relativity and Relativistic Astrophysics} 
                              ed. G. Kunstatter, D. Vincent and J. Williams (World Scientific, Singapore, 1992),
%
                              reprinted as Int. J. Mod. Phys. Proc. Suppl. {\bf D20} 3 (2011). 
							 							  
\bibitem{I93}                 C.J. Isham, ``Canonical Quantum Gravity and the Problem of Time", in {\it Integrable Systems, Quantum Groups and Quantum Field Theories} 
                              ed. L.A. Ibort and M.A. Rodr\'{\i}guez (Kluwer, Dordrecht 1993), gr-qc/9210011.
							  
\bibitem{B94I}                J.B. Barbour, ``The Timelessness of Quantum Gravity. I. The Evidence from the Classical Theory",               Class. Quant. Grav. {\bf 11} 2853 (1994).

\bibitem{Husemoller}          D. Husemoller, {\it Fibre Bundles} (Springer, New York 1994).							  

\bibitem{Kuchar94}            K.V. Kucha\v{r}, ``Geometrodynamics of Schwarzschild Black Holes", Phys. Rev. {\bf D50} 3961 (1994), arXiv:gr-qc/9403003. 

\bibitem{Buckets}             {\it Mach's principle: From Newton's Bucket to Quantum Gravity} ed. J.B. Barbour and H. Pfister (Birkh\"{a}user, Boston 1995).
 	
\bibitem{LR95}                R.G. Littlejohn and M. Reinsch, ``Internal or Shape Coordinates in the $N$-body Problem", Phys. Rev. {\bf A52} 2035 (1995).

\bibitem{Cos}                 P. Coles and F. Lucchin, {\it Cosmology. The Origin and Evolution of Cosmic Structure} (Wiley, Chichester 1995).  

\bibitem{Page1}               D.N. Page, ``Sensible Quantum Mechanics: Are Probabilities only in the Mind?", Int. J. Mod. Phys. D5 583 (1996), gr-qc/9507024.

\bibitem{Grenander96}         U. Grenander, {\it Elements of Pattern Theory}  (Johns Hopkins Press, Baltimore 1996).  

\bibitem{Albouy}              A. Albouy, {\it The Symmetric Central Configurations of Four Equal Masses}, Contemp. Math. {\bf 198} 131 (1996).  
							  							  
\bibitem{Small}               C.G.S. Small, {\it The Statistical Theory of Shape} (Springer, New York, 1996).  

\bibitem{FM96}                A.E. Fischer and V. Moncrief, ``A Method of Reduction of Einstein's Equations of Evolution and a Natural Symplectic Structure 
                              on the Space of Gravitational Degrees of Freedom", Gen. Rel. Grav. {\bf 28}, 207 (1996).
							  
\bibitem{LR97}                R.G. Littlejohn and M. Reinsch, ``Gauge Fields in the Separation of Rotations and Internal Motions in the $N$-Body Problem", 
                              Rev. Mod. Phys. {\bf 69} 213 (1997).  

\bibitem{Graphs-1}            V.K. Balakrishnan, {\it Graph Theory} (McGraw--Hill, New York 1997).
						  						  
\bibitem{Krasinski}			  A. Krasinski, {\it Inhomogeneous Cosmological Models} (Cambridge University Press, Cambridge 1997).  
						  
\bibitem{Sheaves}             G.E. Bredon, {\it Sheaf Theory} (McGraw--Hill, New York 1997).
 							  							  
\bibitem{Sparr98}             G. Sparr, ``Euclidean and Affine Structure/Motion for Uncalibrated Cameras from Affine Shape and Subsidiary Information", 
                              in Proceedings of SMILE Workshop on Structure from Multiple Images, Freiburg (1998).  

\bibitem{Graphs-2}            B. Bollob\'{a}s, {\it Modern Graph Theory}, (Springer-Verlag, New York 1998).		
							  							  
\bibitem{EoT}                 J.B. Barbour, {\it The End of Time} (Oxford University Press, Oxford 1999).
							  
\bibitem{ML99}                K.A. Mitchell and R.G. Littlejohn,  ``Boundary Conditions on Internal Three-Body Wave Functions", physics/9908037.  
							  
\bibitem{Kuchar99}            K.V. Kucha\v{r}, ``The Problem of Time in Quantum Geometrodynamics",  in {\it The Arguments of Time}, ed. J. Butterfield 
                              (Oxford University Press, Oxford 1999).

\bibitem{H99}                 J.J. Halliwell, ``Somewhere in the Universe: Where is the Information Stored When Histories Decohere?", 
                              Phys. Rev. {\bf D60} 105031 (1999), quant-ph/9902008.

\bibitem{Roberts}	          G.E. Roberts, {\it A Continuum of Relative Equilibria in the Five-Body Problem}, 
                              Physica D: Nonlinear Phenomena {\bf 127} 141 (1999).
							  
\bibitem{Kendall}             D.G. Kendall, D. Barden, T.K. Carne and H. Le, {\it Shape and Shape Theory} (Wiley, Chichester 1999).  
							  							  
\bibitem{IshamBook}           C.J. Isham, {\it Modern Differential Geometry for Physicists} (World Scientific, Singapore 1999).


\bibitem{Cohn}                P.M. Cohn, {\it Classic Algebra} (Wiley, Chichester 2000).  

\bibitem{JM00}                K.V. Mardia and P.E. Jupp, {\it Directional Statistics} (Wiley, Chichester 2000).

\bibitem{Gergely}             L.\'{A} Gergely, ``The Geometry of the Barbour-Bertotti Theories I. The Reduction Process", 
                              Class. Quant. Grav. {\bf 17} 1949 (2000), gr-qc/0003064. 							  
							  
\bibitem{RWR}                 J.B. Barbour, B.Z. Foster and N. \'{o} Murchadha, ``Relativity Without Relativity", 
                              Class. Quant. Grav. {\bf 19} 3217 (2002), gr-qc/0012089.
										
\bibitem{Hatcher}             A. Hatcher, {\it Algebraic Topology} (Cambridge University Press, Cambridge 2001).

\bibitem{Pflaum}              M.J. Pflaum, {\it Analytic and Geometric Study of Stratified Spaces}, Lecture Notes in Mathematics {\bf 1768} (Springer, Berlin 2001).  

\bibitem{ArchRat}             R. Montgomery, ``Infinitely Many Syzygies", Arch. Rat. Mech. Anal. {\bf 164} 311 (2002). 

\bibitem{B03}                 J.B. Barbour, ``Scale-Invariant Gravity: Particle Dynamics", Class. Quant. Grav. {\bf 20} 1543 (2003), gr-qc/0211021.

\bibitem{MacFarlane}          A.J. MacFarlane, ``Complete Solution of the Schr\"{o}dinger Equation of the Complex Manifold $\mathbf{CP}^2$", 
                              J. Phys. A: Math. Gen. {\bf 36} 7049 (2003).
			
\bibitem{MacCallum}           H. Stephani, D. Kramer, M.A.H. MacCallum, C.A. Hoenselaers, and E. Herlt, 
                              {\it Exact Solutions of Einstein's Field Equations} 2nd Edition (Cambridge University Press, Cambridge 2003).
			
\bibitem{Geom-Search-Engine}  T. Funkhouser, P. Min, M. Kazhdan, J. Chen, A. Halderman, D. Dobkin and D. Jacobs, 
                             ``A Search Engine for 3D Models" ACM Transactions on Graphics, {\bf 5} 202002 (2003), https://www.cs.princeton.edu/~funk/tog03.pdf .
							 
\bibitem{I03}                 C.J. Isham, ``A New Approach to Quantising Space-Time: I. Quantising on a General Category", Adv. Theor. Math. Phys. {\bf 7} 331, gr-qc/0303060; 
%
                                          ``A New Approach to Quantising Space-Time: II. Quantising on a Category of Sets" {\bf 7} 807 (2003), gr-qc/0304077; 
%
	                                      ``A New Approach to Quantising Space-Time: III. State Vectors as Functions on Arrows" {\bf 8} 797 (2004), gr-qc/0306064; 
%
	 						               ``Quantising on a Category", quant-ph/0401175.  
							 
\bibitem{KieferBook}          C. Kiefer, {\it Quantum Gravity} (Clarendon, Oxford 2004).  

\bibitem{RovelliBook}         C. Rovelli, {\it Quantum Gravity} (Cambridge University Press, Cambridge 2004).  
  
\bibitem{ABFKO}               E. Anderson, J.B. Barbour, B.Z. Foster, B. Kelleher and N. \'{o} Murchadha, ``The Physical Gravitational Degrees of Freedom", 
                              Class. Quant. Grav {\bf 22} 1795 (2005), gr-qc/0407104.   

\bibitem{Montgomery2}         R. Montgomery, ``Fitting Hyperbolic Pants to a 3-Body Problem", Ergod. Th. Dynam. Sys. {\bf 25} 921 (2005), math/0405014.

\bibitem{MP03}                K.V. Mardia and V. Patrangenaru, ``Directions and Projective Shapes", Annals of Statistics {\bf 33} 1666 (2005), math/0508280. 

\bibitem{Younes}              M.I Miller, A. Trouv\'{e} and L. Younes, ``Geodesic Shooting for Computational Anatomy", 
                              Journal of Mathematical Imaging and Vision {\bf 24} 209 (2006).  

\bibitem{Grenander07}         U. Grenader, Pattern Theory: From Representation to Inference (Oxford University Press, Oxford 2007).  
							  
\bibitem{Giu06}	    		  D. Giulini, ``Some Remarks on the Notions of General Covariance and Background Independence", 
                              in {\it An Assessment of Current Paradigms in the Physics of Fundamental Interactions} 
							  ed. I.O. Stamatescu, Lect. Notes Phys. {\bf 721} 105 (2007), arXiv:gr-qc/0603087.
 							  
\bibitem{FORD}                E. Anderson, ``Foundations of Relational Particle Dynamics", Class. Quant. Grav. {\bf 25} 025003 (2008), arXiv:0706.3934.

\bibitem{Diff-Med-Imaging-1}  L. Wang, F. Beg, T. Ratnanather, C. Ceritoglu, L. Younes, J.C. Morris, J.G. Csernansky and M.I. Miller, 
                              ``Large Deformation Diffeomorphism and Momentum Based Hippocampal Shape Discrimination in Dementia of the Alzheimer Type", 
         					  IEEE Transactions on Medical Imaging {\bf 26} (2007).	
							  
\bibitem{Giu09}               D. Giulini, ``The Superspace of Geometrodynamics", Gen. Rel. Grav. {\bf 41} 785 (2009) 785, arXiv:0902.3923.  

\bibitem{LS09}                T.-L. Lee and M. Santoprete, {\it Central Configurations of the Five-Body Problem with Equal Masses}, arXiv:0906.0148.

\bibitem{AF}                  E. Anderson and A. Franzen, ``Quantum Cosmological Metroland Model", Class. Quant. Grav. {\bf 27} 045009 (2010), arXiv:0909.2436. 

\bibitem{+Tri}                E. Anderson, ``Shape Space Methods for Quantum Cosmological Triangleland", Gen. Rel. Grav. {\bf 43} 1529 (2011), arXiv:0909.2439.  
 
\bibitem{WCP}                 V. Guillemin, A. Uribe, and Z. Wang, ``Geodesics on Weighted Projective Spaces", Ann. Global Analysis Geom. {\bf 36} 205 (2009). 

\bibitem{GT09}                D. Groisser, and H.D. Tagare, ``On the Topology and Geometry of Spaces of Affine Shapes", 
                              Journal of Mathematical Imaging and Vision {\bf 34} 222 (2009).  
							  
							 
\bibitem{Younes10}            L. Younes, {\it Shapes and Diffeomorphisms} (Springer, New York 2010).  

\bibitem{Cones}               E. Anderson, ``Relational Mechanics of Shape and Scale", arXiv:1001.1112.

\bibitem{ScaleQM}             E. Anderson, ``Quantum Cosmological Relational Model of Shape and Scale in 1-$d$", Class. Quantum Grav. {\bf 28} 065011 (2011), arXiv:1003.4034. 

\bibitem{Midi}                J.F. Barbero and E.J.S. Villase$\tilde{\mn}$or, ``Quantization of Midisuperspace Models", Living Rev. Rel. {\bf 13} 6 (2010), arXiv:1010.1637.

\bibitem{APoT1}               E. Anderson, ``The Problem of Time in Quantum Gravity", in {\it Classical and Quantum Gravity: Theory, Analysis and Applications} 
                              ed. V.R. Frignanni (Nova, New York 2012), arXiv:1009.2157.    

\bibitem{Morphometrics}	      R. Sparks and A. Madabhushi, 
                              ``Novel Morphometric based Classification via Diffeomorphic based Shape Representation using Manifold Learning",
							  Med. Image Comput. Comput. Assist Interv. {\bf 13} 658 (2010).
							  						  
\bibitem{Kreck}               M. Kreck, {\it Differential Algebraic Topology: From Stratifolds to Exotic Spheres} (American Mathematical Society, Providence 2010).							 
 							  
\bibitem{FileR}               E. Anderson, ``The Problem of Time and Quantum Cosmology in the Relational Particle Mechanics Arena", arXiv:1111.1472.  

\bibitem{Lee1}                J.M. Lee, {\it Introduction to Topological Manifolds} (Springer, New York 2011).

\bibitem{Bhatta}              A. Bhattacharya and R. Bhattacharya, {\it Nonparametric Statistics on Manifolds with Applications to Shape Spaces} 
                             (Cambridge University Press, Cambridge 2012).

\bibitem{QuadI}               E. Anderson, ``Relational Quadrilateralland. I. The Classical Theory", Int. J. Mod. Phys. {\bf D23} 1450014 (2014), arXiv:1202.4186.

\bibitem{APoT2}               E. Anderson, ``Problem of Time in Quantum Gravity", Annalen der Physik, {\bf 524} 757 (2012),  arXiv:1206.2403.    

\bibitem{Diff-Med-Imaging-2}  A. Sotiras, C. Davatzikos and N. Paragios, ``Deformable Medical Image Registration: A Survey", 
                              in IEEE Transactions on Medical Imaging {\bf 32} (2013).   
							  
\bibitem{QuadII}              E. Anderson and S.A.R. Kneller, ``Relational Quadrilateralland. II. The Quantum Theory", Int. J. Mod. Phys. {\bf D23} 1450052 (2014), arXiv:1303.5645.

\bibitem{ABeables}            E. Anderson, ``Beables/Observables in Classical and Quantum Gravity", SIGMA {\bf 10} 092 (2014), arXiv:1312.6073. 

\bibitem{Sniatycki}           J. \'{S}niatycki, {\it Differential Geometry of Singular Spaces and Reduction of Symmetry} 
                             (Cambridge University Press, Cambridge 2013).
							    
\bibitem{AKendall}			  E. Anderson, ``Kendall's Shape Statistics as a Classical Realization of Barbour-type Timeless Records Theory approach to Quantum Gravity", 
                              Stud. Hist. Phil. Mod. Phys. {\bf 51} 1 (2015), arXiv:1307.1923. 

\bibitem{BI}                  E. Anderson, ``Background Independence", arXiv:1310.1524.
							  
\bibitem{Mercati}             F. Mercati, ``A Shape Dynamics Tutorial", arXiv:1409.0105.  
							  
\bibitem{APoT3}               E. Anderson, ``Problem of Time and Background Independence: the Individual Facets", arXiv:1409.4117.  

\bibitem{Gowdy}               R.H. Gowdy, ``Gowdy Spacetimes", Scholarpedia {\bf 9} 31673 (2014),  
                              http://www.scholarpedia.org/article/Gowdy$\u{\m }$Spacetimes .

\bibitem{ASoS}                E. Anderson, ``Spaces of Spaces", arXiv.1412.0239.
								  					  
\bibitem{MIT}                 A. Edelman and G. Strang, ``Random Triangle Theory with Geometry and Applications", Foundations of Computational Mathematics (2015), arXiv:1501.03053.   

\bibitem{AConfig}             E. Anderson, ``Configuration Spaces in Fundamental Physics", arXiv:1503.01507. 

\bibitem{AMech}               E. Anderson, ``Six New Mechanics corresponding to further Shape Theories", Int. J. Mod. Phys. {\bf D 25} 1650044 (2016), arXiv:1505.00488. 

\bibitem{ASphe}               E. Anderson, ``Spherical Relationalism", arXiv:1505.02448.

\bibitem{ABeables2}           E. Anderson, ``Explicit Partial and Functional Differential Equations for Beables or Observables" arXiv:1505.03551.
	
\bibitem{PE16}                V. Patrangenaru and L. Ellingson, ``Nonparametric Statistics on Manifolds and their Applications to Object Data Analysis" 
                             (Taylor and Francis, Boca Raton, Florida 2016).  

\bibitem{KKH16}               F. Kelma, J.T. Kent and T. Hotz, ``On the Topology of Projective Shape Spaces", arXiv:1602.04330. 

\bibitem{ABeables3}           E. Anderson, ``On Types of Observables in Constrained Theories", arXiv:1604.05415.
							
\bibitem{TQ-17}  			  J. Cantarella, T. Needham, C. Shonkwiler and G. Stewart {\it Random Triangles and Polygons in the Plane}, arxiv:1702.01027 
							 
\bibitem{ABook}               E. Anderson, {\it Problem of Time. Quantum Mechanics versus General Relativity}, (Springer International 2017), Found. Phys. {\bf 190};
                              free access to its extensive Appendices is at https://link.springer.com/content/pdf/bbm$\%$3A978-3-319-58848-3$\%$2F1.pdf .

\bibitem{II}                  E. Anderson, ``The Smallest Shape Spaces. II. 
                              4 Points on a Line Suffices for a Complex Background-Independent Theory of Inhomogeneity", arXiv:1711.10073.

\bibitem{III}                 E. Anderson, ``The Smallest Shape Spaces. III. Triangles in the Plane and in 3-$d$", arXiv:1711.10115.

\bibitem{A-Pillow}            E. Anderson, ``Alice in Triangleland: Lewis Carroll's Pillow Problem and Variants Solved on Shape Space of Triangles", arXiv:1711.11492.

\bibitem{2-Herons}            E. Anderson, ``Two New Perspectives on Heron's Formula",  arXiv:1712.01441.  
							  
\bibitem{Ineq}                E. Anderson, ``Shape (In)dependent Inequalities for Triangleland's Jacobi and Democratic-Linear Ellipticity Quantitities", arXiv:1712.04090.
							  
\bibitem{Max-Angle-Flow}      E. Anderson, ``Maximal Angle Flow on the Shape Sphere of Triangles", arXiv:1712.07966 
								   
\bibitem{A-Monopoles}         E. Anderson, ``Monopoles of Twelve Types in 3-Body Problems", arXiv:1802.03465.

\bibitem{ATorus}              E. Anderson, ``Background Independence and Shape Theory on the Torus, forthcoming 2018.   

\bibitem{Project-1}           E. Anderson, ``Background Independence {\sl Includes} Shape Theory", forthcoming 2018.

\bibitem{A-Sylvester}         E. Anderson, ``Shape Theory's answer to Background-Independent Vertex-Primary Sylvester Prob(Quadrilateral is Convex)", forthcoming 2018. 

\bibitem{A-Coolidge}          E. Anderson, ``While Heron's Form is Shape-Theoretic, Coolidge's is not", forthcoming 2018. 

\bibitem{A-Quad-Ineq}         E. Anderson, ``Shape-Independent Inequalities for Quadrilaterals", forthcoming 2018.  

\bibitem{IV}                  E. Anderson, ``The Smallest Shape Spaces. IV. Quadrilaterals in the Plane", forthcoming 2018.
    
\bibitem{Affine-Shape-1}      E. Anderson, ``Affine Shape Apace. I. Topological Structure", forthcoming 2018.

\bibitem{Affine-Shape-2}      E. Anderson, ``Affine Shape Apace. II. Riemannian Metric Structure", forthcoming 2018.

\bibitem{A-Generic}           E. Anderson  ``Absolute versus Relational Motion Debate: a Modern Global Version", forthcoming 2018.  

\bibitem{A-Perimeter}         E. Anderson, ``Geometrically Significant Functions for Triangles Plotted over the Shape Sphere of Triangles", forthcoming 2018.

\bibitem{Project-2}           E. Anderson, ``Background Independence Leads to Study of Stratified Manifold--Sheaf Pairs", forthcoming 2018. 

\bibitem{QLS}                 E. Anderson, ``Quantizing the Smallest Shape Spaces", forthcoming 2018. 

\bibitem{Project-3}           E. Anderson, ``Constraint Closure and the Problem of Observables: Outline of Sheaf-Theoretic Inputs", forthcoming 2018. 

\bibitem{Quantum-Triangles}   E. Anderson, ``Quantum Triangles", forthcoming 2018.  

\bibitem{Forthcoming}         E. Anderson, various forthcoming works on Classical-level Topologenesis.  

\end{thebibliography}
\end{document}